\documentclass[11pt,a4paper]{article}

\usepackage{jheppub}
\usepackage{caption, subcaption}  
\usepackage{color,latexsym, array,multirow,  verbatim, enumerate,cancel,graphicx}
\usepackage{slashed}
\usepackage{comment}
\usepackage{bm}

\def\met{\slashed{E}_T}

\def\beq {\begin{equation}}
\def\eeq {\end{equation}}
\def\bi {\begin{itemize}}
\def\ei {\end{itemize}}
\def\bea {\begin{eqnarray}}
\def\eea {\end{eqnarray}}

\numberwithin{equation}{section} 

\title{Triggering long-lived particles in HL-LHC and the challenges in the first stage of the trigger system}

\author[a]{Biplob Bhattacherjee}
\author[b]{Swagata Mukherjee}
\author[a]{Rhitaja Sengupta}
\author[a]{Prabhat Solanki}

\affiliation[a]{Centre for High Energy Physics, Indian Institute of Science,\\
Bangalore 560012, India}

\affiliation[b]{III. Physikalisches Institut A, RWTH Aachen University,\\
Otto-Blumenthal-Str. 16, 52074 Aachen, Germany}

\emailAdd{biplob@iisc.ac.in}
\emailAdd{mukherjee@physik.rwth-aachen.de}
\emailAdd{rhitaja@iisc.ac.in}
\emailAdd{prabhats@iisc.ac.in}

\abstract{Triggering long-lived particles at the first stage of the trigger system is very crucial in LLP searches to ensure that we do not miss them at the very beginning.
The future High Luminosity runs of the Large Hardron Collider will have increased number of pile-up events per bunch crossing. There will be major upgrades in hardware, firmware and software sides, like tracking at level-1 (L1) as well as inclusion of the MIP timing detector. The L1 trigger menu will also be modified to cope with pile-up and maintain the sensitivity to physics processes. 
In our study we found that the usual level-1 triggers, mostly meant for triggering prompt particles, will not be very efficient for LLP searches in the 140 PU environment of HL-LHC, thus pointing to the need to include dedicated L1 triggers in the menu for LLPs.
 We consider the decay of the LLP into jets and develop dedicated jet triggers using the track information and if available, the regional timing information at L1 to select LLP events. We show in our work that these triggers give promising results in identifying LLP events with moderate trigger rates.}

\begin{document}
\maketitle
\flushbottom


\section{Introduction}
\label{sec:intro}

The RunII of LHC has ended in 2018 and no clear hint of new physics is found yet. LHC and its two general purpose detectors were mainly built to look for signatures of new physics involving prompt particles, which was widely searched for in LHC during both RunI and RunII. In the absence of any clear signal in that sector, the focus of LHC searches is shifting towards long-lived particles(LLP) and in upcoming LHC runs LLP searches will be one of the priorities of the LHC experiments. The presence of a long-lived new particle can be a salient trait of several beyond standard model scenarios. A plethora of BSM models predict the presence of long-lived particles. Supersymmetry (SUSY) in different incarnations, several dark matter models, or portal models predicting interactions between a possible hidden sector and the SM are potential sources of LLP~\cite{Lee:2018pag}. Among the SUSY models predicting LLPs \cite{
Giudice:1998bp,Giudice:2004tc,Barbier:2004ez,ArkaniHamed:2004yi,Burdman:2006tz,
deAlwis:2008aq,Meade:2010ji,Biswas:2010yp,Fan:2011yu,Ibe:2012hu,
Graham:2012th,Bhattacherjee:2012ed,Arvanitaki:2012ps,Cerdeno:2013oya,
Rolbiecki:2015gsa,Banerjee:2018uut,Dercks:2018eua,Fukuda:2019kbp}, some of the most compelling ones 
are the R-parity violating SUSY, gauge and anomaly mediated SUSY breaking scenarios, split SUSY, and stealth SUSY. LLPs can also be present in many dark matter models \cite{Hall:2009bx,TuckerSmith:2001hy,Co:2015pka,Hessler:2016kwm,
Belanger:2018sti,Goudelis:2018xqi,Bae:2020dwf}.
In hidden valley scenarios \cite{Strassler:2006im,Strassler:2006ri,Strassler:2006qa}, particles of the hidden sector such as long-lived dark photons or dark hadrons may decay to standard model particles. Axions or axion-like particles~\cite{Curtin:2018mvb}, which provides a nice resolution to the strong CP Problem, can also have long lifetimes and can be probed in LHC in different final states. 

Several LLP searches have been performed at the ATLAS and CMS experiments \cite{Aad:2014gfa,Aaboud:2017mpt,Aaboud:2017iio,Aad:2019tua,Aaboud:2019opc,
Aad:2019tcc,Khachatryan:2016sfv,Sirunyan:2017jdo,Sirunyan:2017sbs,
Sirunyan:2019gut,Sirunyan:2019wau}, even though the detectors were a priori designed and optimised for promptly produced particles. For neutral LLPs that decay to charged particles, displaced vertex searches were carried out, while for charged LLP, several unique signatures like disappearing track, kinked track, long time of flight in muon chamber, high ionisation energy loss per path length of a track (dE/dx) etc, were looked for. For a comprehensive review of various LLP models and experimental searches, the readers are referred to \cite{Alimena:2019zri} and the references therein.

Depending on lifetime of LLP, the signature in detector can widely vary, thus inclusive searches are often not possible, and dedicated search strategies need to be designed. While LHC experiments have performed extensive searches to look for LLPs, there are some other interesting signatures that are yet to be searched for in the experiments, for example, difference in energy deposition patterns in the hadron calorimeter (HCAL) for displaced jets or backward moving objects originating from a heavy slow-moving or stopped LLP \cite{Banerjee:2017hmw,Bhattacherjee:2019fpt} compared to the energy deposition pattern of prompt jets. Many of these signatures are experimentally challenging and usage of modern machine learning techniques would be beneficial to have better discovery potential.

When it comes to LLP search in LHC experiments, trigger poses a critical challenge. 
To overcome challenges of triggering LLP signatures, several novel ideas have been devised by the experimental collaborations. For example, the ``CalRatio'' trigger~\cite{Aad:2013txa}, which is a trigger which makes use of the ratio of energy deposited in HCAL and electromagnetic calorimeter (ECAL) for a particular jet. 
Another idea already used was to trigger on trackless jet signature~\cite{Aad:2013txa} at high-level trigger, which provides a robust way to identify displaced jets coming from LLP that decayed after crossing the tracker. For LLP decays taking place just before or inside muon spectrometers, a special trigger, which looks for large number of charged hadrons traversing a narrow region of the muon spectrometer, was used by the ATLAS collaboration \cite{Aad:2013txa,Aaboud:2018aqj}.

In high-luminosity LHC (HL-LHC) \cite{Apollinari:2017cqg} the triggering challenge will be more severe because of high pile up (PU). 
High PU makes it difficult to reconstruct
the hard interaction, because a huge number of soft charged particles give rise to many tracks, which overlap spatially. Moreover, the calorimeter energy
deposits can also overlap due to particles coming from many pp collisions. This can increase the rate of false triggers.
It was found that the trackless jet and the CalRatio triggers are particularly sensitive to number of PU
interactions and showed significant reduction in efficiency with increasing PU \cite{Aad:2013txa}. 
So it is understood that the efficiency and rate of the novel LLP triggers already used in experimental searches will be severely affected by high PU in HL-LHC, and in order to keep the rate under control the trigger level cuts might need to be tightened. However, it would be good to apply dedicated cuts rather than tighter cuts to keep the trigger rates in acceptable limits as well as to select LLP events efficiently. Since the cross sections of the LLP processes can be really small $\sim\mathcal{O}({\rm fb})$ or even less, the goal must be to select as many LLP events as possible at L1. If the event does not pass the L1 trigger, we will lose it forever, and due to this the prospect of discovering LLPs at the LHC will be severely hampered. 

In our paper, we will present the results in the context of CMS detector and its upgrade in HL-LHC.
The improved detector systems of the CMS Phase-2 upgrades \cite{Contardo:2020886} will 
help in maintaining the necessary performance of object reconstruction and identification under the arduous conditions of high PU at the HL-LHC. 
A key component of the upgrade of CMS experiment for HL-LHC is 
a unique arrangement of the outer part of the tracker. Utilizing the strong magnetic field of the CMS detector, it will permit the use of tracks in the L1 trigger stage, with input event rate of  40 MHz, arising from pp bunch crossings~\cite{James:2647214}. This distinctive attribute, along with the opportunities rendered by advancement in FPGA processing capability and bandwidth, will permit more elegant L1 triggers.
The new track trigger facility will be able to identify tracks with $p_T > 2 {\rm~GeV}$ at the hardware-based first-level trigger, known as L1\footnote{Tracking displaced charged particles at L1 is challenging. There are some works going on to improve tracking at L1 to include LLP scenarios that will increase the sensitivity of LLP searches \cite{Martensson:2019sfa}.}. 
Another important upgrade is of CMS detector is a new timing layer~\cite{MTD:TDR} which will be built to measure
minimum ionizing particles (MIPs) with an excellent time resolution of about 30 ps. The addition of the timing layer, which is able to identify charged particles, will remarkably suppress the effect of PU. While full readout of the 
timing detector at level-1 trigger will not be possible, because of bandwidth constraints, it may still 
be possible for this sub-detector to play a pivotal role in the L1 trigger, by utilising data from only the regions
of interest. 

There are certain optimistic scenarios where PU won't be a major problem for LLP search and dedicated LLP triggers might not be essential. 
For example, when the LLP is very heavy and its hadronic decay products deposit enough energy in calorimeters to push the events pass the usual $H_T$ trigger thresholds, or when the LLP decays to muons and muon chambers being the last layer of the detector is least affected by PU, so the displaced muon signature will still be prominent even in high PU scenario. However, there are some pessimistic scenarios where LLP is not very heavy and its hadronic decay products are less energetic.
Both the upgrades of CMS detector, track-trigger and timing layer, might be very useful for LLP searches, in situations where the LLP is light and decays to jets. In order to be sensitive to models where LLP decays to 
low $p_T$ jets, it is extremely important to have a good handle on PU mitigation. In this paper we have explored how the upgraded CMS detector might perform in the high PU environment of HL-LHC to differentiate between prompt and displaced jets. We design some dedicated LLP triggers for selecting LLP events efficiently keeping the background rates within the acceptable limit, using the L1 tracking and timing information from the MIP timing detector.

The rest of the paper is outlined as: in section \ref{sec:hllhc_llp}, we discuss the HL-LHC upgrades and L1 triggers in detail and the LLP scenarios that might not get selected with the standard triggers. In section \ref{sec:first_LLP}, we define a particular LLP scenario, where LLPs are pair-produced directly and then decay to jets, and study the effect of PU on different distributions and also the performance of standard L1 jet triggers in selecting LLP events. In section \ref{sec:triggers}, we develop some dedicated triggers for LLPs using the tracking information and timing information from the MTD at L1, which can efficiently identify jets coming from the decay of light LLPs, unlike the standard jet triggers. In section \ref{sec:other_bench}, we briefly discuss a few other LLP scenarios, each highlighting a different aspect of triggering LLPs at L1. Finally, in section \ref{sec:concl}, we conclude. We also present in appendix \ref{app:magnetic} a description of propagation of both prompt and displaced charged particles in the magnetic field.

\section{The high luminosity LHC and triggering long-lived particles}
\label{sec:hllhc_llp}

In this section, we discuss the major upgrades for the high luminosity LHC and also some of the triggers at level-1 (L1) relevant for this work. Although these triggers are optimized for standard prompt scenarios, we do not know how they perform for non-prompt cases. In the later part of this section, we briefly discuss which LLP scenarios will be easier to trigger using standard triggers or have some other sensitivity and which scenarios will be more difficult to trigger.

\subsection{The Phase-II HL-LHC upgrade and level-1 triggers}
\label{ssec:descr-hllhc}

We start with a brief discussion of the major upgrades for the high luminosity runs of the LHC.
We will also discuss the standard level-1 (L1) triggers for the HL-LHC runs.
Peak instantaneous luminosity at HL-LHC will increase up to $\sim 7.5\times10^{34}{\rm~cm}^{-2}{\rm s}^{-1}$ which is much higher than the current LHC design value of $1.0\times10^{34}{\rm~cm}^{-2}{\rm s}^{-1}$ \cite{Apollinari:2017cqg}.
The increase in the instantaneous luminosity of the HL-LHC runs also mean an increase in the average pile-up rate, estimated to be close to $140$ for a peak luminosity of $\sim 5\times10^{34}{\rm~cm}^{-2}{\rm s}^{-1}$ (and $200$ for peak luminosity $\sim 7.5\times10^{34}{\rm~cm}^{-2}{\rm s}^{-1}$) average PU vertices in each bunch crossing at the start(end) of the HL-LHC compared to the present number of $30$-$50$ PU events per bunch crossing. To handle such a large amount of PU without compromising with its physics reach, the HL-LHC runs will witness many detector upgrades as proposed in the Phase-II upgrade of the detectors.   

The Phase-II upgrade of the CMS tracker will include installation of a new silicon tracker (aka outer tracker) and a new pixel detector (aka inner tracker), which will have the capability to read out important information related to tracking, though limited, at 40 MHz to the hardware-based L1 trigger system. It can read out the hits left only by high momentum tracks ($p_T>2{\rm~GeV}$), out of the huge number of silicon hits, all of which cannot be read due to the latency limitation at L1. Inclusion of L1 tracking can help reduce the PU rate at the L1 triggers keeping the efficiency of interesting physical processes at par with those obtained for the present LHC runs. 
This will become clear when we discuss the standard L1 multijet triggers for CMS at HL-LHC. Tracker modules, made out of two sensors that are placed side by side, and an integrated circuit that harmonize the hits in them, provides both spatial and transverse momentum ($p_T$) measurements. The pair of hits in such a module are called stubs. A track will produce a stub if hits in the sensors of a $p_T$ module fall in a specified window corresponding to the $p_T$ threshold. Tracks having $p_T$ below this threshold will be
rejected. However, the $p_T$ measurement assumes that the track originated from the beamline, and therefore this will lead to wrong estimation of track parameters for displaced tracks coming from decays of long-lived particles, as has been discussed in \cite{Gershtein:2017tsv}. 

Also, the Phase-II upgrade of CMS is proposed to have a MIP (minimum ionizing particles) timing detector (MTD) \cite{MTD:TDR}. 
Recently there has been some works which study the prospects of MTD in long-lived particle searches and determination of LLP properties like mass and lifetime \cite{Liu:2018wte,Mason:2019okp,Kang:2019ukr,Klimek:2019cny,Du:2019mlc,
Banerjee:2019ktv}. Heavy LLPs, which move slowly, with proper mean decay lengths $\sim 1\text{m}$, decay (usually late, but within tracker volume) to SM particles which will reach the MTD with significant time difference compared to SM prompt particles. This is the idea explored in these works where putting a cut on time of arrival of a particle (or jet) at the MTD can separate LLP events from non-LLP ones.

Before discussing the HL-LHC L1 multijet triggers, let us briefly discuss how jets are usually formed at L1.
In the current LHC runs, the L1 jets are formed using trigger towers, which are segmented as $\Delta\eta\times\Delta\phi$ size of $0.087\times0.087$ in the barrel (up to $|\eta|\sim1.5$) and this size varies up to $\sim 0.17\times0.17$ as we move to $|\eta|=2.5$ (for the CMS detector) as given in the \texttt{Delphes} card for CMS. For the HL-LHC CMS upgrade, there is a proposition for High Granularity Calorimeters (HGCAL) at the  endcaps of the detector, both for the electromagnetic and hadronic calorimeters which will have much smaller segmentations (with a size of $0.02\times0.02$) \cite{Collaboration:2293646}. In this work, we do not use the HGCAL segmentation and use the barrel segmentation only, even for the endcaps, up to $|\eta|\sim3$ and then use the HF (hadronic forward calorimeter) segmentation. L1 Calo jets are usually clustered by taking only towers having transverse energy ($E_T$) greater than $2{\rm~GeV}$ and up to $|\eta|=2.5$. At L1, standard sequential jet formation algorithms like anti-$k_T$\cite{Cacciari:2008gp} cannot be used due to limited time available for making trigger decisions. Therefore, jets are formed using sliding windows of size $9\times9$ around a tower with the maximum $E_T$, which has to be greater than $4{\rm~GeV}$. This reasonably matches with the anti-$k_T$ clustering with $R=0.4$ as shown in \cite[fig.4]{Zabi:2016ljo}, and takes much lesser time than the sequential clustering algorithms. We believe that in HL-LHC runs as well, the L1 jets formed with some particular sliding window size will resemble jets formed with anti-$k_T$ with the same cone size. Therefore, for this work, we use anti-$k_T$ algorithm to cluster jets.


Let us now discuss the L1 triggers designed for processes involving jets at the HL-LHC runs of CMS, and how they can control pile-up (PU).
The trigger menu involving more than one jets at the HL-LHC (dijet, quad jet, $H_T$ triggers, etc.) require all the jets contributing to the trigger selection to come from the same $z-$vertex. 
The same $z-$vertex condition is defined in terms of the L1 tracks associated with a jet. 
Tracks with $p_T>2{\rm~GeV}$, within $|\eta|<2.4$ having transverse displacement ($L_{xy}$) less than $1{\rm~cm}$ and within longitudinal spread of $|z_0|<30{\rm~cm}$ are selected. The L1 tracks, satisfying above requirements, are associated with a L1 jet by taking all tracks within a cone of $\Delta R=0.4$ around the jet axis. The jet $z$-vertex is defined as the $p_T$ weighted average $z_0$ of all the tracks associated with the jet. The same $z$-vertex condition implies that the $z-$vertex of all jets that contribute to the trigger decision should satisfy, $\Delta z \leq 1{\rm~cm}$ \cite{Contardo:2020886}.

Based on this, the following jet triggers will be used at L1 of HL-LHC \cite{Contardo:2020886}:

\begin{itemize}
\item \textbf{Single jet:} at least one jet with $p_T>173{\rm~GeV}$; no tracking information needed \footnote{The single jet trigger has a very high $p_T$ threshold and therefore, is not affected much by pile-up jets. Therefore, this trigger can be used even without the tracking information at L1.}.
\item \textbf{Dijet:} at least two jets with $p_T>136{\rm~GeV}$; both these jets should satisfy the same $z-$vertex condition.
\item \textbf{Quad jet:} at least four jets with $p_T>72{\rm~GeV}$; all these jets should satisfy the same $z-$vertex condition.
\end{itemize}

In addition to these triggers, there are the $H_T$ and MET triggers, which also use the tracking information. The $H_T$ trigger requires 
the total scalar $p_T$ sum of all the jets, $H_T>350{\rm~GeV}$. Also, all the jets contributing to $H_T$ should satisfy the same $z-$vertex condition. 
The MET trigger also requires all the objects (like electrons, muons and jets) which help pass the trigger come from the same vertex. All the above discussed thresholds are for 140 PU scenario. For the 200 PU case, these might be increased to keep the trigger rates within the allowed trigger bandwidth.

The same $z-$vertex condition added in the L1 triggers help in controlling the pile-up rate because pile-up jets are most likely to come from different $z$ positions. For instance, the dijet and quad jet trigger rates increase from 26 and 12 kHz to 52 and 185 kHz respectively, on removing the same $z$-vertex condition \cite{CMSrates}. The same $z$-vertex condition, therefore, helps a lot to maintain the trigger rates for the above discussed thresholds, and as we can see the effect of PU is more in cases of triggers with low $p_T$ thresholds. 

However, we don't know how this condition will affect triggering of events with LLPs, as these triggers are mostly optimized for prompt decays of particles. It is important to study the performance of standard L1 triggers for different LLP scenarios, to make sure we are not missing interesting events at the very first stage of the trigger system. Studying the performance of standard triggers can also guide towards designing dedicated triggers for long-lived particles at L1. We want to address all these issues in this work by studying how efficiently can we select LLP events in the 140 PU environment of the HL-LHC with standard triggers and also how some dedicated LLP triggers might help at L1.

We will now present a brief discussion of various kinds of LLP models in the next section and also which of them will surely be selected efficiently by standard L1 triggers, as discussed above. In this work, we will mostly concentrate on LLP scenarios where we are not sure of the performance of the standard triggers. 

\subsection{Long-lived particle scenarios sensitive to standard L1 triggers}
\label{ssec:llp_models}

Long-lived particles can either be pair-produced directly or can come from the decay of a resonance. 
The standard L1 triggers, as we have discussed in the previous section, will be able to trigger events with hard prompt particles, which can pass the $p_T$ thresholds as well as have enough associated tracks reconstructed at L1. The latter helps the final state objects to pass the same $z$-vertex condition. Therefore, if the LLP is produced along with prompt particles with high energies, it will get selected by the standard triggers. For instance, if the LLP is produced in the following process:

$$pp\rightarrow AA,\, A\rightarrow q\bar{q}X$$

where $A$ decays promptly, $X$ is long-lived, and if the mass difference between $A$ and $X$ is large, then there will be hard prompt jets in the events which can help trigger such events.

Scenarios where the LLPs are pair-produced without any other prompt particles, like 

$$pp\rightarrow A,\, A\rightarrow XX, ~~~{\rm~ or} ~~~ pp\rightarrow XX$$

might suffer due to the same $z$-vertex condition. Since the L1 tracks can be reconstructed with high efficiency up to a transverse distance of only 1 cm from the beamline ($L_{xy}<1{\rm~cm}$) \cite{James:2647214}, if the LLPs decay after that, their decay products mostly won't have any associated tracks and therefore, won't be able to satisfy the same $z$-vertex condition.

However, if the LLP is really massive and decays to quarks or gluons, its decay products will have very high transverse momenta, which can be easily triggered by standard single jet trigger which does not use the L1 tracking information. Say, we have events where LLPs of mass 500 GeV are pair produced and they decay to jets, then at least one of these jets will have $p_T>173{\rm~GeV}$ and pass the single jet trigger \footnote{If the LLP decays mostly after the ECAL and before the HCAL, one can use the CalRatio trigger for selecting such events, where the ratio of the ECAL to HCAL energy deposition is low. Such triggers have been proposed by the ATLAS collaboration\cite{Aad:2013txa,Aaboud:2018aqj}, and they will be mostly sensitive to LLPs with high decay length. However, the performance of such a trigger in the high PU environment of HL-LHC is yet to be studied, and therefore, we do not discuss this further in the present work.}.

The decay mode of the LLP is also an important factor in its searches. If the LLP decays to final states involving electrons, muons or photons, the final states are cleaner and also easier to trigger than if the LLP decays to quarks (or gluons) which hadronize to form jets. Also, if regional timing is available at L1, then these displaced electrons/muons/converted photons can be identified very easily. Jets are complex objects consisting of many particles, and their timing is defined using some statistical measure (like median time of all hits), which can be easily biased by pile-up hits on the MTD (as we will find in the later sections).

Based on discussions till now, we find that LLP scenarios in increasing order of the difficulty to trigger them at L1 are $-$
\begin{itemize}
\item LLPs with associated hard and prompt particles,
\item LLPs without associated hard and prompt particles
\begin{itemize}
\item massive LLPs decaying to jets,
\item LLPs decaying to electrons, photons, muons
\item LLPs with moderate masses and moderate decay lengths decaying to jets
\item very light LLPs decaying to jets
\end{itemize}
\end{itemize}
where by moderate LLP masses, we mean cases where the decay products don't have enough $p_T$ to pass the trigger threshold, and by moderate decay lengths, we mean significant amount of decays happening within the tracker volume. By very light LLPs, we mean a ballpark of a few GeV, which can never lead to significant energy deposition in the calorimeters and therefore, will mostly require associated hard and prompt particles for passing the level-1 trigger. They can also be triggered if they are produced with a hard ISR jet, but this will again reduce the cross-section of the process (we discuss this later in section \ref{sec:other_bench}).

In this work, we concentrate on the second most difficult possibility where the LLPs are not too heavy (around 10-100 GeV), they are produced without any hard prompt particles, their decay length is moderate such that they mostly decay within the tracker and they decay into final states with jets with 100\% branching.
We discuss four scenarios based on this $-$ the first two with direct pair-production of LLPs and the other two where the LLP comes from the decay of the SM Higgs boson and a heavy resonance respectively. The first two scenarios differ in the decay mode of the LLP $-$ one where the LLP decays to jets only, and the second where it decays to jets and an invisible particle. The motivation for the latter decay mode is to discuss the possibility of using a missing transverse energy trigger for such LLPs. We present our analyses with some benchmark points from the first scenario defined in the next section and discuss briefly the other scenarios in section \ref{sec:other_bench}. 

\section{Scenario with direct pair-production of LLPs and their decay to jets in the HL-LHC}
\label{sec:first_LLP}

In this section, we calculate the signal efficiencies for some benchmarks from the scenario, where LLPs are pair-produced directly and then decay to jets, using the single jet, dijet and quad jet triggers. Later, we also discuss how adding 140 PU affects the jet distributions in both LLP events as well as background events. In this work, we study only the dominant background $-$ QCD dijet events.

\subsection{Efficiency of standard Phase-II L1 triggers for LLPs}
\label{ssec:eff_L1_std}

In this section, we define our first LLP scenario (A), discuss the fractions of decays that we expect in different parts of the detector for various benchmark points (corresponding to different LLP masses and $c\tau$ values) in this scenario and also quote their L1 trigger efficiency when the LLP decays into jets and when the standard jet triggers are used at the HL-LHC.

We start with the following LLP scenario, which is just the direct pair-production of LLPs. Later in section \ref{sec:other_bench}, we present a discussion about some other LLP scenarios.

$$\mathbf{{\rm (A)~~~}p p \rightarrow X X,\ X \rightarrow j j}$$

We generate, using \texttt{PYTHIA6}\cite{Sjostrand:2006za}, LLPs from a quark initiated $Z$-mediated process \footnote{We use a supersymmetric LQD-type R-parity violating coupling, where a pair of sneutrinos are directly produced and they decay to two quarks, to simulate this type of LLP scenario.}, where they are directly produced and then decay to two quarks, which hadronize to give jets.
We study three different mass points, having two different mean proper decay lengths each $-$ $50{\rm~GeV}$, $100{\rm~GeV}$, and $200{\rm~GeV}$ having mean proper decay lengths \footnote{Hereafter, we will mostly refer to the mean proper decay length as just decay length, unless stated otherwise.} of $10{\rm~cm}$ and $100{\rm~cm}$ each. The LLP is assumed to decay to light jets with $100\%$ branching. We denote each benchmark point as ``(M\underline{\hspace{0.3cm}},$c\tau$\underline{\hspace{0.3cm}})'' with the mass of the LLP in GeV and the decay length $c\tau$ in cm.

Jets are defined as clusters of ECAL and HCAL energy deposition towers using anti-$k_T$ jet clustering algorithm with $R=0.4$. We use \texttt{Delphes-3.4.1}\cite{deFavereau:2013fsa} for detector simulation in our work with some modification such that the displaced stable particles deposit energy in the $\eta$ and $\phi$ bins corresponding to the detector segmentation rather than their actual $\eta-\phi$. LLPs decaying beyond the radial and half-length values from where the HCAL starts won't have any energy deposits in our setup\footnote{In experiment, the HCAL has longitudinal segmentation, and therefore, still there will be jets formed even if the LLP decays after the HCAL starts, until the LLP decays within the HCAL. In \texttt{Delphes} however, there is no longitudinal segmentation and the energy of all hadrons are deposited where the HCAL starts. Therefore, we won't get jets once the decay happens after this radial and half-length value.}.
Before studying the trigger efficiency of these benchmarks, we need to quantify the amount of decays occurring in various detector parts. This fraction depends on the boost (which is related to the mass and the production mode, whether direct or from a resonance) and the proper lifetime of the LLP as well as the volume of each detector part. The decay length of the LLP in the lab frame, $d$, is related to the boost of the LLP, $\beta\gamma$, and mean lifetime of the LLP in its rest frame, $\tau$, as 

\begin{equation}
d=\beta\gamma c\tau
\label{eq:decay_length}
\end{equation}

where $c$ is the velocity of light. We divide the decays into the following exclusive regions, where we have followed the CMS detector geometry \cite{CMS_geo}:

\begin{itemize}
\item \underline{Reco as L1 tracks}: corresponds to transverse distance ($L_{xy}$) of $1{\rm~cm}$ from the beamline and half-length ($|z_0|$) of $30{\rm~cm}$; if the LLP decays within this region, the tracks of its decay products are expected to be reconstructed at L1, typically with $\sim95\%$ efficiency and hence it stands a chance to pass the L1 trigger with the same $z$-vertex condition.
\item \underline{Before MTD}: corresponds to transverse distance of $1.161{\rm~m}$ and half-length of $2.6{\rm~m}$; for this fraction of LLP events, we can use the timing information, however, the decay products of these LLPs may not have L1 reconstructed tracks.
\item \underline{Before ECAL}: corresponds to transverse distance of $1.29{\rm~m}$ and half-length of $3{\rm~m}$; the MTD timing information cannot be used \footnote{The calorimeters have some time information, which has a resolution of few 100 ps \cite{CMS-PAS-EXO-12-035}. We do not use it in this work.} but still the jets from the LLPs decaying in this region won't have many tracks reconstructed at L1.
\item \underline{Before HCAL}: corresponds to transverse distance of $1.811{\rm~m}$ and half-length of $3.9{\rm~m}$; timing information cannot be used but still the jets from the LLPs decaying in this region won't have tracks reconstructed at L1 \footnote{In \texttt{Delphes}, the electrons and photons deposit energy only in the ECAL, therefore, jets after the ECAL won't have energy contribution from photons and electrons in our setup.}.
\item \underline{Before MS}: decays before the muon spectrometer (MS); corresponds to transverse distance of $4.020{\rm~m}$ and half-length of $5.68{\rm~m}$; from here on, we won't have any calorimeter energy deposit and hence, no jets. 
\item \underline{Inside MS}: decays inside the muon spectrometer (MS); corresponds to transverse distance of $7.38{\rm~m}$ and half-length of ${10.86\rm~m}$.
\item \underline{Outside detector}: the LLP decays outside the detector.
\end{itemize}

\begin{table}[hbt!]
\centering
\begin{tabular}{|c||c|c|c|c|c|c|c||}
\hline
Mass [GeV], & Reco as & Before & Before & Before & Before & Inside & Outside \\
Decay Length [cm] & L1 tracks & MTD & ECAL & HCAL & MS & MS & detector\\
\hline\hline
50, 10 & 13.20 & 80.51 & 1.26 & 1.85 & 1.72 & 1.19 & 0.26\\
50, 100 & 1.38 & 46.55 & 3.38 & 7.58 & 11.02 & 12.55 & 17.55\\
100, 10 & 10.59 & 86.81 & 0.70 & 0.91 & 0.68 & 0.28 & 0.02\\
100, 100 & 1.15 & 49.46 & 3.71 & 8.84 & 13.66 & 12.31 & 10.88\\
200, 10 & 10.83 & 88.42 & 0.27 & 0.31 & 0.14 & 0.02 & 0.002\\
200, 100 & 1.12 & 56.27 & 3.79 & 9.66 & 14.27 & 9.72 & 5.17\\
500, 10 & 11.94 & 87.99 & 0.04 & 0.04 & 0.04 & 0 & 0\\
500, 100 & 1.32 & 66.65 & 3.75 & 9.54 & 12.80 & 4.86 & 1.10\\
\hline\hline
\end{tabular}
\caption{Percentage of decays for different benchmark points from scenario (A) in various detector parts using the CMS detector dimensions.}
\label{tab:decay_frac_BP1}
\end{table}

Table \ref{tab:decay_frac_BP1} shows the percentage of the LLP decays in different detector parts for different benchmarks from scenario (A).
From table \ref{tab:decay_frac_BP1}, we notice a general trend that for a particular production mode of the LLP, with decreasing mass (increasing boost) and increasing proper decay length, the decay length in the lab frame shifts to higher values, as is expected from Eq. \eqref{eq:decay_length}. Therefore, we get relatively higher fraction of decays in the later parts of the detector. However, the $50{\rm~GeV}$ point in this scenario is different because at LLP mass values of $40-50{\rm~GeV}$, they will preferentially be produced by the decay of an on-shell $Z$-boson, since it is allowed in the way we are generating the LLPs through a $Z$ mediated process, and hence they will have very little boost and smaller decay lengths. This dependence of the fraction of decays of the LLP, in various regions of the tracker, on its mass and $c\tau$ is also illustrated in \cite{Banerjee:2019ktv}.

\begin{figure}[hbt!]
\centering
\includegraphics[scale=0.19]{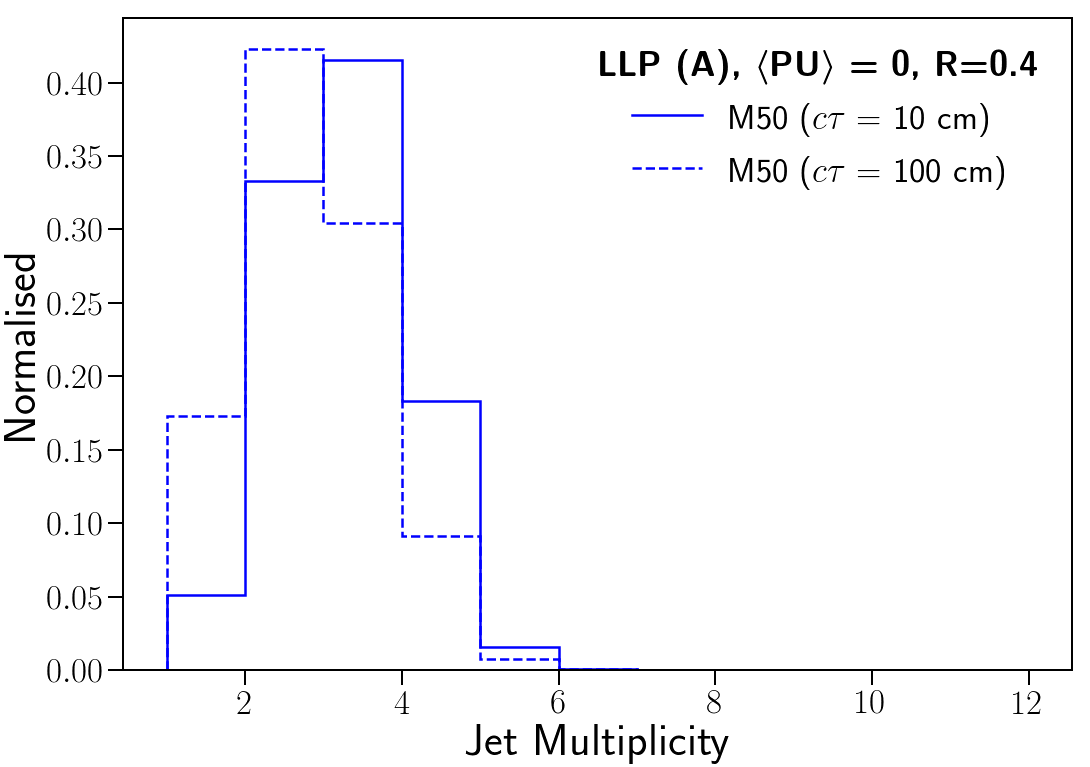}
\includegraphics[scale=0.19]{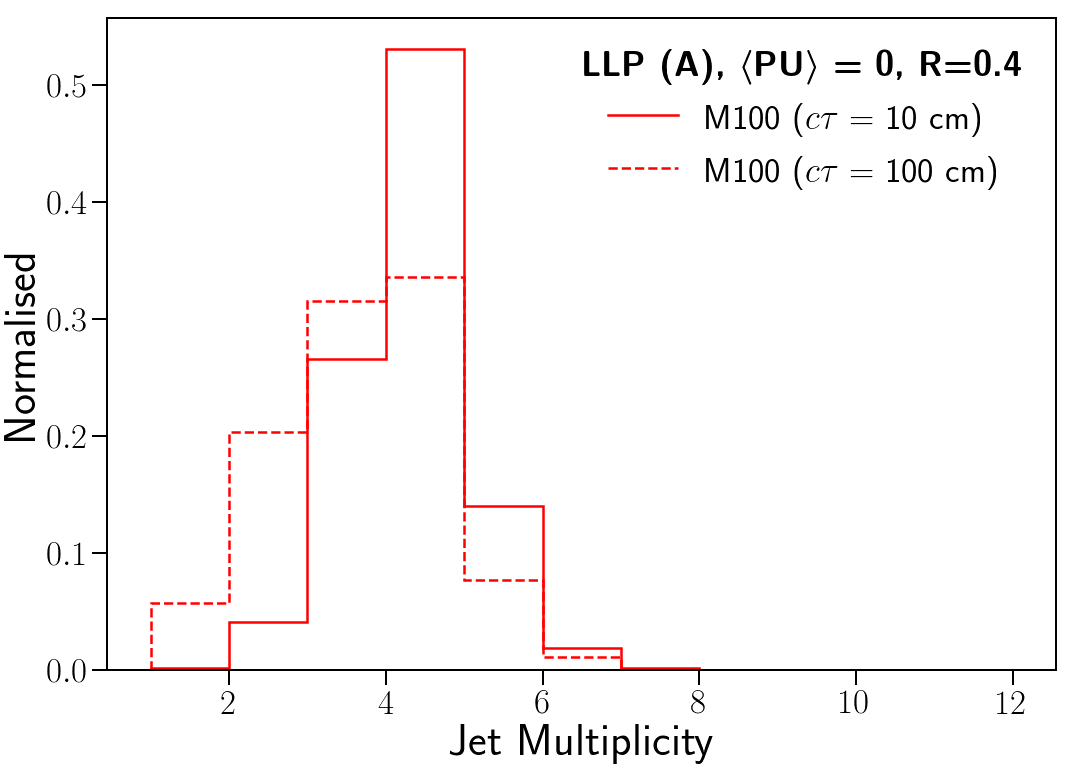}\\
\includegraphics[scale=0.19]{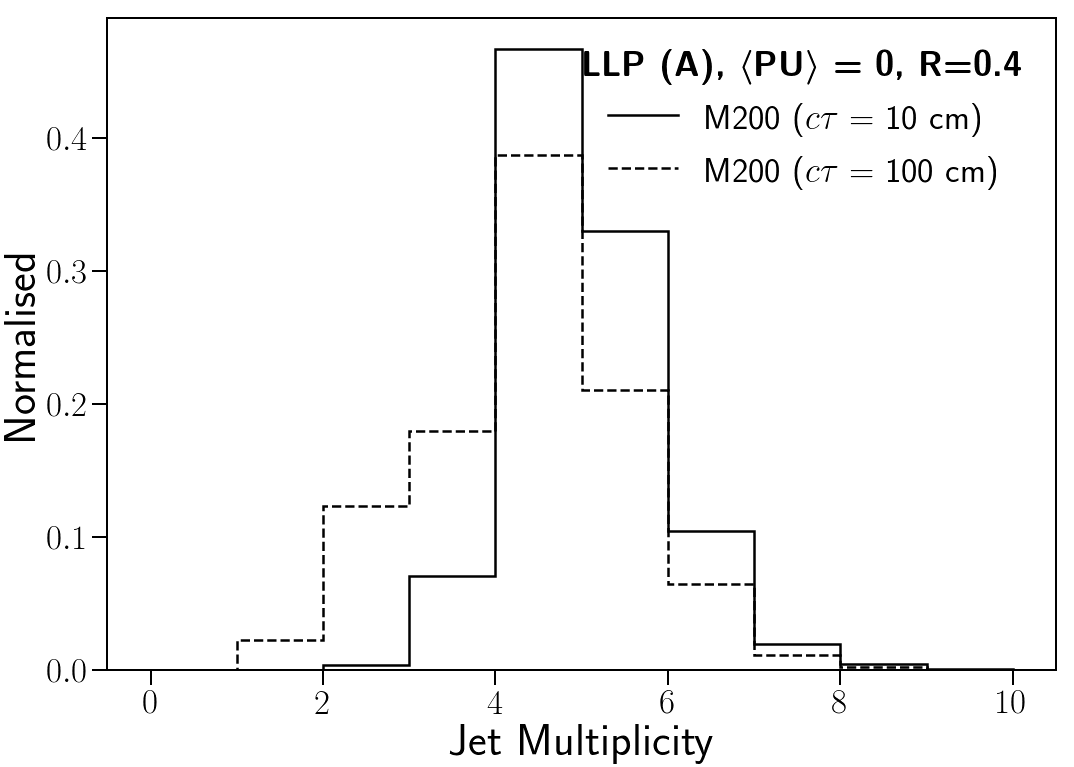}
\includegraphics[scale=0.19]{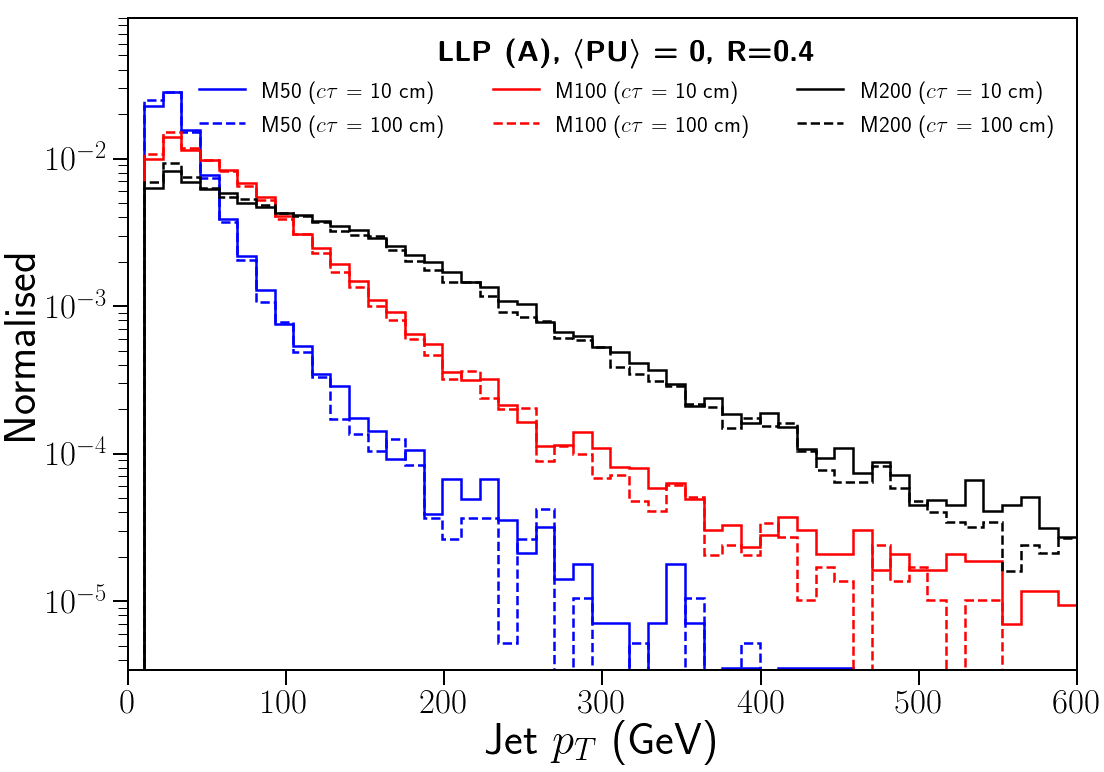}~
\caption{Multiplicity ({\it top panel and bottom left}) and $p_T$ ({\it bottom right}) distribution of jets with $p_T>15{\rm~GeV}$ coming from the decay of LLPs with different masses and decay lengths from scenario (A) as described above. The distributions are normalised such that their area is 1.}
\label{fig:LLP_BP1_pt}
\end{figure}

Fig.\ref{fig:LLP_BP1_pt}~ shows the multiplicity ({\it left}) and $p_T$ ({\it right}) distribution of jets with transverse momentum greater than 15 GeV from decays of LLPs from scenario (A) characterized by different masses and decay lengths. The jet multiplicity increases with mass of the LLP and decreases with increasing decay length, as is expected since massive LLPs and LLPs with smaller decay lengths decay more often within the starting of the HCAL (see table \ref{tab:decay_frac_BP1}). For highly displaced LLP decays, like when the LLP decays just before the HCAL, the two jets from the decay can be close enough to get identified as a single jet, and this also affects the multiplicity.

The jet $p_T$ distribution becomes harder with increasing mass of the LLP.
With increasing decay length, the $p_T$ distribution becomes slightly softer due to the fact that higher $p_T$ jets will most probably come from LLPs with larger boosts, and these will mostly decay after the HCAL starts, and hence are not identified as jets.

\begin{figure}[hbt!]
\centering
\includegraphics[scale=0.25]{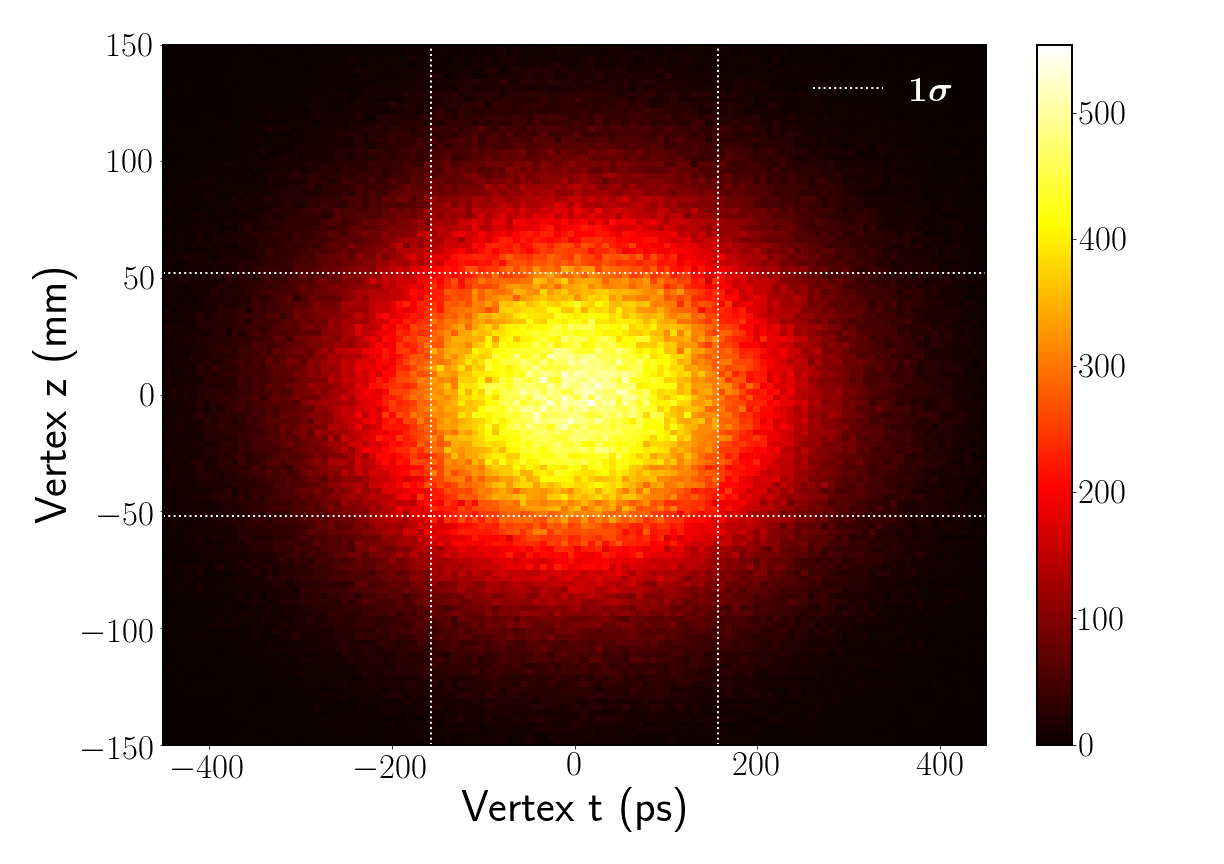}
\caption{The spread of $z$ positions and timings of 140 average PU vertices for 10,000 events. The color bar indicates the number of vertices having a particular $z$ position and time $t$. The vertices follow a two-dimensional Gaussian distribution, with $z=0$ and $t=0$ having the maximum probability of having a vertex. The total spread is of 25 cm and 800 ps respectively in $z$ and $t$ directions. The dotted grey lines correspond to the $1\sigma$ values of the Gaussian distributions in the $z$ and $t$ directions, with $\sigma_z=5.3{\rm~cm}$ and $\sigma_t=160{\rm~ps}$ respectively.}
\label{fig:PU_spread}
\end{figure}

We now move towards finding the trigger efficiency of our LLP benchmarks when standard jet triggers are used at L1.
To find trigger efficiency, we need to merge our LLP samples with pile-up.
We generate 1 million soft QCD events using \texttt{PYTHIA8}\cite{Sjostrand:2007gs} and use this as the pile-up. We merge this PU with the hard process using the \texttt{PileUpMerger} \footnote{We had to modify the \texttt{PileUpMerger} code of \texttt{Delphes} slightly as it was cancelling offset in the $z$ and $t$ values with respect to the first stable particle, which will clearly be a problem when our first stable particle comes from the decay of the LLP.} of \texttt{Delphes-3.4.1} with each event having an average of 140(200) PU vertices in addition to a hard $pp$ collision event. Fig.\ref{fig:PU_spread} shows the spread of PU vertices in 10,000 $pp\rightarrow\nu\nu$ events merged with an average of 140 PU events per hard collision. The number of PU events merged with each hard collision follows a Poisson distribution with average value of 140. The details of how the PU vertices are spread in the $z$ and time directions are described in the caption of fig.\ref{fig:PU_spread}.  

\begin{table}[hbt!]
\centering
\begin{tabular}{|c|c||c|c|c|c|c|c|c||}
\hline
Number of & Mass [GeV], & Single & Dijet & Dijet & Quad & Quad \\
$\langle PU\rangle$ & Decay Length [cm] & jet &  & (trk) & jet & jet (trk)  \\
\hline\hline
\multirow{6}{*}{$\langle PU\rangle=0$} & 50, 10 &$0.92\%$  &$0.83\%$  &$0.03\%$ &$0.01\%$  &$0.00\%$    \\
& 50, 100 & $0.50\%$ & $0.17\%$  & $0.02\%$ & $0.00\%$ & $0.00\%$    \\
& 100, 10 & $11.39\%$ & $10.20\%$  & $0.83\%$ & $1.24\%$ & $0.05\%$   \\
& 100, 100 & $8.26\%$  & $3.44\%$ & $0.53\%$ & $0.18\%$ &$0.03\%$     \\
& 200, 10 & $54.40\%$  & $47.73\%$  & $7.08\%$ & $13.40\%$ & $0.67\%$    \\
& 200, 100 & $43.73\%$ & $25.72\%$  & $4.36\%$ & $3.99\%$ & $0.28\%$    \\
\hline
\multirow{4}{*}{$\langle PU\rangle=140$} & 50, 10 & $2.72\%$ & $2.81\%$ & $0.47\%$ & $59.65\%$ & $10.99\%$ \\
& 50, 100 & $1.78\%$ & $0.94\%$ & $0.25\%$ & $55.94\%$ & $10.57\%$  \\
& 100, 10 & $27.43\%$ & $28.03\%$ & $5.83\%$ & $75.85\%$ &  $14.21\%$  \\
& 100, 100 & $20.48\%$ & $12.59\%$ & $2.93\%$ & $66.94\%$ &   $12.26\%$  \\
& 200, 10 & $81.04\%$ & $77.71\%$ & $23.68\%$ & $89.19\%$ &   $18.23\%$  \\
& 200, 100 & $67.98\%$ & $50.11\%$ & $13.98\%$ & $77.18\%$ &  $14.86\%$  \\
\hline\hline
\end{tabular}
\caption{Efficiency of selecting LLP events for benchmark points from scenario (A) with the standard Phase-II L1 jet triggers both for zero PU and 140 PU scenarios. For the multijet triggers, efficiencies for both, without and with (columns with `(trk)') the same-$z-$vertex condition are quoted.}
\label{tab:std_eff_llp}
\end{table}

Table \ref{tab:std_eff_llp} shows the signal efficiencies when standard single jet, dijet and quad jet triggers are used on LLP events from benchmark points in scenario (A). We present the efficiencies before (0 PU) and after merging with 140 PU. For comparison, we also present the dijet and quad jet efficiencies without the same $z$-vertex condition, which does not need the L1 track information (columns without ``(trk)'').
Adding PU increases the trigger efficiency for the signal benchmarks, as we can see from table \ref{tab:std_eff_llp} that the trigger efficiencies for the quad jet (trk) trigger is increased from $\sim 0$ to $\sim 10\%$ in the presence of PU for the 50 GeV benchmark point. Energy depositions coming from PU might increase the $p_T$ of the LLP jets. However, even if they satisfy the $p_T$ threshold, these jets will have very less probabilities of having tracks reconstructed at L1 (see table \ref{tab:decay_frac_BP1}), and therefore, they won't satisfy the same $z$-vertex condition. Jets from PU processes, on the other hand, can satisfy the $p_T$ threshold as well as the same $z$-vertex condition together in some cases and therefore, help trigger some of the LLP events. One way to see that these events are triggered indeed due to the PU jets is the fact that quad jet (trk) trigger efficiencies don't depend as much on the decay length of the LLP, as their percentage of decaying within the region where the L1 tracks can be reconstructed efficiently depends (see the second column of table \ref{tab:decay_frac_BP1}). Also, this effect of PU is dominant for the quad jet triggers than the single and double jet ones due to the lower $p_T$ threshold of the former.

We also find that removing the same vertex condition is helpful in improving the trigger efficiency, but it again comes with the cost of increasing the rates due to PU to very high values and decreasing the purity of the LLP sample. As we have discussed earlier in section \ref{ssec:descr-hllhc}, the quad jet rate increases from $12$ kHz to $185$ kHz, on lifting up the same $z$-vertex condition. Since we are limited by the trigger bandwidth, we need additional conditions after lifting up the same vertex condition. These conditions should be such that they are able to remove most of the PU contribution and can then differentiate between hard processes involving long-lived particles and QCD ones.

To achieve this, we first need to study how pile-up affects the jet distributions for the LLP events, which is our signal, and QCD dijet events, which is the dominant background. 

\subsection{Effect of high pile-up on jet distributions}
\label{ssec:narrow_jets}

The same $z$-vertex condition, though is a great way to control PU rates, it affects the signal efficiency of LLP events adversely as we have found in the previous section. However, if we want to lift that up, we have to study how the pile-up contribution affects the signal and background events.
As discussed in the previous section, we will study the dominant background coming from QCD dijet events in this work. We generate the QCD dijet events with different exclusive $p_T^{gen}$ cuts at the generation level of the partons $-$ $p_T^{gen}\in$(50,100) GeV, $p_T^{gen}\in$(100,150) GeV, $p_T^{gen}\in$(150,200) GeV and $p_T^{gen}>$200 GeV, using \texttt{PYTHIA6}.
Merging the hard processes with PU affects the jet multiplicity as well as other distributions, and since we have around 140 PU events per bunch crossing, the PU contribution will be dominant. We start by looking at the jet multiplicity distribution for the signal and the background with different average PU ($\langle{\rm PU}\rangle$) values.

\begin{figure}[hbt!]
\centering
\includegraphics[scale=0.25]{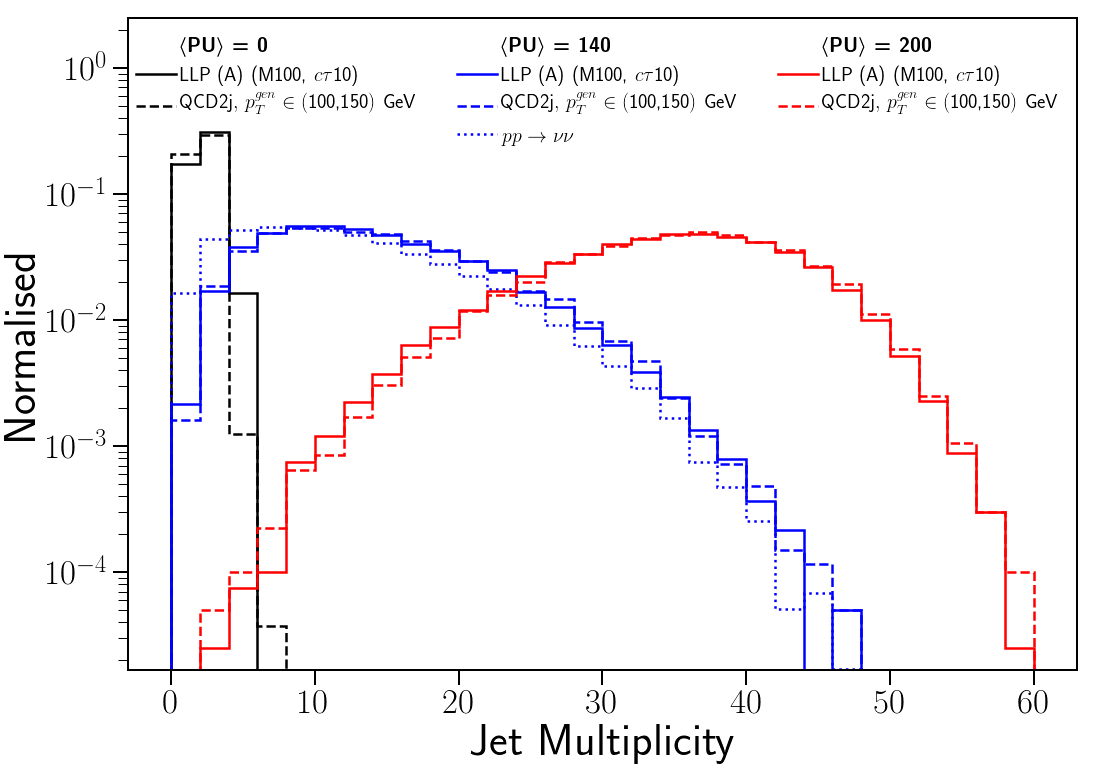}
\caption{Number of jets coming from QCD dijet processes (background) and LLP process corresponding to scenario (A), with $M_X = 100{\rm~GeV}$ and $c\tau = 10{\rm~cm}$ (signal) with zero PU, 140 PU, and 200 PU. Number of jets coming from a process having just 140 PU (generated by merging 140 PU with $pp\rightarrow Z\rightarrow\nu\nu$ process) is also shown.}
\label{fig:jet_multiplicity_PU}
\end{figure}

We have to study jets having transverse momentum above some threshold, and if this is very low, it will be significantly affected by the pile-up, and if this is too high, our signal jets from lighter LLPs won't be selected. 
Since our background  QCD dijet events have been generated starting from a $p_T^{gen}$ cut of 50 GeV, we consider jets with a transverse momentum greater than 60 GeV.  
Fig.\ref{fig:jet_multiplicity_PU} shows the number of jets having $p_T>60 {\rm~GeV}$ and $|\eta|<2.5$ coming from signal (a LLP benchmark point from scenario (A)) and background (QCD with $p_T^{gen}\in(100,150){\rm~GeV}$) with 0 PU, 140 PU and 200 PU. 
We find that increasing the amount of PU shifts the average number of jets from 3-4 (for 0 PU) to about $\sim 10$ and $\sim 40$ for 140 and 200 PU scenarios respectively, for both the LLP benchmark point and QCD dijet events.

Clustering jets with $R=0.4$ on the 140 PU (200 PU) sample, therefore, affects the distributions coming from just the hard processes (0 PU sample). 
Also, this bias is the same for both QCD and LLP processes and it will make both of their distributions look alike, even if they were different to start with, since the PU jets dominate both the samples. Therefore, it becomes hard to differentiate between them. 

Therefore, if we want to lift up the same $z$-vertex condition for improving the LLP signal efficiencies, we have to use some other ways for reducing the pile-up contribution.
The pile-up contribution is mostly uniform throughout the detector and therefore, its effect will depend on the area of the jet. 
Narrow jets with less area in the $\eta\times\phi$ plane will, therefore, have smaller PU contribution and also, it is more unlikely for a PU jet to contain more than $60{\rm~GeV}$ transverse momentum within a smaller region of $\eta\times\phi$. Therefore, we can reduce the multiplicity of jets coming from just PU collisions.

The cone size of hadronic objects need to be motivated by the physical size of the spread of the object in $\eta-\phi$. The cone size should be such that it is big enough to fully or at least mostly contain the hadronic activity of the jets from signal events, and small enough to not include a huge amount of PU contribution, while it is understood that some PU contamination would be unavoidable. 
If most of the energy of the signal jets are not contained within the smaller cone size, they won't be able to satisfy the $p_T$ threshold, and that will decrease the signal efficiency.
For jets coming from the decay of long-lived particles, we have another advantage of considering narrow jets. Displaced jets from LLP decays will have energy deposition contained in smaller region as has been pointed out in \cite{Aaboud:2019opc} and also discussed in \cite{Bhattacherjee:2019fpt}. Therefore, their $p_T$ might be affected less on considering narrow jet cone sizes. 

\begin{figure}[hbt!]
\centering
\includegraphics[scale=0.19]{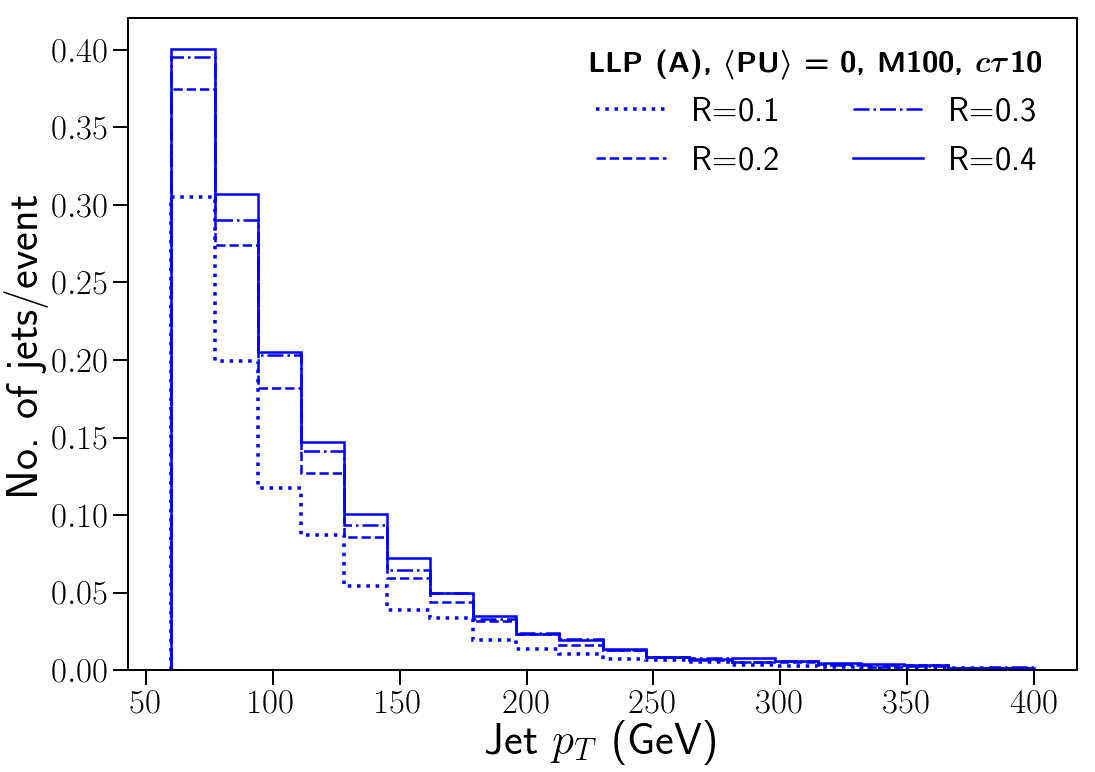}~~
\includegraphics[scale=0.19]{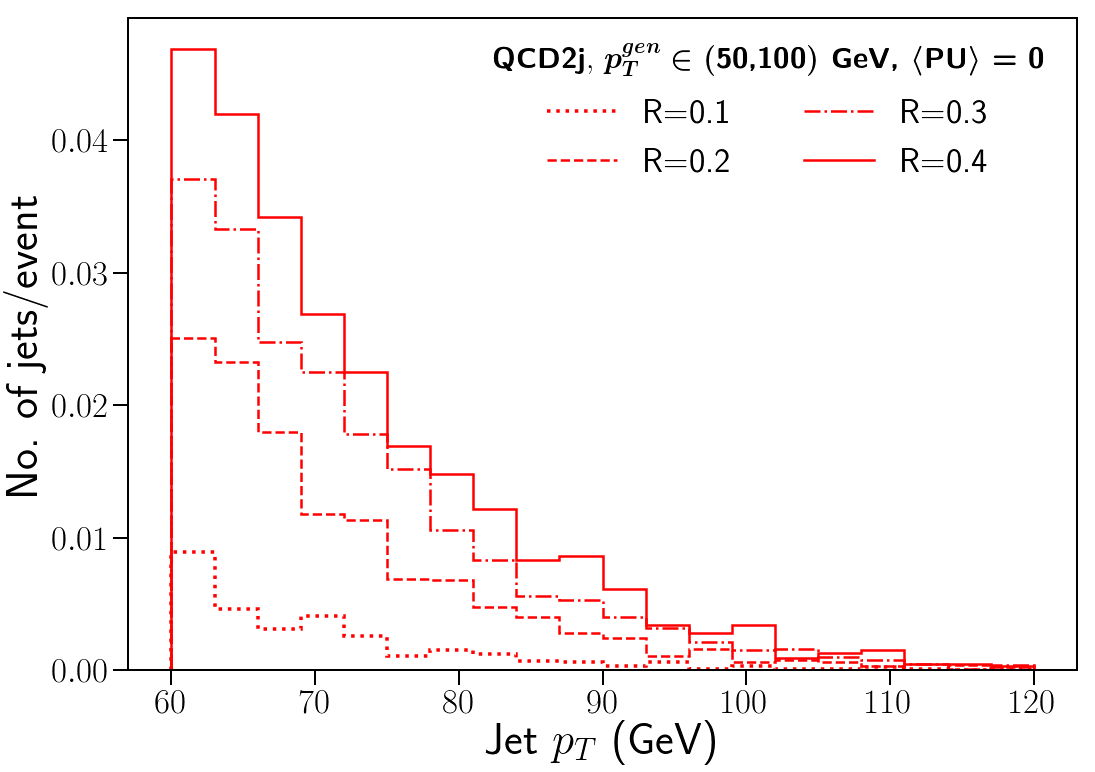}
\caption{Comparison of jet $p_T$ distributions for jets clustered using anti-$k_T$ with cone sizes $R=0.1$, $R=0.2$, $R=0.3$ and $R=0.4$ coming from LLPs with mass of the LLP 100 GeV and decay length 10 cm in scenario (A) ({\it left}) and from QCD dijet events with $p_T^{gen}\in(50,100){\rm~GeV}$ without any PU. The distributions are drawn for 10,000 events.}
\label{fig:jetpt_var_r}
\end{figure} 

We compare the jet $p_T$ distributions with different cone sizes ($R=0.1,\, 0.2,\, 0.3,\, 0.4$), for one benchmark point of signal from scenario (A), (M100,$c\tau10$) ({\it left}) and for the background QCD dijet events with $p_T^{gen}\in(50,100){\rm~GeV}$ ({\it right}) without PU in fig.\ref{fig:jetpt_var_r} for all jets with $p_T>60{\rm~GeV}$ and $|\eta|<2.5$ cuts in 10,000 events. We find that the $p_T$ distribution becomes softer as is expected with reducing cone size $R$. We find that the QCD jets with $p_T^{gen}\in(50,100){\rm~GeV}$ are affected much more than the LLP benchmark, which is expected since they start from the beamline and spread more in the physical region till they reach the calorimeters and deposit energy in a cone of radius $R=0.4$ mostly. 
As the $p_T$ of the QCD jet increases, the jet constituents will be more collimated and hence, narrow jets won't affect the $p_T$ distribution much.

\begin{figure}[hbt!]
\centering
\includegraphics[scale=0.19]{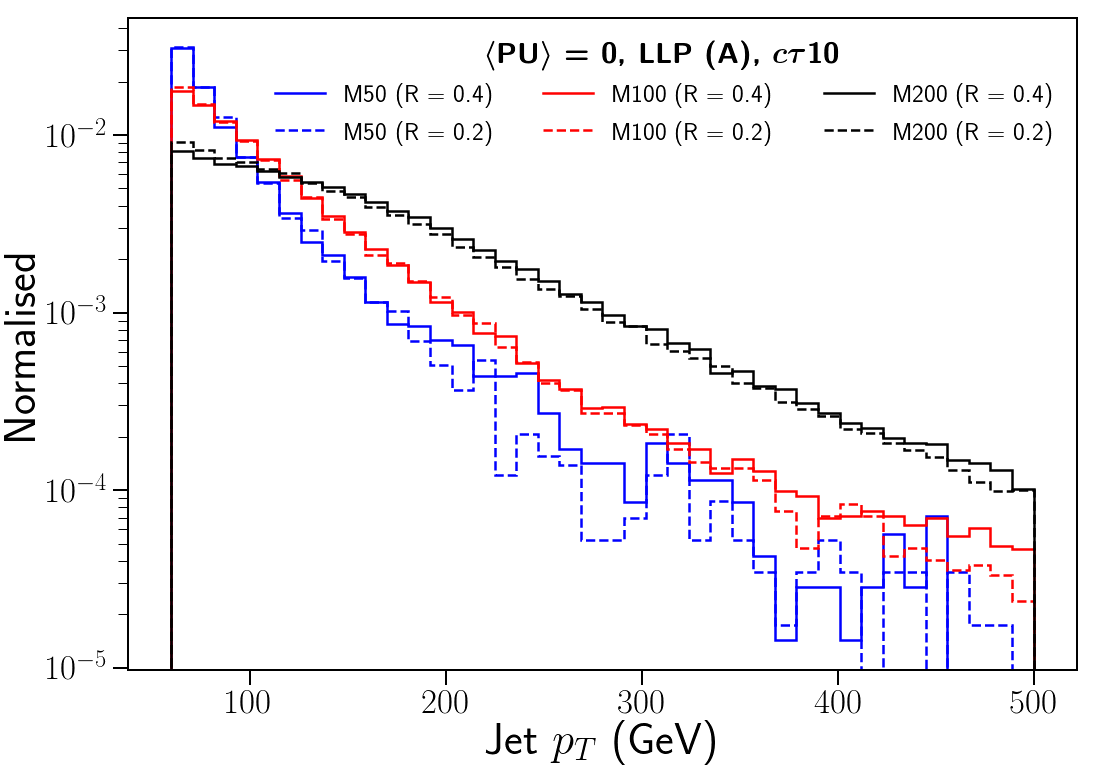}~~
\includegraphics[scale=0.19]{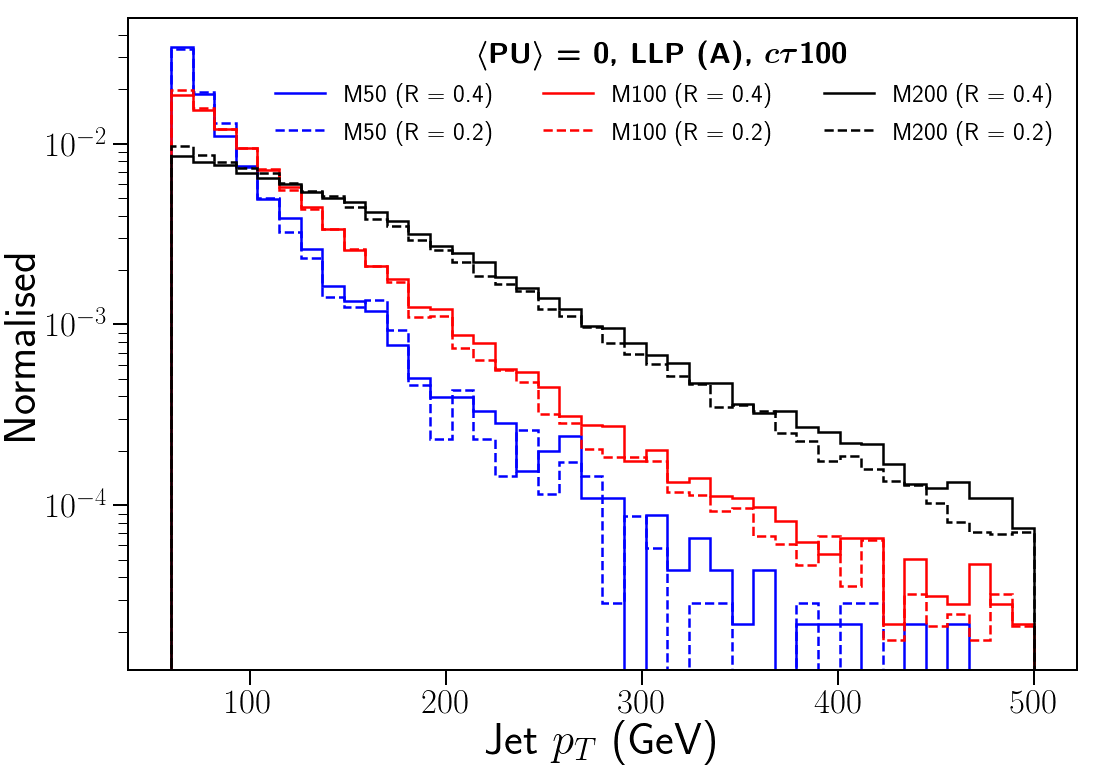}
\caption{Comparison of jet $p_T$ distributions for jets clustered using anti-$k_T$ with $R=0.2$ and $R=0.4$ coming from LLP benchmark points from scenario (A) with different masses having a decay length of 10 cm ({\it left}) and 100 cm ({\it right}) with zero PU.}
\label{fig:jetpt_llp_r02vsr04}
\end{figure}

Our finding is also supported by experimental collaborations, which have used small cone size when it is known that the hadronic object of interest is expected to be narrow. For example, ATLAS collaboration have used 0.2 cone size for tau jet triggers \cite{ATLAS:2017hlo}, as hadronic taus are narrower than their QCD counterparts. 
We decide to use $R=0.2$ jets because the signal $p_T$ distribution is affected little till that cone size value.
Fig.\ref{fig:jetpt_llp_r02vsr04} compares the jet $p_T$ distribution for the LLP benchmark points from scenario (A) with decay length 10 cm ({\it left}) and 100 cm ({\it right}) using the usual $R=0.4$ jets with narrow jets ($R=0.2$).
We find that using narrow jets does not change the jet $p_T$ distribution for the LLP hard process much. 

\begin{figure}[hbt!]
\centering
\includegraphics[scale=0.19]{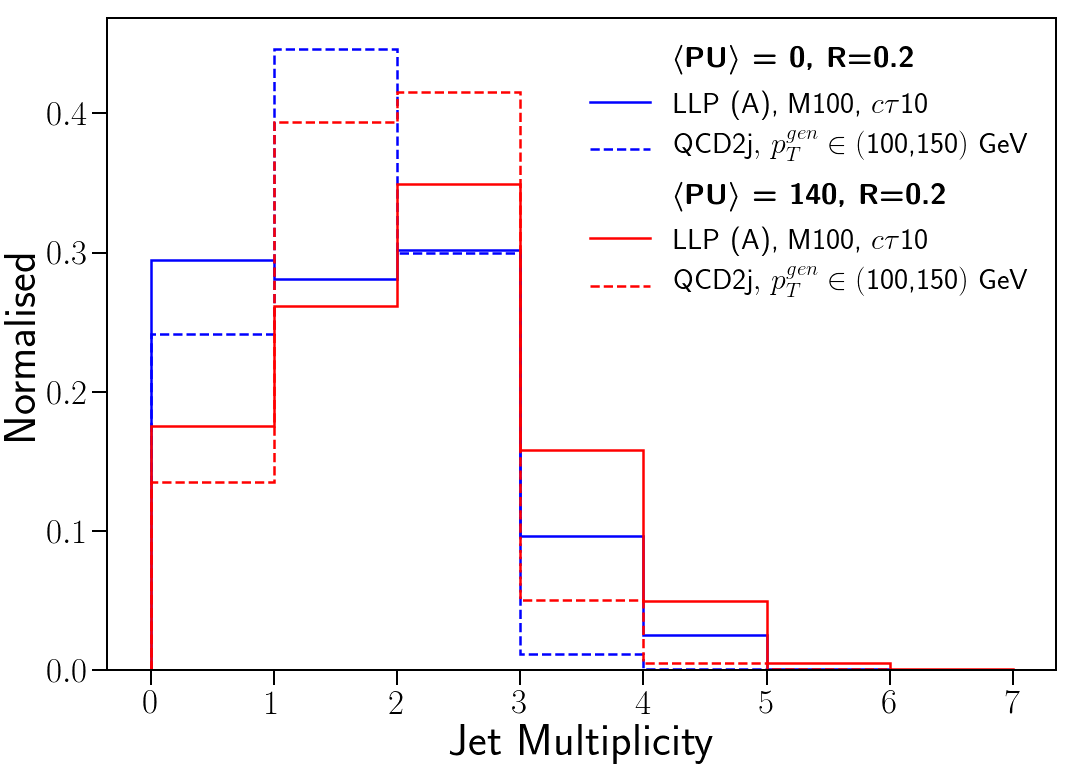}~~
\includegraphics[scale=0.19]{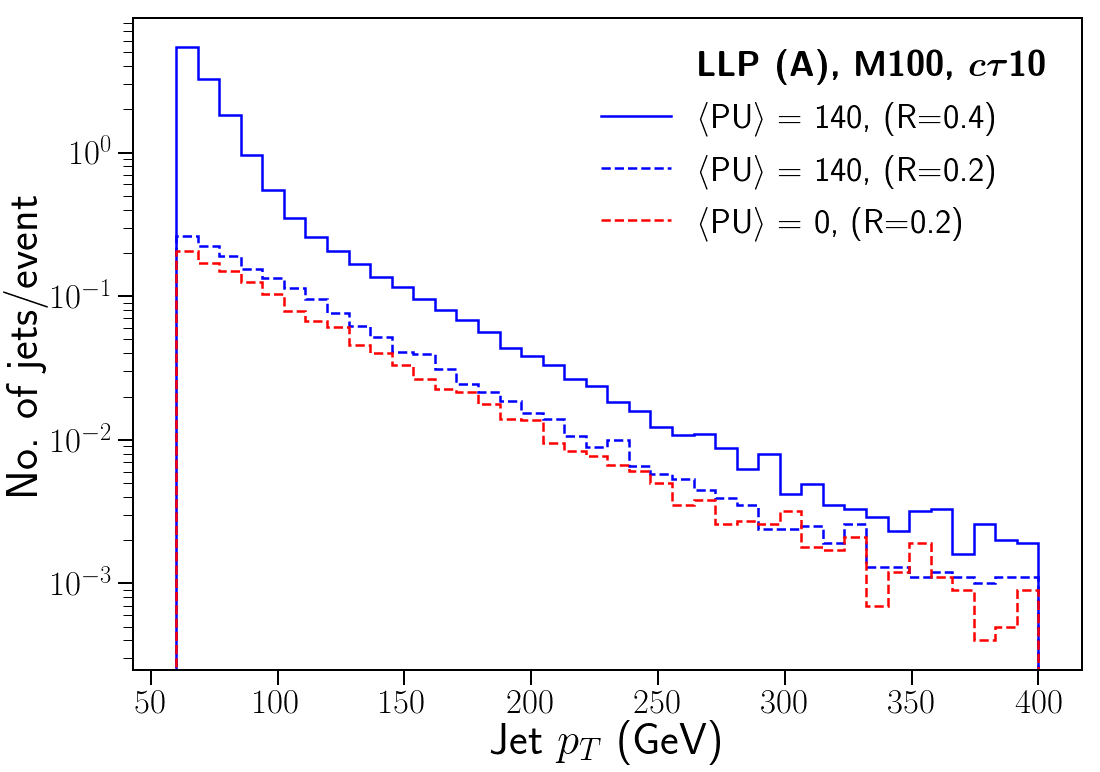}
\caption{{\it Left:} Jet multiplicity for jets clustered with anti-$k_T$ $R=0.2$ having $p_T>60{\rm~GeV}$ from LLP benchmark (M100,$c\tau$10) from scenario (A) and QCD dijet events with $p_T^{gen}\in(100,150){\rm~GeV}$ merged with no PU and 140 PU; {\it Right:} $p_T$ distribution for jets clustered using anti-$k_T$ with $R=0.2$ and $R=0.4$ coming from LLP benchmark ($M_X = 100{\rm~GeV}$ and $c\tau = 10{\rm~cm}$) from scenario (A) with 140 PU compared with the zero PU jet $p_T$ distribution, all drawn for 10,000 events.}
\label{fig:jetpt_llp_PU}
\end{figure}

Let us now add PU and see how considering narrow jets help. The {\it left} panel of fig.\ref{fig:jetpt_llp_PU} shows the comparison of the jet multiplicity distribution coming from a LLP benchmark point from scenario (A) and from QCD dijet events with 0 PU and 140 PU clustered with $R=0.2$, and the {\it right} panel of fig.\ref{fig:jetpt_llp_PU} shows the comparison of the $p_T$ distribution of jets coming from a LLP benchmark point from scenario (A) with 0 PU (with $R=0.2$) and with 140 PU for jets with $R=0.2$ and $R=0.4$.
We find that the multiplicity of jets with $p_T>60{\rm~GeV}$ decreases when we consider $R=0.2$ jets from that when we had $R=0.4$ jets. The multiplicity distribution does not exactly match the zero PU multiplicities, but we are able to get rid of many PU jets. The $p_T$ distribution is for all the jets with $p_T>60{\rm~GeV}$ and $|\eta|<2.5$ in 10,000 events for each case. When jets are clustered using anti-$k_T$ with $R=0.4$, processes merged with 140 PU have very high jet multiplicities (fig.\ref{fig:jet_multiplicity_PU}), most of which are PU jets with low $p_T$ values. Therefore, the $p_T$ distribution is biased to low values.
Jets with $R=0.2$ restore the hard process $p_T$ distribution reasonably well, though there is still some PU contribution. 
Hereafter, we will mostly consider jets with $R=0.2$, $p_T>60{\rm~GeV}$ and $|\eta|<2.5$ for our analyses, unless stated otherwise.

\section{Dedicated LLP triggers}
\label{sec:triggers}

In the previous section, we have discussed how the standard L1 triggers are not very efficient in selecting events from LLP scenario (A), where a pair of LLPs is directly produced and they decay into jets, especially when the LLP is light. In this section, we discuss some dedicated triggers based on information available at L1 for long-lived particles that can improve their signal efficiencies.


\subsection{Trigger based on L1 tracking}

For long-lived particles,
the trackless jet trigger was used in \cite{Aad:2013txa} as a High Level Trigger. Since jets coming from LLP decays will be displaced, these jets will have very few reconstructed tracks and will be mostly trackless. Their trigger design was to select events with jets having $E_T>25{\rm~GeV}$, which have no reconstructed tracks with $p_T>0.8 {\rm~GeV}$ within $\Delta R=\sqrt{\Delta\eta^2+\Delta\phi^2}=0.2$ around the jet axis \footnote{In this work, we denote the cone-size of an anti-$k_T$ clustered jet as $R$, and we use $\Delta R$ to denote some $\Delta\eta\times\Delta\phi$ region around the jet axis.}. They need an associated muon with $p_T>10{\rm~GeV}$ to pass the L1 trigger.

\begin{figure}[hbt!]
\centering
\includegraphics[scale=0.19]{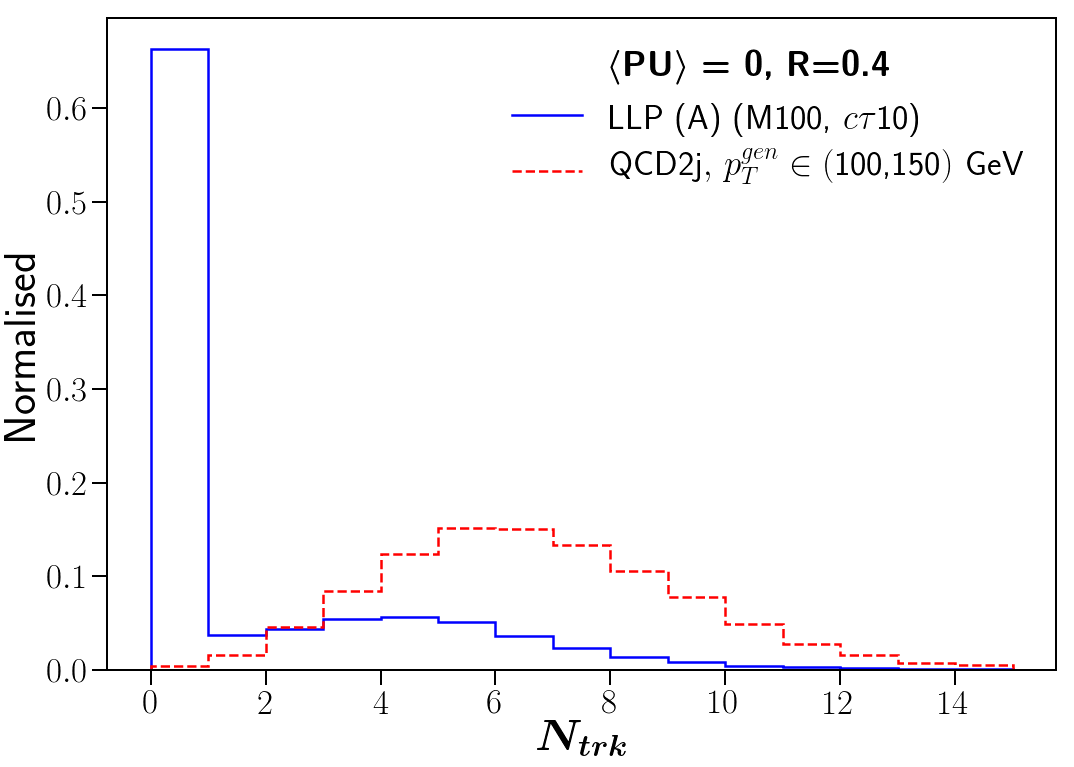}~~
\includegraphics[scale=0.19]{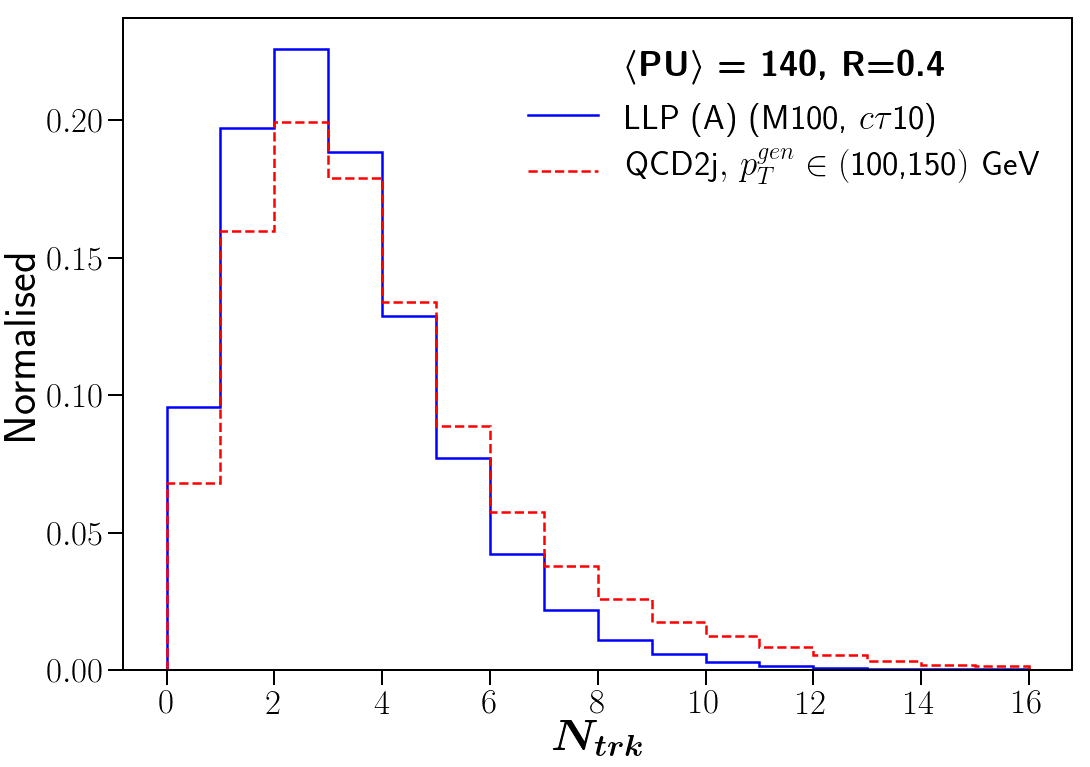}\\
\includegraphics[scale=0.19]{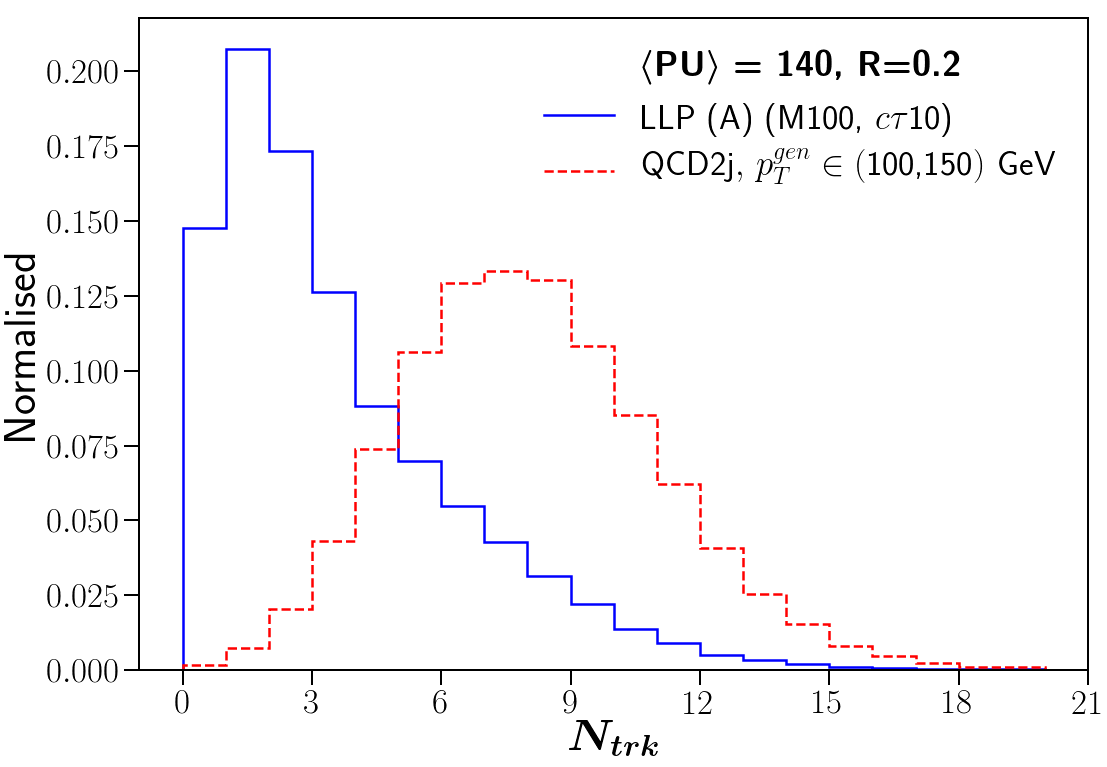}
\caption{Distribution of number of tracks associated with jets clustered using anti-$k_T$ with $R=0.4$ coming from LLP benchmark ($M_X = 100{\rm~GeV}$ and $c\tau = 10{\rm~cm}$) from scenario (A) with zero PU (\textit{top left}) and 140 PU (\textit{top right}) compared with the number of tracks associated with QCD jets. In the {\it bottom} plot, we have shown the distributions for jets with $R=0.2$ merged with 140 PU.}
\label{fig:Ntracks}
\end{figure}

However, with the Phase-II upgrade, since tracking will be available at L1, we can think of designing some trigger for LLPs based on the tracking information at L1. However, we would only be able to have tracks for particles with $p_T>2{\rm~GeV}$ that are produced within a radial distance of 1 cm and half-length of 30 cm. Long-lived particles, decaying after a radial distance of 1 cm or half-length of 30 cm won't have jets with tracks reconstructed at L1 unlike the QCD jets. Fig. \ref{fig:Ntracks} ({\it top left}) shows the number of L1 tracks associated with a $R=0.4$ jet with $p_T>60{\rm~GeV}$, i.e., tracks within $\Delta R=0.4$ of the jet axis ($N_{\text{trk}}$), for jets from LLP benchmark ($M_X = 100{\rm~GeV}$ and $c\tau = 10{\rm~cm}$) from scenario (A) and QCD jets, when there is no PU. We notice that they have significant difference.

However, with increasing the amount of PU to 140, 
the PU contribution dominates in both samples. As we can see from fig. \ref{fig:jet_multiplicity_PU}, the jet multiplicity in an event increases from around 3-4 to 10 on adding PU. Therefore, most of the jets in an event are PU jets and they control all the distributions in both signal and background. As a result, there is no difference between the $N_{\text{trk}}$ distribution for jets from LLP events and QCD ones. But once we consider narrow jets of R=0.2, as we have already discussed and demonstrated earlier, the number of jets coming only from PU decreases (compare figs. \ref{fig:jet_multiplicity_PU} and \ref{fig:jetpt_llp_PU}). Therefore, $N_{\text{trk}}$ of a narrow ($R=0.2$) jet again has some difference between LLP and QCD events, with the former shifted towards lesser number of tracks than the latter. The {\it top right} and {\it bottom} plots of fig.\ref{fig:Ntracks} show the $N_{\text{trk}}$ distributions in the 140 PU scenario with $R=0.4$ and $R=0.2$ jets respectively.

There is still some amount of PU contamination in the $N_{\text{trk}}$ distribution of $R=0.2$ jets which is evident from the comparison of the {\it top left} and {\it bottom} panels of fig.\ref{fig:Ntracks}. Therefore, just using the $N_{\text{trk}}$ variable to discriminate between displaced and prompt jets won't be very efficient. We study many possible variables that can be constructed out of the tracking information at L1, and discuss how these variables and their correlations can help identifying displaced jets from prompt ones.

\subsubsection{Training a BDT to classify displaced and prompt jets based on tracking information}
\label{sssec:train_track}

We consider the following variables in addition to $N_{\text{trk}}$, which can help distinguish between jets coming from QCD dijet and LLP events:

\begin{itemize}

\item $\mathbf{\sum p_T}$ {\bf of all tracks associated with a jet:} For jets coming from LLP decays, the tracks associated with the jet will mostly be coming from PU. Therefore, we expect weaker correlation between the jet $p_T$ and the sum of $p_T$ of all the tracks associated with the jet ($\sum p_T=\sum_{i=1}^{N_\text{trk}} p_{T,i}$) for an LLP jet than in a QCD prompt jet, where most of the tracks will actually be the constituents of the jet (see the correlation matrix of some of the track variables for signal and background in Appendix \ref{app:corr_matrix}). 

\item {\bf Jet $\mathbf{z}$ vertex $\mathbf{z_{\text{j\_vtx}}}$:} The $z$-position of the jet vertex is defined as the $p_T$ weighted average of $z$-values of all the tracks associated with a jet ($z_{\text{j\_vtx}}=\sum_{i=1}^{N_\text{trk}} p_{T,i}\times z_i/\sum_{i=1}^{N_\text{trk}} p_{T,i}$) \cite{Contardo:2020886}.

\item $\mathbf{\Delta z_{\text{j\_vtx}}: }$ We also calculate the $p_T$ weighted average of the difference between the $z$-values of tracks and the $z$-position of the jet vertex ($\Delta z_{\text{j\_vtx}}=\sum_{i=1}^{N_\text{trk}} p_{T,i}\times |z_i-z_{\text{j\_vtx}}|/\sum_{i=1}^{N_\text{trk}} p_{T,i}$). This might give us an idea of how much the jet vertex is shifted along the $z$-direction due to stray PU tracks associated with an LLP or QCD jet, and we expect this to be smaller for the latter than the former.

\item $\mathbf{p_{T\text{(vtx)}}^{\text{miss}}:}$ We calculate a variable called $p_{T\text{(vtx)}}^{\text{miss}}$, which is the missing transverse momentum calculated for the vertex from which a jet is coming from. Say, a jet has the $z$ vertex position at $z_{\text{j\_vtx}}$, and there are $n$ jets coming from the same $z$ vertex, within 1 cm (including the starting jet), then $p_{T\text{(vtx)}}^{\text{miss}} = \sqrt{(\sum_n p_x^i)^2+(\sum_n p_y^i)^2}$ where $p_x^i$ and $p_y^i$ are the $x$ and $y$ components of the momentum of the $i^{\rm th}$ jet coming from the same vertex.

\item $\mathbf{n_{z_{\text{trk\_max}}}:}$ Next, we group all tracks associated with a jet having same $z$-positions (within $1{\rm~mm}$) and label these different $z$-values as $z_a~(a=1,2,...)$, each having $n_{z_a}$ number of tracks, and $\sum_{z_a} n_{z_a}=N_{\text{trk}}$. For each jet, the maximum number of tracks coming from the same $z$-position ($n_{z_{\text{trk\_max}}}$), is stored. 
 
\item $\mathbf{\Delta z_{\text{trk\_max}}:}$ We calculate the average of the difference between the $z$-position with maximum number of tracks and all the other $z$-values weighted by the number of tracks coming from that particular $z$-value ($\Delta z_{\text{trk\_max}}=\sum_{z_a} n_{z_a}\times |z_a-z_{\text{trk\_max}}|/N_{\text{trk}}$). 

\item $\mathbf{\sum p_T^{z_{\text{trk\_max}}},\, \sum p_T^{z_a\neq z_{\text{trk\_max}}},\, \sum p_T^{z_{\text{trk\_max}}}/\sum p_T:}$ Also, the sum of $p_T$ of tracks coming from the $z$-position with maximum tracks and that of tracks coming from all other $z$-vertices are calculated separately as ($\sum p_T^{z_{\text{trk\_max}}}=\sum_{i=1}^{n_{z_{\text{trk\_max}}}} p_{T,i}$) and ($\sum p_T^{z_a\neq z_{\text{trk\_max}}}=\sum p_T-\sum p_T^{z_{\text{trk\_max}}}$) respectively. The fraction of $p_T$ coming from the $z$-vertex with maximum tracks in the total $p_T$ sum of all tracks associated with a jet ($\sum p_T^{z_{\text{trk\_max}}}/\sum p_T$) is also a variable of interest, which is expected to be closer to 1 for QCD prompt jets than LLP jets, because most of the tracks associated with a prompt QCD jet will come from the same $z$-position (within 1 mm).

\item {\bf Entropy variables:} To quantify the spread of $z$-values of tracks associated with a jet, we also consider the Shannon entropy \cite{shannon} of the $z$-position ($S(z_i) = -\sum_{i=1}^{N_\text{trk}} P(z_i)\times\\\text{log}_{N_{\text{trk}}} P(z_i)$) as well as the $p_T$ ($S(p_{T,i}) = -\sum_{i=1}^{N_\text{trk}} P(p_{T,i})\text{log}_{N_{\text{trk}}} P(p_{T,i})$) of the tracks associated with a jet. We consider different variants of these entropy variables $-$ $S(|z_i|/\sum |z_i|)$, $S(z_i+301)$ and $S(z_i+301/\sum (z_i+301))$ for the starting $z$ values of the tracks; and $S(p_{T,i})$ and $S(p_{T,i}/\sum p_{T,i})$ with the track $p_T$ values. 

If the input of the Shannon entropy is a set of numbers with very less variation, the entropy value is close to 0, and if the variation is maximum, the entropy value is closer to 1. We expect for QCD prompt jets, most of the L1 tracks to come from the same $z$ positions unlike for displaced jets, which will mostly have tracks from PU associated with them, which will have different $z$ positions. The same might be true for $p_T$ of the tracks which will be more correlated for prompt jets.

\item {\bf Variables with tracks within the narrow jet:} We also calculate all the variables discussed above with tracks within $\Delta R=0.2$ of the jet axis (variables with a superscript saying $0.2$). 

\item $\mathbf{N_{\text{trk}}/N_{\text{trk}}^{(0.2)},\, \sum p_T/\sum p_T^{(0.2)}:}$ Finally, we also calculate the ratios $\frac{N_{\text{trk}}}{N_{\text{trk}}^{(0.2)}}$ and $\frac{\sum p_T}{sum p_T^{(0.2)}}$. 

\end{itemize}

\begin{figure}[hbt!]
\centering
\includegraphics[scale=0.175]{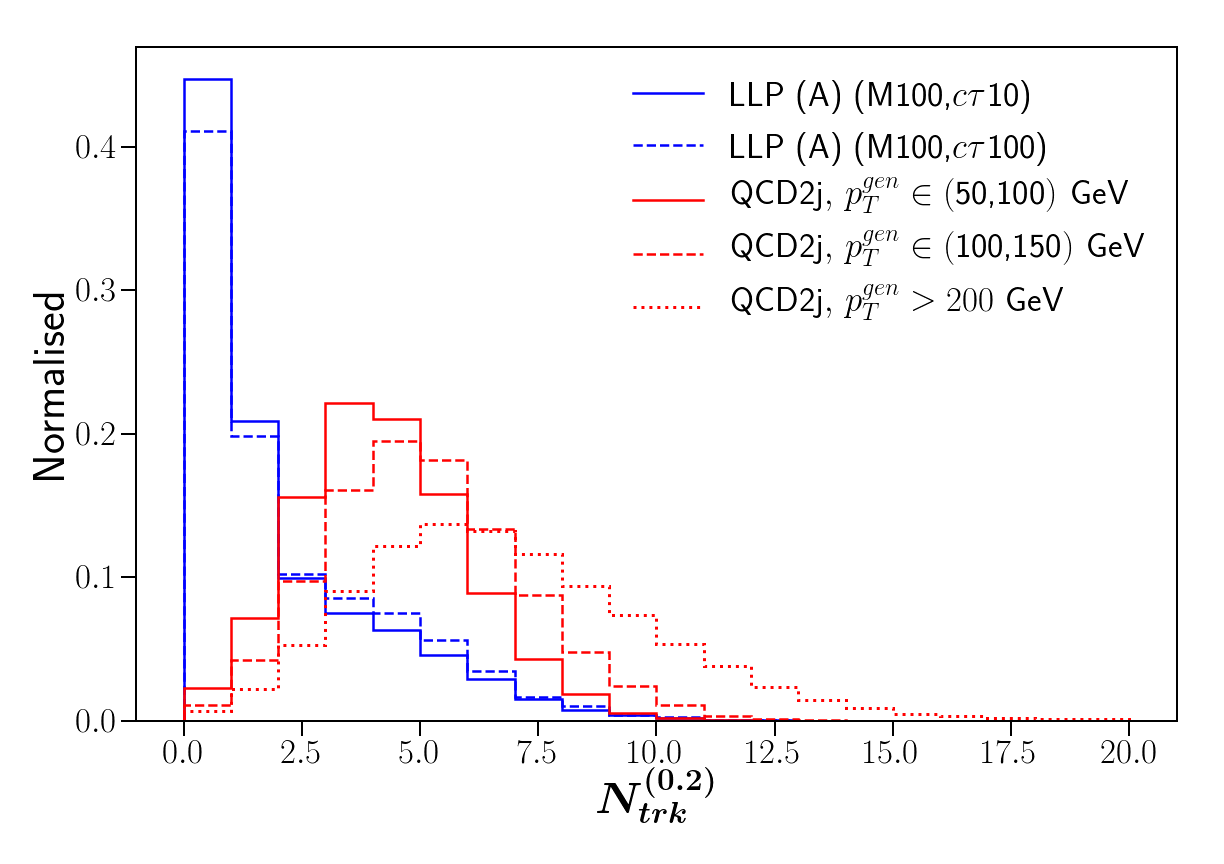}~
\includegraphics[scale=0.175]{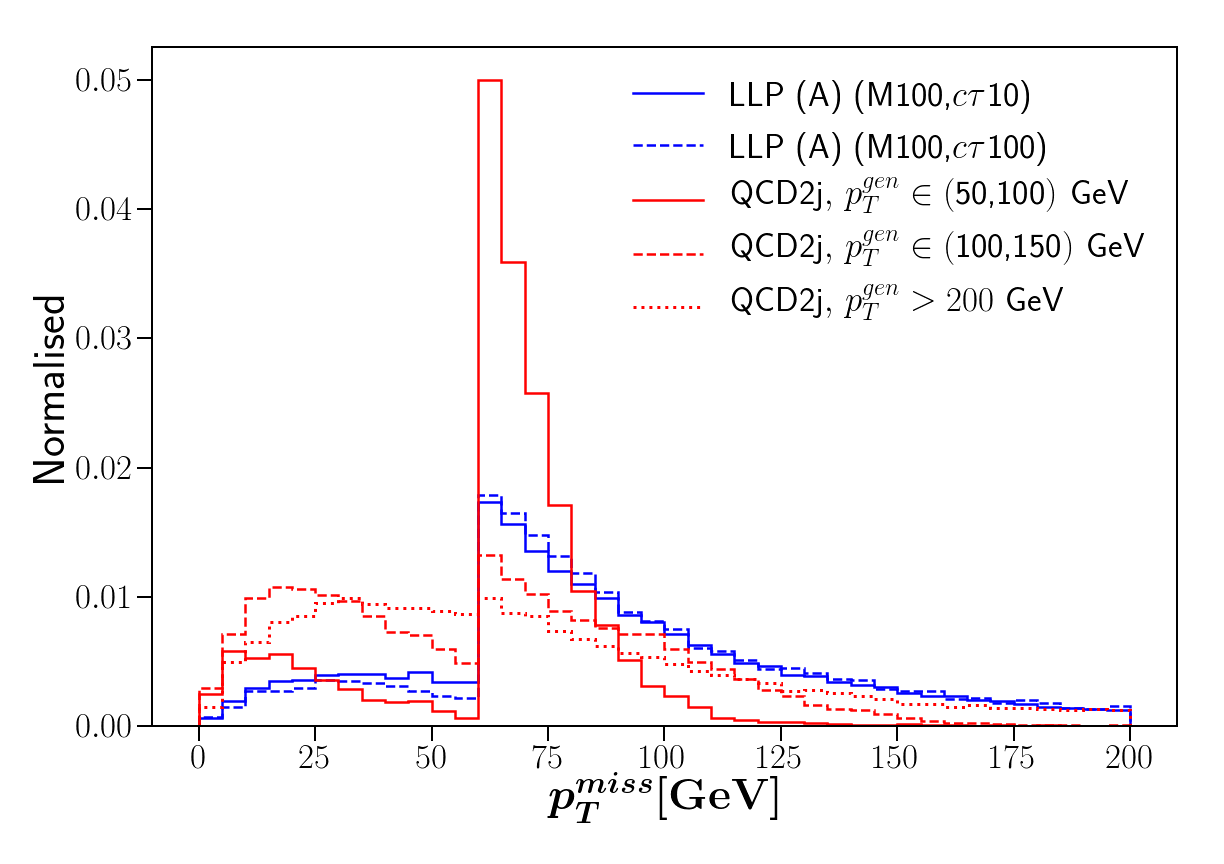}\\
\includegraphics[scale=0.175]{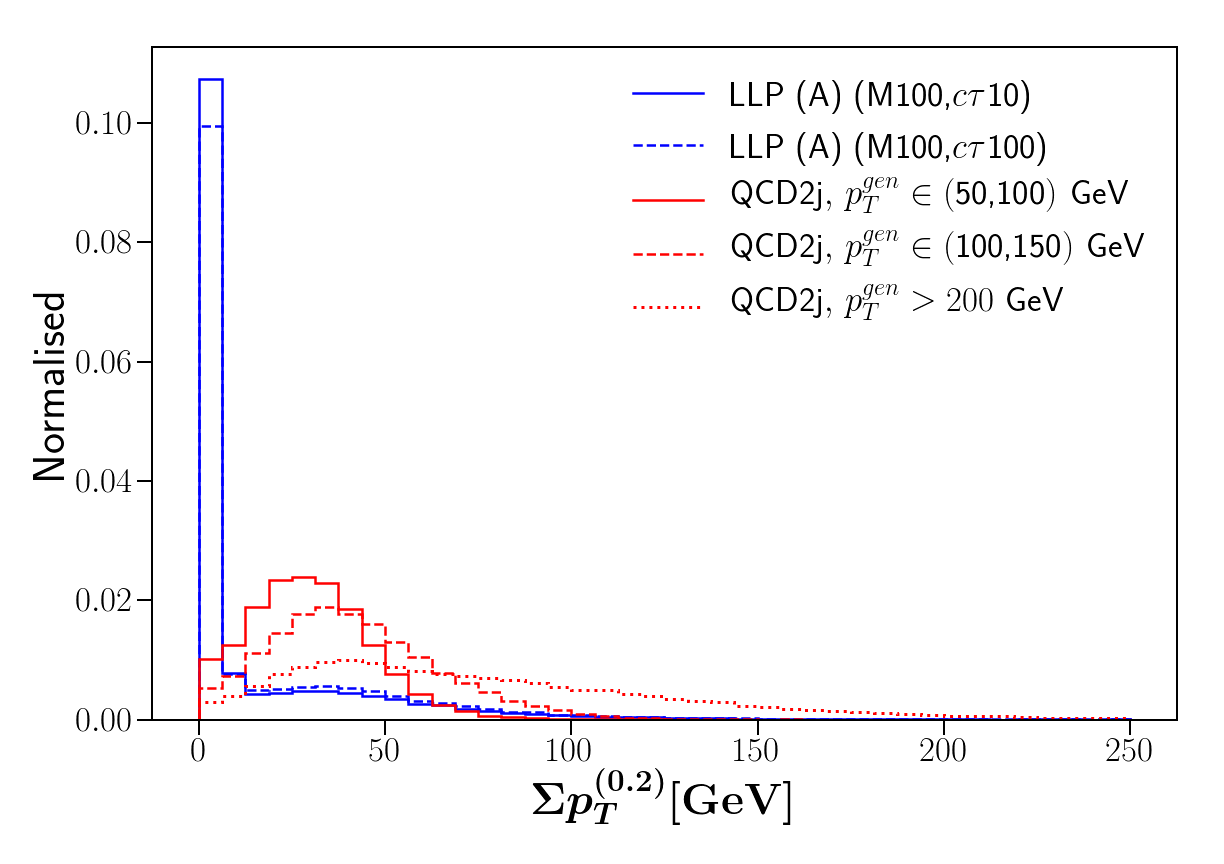}~
\includegraphics[scale=0.175]{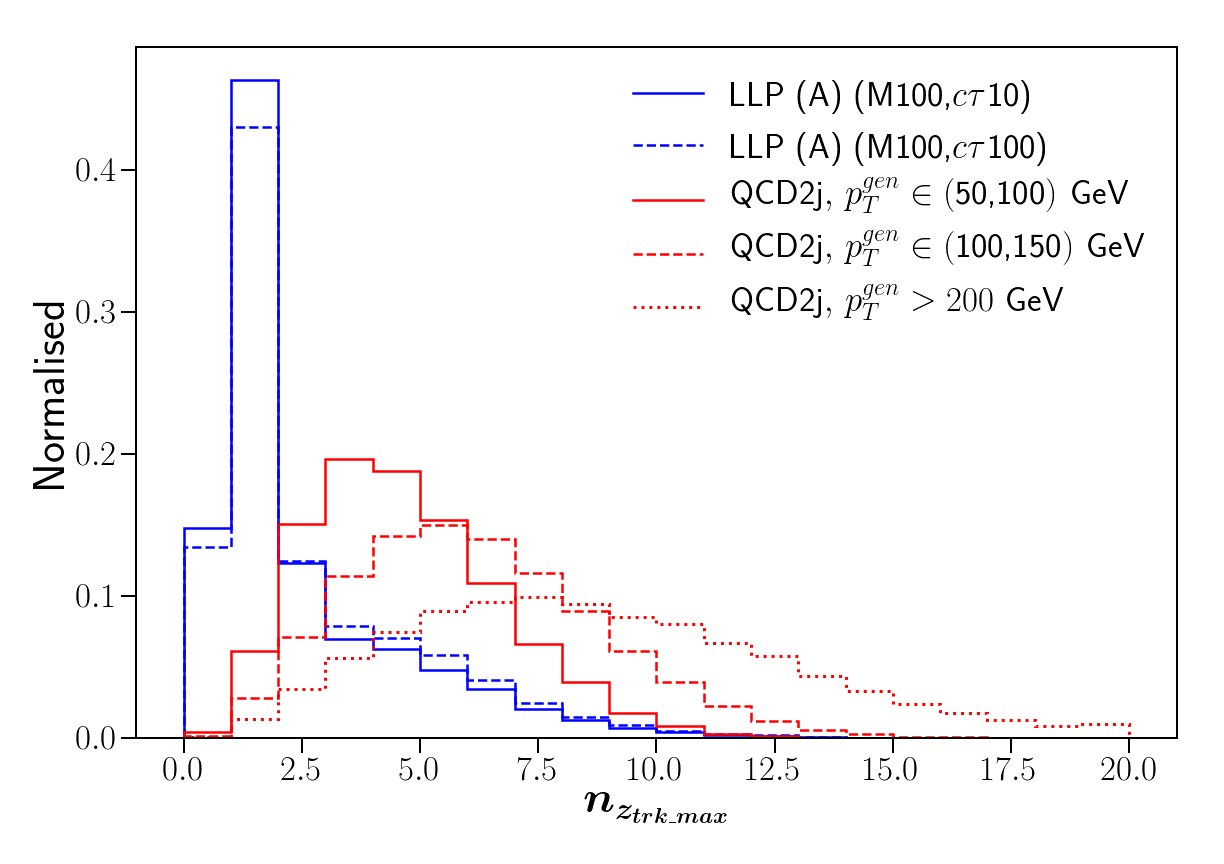}\\
\includegraphics[scale=0.175]{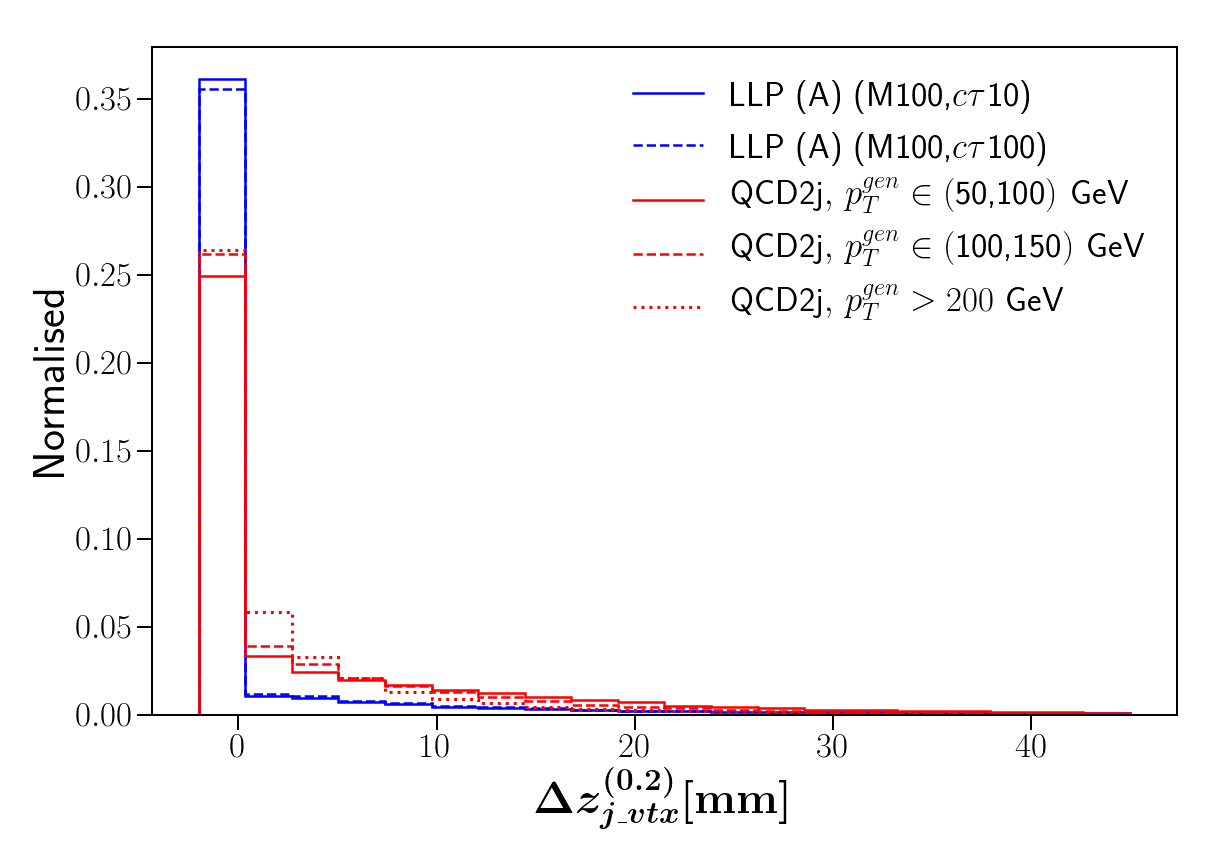}~
\includegraphics[scale=0.175]{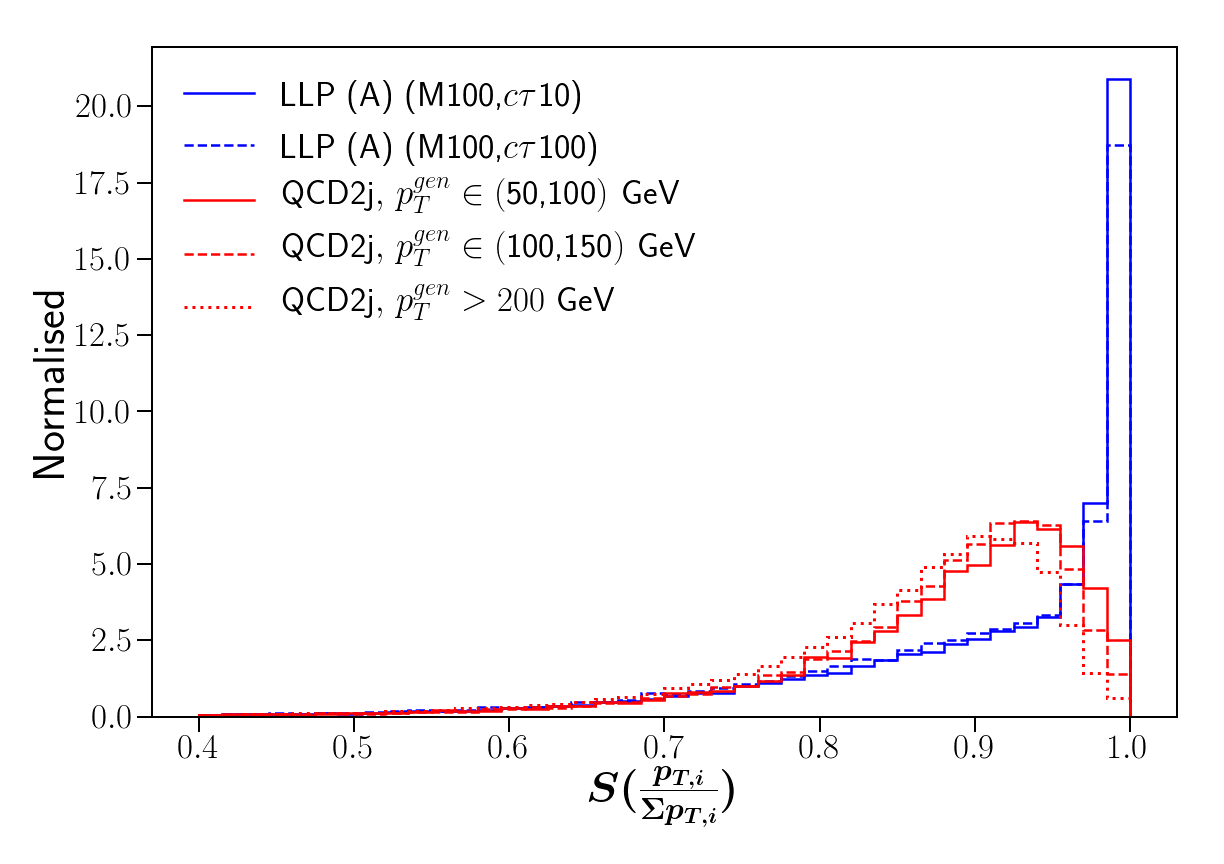}
\caption{Distributions for some of the track variables for two signal points from LLP scenario (A) and QCD dijet processes with different $p_T$ cuts at the parton level.}
\label{fig:track_vars}
\end{figure}

Fig.\ref{fig:track_vars} shows the distributions of some of the track variables for two signal benchmark points from scenario (A) and for the background QCD dijet processes for three different $p_T$ regions at the generation level. We have shown distributions of all other variables in the appendix \ref{app:track_vars}. The $N_{\text{trk}}^{(0.2)}$ variable is similar to the $N_{\text{trk}}$ variable, only with lesser number of tracks. The $p_{T\text{(vtx)}}^{miss}$ variable has a longer tail for the LLP jets. This is due to the fact that the LLP jets won't have much associated L1 tracks and they are not identified to come from the same $z$-vertex. Therefore, a vertex will have mostly a single jet associated to it, and there will be missing transverse energy at the vertex. The $p_T$ threshold for jets is $60{\rm~GeV}$ in our case and therefore, there is a peak around that value, since if a single jet is identified with a vertex and its $p_T$ is $60{\rm~GeV}$, the missing transverse momentum associated with that vertex will be $60{\rm~GeV}$ only. The entropy variable $S(p_{T,i}/\sum p_{T,i})$ peaks around 1 for the signal since the LLP jets have mostly PU tracks which have a wide distribution of $p_T$ values.

As we can see from fig.\ref{fig:track_vars}, no single variable is very powerful in discriminating between the signal and the background, however, their correlations do have some differences, as we can see in appendix \ref{app:corr_matrix}.
We, therefore, train a BDT classifier using the \texttt{TMVA}\cite{hoecker2007tmva} framework of \texttt{ROOT}\cite{Brun:1997pa} with all these variables along with the $p_T$ and $\eta$ of the jets for the LLP benchmark points from scenario (A) as signals and the background being a merged sample of QCD dijet processes with $p_T$ between 50-100 GeV, 100-150 GeV, 150-200 GeV and greater than 200 GeV at the generation level, with appropriate weights according to their cross-sections respectively. Out of all the variables that we use for training, following is the list of the top five variables, which are most efficient in discriminating between the signal and the background:

$$p_{T\text{(vtx)}}^{\text{miss}},\, n_{z_{\text{trk\_max}}},\, N_{\text{trk}},\, \sum p_T^{(0.2)},\, \Delta z_{\text{j\_vtx}}^{(0.2)}$$ 

\begin{figure}[hbt!]
\centering
\includegraphics[scale=0.2]{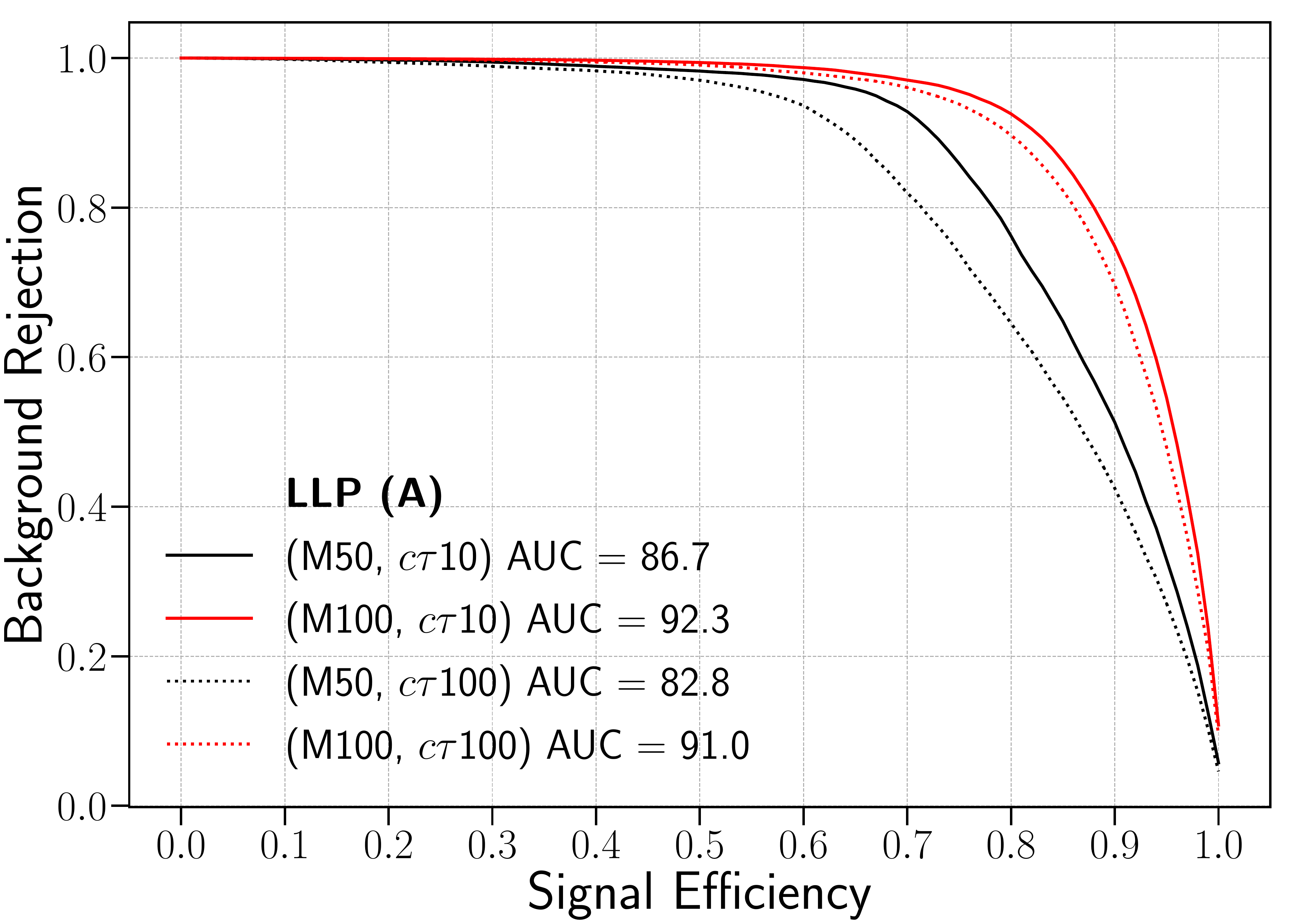}\\\vspace{0.5cm}
\includegraphics[scale=0.19]{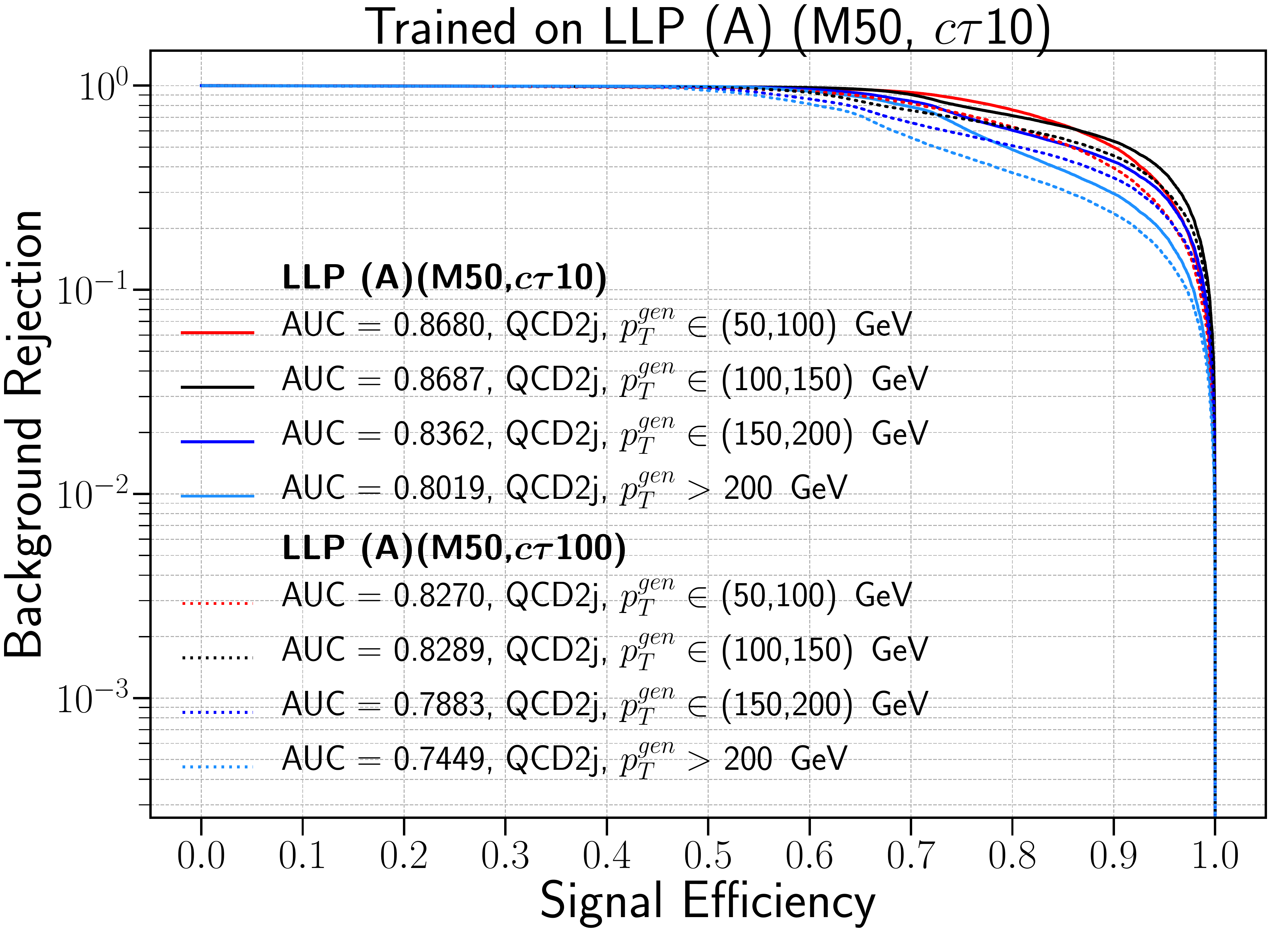}~
\includegraphics[scale=0.19]{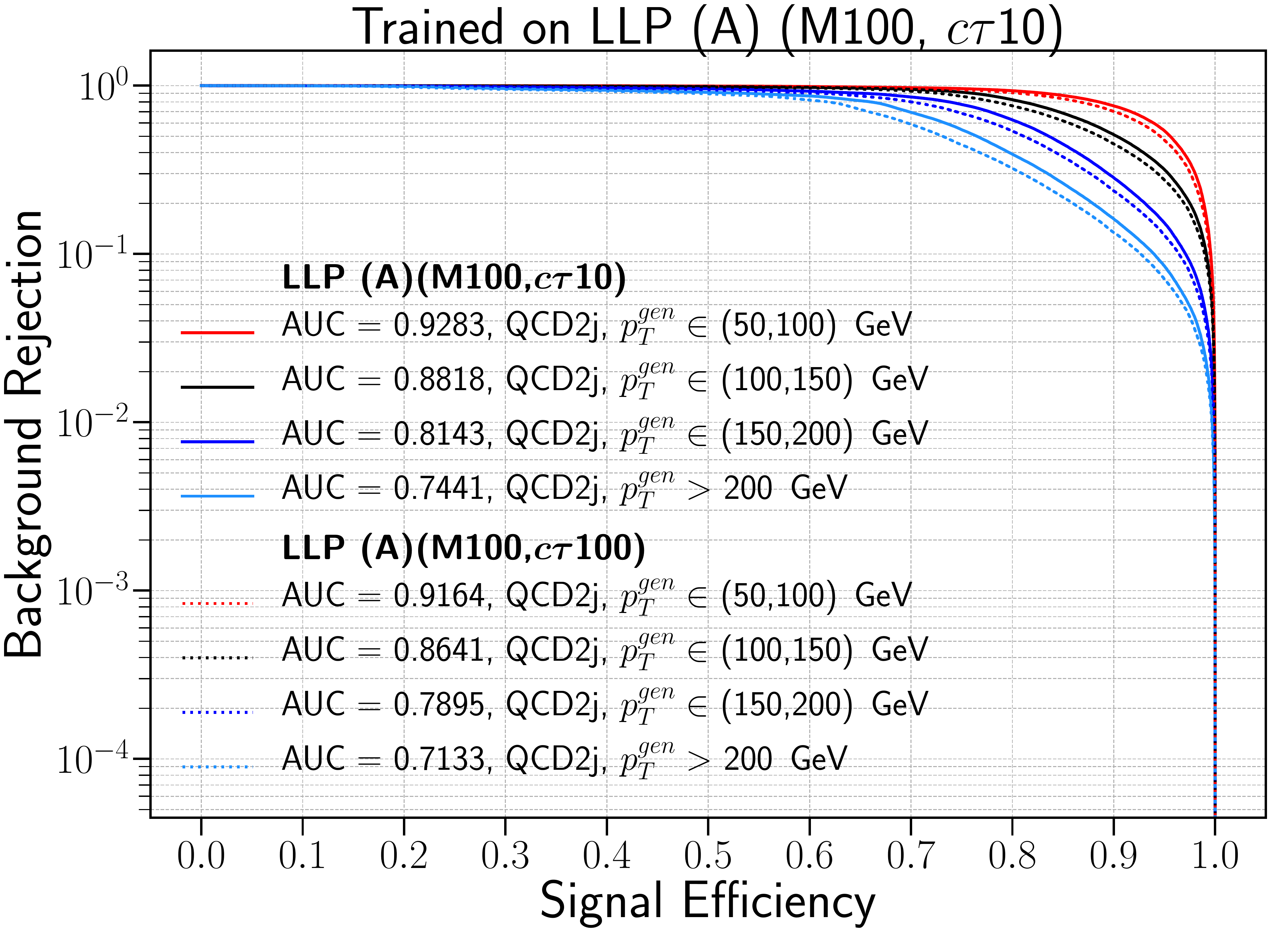}
\caption{ROC curves for selecting LLP jets for two points from BP1 $-$ $M_X=100{\rm~GeV}$, $c\tau=10{\rm~cm}$ (M100,DL10) and $M_X=50{\rm~GeV}$, $c\tau=10{\rm~cm}$ (M50,DL10) from QCD dijet samples with different $p_T$ cuts at the parton level.}
\label{fig:ROC_track}
\end{figure}

The {\it top panel} of fig.\ref{fig:ROC_track} shows the ROC curves for classifying the signal and background using tracking variables. We now train the BDT on a particular mass value and decay length and apply it on that benchmark point as well as the benchmark point with the same mass and a different decay length. 
The {\it bottom} plots of fig.\ref{fig:ROC_track} show the application ROC curves for the benchmark points from scenario (A) when trained on (M50,$c\tau$10) ({\it left}) and (M100,$c\tau$10) ({\it right}) and applied on (M50,$c\tau$100) and (M100,$c\tau$100) respectively. The application ROCs are shown for jets from each of the $p_T^{gen}$ bins of the background QCD dijet events. 

\begin{figure}[hbt!]
\centering
\includegraphics[scale=0.4]{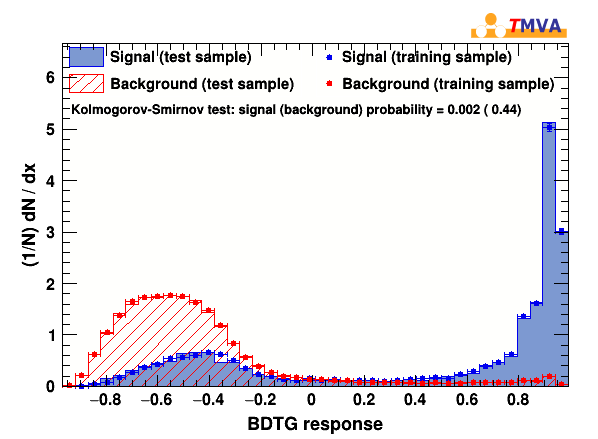}
\caption{Distributions of the BDT scores of signal jets coming from LLP benchmark point with mass 50 GeV and decay length 10 cm from scenario (A) and background jets from merged dijet process with different $p_T^{gen}$ bins.}
\label{fig:BDTscore_gt50_50-10}
\end{figure}

Even after considering $R=0.2$ jets, we have not been able to eliminate the PU contribution completely. There might be jets with $p_T>60{\rm~GeV}$ which come from the PU in the LLP events. These jets are similar to prompt QCD jets, and we therefore expect that the BDT classifier won't be able to discriminate between these jets.
In fig.\ref{fig:BDTscore_gt50_50-10}, we show the BDT score distributions for LLP benchmark point with mass 50 GeV and decay length 10 cm from scenario (A) being the signal and the merged sample of dijet processes with different $p_T^{gen}$ cuts as the background. 
We find that a small part of the signal BDT score distribution is similar to that of the background which are the remnant PU jets in the signal sample, and therefore, have similar BDT scores as the QCD jets background.

\subsubsection{Triggers based on the BDT training using variables from tracking}

We now discuss some triggers using the above BDT classification and quote the signal efficiency and background rates for each of them.
In the Phase-II upgrade, the L1 trigger hardware will have FPGAs which will be able to handle small scale machine learning (ML) applications, like BDT classification \cite{Duarte:2018ite,Summers:2020xiy}.
We choose three points from the ROC curves, corresponding to $98\%$, $90\%$ and $70\%$ background rejections. Following are the different trigger cuts we apply for selecting the signal LLP events from the background QCD prompt jets:

\begin{itemize}
\item $\mathbf{T_1}$: at least one $R=0.2$ jet with $p_T>60{\rm~GeV}$;
\item $\mathbf{T_2^{a}}$: $T_1$ + that jet passes the BDT threshold corresponding to a background rejection of $98\%$ ($a=0$), $90\%$ ($a=1$) and $70\%$ ($a=2$);
\item $\mathbf{T_3^{a}}$: $T_1$ + no other jet from the same $z$-vertex (i.e., $\Delta z$ with all other jets is greater than 1 cm) + $T_3^{a}$;
\item $\mathbf{T_{4b}^{a}}$:  at least one jet with $p_T\in(60,100)$ ($b=1$)/ $p_T\in(60,120)$ ($b=2$) + no other jet from the same $z$-vertex (i.e., $\Delta z$ with all other jets is greater than 1 cm) + $T_3^{a}$;
\end{itemize}

$T_1$ is the efficiency of getting events with at least one jet with $p_T$ greater than 60 GeV. Since the LLP jets are narrower than QCD jets (due to late decays), their $p_T$ distribution is not affected much when taking $R=0.2$ jets. However, the low $p_T$ QCD jets will be affected as we have discussed earlier. We then choose a point from the ROC curve corresponding to a particular background rejection, and use the BDT threshold corresponding to that point as a cut on the jet's BDT score. If an event has at least one jet satisfying the $p_T$ condition and having a BDT score above the threshold, we select the event as a signal and this is our $T_2^{a}$ trigger. For $T_3^{a}$ trigger, we have an added condition that our trigger jet (passing $T_1$ and $T_2^{a}$) shall not have any other jet coming from a $z$-vertex which is within 1 cm of the trigger jet's $z$-vertex position. We expect this condition to affect the QCD jets more since they usually have a same $z$-vertex unlike the LLP jets which mostly won't have much reconstructed L1 tracks, and hence mostly not come from the same $z$-vertex (as we already saw in the standard same vertex multijet trigger efficiencies quoted for zero PU in table \ref{tab:std_eff_llp}). Finally, we define another trigger $T_{4b}^{a}$ which demands the transverse momentum of jets to be within particular $p_T$ ranges. As seen from fig. \ref{fig:ROC_track}, the ROC performance degrades when QCD jets with higher $p_T^{gen}$ bins are used, and therefore, putting an upper limit on the jet $p_T$ can reduce background from higher $p_T^{gen}$ bins. 

An efficient LLP trigger design must be such that it has high signal efficiency and the rates are comparable to that of other triggers, ensuring that the total rate does not exceed the allowed trigger bandwidth. We now discuss the calculation of trigger rates, which will be dominated by background processes due to its huge cross section.

The trigger rate (in Hz) from background is calculated as follows:

\begin{equation}
\mathcal{R}_B=\sigma {\rm~(nb)}\times\mathcal{L}{\rm~(nb^{-1}Hz)}\times\epsilon_B
\label{eq:rate}
\end{equation}
where $\sigma$ is the background cross section (in nb), $\mathcal{L}$ is the peak instantaneous luminosity (in ${\rm nb^{-1}Hz}$) and $\epsilon_B$ is the efficiency of the triggers to select a background event. For the HL-LHC, $\mathcal{L}=5.6\times 10^{34}{\rm~cm^{-2}s^{-1}} = 56{\rm~nb^{-1}Hz}$ for the 140 PU scenario. For example, the QCD dijet rate for $p_T^{gen}\in(50,100){\rm~GeV}$ is:
\begin{equation}
\mathcal{R}_B=1.868\times10^{4}{\rm~nb}\times56{\rm~nb^{-1}Hz}\sim1046{\rm~kHz}
\end{equation}
We use the leading order \texttt{PYTHIA} cross sections for calculating the rates. The higher order corrections can contribute to another factor of $\sim2$ \footnote{The other single jet backgrounds can be $Z$+jets or $t\bar{t}$, where the cross sections are few thousand pb, which is very small compared to the dijet cross sections for low $p_T^{gen}$ bins.}. We have to multiply this huge rate by $\epsilon_B$, which is the trigger efficiencies for background events, and which will quantify how much of this rate can we reject using our triggers.

\begin{table}[hbt!]
\centering
\resizebox{\textwidth}{!}{ 
\begin{tabular}{|c|c||c|c|c|c|c||}
\hline
\multirow{3}{*}{LLP (A)} & QCD2j & $T_1$ & $T_2^0$ & $T_3^0$ & $T_{41}^0$ & $T_{42}^0$\\
 & $p_T^{gen}$ [GeV] & $\mathcal{R}_B$ [kHz] & $\mathcal{R}_B$ [kHz] & $\mathcal{R}_B$ [kHz] & $\mathcal{R}_B$ [kHz] & $\mathcal{R}_B$ [kHz]\\
 & ($\mathcal{R}_B$ [kHz]) & $(\epsilon_S~[\%])$  & $(\epsilon_S~[\%])$ & $(\epsilon_S~[\%])$ & $(\epsilon_S~[\%])$ & $(\epsilon_S~[\%])$\\
\hline\hline
&50,100 (1046) & 301.5(23.43) & 7.2(13.29) & 7(13.18)& 6.4(10.68)& 6.7(11.91) \\
$M=50{\rm~GeV}$ & 100,150 (53.4) & 46.4(23.43) & 1.5(14.84) & 1.3(14.64)  & 0.7(12.21) & 0.9(13.39)  \\
$c\tau=10{\rm~cm}$ & 150,200 (7.5) & 7.3(23.43) & 0.3(14.15)  & 0.2(14.01) & 0.06(11.54)  & 0.08(12.75)  \\
& $>$200 (2.7) & 2.7(23.43) & 0.1(13.97) & 0.08(13.84) & 0.02(11.37) & 0.02(12.58)	\\
\hline\hline
&50,100 (1046) & 301.5(18.34) & 7.2(8.55) & 7(8.48) & 6.4(6.69) & 6.7(7.47) \\
$M=50{\rm~GeV}$ & 100,150 (53.4) & 46.4(18.34) & 1.5(9.92) & 1.3(9.79) & 0.7(7.90) & 0.9(8.71)  \\
$c\tau=100{\rm~cm}$ & 150,200 (7.5) & 7.3(18.34) & 0.3(9.33) & 0.2(9.23) & 0.06(7.39) & 0.08(8.19)  \\
& $>$200 (2.7) & 2.7(18.34) & 0.1(9.15)	& 0.08(9.06) & 0.02(7.23) & 0.02(8.03)	\\
\hline\hline
&50,100 (1046) & 301.5(82.38) & 7(61.80) & 6.7(60.20) & 5.4(37.15) & 5.9(46.59) \\
$M=100{\rm~GeV}$ & 100,150 (53.4) & 46.4(82.38) & 1.4(53.89) & 1.2(52.64) & 0.3(28.14) & 0.5(38.12) \\
$c\tau=10{\rm~cm}$ & 150,200 (7.5) & 7.3(82.38) & 0.3(35.40) & 0.2(34.66) & 0.0(7.41) & 0.01(16.71)    \\
& $>$200 (2.7) & 2.7(82.38)	& 0.1(25.46) & 0.08(25.05)	& 0.0(1.95) & 0.0(6.25)	\\
\hline\hline
&50,100 (1046) & 301.5(68.84) & 7(48.32) & 6.7(46.40) & 5.4(26.31) & 5.9(33.14) \\
$M=100{\rm~GeV}$ & 100,150 (53.4) & 46.4(68.84) & 1.4(41.10) & 1.2(39.54) & 0.3(19.11) & 0.5(26.15) \\
$c\tau=100{\rm~cm}$ & 150,200 (7.5) & 7.3(68.84) & 0.3(25.54) & 0.2(24.67) & 0.0(4.66) & 0.01(10.71)    \\
& $>$200 (2.7) & 2.7(68.84) & 0.1(17.91) & 0.08(17.43) & 0.0(1.36) & 0.0(4.07)	\\
\hline\hline
\end{tabular}
}
\caption{Efficiency of selecting QCD and LLP events for LLP scenario (A) benchmark points with the modified trackless trigger at L1 for Phase-II. BDT cut is applied considering 98\% background rejection. Quantity in parenthesis represents corresponding signal efficiency.}
\label{tab:mod0_eff_llp}
\end{table}

\begin{table}[hbt!]
\centering
\resizebox{\textwidth}{!}{ 
\begin{tabular}{|c|c||c|c|c|c||}
\hline
\multirow{3}{*}{LLP (A)} & QCD2j & $T_2^2$ & $T_3^2$ & $T_{41}^2$ & $T_{42}^2$\\
 & $p_T^{gen}$ [GeV] & $\mathcal{R}_B$ [kHz] & $\mathcal{R}_B$ [kHz] & $\mathcal{R}_B$ [kHz] & $\mathcal{R}_B$ [kHz]\\
 & ($\mathcal{R}_B$ [kHz]) & $(\epsilon_S~[\%])$  & $(\epsilon_S~[\%])$ & $(\epsilon_S~[\%])$ & $(\epsilon_S~[\%])$\\
\hline\hline
& 50,100 (1046) & 103.2(19.79) & 95(19.28) & 86.1(16.77) & 93.1(18.11)    \\
$M=50{\rm~GeV}$ & 100,150 (53.4) & 19.2(19.36) & 13.4(18.87) & 5.7(16.34) & 10.6(17.69)  \\
$c\tau=10{\rm~cm}$ & 150,200 (7.5) & 3.3(18.06) & 1.6(17.67) & 0.2(15.08) & 0.6(16.46) \\
& $>$200 (2.7) & 1.2(17.58) & 0.4(17.23) & 0.05(14.61) & 0.08(16.01)	\\
\hline\hline
& 50,100 (1046) & 103.2(14.48) & 95(14.03) & 86.1(11.63) & 93.1(12.78)    \\
$M=50{\rm~GeV}$ & 100,150 (53.4) & 19.2(14.08) & 13.4(13.65) & 5.7(11.24) & 10.6(12.40)  \\
$c\tau=100{\rm~cm}$ & 150,200 (7.5) & 3.3(12.94) & 1.6(12.58) & 0.2(10.15) & 0.6(11.31) \\
& $>$200 (2.7) & 1.2(12.50) & 0.4(12.17) & 0.05(9.76)	& 0.08(10.89)	\\
\hline\hline
& 50,100 (1046) & 100.5(77.73) & 87(72.90) & 77.6(52.11) & 85.2(61.72)    \\
$M=100{\rm~GeV}$ & 100,150 (53.4) & 19.4(73.28) & 11.5(69.56) & 3.7(47.56) & 8.6(57.81)  \\
$c\tau=10{\rm~cm}$ & 150,200 (7.5) & 3.6(69.24) & 1.3(66.49) & 0.1(44.18) & 0.2(53.94) \\
& $>$200 (2.7) & 1.4(64.27) & 0.3(62.41) & 0.02(39.85) & 0.03(49.09)	\\
\hline\hline
& 50,100 (1046) & 100.5(64.02) & 87(59.53) & 77.6(39.17) & 85.2(47.27)    \\
$M=100{\rm~GeV}$ & 100,150 (53.4) & 19.4(59.83) & 11.5(56.16) & 3.7(35.01) & 8.6(43.47)  \\
$c\tau=100{\rm~cm}$ & 150,200 (7.5) & 3.6(55.60) & 1.3(52.72) & 0.1(31.97) & 0.2(39.52) \\
& $>$200 (2.7) & 1.4(50.09) & 0.3(48.53) & 0.02(28.49) & 0.03(35.28)	\\
\hline\hline
\end{tabular}
}
\caption{Efficiency of selecting QCD and LLP events for LLP scenario (A) benchmark points with the modified trackless trigger at L1 for Phase-II. BDT cut is applied considering 70\% background rejection. Quantity in parenthesis represents corresponding signal efficiency.}
\label{tab:mod2_eff_llp}
\end{table}

Tables \ref{tab:mod0_eff_llp} and \ref{tab:mod2_eff_llp} show the trigger efficiencies for different LLP signal benchmark points from scenario (A) and the background rates for different $p_T^{gen}$ bins for a background rejection of 98\% and 70\% (in each of the $p_T^{gen}$ bins) respectively. A similar table for a background rejection of 90\% has been given in appendix \ref{app:trigger_eff}. We find that the trigger rates for $p_T^{gen}\in(50,100){\rm~GeV}$ can be reduced to about 5.4-6.4 kHz ($T_{41}^0$ column of table \ref{tab:mod0_eff_llp}) from 1046 kHz for a signal efficiency of $\sim11\%$ for (M50,$c\tau$10) and $\sim37\%$ for (M100,$c\tau$10). The actual rates will be twice if we consider higher orders, and therefore, the final background rates reduce to about 10-12 kHz.

For 70\% background rejection, the background rates are really high ($\sim 2\times86 {\rm~kHz}$), and the signal efficiencies increase by few percentage. So if we are limited by rates, choosing a 98\% or 90\% background rejection point from the ROC will be more useful, because they have low background rates and there is little compromise on the signal efficiencies. We will again discuss the 70\% background rejection point when discussing the timing variables, to see whether timing can give extra background rejection.

The BDT training, here, was done using a merged background sample of QCD dijet events with different $p_T^{gen}$ cuts weighted according to their cross-section. 
The same could have been done using a categorical training with each $p_T$ bin as separate backgrounds. That will improve the performance of the application ROCs ({\it bottom panel} plots of fig.\ref{fig:ROC_track}) for the higher $p_T^{gen}$ bins, which are now poorer than the $p_T^{gen}\in(50,100){\rm~GeV}$ bin because it has the highest cross section and the highest weight in the training. We here present a conservative analysis, and the rates of the higher $p_T$ bins are very low to start with. Also, using the upper cut on $p_T$ values of jets also suppress the dijet background from higher $p_T^{gen}$ bins.

The total signal efficiency for the benchmark point (M50,$c\tau$10) is $\sim 13\%$ from the standard single jet, dijet and quad jet triggers (from table \ref{tab:std_eff_llp}). From the track-based triggers that we have defined in this section, we get at best a signal efficiency of $19.79\%$ for $70\%$ background rejection for the same benchmark point. This might not seem much of an improvement unless we are selecting different events from the standard triggers and the one defined by us. 

It will be, therefore, interesting to check the exclusiveness of the events selected by these dedicated LLP triggers, that we define here, and the ones selected by standard L1 single jet, dijet and quad jet triggers. The overlap between the events selected with standard triggers and the dedicated triggers defined in this work is just 5\% for the benchmark point (M50,$c\tau$10) in scenario (A). Therefore, they are mostly selecting different events, and that is because, the standard triggers are mostly triggered by pile-up jets, and our defined triggers are mostly triggered by jets from LLPs. Therefore, if we add this type of dedicated triggers, we can select an additional $\sim 14\%$ of signal events even for the LLP of mass 50 GeV and decay length 10 cm, making the total signal efficiency close to $\sim 27\%$.

Also, for lower mass LLPs, the signal efficiency is mostly affected due to the first cut only, $T_1$. It would be important to lower the $p_T$ threshold to increase sensitivity for such lighter LLPs. However, decreasing the $p_T$ threshold will increase the PU contamination drastically, even when we consider narrow jets of $R=0.2$. In that case, we will need much more sophisticated techniques to deal with PU at lower $p_T$ thresholds, which is outside the scope of the present work.

\subsection{Trigger using the MIP Timing Detector}
\label{ssec:time_L1}

As discussed earlier, the HL-LHC CMS upgrade has a scheme for a MIP Timing Detector, abbreviated as MTD.  
The proposed design of the MTD is such that it will be capable of measuring timing of all electrically charged particles which have transverse momentum more than
$0.7{\rm~GeV}$ in barrel ($|\eta|<1.5$) and momentum more than 
$0.7{\rm~GeV}$ in endcap ($1.5<|\eta|<3$) with timing precision of around $30{\rm~ps}$. 
This timing layer will be positioned at 
$1.161$ meters away from the beam pipe of CMS: in the small gap between the tracker and the ECAL with a half-length of $2.6{\rm~m}$ \cite{MTD:TDR}. 

One of the main purpose for adding the MTD is to reduce the huge amount of pileup (PU) by using 4-dimensional vertex reconstruction. However, for this 4-D vertex reconstruction, one would need the full timing information which might not be available at the L1. However, regional timing information can be used at L1 as has been mentioned in the MTD TDR \cite{MTD:TDR}. Therefore, in this work we explore the prospects of availability of regional timing at L1. Before discussing that, let us discuss some aspects of timing in general.

If $T_{arrival}$ is the time of arrival of the particle at the outer tracker as measured by the MTD and $T_{light}$ is the time of arrival of the particle if it were travelling at the speed of light in the same direction as the former one starting from $z=0$, we calculate the time delay of a particle following the formula given in Ref. \cite{Mason:2019okp}:
\begin{equation}
\Delta T = T_{arrival}- T_{light}
\label{eq:delT}
\end{equation}
where $T_{light}$ is calculated as: 

\begin{equation}
T_{light} = \frac{R}{c~{\rm sin}(\theta)}
\label{eq:tlight}
\end{equation}
where $\theta$ is the angle made by the particle from the positive $z$-direction ($\theta \in (0,\pi)$) and $R$ is the radial distance where the particle hits the MTD.


For stable charged particles starting from $z=0$, the time delay $\Delta T$ (Eq. \ref{eq:delT}), without any magnetic field and time smearing, is,

\begin{equation}
\Delta T = \frac{R}{c {\rm~sin}(\theta)}\left(\frac{E}{p}\right) - \frac{R}{c {\rm~ sin}(\theta)} = \frac{R}{c}\left(\frac{\sqrt{p_T^2/{\rm sin}^2(\theta)+m^2}}{p_T}-\frac{1}{{\rm sin}(\theta)}\right)
\label{eq:delT_theta}
\end{equation}

$\Delta T$ is maximum for $\theta = \pi/2$ and decreases with $\theta$ on either sides of $\pi/2$ for fixed $p_T$. Pseudorapidity, $\eta$ is related to $\theta$ \footnote{$\eta = -{\rm ln}\left({\rm tan}\frac{\theta}{2}\right)$, $\eta \in (-\infty, \infty)$}~ such that $\eta = 0.0$ for $\theta=\pi/2$. We, therefore, expect $\Delta T$ to be maximum in the central region of the detector and falling off with increasing $|\eta|~$ up to $|\eta|=1.5$ where the threshold is determined by $p_T$ \footnote{Note that $T_{arr}$ and $T_{light}$, both, increase with increasing $|\eta|$ as is expected. The difference of the two shows the opposite trend.}~. Also, with increasing $p_T$, the delay shall decrease. The minimum $p_T$ threshold of the MTD, $p_T=0.7{\rm~GeV}$, therefore, gives the maximum time delay for each particle if it starts from $z=0$ in the barrel region. 

However, in the endcap region, the MTD threshold depends on the absolute momentum of the charged particles instead of their transverse momentum. Therefore, their transverse momentum can still be very small and since they hit the MTD at higher $|\eta|$ values, they will take much longer time.

\begin{figure}[hbt!]
\centering
\includegraphics[width=0.5\textwidth]{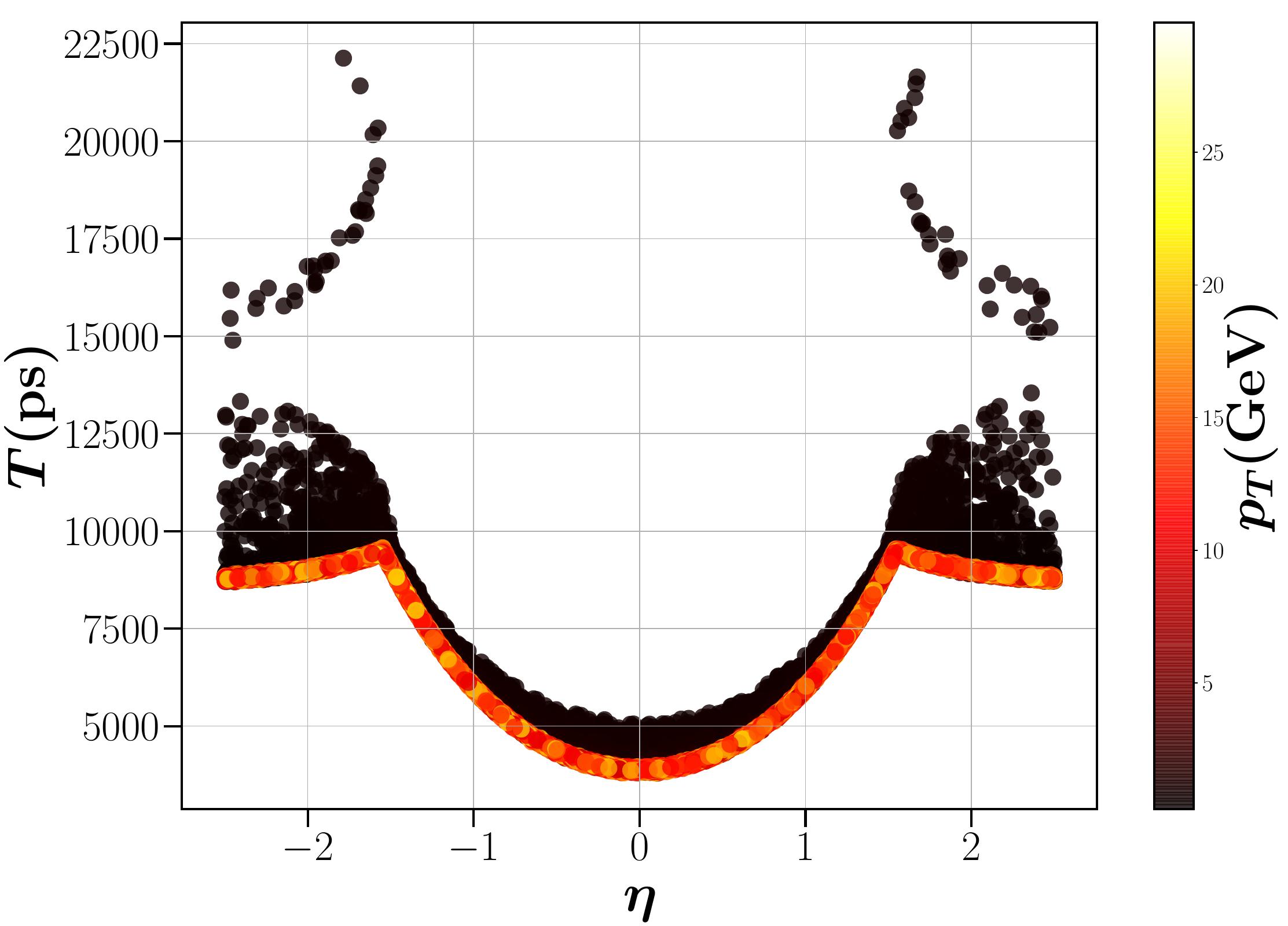}~
\includegraphics[width=0.5\textwidth]{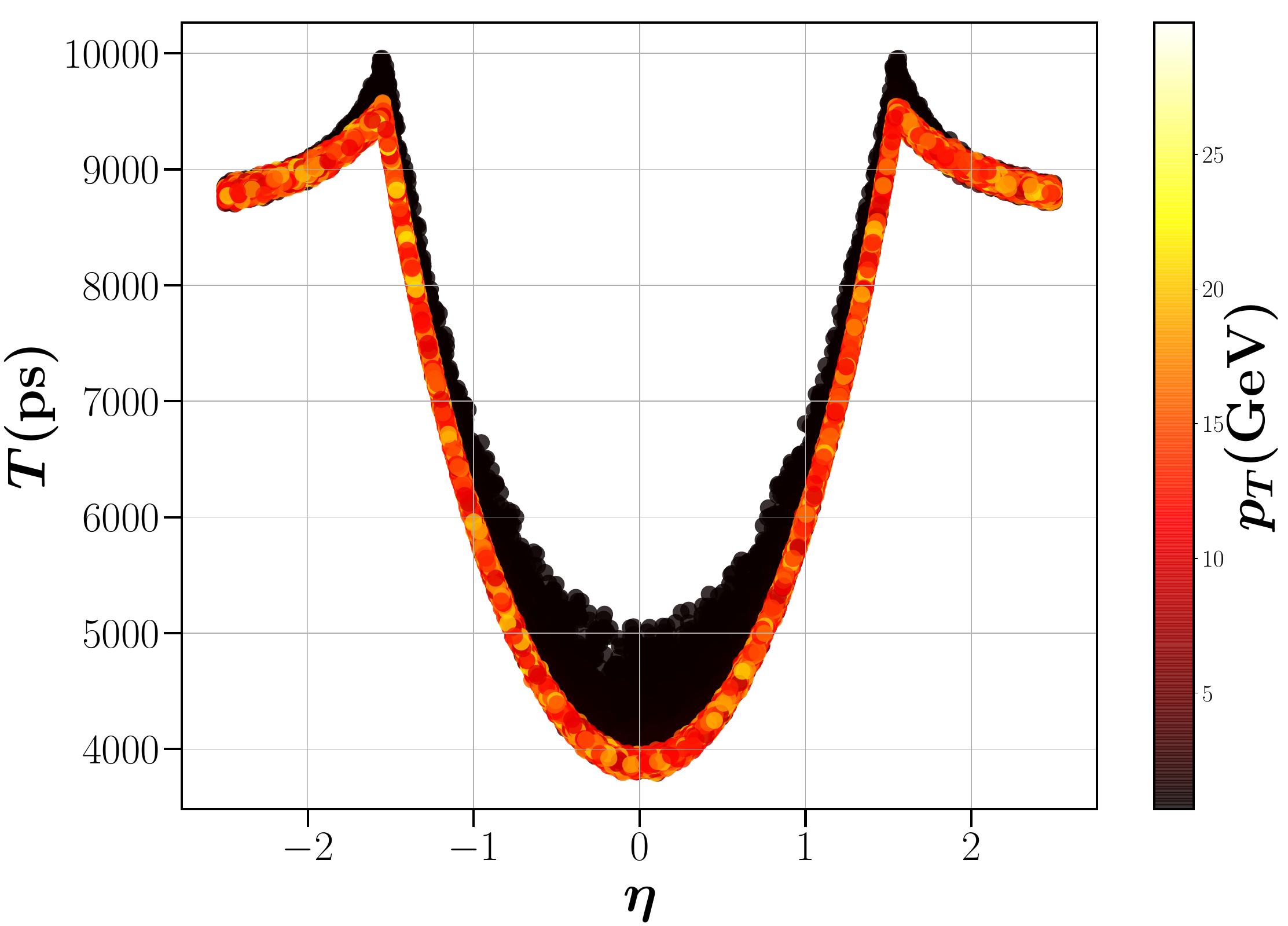}\\
\includegraphics[width=0.5\textwidth]{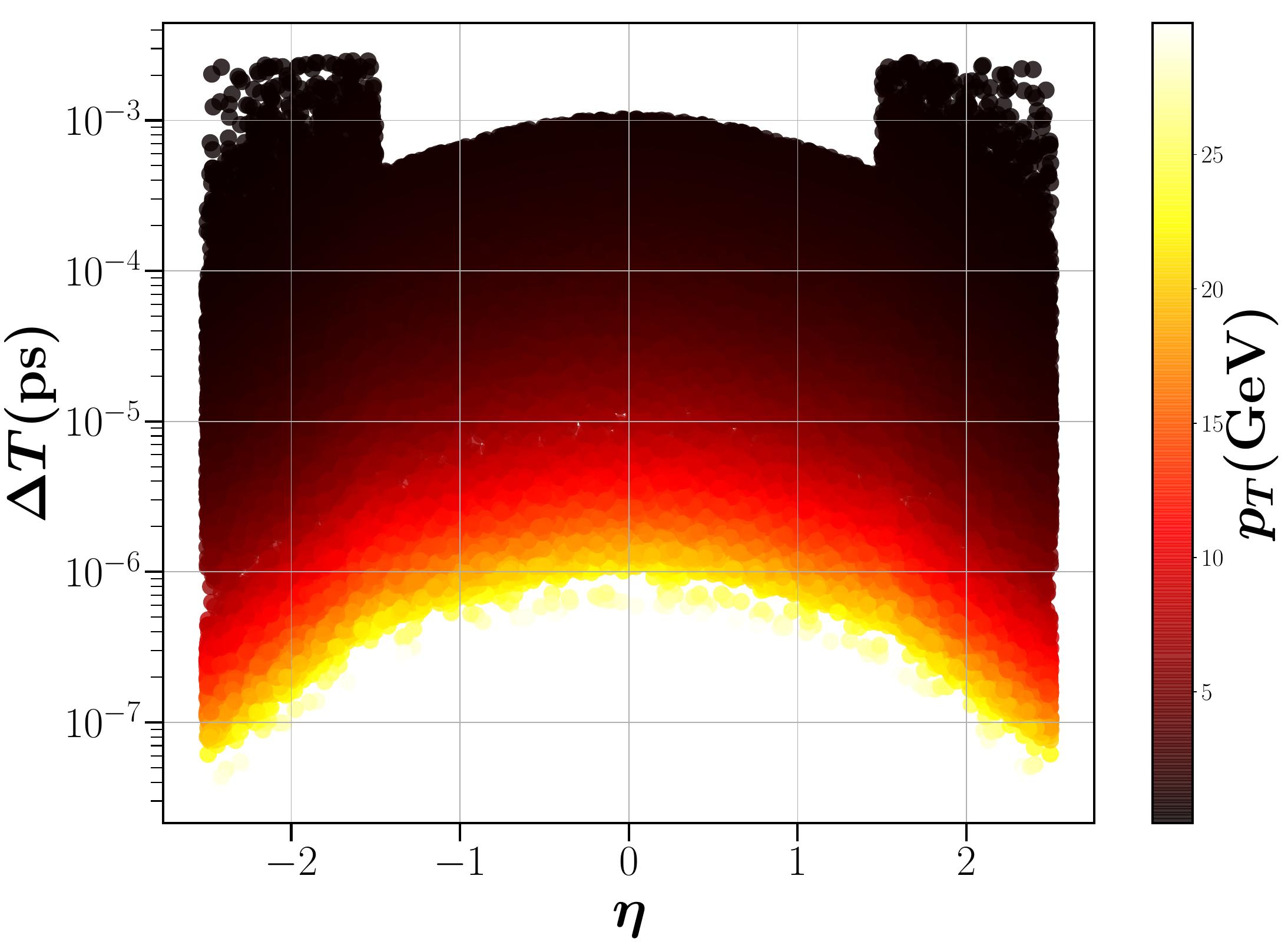}~
\includegraphics[width=0.5\textwidth]{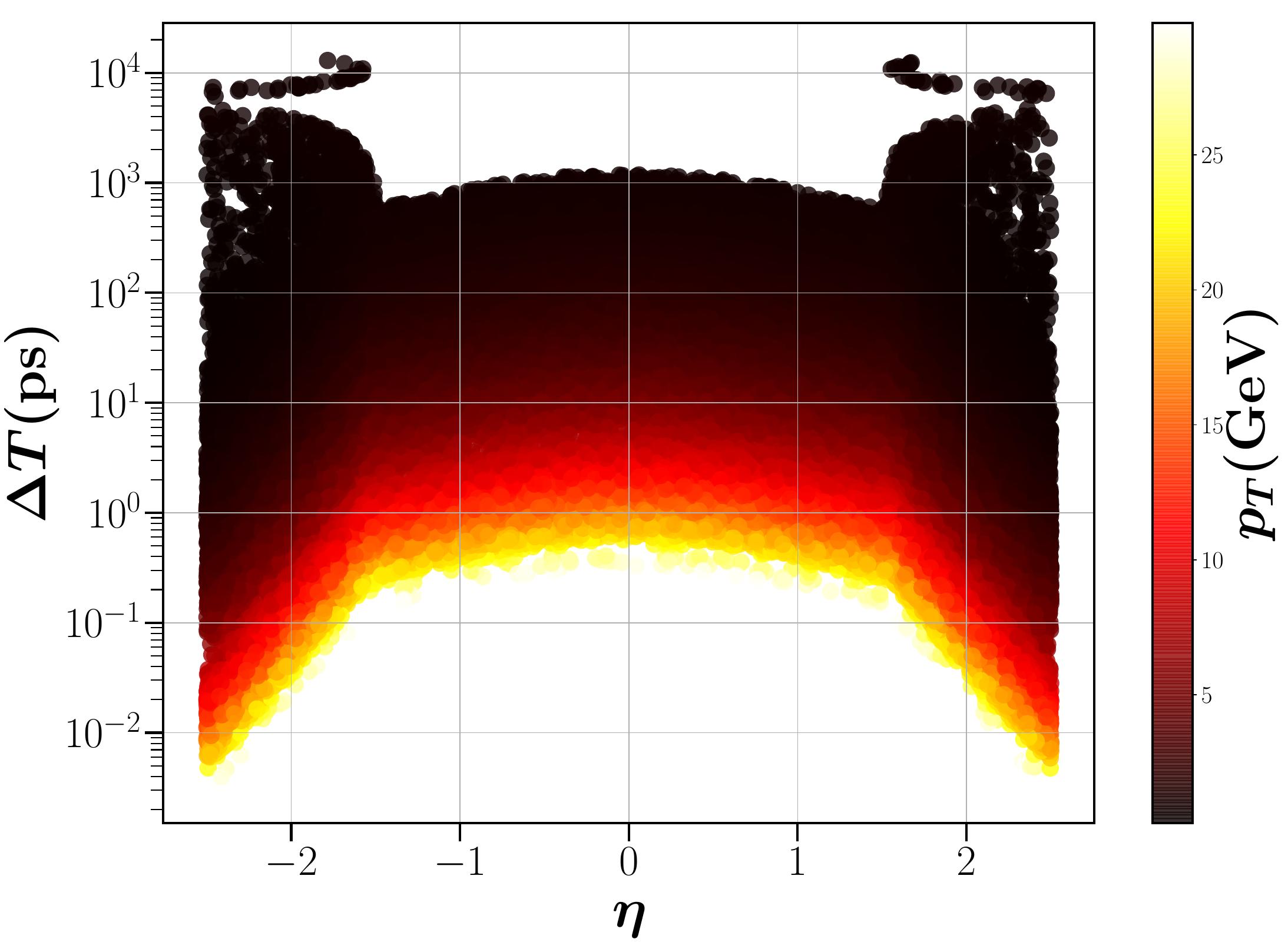}\\
\includegraphics[width=0.5\textwidth]{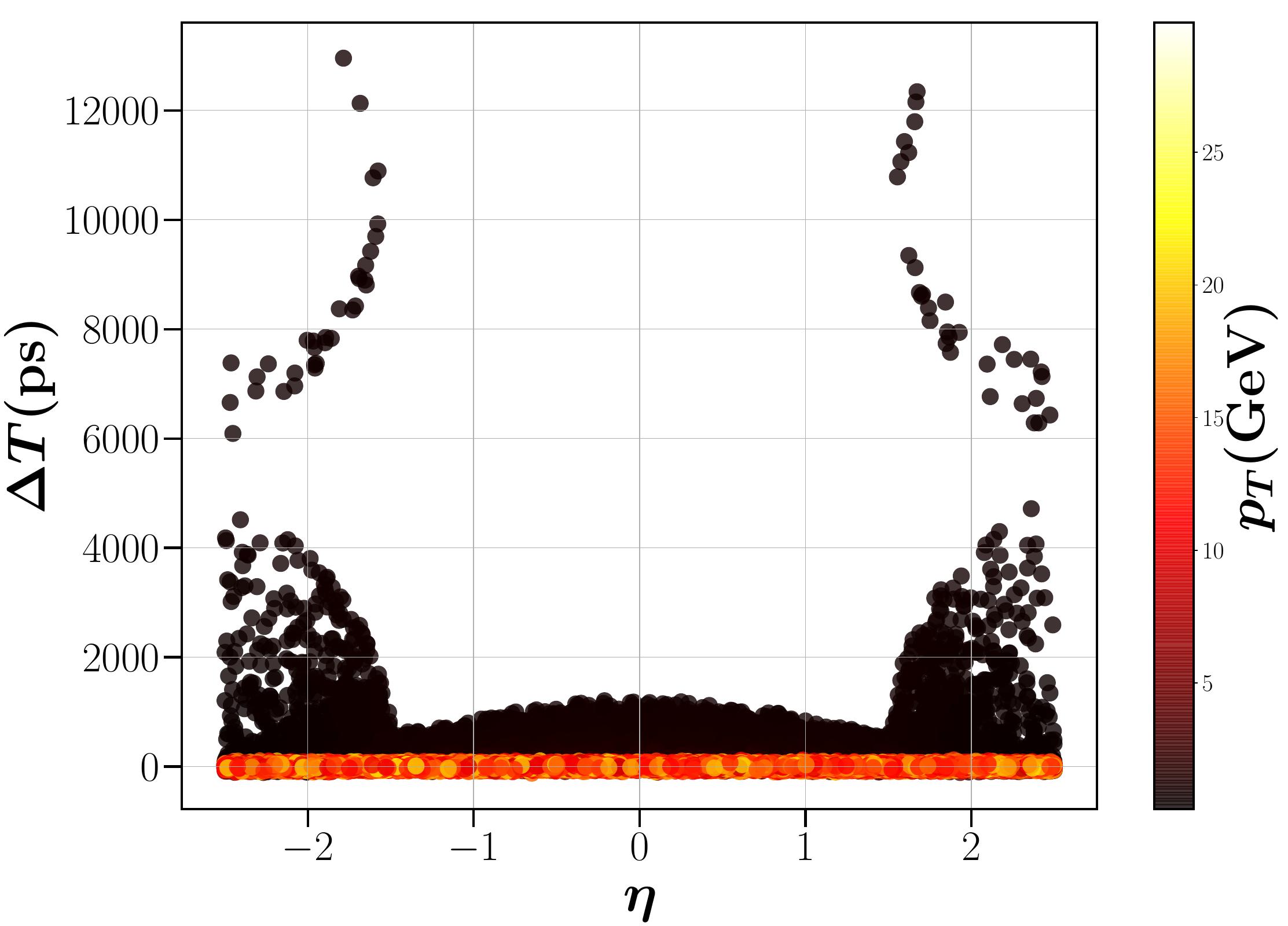}
\caption{$T$ as a function of $\eta$ and $p_T$ of electrons with both magnetic field and $30{\rm~ps}$ time smearing for all particles (\textit{top left}) and for particles with $p_T>0.7{\rm~GeV}$ (\textit{top right}); $\Delta T$ as a function of $\eta$ and $p_T$ of electrons with no magnetic field and smearing of time (\textit{centre left}), magnetic field but no smearing (\textit{centre right}), both magnetic field and $30{\rm~ps}$ time smearing (\textit{bottom}).
}
\label{fig:delT_basic}
\end{figure}

The time delay of charged particles will be affected by the magnetic field which bends the charged particle trajectory according to the $p_T$ and mass of the particle. We again use the \texttt{ParticlePropagator} code of \texttt{Delphes} to propagate the particles in a magnetic field of $3.8 {\rm~T}$ along the $z$-direction (the beam axis) which corresponds to the value of $B_z$ for the CMS detector. A charged particle will follow a helical trajectory inside such a magnetic field and we need to find out the (minimum) time when this helix crosses the MTD.
We briefly describe how time of a charged particle hitting the MTD is found out when the particle travels inside a magnetic field in appendix \ref{app:magnetic}~\footnote{We found a possible bug in \texttt{ParticlePropagator} code of \texttt{Delphes} when calculating the minimum time when the particle hits the MTD which we discuss in appendix \ref{app:magnetic}~.}~.
This discussion is presented to have a documentation of how the magnetic field affects the propagation of a charged particle.

In the magnetic field, if the charged particle has enough $p_T$ such that its radius of curvature is greater than the radial position of the MTD, the particle will hit the MTD radially. However, if the $p_T$ is really small, the particle moves in a helix of very small radius and hits the MTD at its half-length. In the latter case, the path length of the particle is very large, and therefore, it takes longer time to reach the MTD. Since in the endcap regions, we select particles with $p>0.7{\rm~GeV}$, they can have very small values of $p_T$, even $p_T<0.7{\rm~GeV}$ is possible and therefore, will have large time values recorded by the MTD. This can be seen by comparing the plots in the {\it top panel} of fig.\ref{fig:delT_basic}, where the former is for particles having all possible $p_T$ values and the latter is particles with $p_T>0.7{\rm~GeV}$.

As mentioned earlier, the MTD is proposed to have a time resolution of $30~\text{ps}$. 
This is the timing resolution needed to bring down the number of PU events in HL-LHC to the order of present LHC PU events. We therefore apply a gaussian smearing of the time that we get from \texttt{Delphes} with a standard deviation of $30~\text{ps}$. This will affect the higher values of time delays for particles with low $p_T$ very little, however, particles having higher $p_T$ values with lesser time delays will be affected by this smearing and we can also have negative delays as well due to smearing (mostly for high $p_T$ particles). 

Fig. \ref{fig:delT_basic} shows $T$ as a function of $\eta$ and $p_T$ of prompt electrons with both magnetic field and $30{\rm~ps}$ time smearing for all particles (\textit{top left}) and for particles with $p_T>0.7{\rm~GeV}$ (\textit{top right}); $\Delta T$ as a function of $\eta$ and $p_T$ of electrons with no magnetic field and smearing of time (\textit{centre left}), magnetic field but no smearing (\textit{centre right}), both magnetic field and $30{\rm~ps}$ time smearing (\textit{bottom}).
The time of flight increases with increasing $\eta$ up to $|\eta|=1.5$ due to increased distance to be travelled before reaching the MTD and decreasing $p_T$ due to decreased velocity. The distance to the MTD, given by $\sqrt{r^2+z^2}$ increases with $\eta$ only up to $|\eta|=1.5$ where $r$ is fixed and $z$ increases with increasing $\eta$. After that, the endcap region starts, where $z$ is constant (the half-length of the MTD) and $r$ decreases with increasing $\eta$. Therefore, we find a dip in the absolute timing value. We find that with increasing $\eta$ and increasing $p_T$, the time delay decreases in the barrel as we had discussed earlier. On switching magnetic field on, we find that the time delay increases significantly for electrons, which is expected due to the increase in its path length in the magnetic field. On adding smearing of time, the resolution is very small compared to large time delays, and therefore, its effect is minimal. However, for the high $p_T$ particles with time delays of the order of few ps, smearing affects the timing and we now will have negative time delays as well due to smearing.

Charged particles coming from LLPs decaying after the MTD can hit the MTD from the other side. \texttt{Delphes} doesn't show time for such particles. The proposed design for MTD consists of a LYSO crystal in the inner-side of the detector with a SiPM in the back side for readout, which will only give the time hit if the particle hits the MTD from the front at the LYSO crystal but not when it hits the same from back side at the SiPM. 
However, if the design of the MTD involves double-sided SiPM crystals, it can show hits for backward moving particles coming from LLP decays, as discussed in \cite{Banerjee:2017hmw}, as well. That could be an interesting possibility and can have some exciting prospects for LLP studies, and would require minor modification of the present \texttt{ParticlePropagator} code. In this work we don't include time of backward moving particles.

\begin{figure}[hbt!]
\centering
\includegraphics[width=0.7\textwidth]{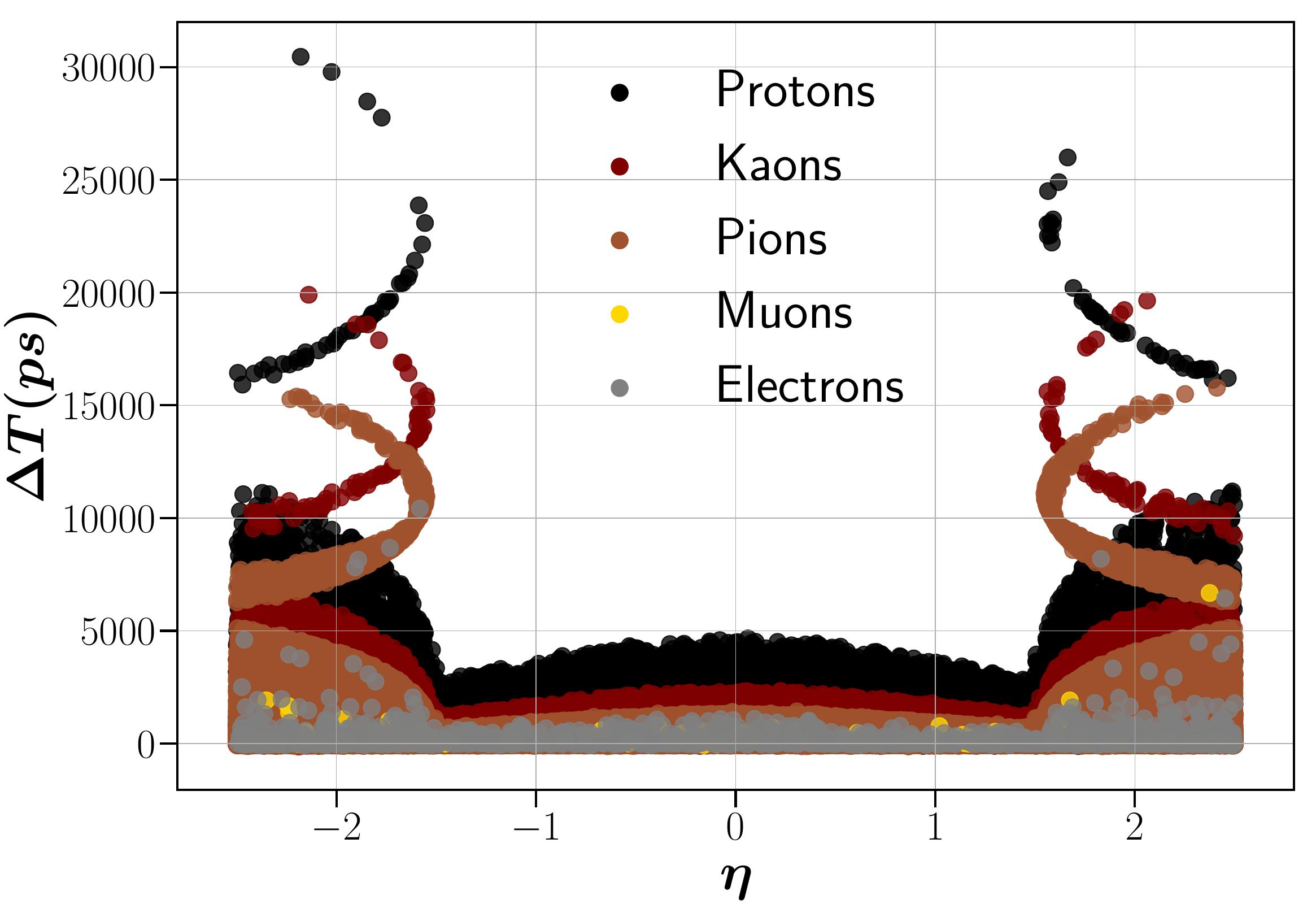}~
\caption{Timing of different particles within a quark jet with $p_T^{gen}\in(50,100){\rm~GeV}$.}
\label{fig:jets_q_g}
\end{figure}

Fig.\ref{fig:jets_q_g}~ shows the typical values of time difference ( w.r.t $z=0$) of various constituents of quark jet with $p_T^{gen}\in(50,100){\rm~GeV}$. We have electrons, muons, pions, kaons and protons as the major charged constituents of a jet. These constituents will all reach the MTD at different times depending primarily on their mass and momentum. 
Association of a timing to a jet is, therefore, a difficult task as jet consists of many components. Mostly the median of the timing of all the hits associated with a jet is used as the timing of a jet \cite{CMS:2016azs}. 

Therefore, timing of a jet can be defined mostly by a statistical measure of all the MTD hits associated with a jet. The statistical measures will be affected by hits on the MTD coming from pile-up which get associated with a jet. It is, therefore, important to study how the number of MTD hits associated with a jet gets affected in the 140 PU scenario and also, how these PU hits affect the timing of a jet.

\subsubsection{Timing of a jet and the effect of PU}

\begin{figure}[hbt!]
\includegraphics[scale=0.18]{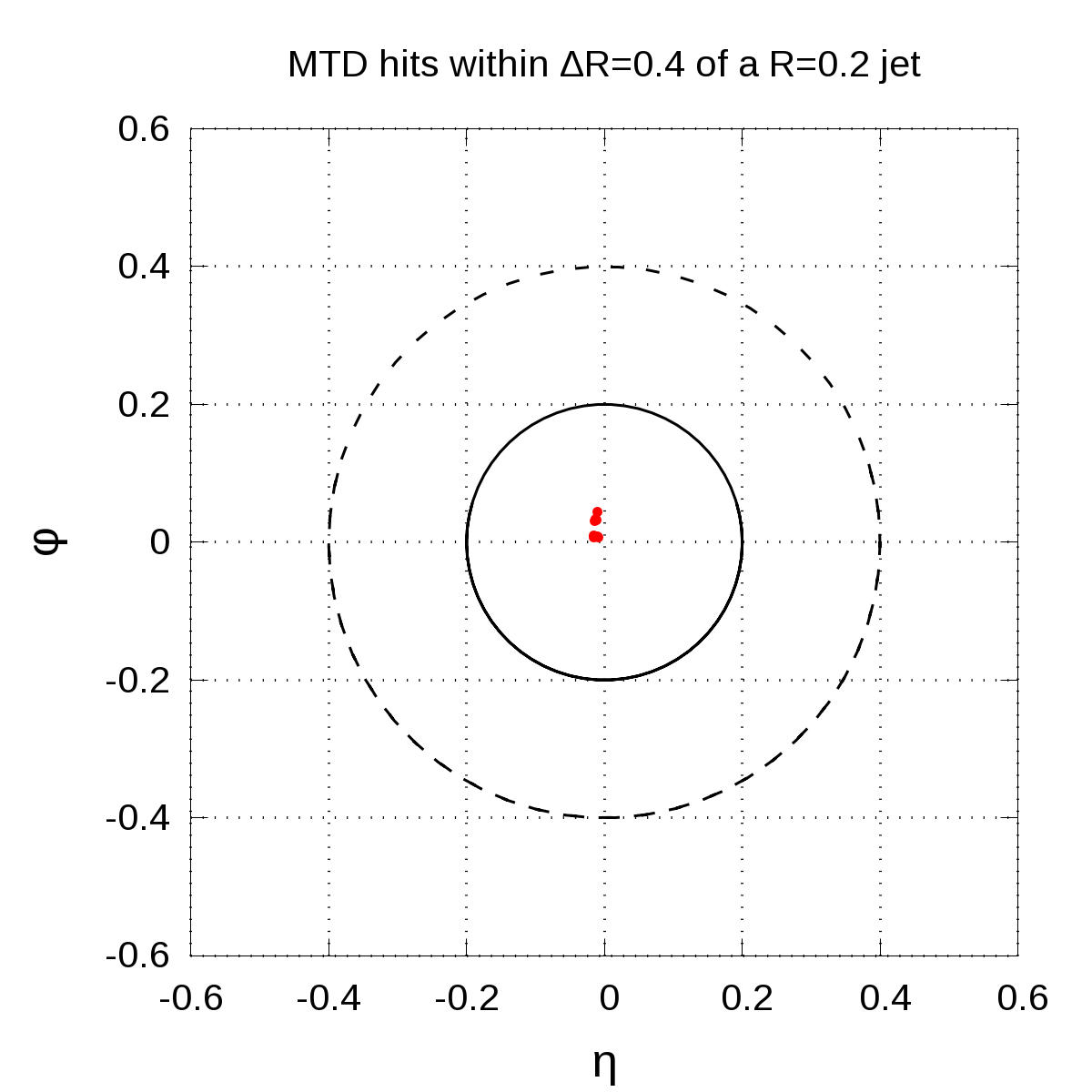}~
\includegraphics[scale=0.18]{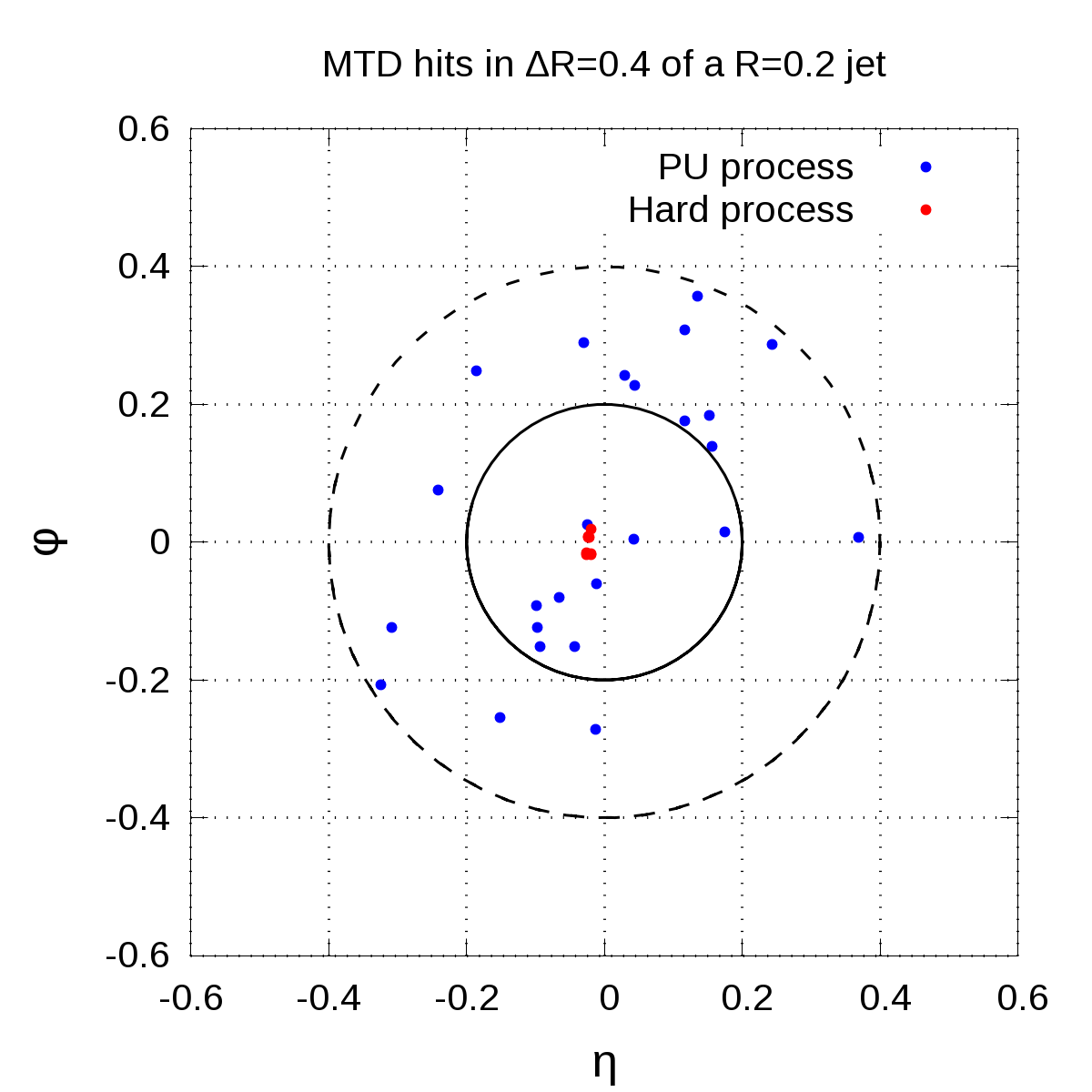}
\caption{MTD hits within $\Delta R=0.4$ of a displaced jet (of cone-size $R=0.2$) which comes from the decay of an LLP of mass 100 GeV which decays just before the MTD without ({\it left}) and with ({\it right}) 140 PU events. Red and blue points show hits from the hard and PU process respectively. The solid and dashed circles mark a region of $\Delta R=0.2$ and $\Delta R=0.4$ around the jet axis (at $\eta=\phi=0$) respectively.}
\label{fig:MTD_hits}
\end{figure}

Timing of a jet, defined as the median of all hits inside the jet, might get contaminated by pile-up hits. To study the extent of this contamination, we
look at the fraction of hits on the MTD which are associated with a jet and coming from PU or the hard collision which produces the LLPs. Fig.\ref{fig:MTD_hits} shows the MTD hits in a $R=0.2$ jet, within $\Delta R=0.4$ of the jet axis, coming from a LLP of mass 100 GeV which decays just before the MTD. Since the LLP decays just before the MTD, the hits from the hard collision are therefore concentrated in a very small physical region. The {\it left} plot is without PU and the {\it right} one is after merging the hard process with 140 PU. We find that even within $\Delta R=0.2$ of the jet axis, there are many PU hits in addition to the hits from the hard process. Pile-up can also change the jet axis as can be seen from fig.\ref{fig:MTD_hits} by comparing the position of the MTD hits from hard collision (in red) with respect to the jet axis ($\eta=\phi=0$) in the zero PU ({\it left plot}) and 140 PU ({\it right plot}) cases. 

\begin{figure}[hbt!]
\centering
\includegraphics[width=\textwidth]{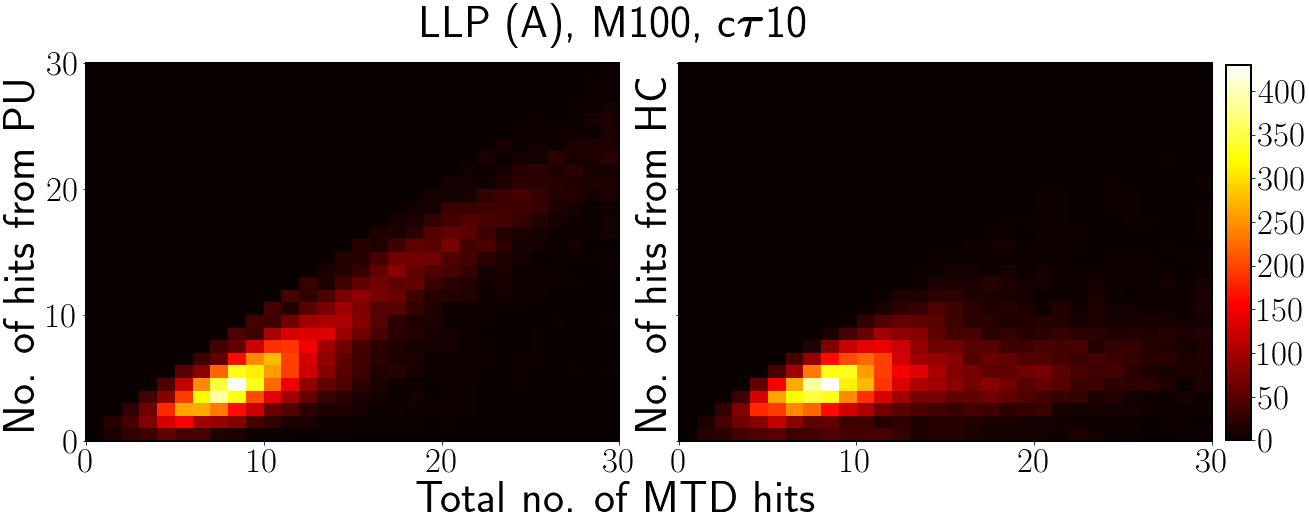}\\\vspace{2mm}
\includegraphics[width=\textwidth]{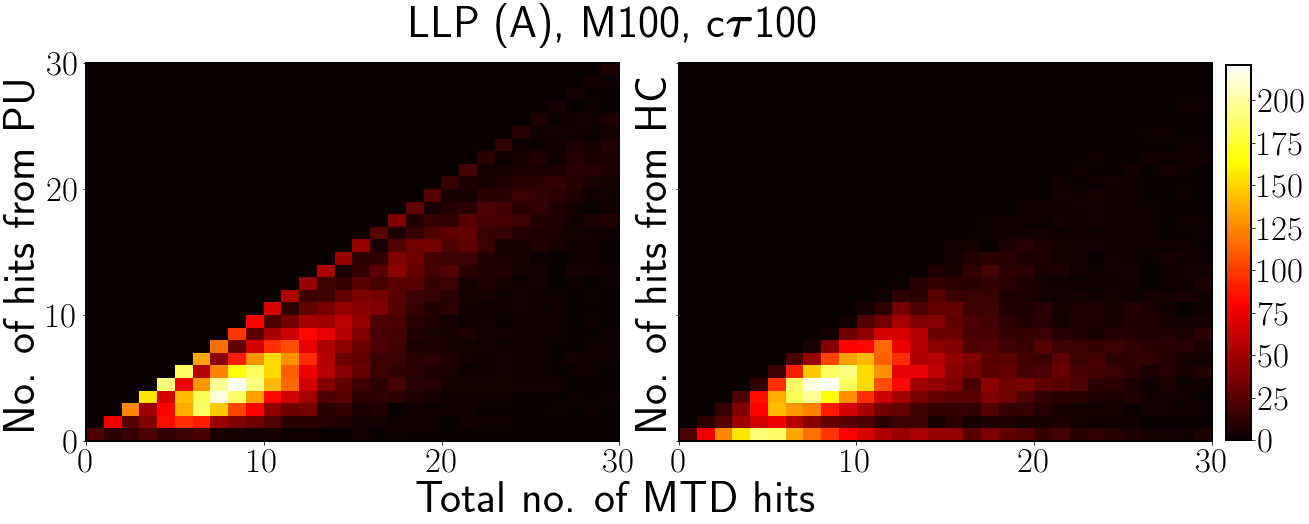}
\caption{Number of total MTD hits and the ones which are coming from ({\it left}) pile-up (PU) and ({\it right}) hard collision (HC) for jets coming from the decay of LLP of mass 100 GeV having decay length 10 cm ({\it top}) and 100 cm ({\it bottom}) from scenario (A).}
\label{fig:frac_isPU}
\end{figure}

The above illustration shows qualitatively that many MTD hits from pile-up processes get associated with a jet. To quantify this number, in fig.\ref{fig:frac_isPU}, we show the number of total MTD hits associated with a $R=0.2$ jet, with the number of hits coming from the actual hard collision which produces the LLPs (denoted as HC) and those coming from PU events, for a LLP of mass 100 GeV having decay lengths 10 cm and 100 cm from scenario (A). The color bar shows the number of jets.
In both the cases, most of the jets have almost equal amounts of PU and HC hits (corresponding to the denser region around total number of MTD hits close to $\sim 10$). However, with increasing number of total MTD hits, the contribution from PU events is more than the hard collision event.
Therefore, the timing of a jet will depend significantly on the timing of particles coming from PU, since they contribute to atleast 50\% of the MTD hits in a $R=0.2$ jet.

If the LLP decays outside the MTD, the jets will have energy deposits but no MTD hits, however, PU hits might still get associated with such jets. These jets will, therefore, have no hits from hard collision and the total number of MTD hits will be equal to the number of PU hits associated with the jet. Since for decay length 100 cm, decays outside the MTD are more probable, we find more jets having MTD hits from HC to be 0 and from PU to be equal to the total number of hits. These correspond to the lines with slopes $1$ (in the {\it left plot}) and $0$ (in the {\it right plot}) of the bottom panel of fig.\ref{fig:frac_isPU}. Therefore, with increasing decay lengths, contribution of PU is more in the displaced jets.

\begin{figure}[hbt!]
\centering
\includegraphics[scale=0.195]{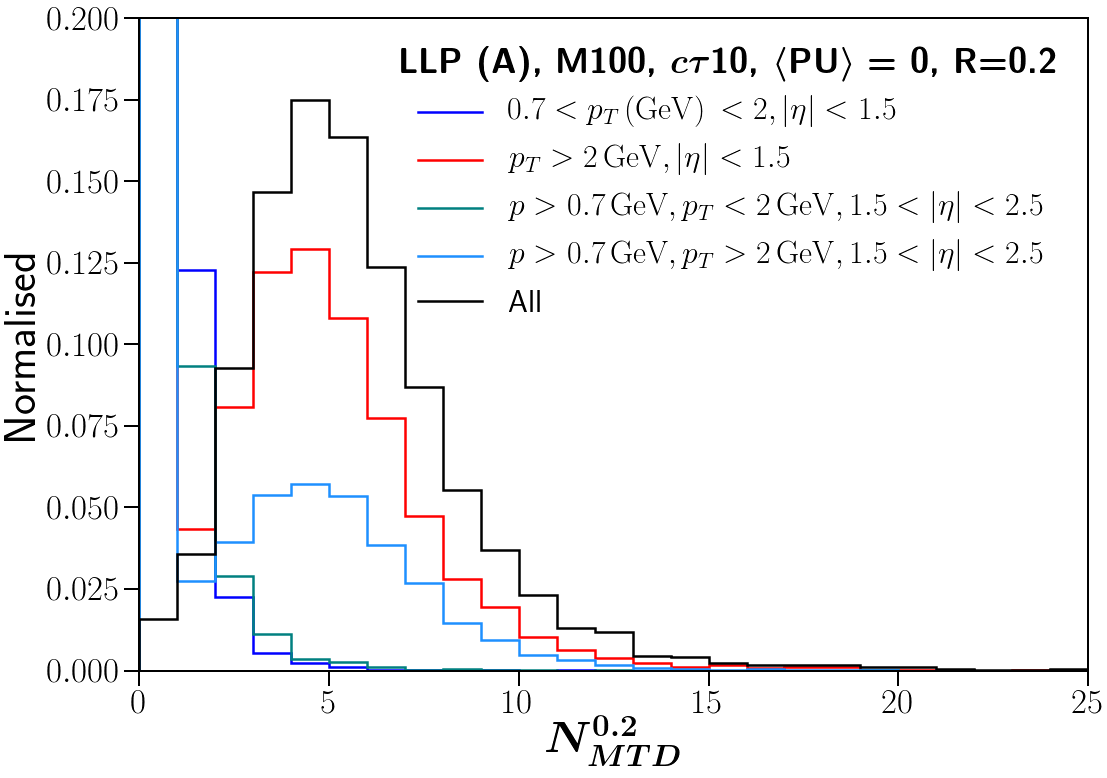}~
\includegraphics[scale=0.195]{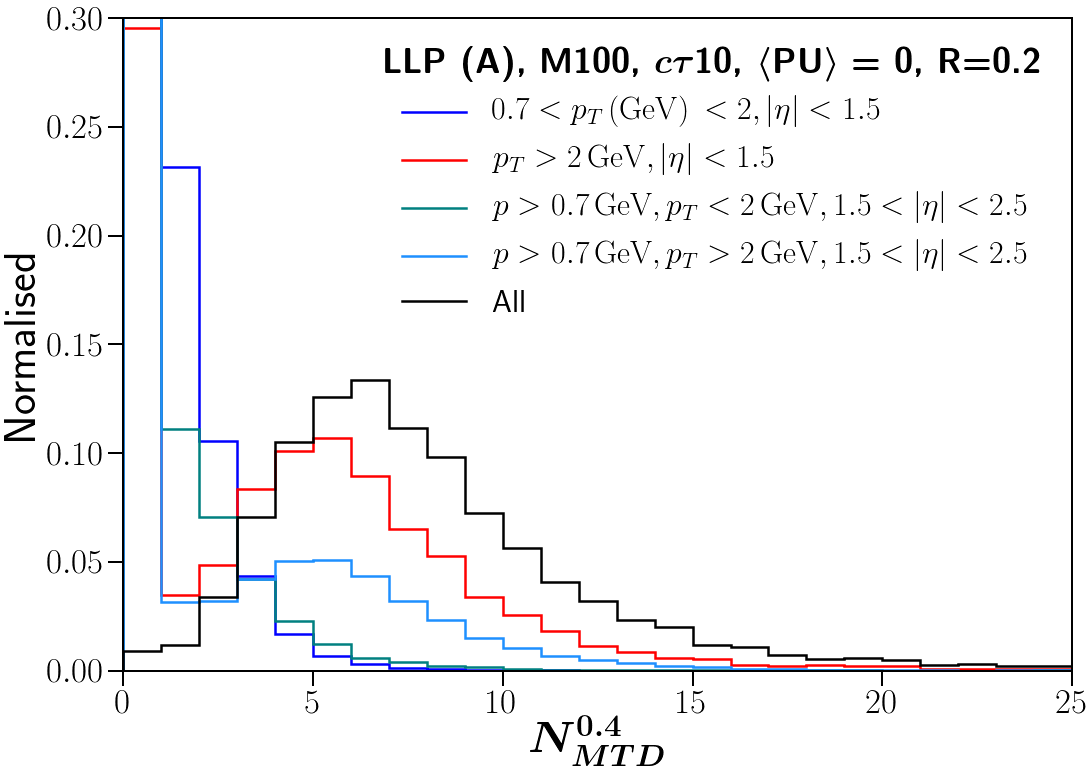}\\
\includegraphics[scale=0.195]{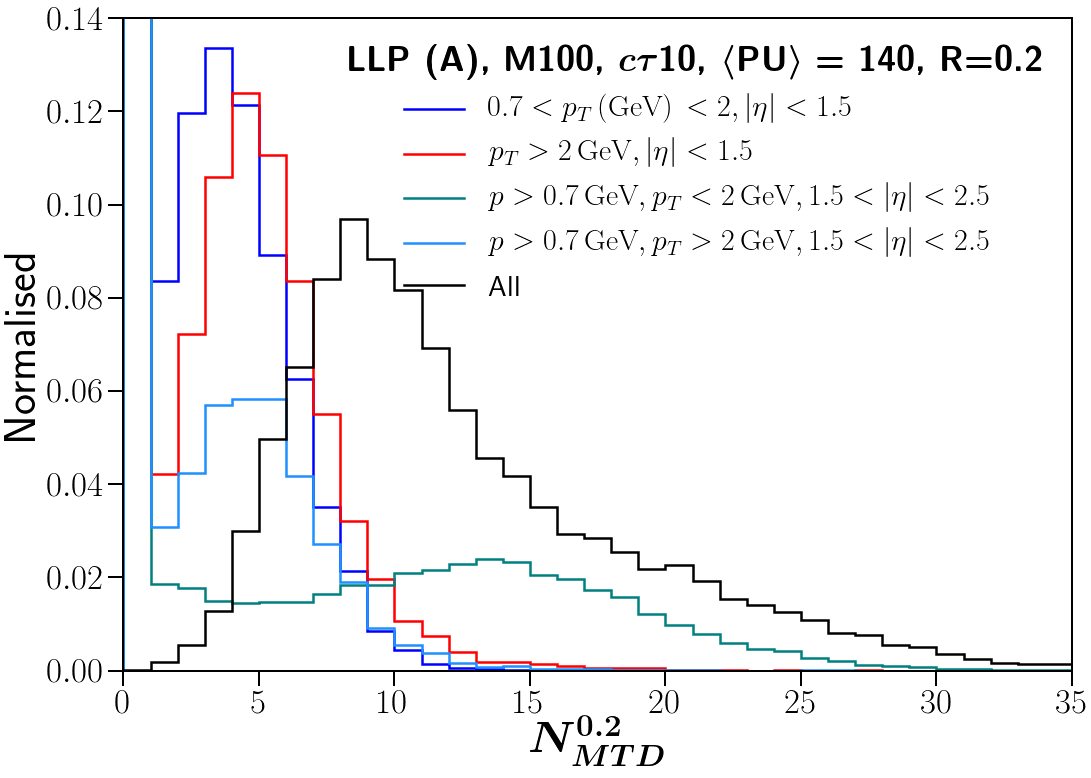}~
\includegraphics[scale=0.195]{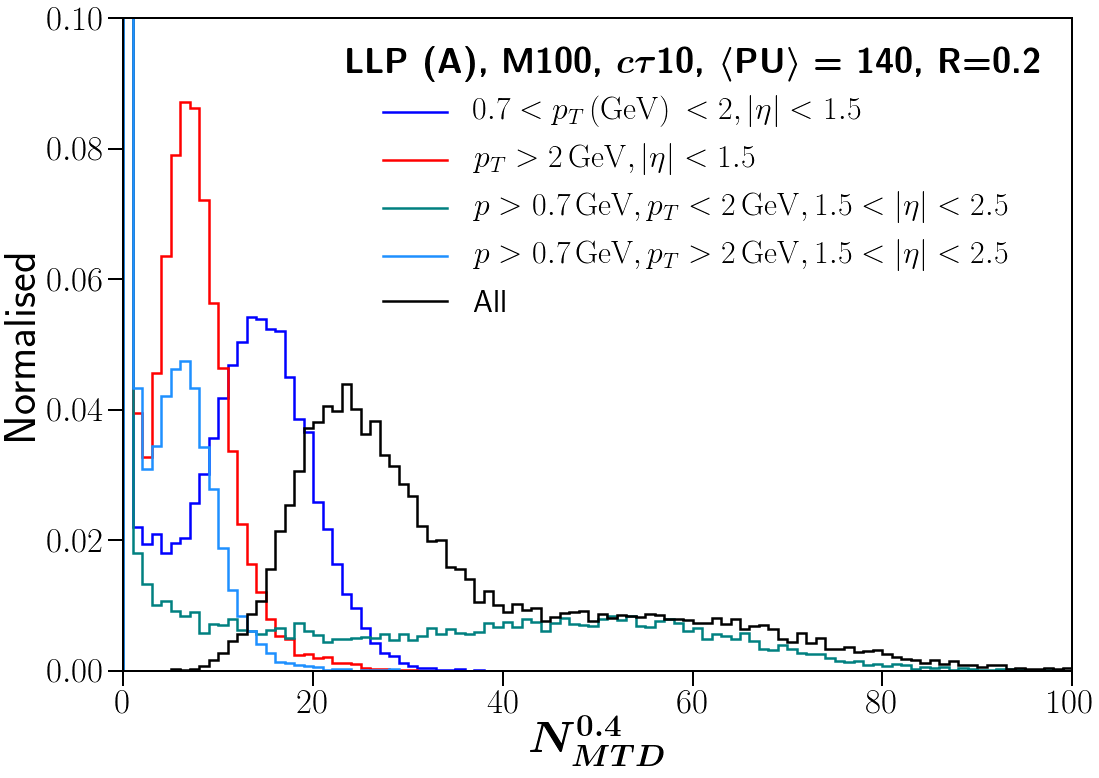}\\
\caption{Distribution of number of MTD hits within $\Delta R=0.2$ ({\it left panel}) and $\Delta R=0.4$ ({\it right panel}) of a $R=0.2$ jet axis without ({\it top panel}) and with ({\it bottom panel}) 140 PU events for LLP benchmark with mass 100 GeV and decay length 10 cm from scenario (A). Each of the plots shows the normalised number of hits in different $p_T$ and $p$ regions, and ``All'' shows the distribution of the total number of hits.}
\label{fig:nmtd_parts_R02}
\end{figure}

The timing of these extra hits coming from PU and associated with a jet will depend mostly on their momentum and the $\eta$ value where it hits the MTD. To quantify this, we now study the number of MTD hits having different $p_T$ and $p$ values and compare them for the zero and 140 PU scenarios.
Fig.\ref{fig:nmtd_parts_R02} shows the normalised distributions of the number of hits with different $p_T$ and $p$ values within $\Delta R=0.2$ ({\it left panel}) and $\Delta R=0.4$ ({\it right panel}) of a $R=0.2$ jet axis for 0 PU ({\it top panel}) and 140 PU ({\it bottom panel}). The following $p_T$ and $p$ regions are considered $-$

\begin{itemize}
\item $0.7 {\rm~GeV} <p_T<2 {\rm~GeV}, |\eta|<1.5$
\item $p_T>2 {\rm~GeV}, |\eta|<1.5$
\item $p>0.7 {\rm~GeV}, p_T<2 {\rm~GeV}, 1.5<|\eta|<2.5$
\item $p>0.7 {\rm~GeV}, p_T>2 {\rm~GeV}, 1.5<|\eta|<2.5$
\end{itemize}

These four conditions correspond to the two different MTD threshold regions $-$ up to barrel ($|\eta|<1.5$) and then for endcaps ($1.5<|\eta|<2.5$) and the possibility that the hits are associated with a L1 track ($p_T>2{\rm~GeV}$) or not. From fig.\ref{fig:nmtd_parts_R02}, we find that when there is no PU, the mean number of total MTD hits ($N_{\text{MTD}}$) is between 5-6 within $\Delta R=0.2$ and between 6-7 within $\Delta R=0.4$ from the jet axis. However, with the addition of 140 PU vertices, the $N_{\text{MTD}}$ distribution develops a longer tail extending up to around 30-35 and 75-80 hits within $\Delta R=0.2$ and $\Delta R=0.4$ respectively. This increase in the number of MTD hits is  mostly due to the increase of hits with $p>0.7 {\rm~GeV}$, $1.5<|\eta|<2.5$ and $p_T<2 {\rm~GeV}$. The contribution of hits from this $p_T$ and $p$ region was very small before adding PU and therefore, we conclude that these are mostly coming from pile-up. These are very low $p_T$ hits in the forward region which don't have any associated tracks at L1. 
Since they have such low $p_T$ values, they will bend more in the magnetic field and take much longer time to arrive at the MTD. The number of such hits dominate the tails of the $N_{\text{MTD}}$ distribution, and therefore, will affect the median time of a jet accordingly.
Another important observation is that the total $N_{\text{MTD}}$ distribution for hits within $\Delta R=0.4$ is affected more than that within $\Delta R=0.2$ due to these increased number of low $p_T$ PU hits in the endcap region. Therefore, we only use hits within $\Delta R=0.2$ of the jet axis hereafter, since they have relatively lesser PU contribution.

\begin{figure}[hbt!]
\centering
\includegraphics[scale=0.175]{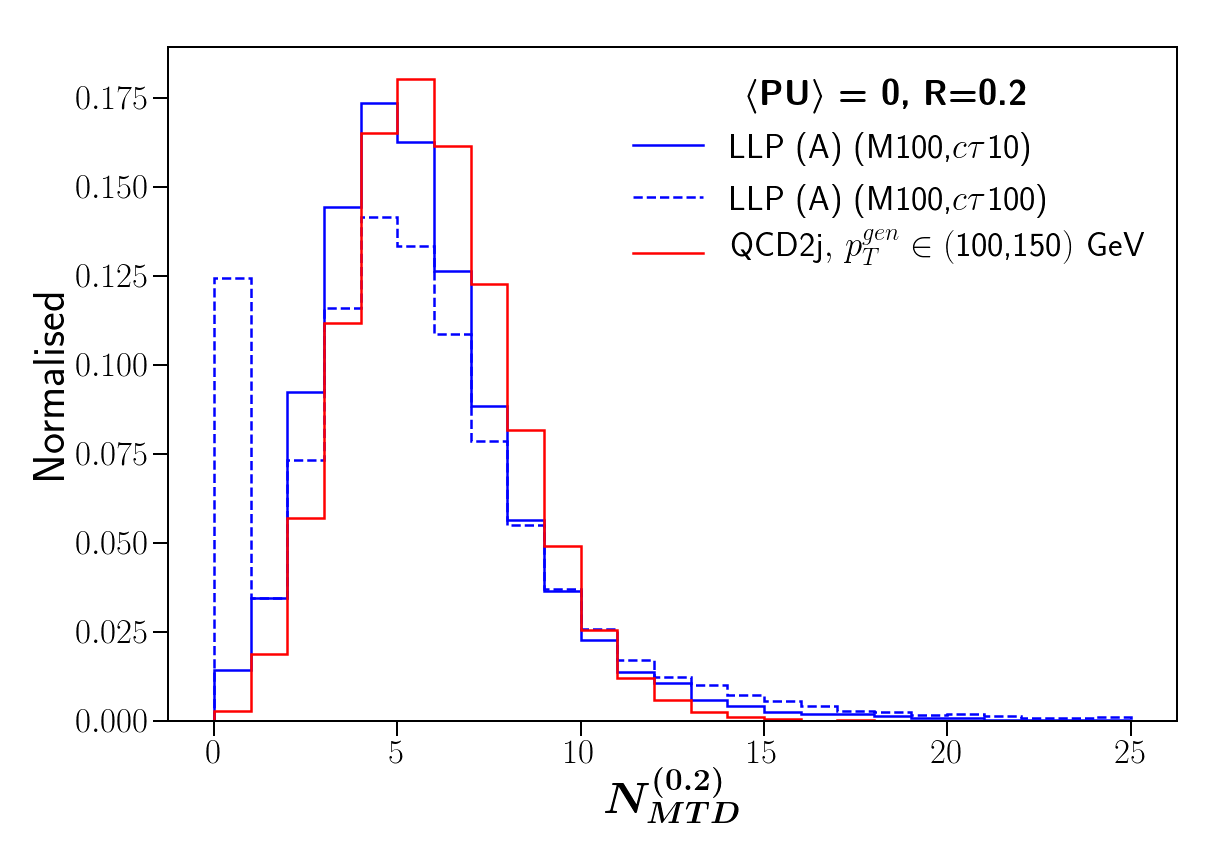}~
\includegraphics[scale=0.175]{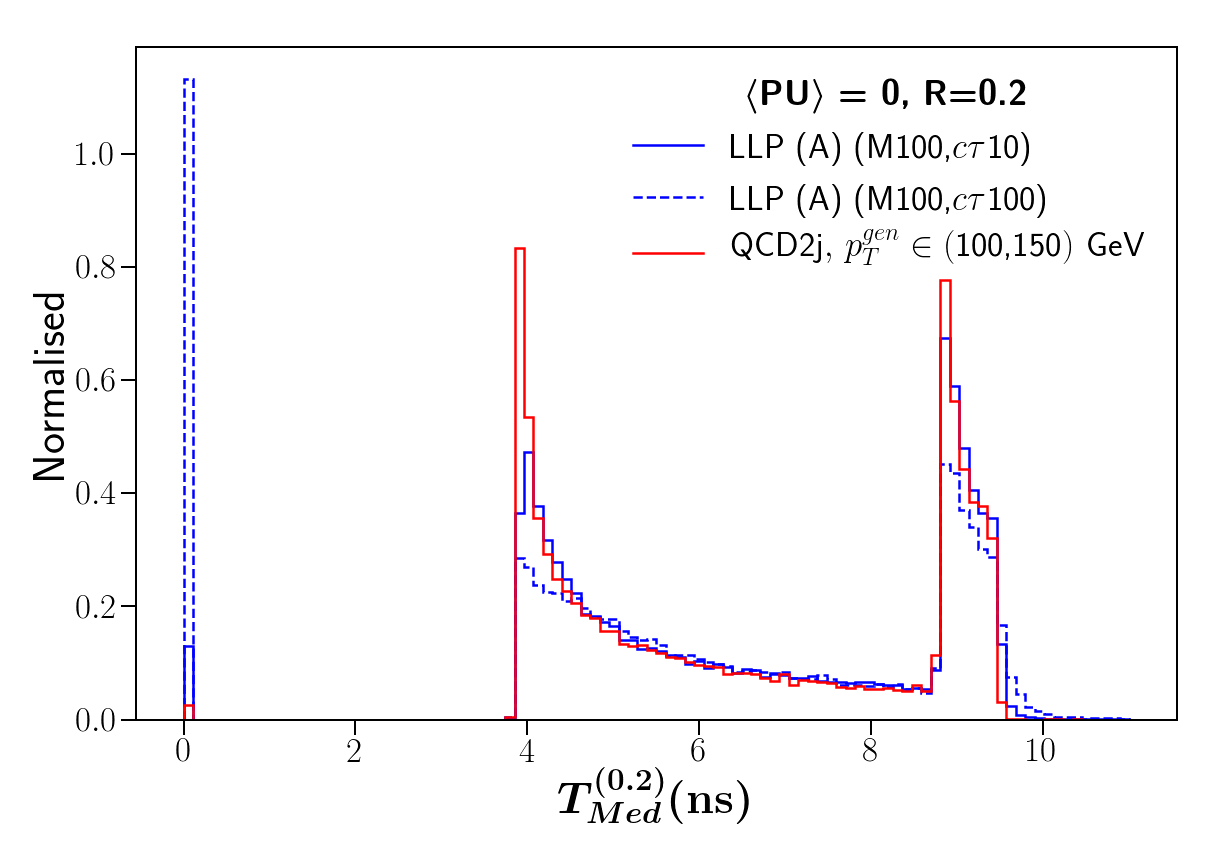}\\
\includegraphics[scale=0.175]{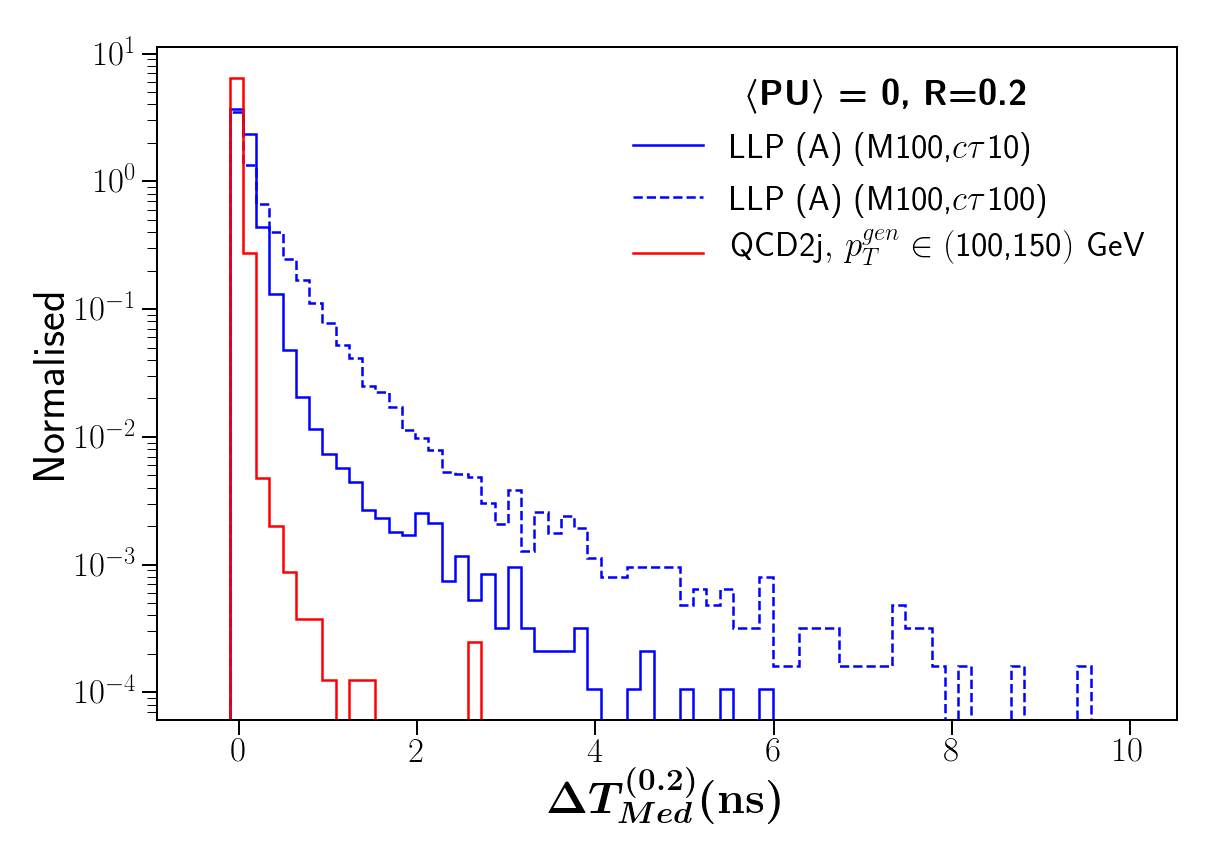}~
\includegraphics[scale=0.175]{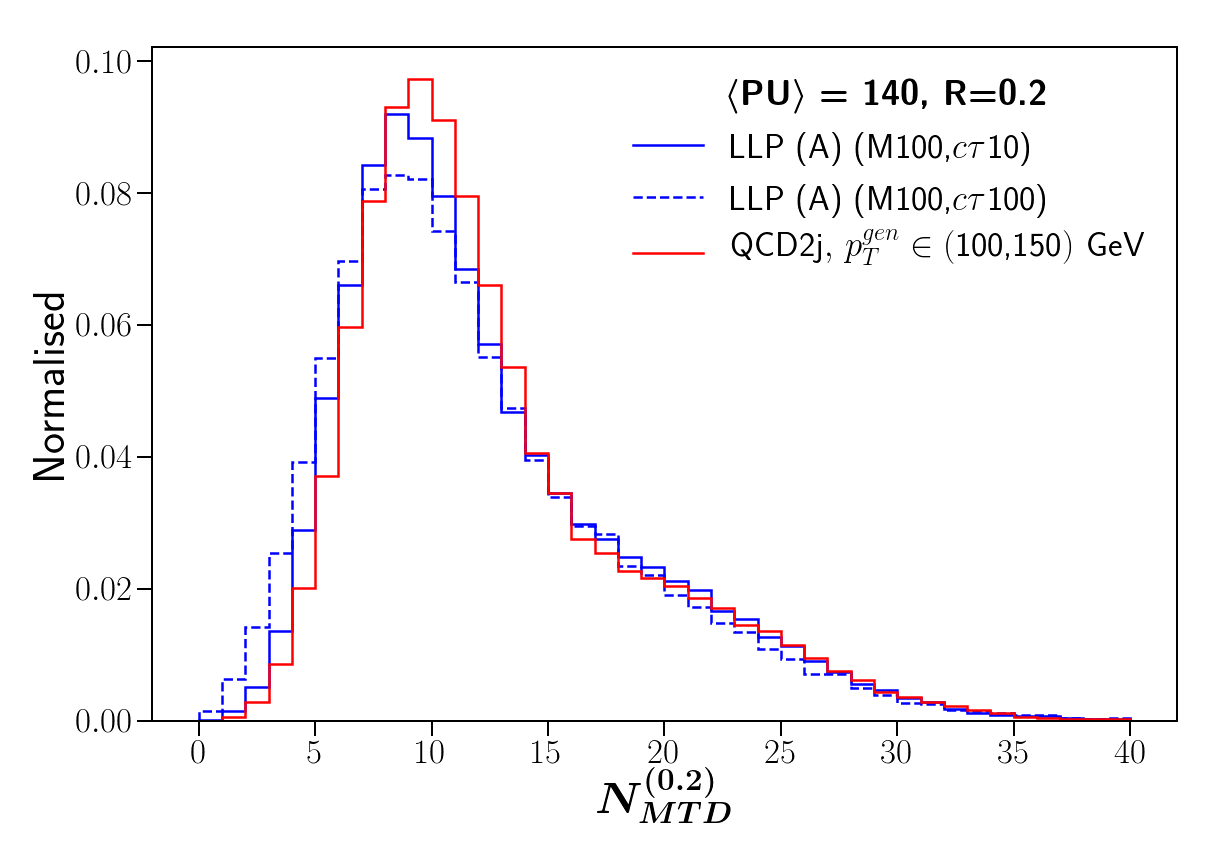}\\
\includegraphics[scale=0.175]{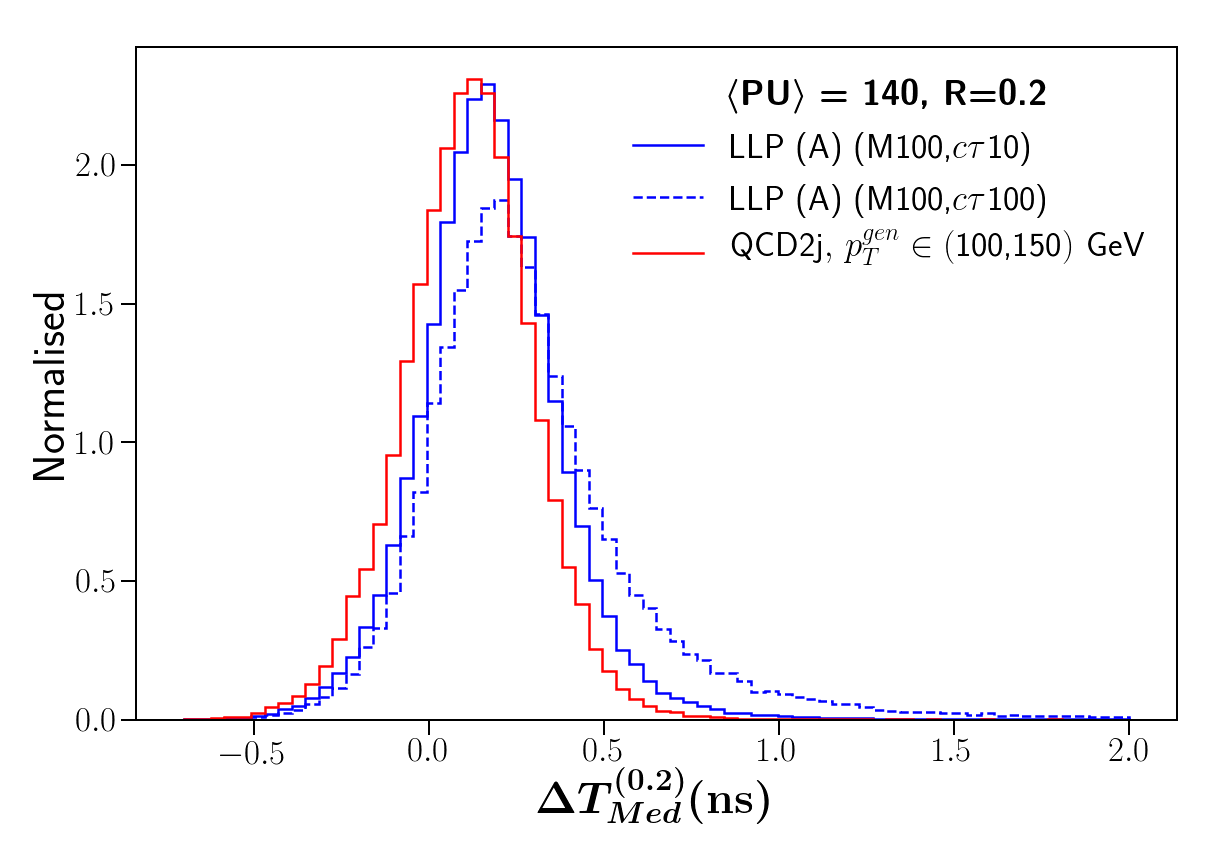}~
\includegraphics[scale=0.175]{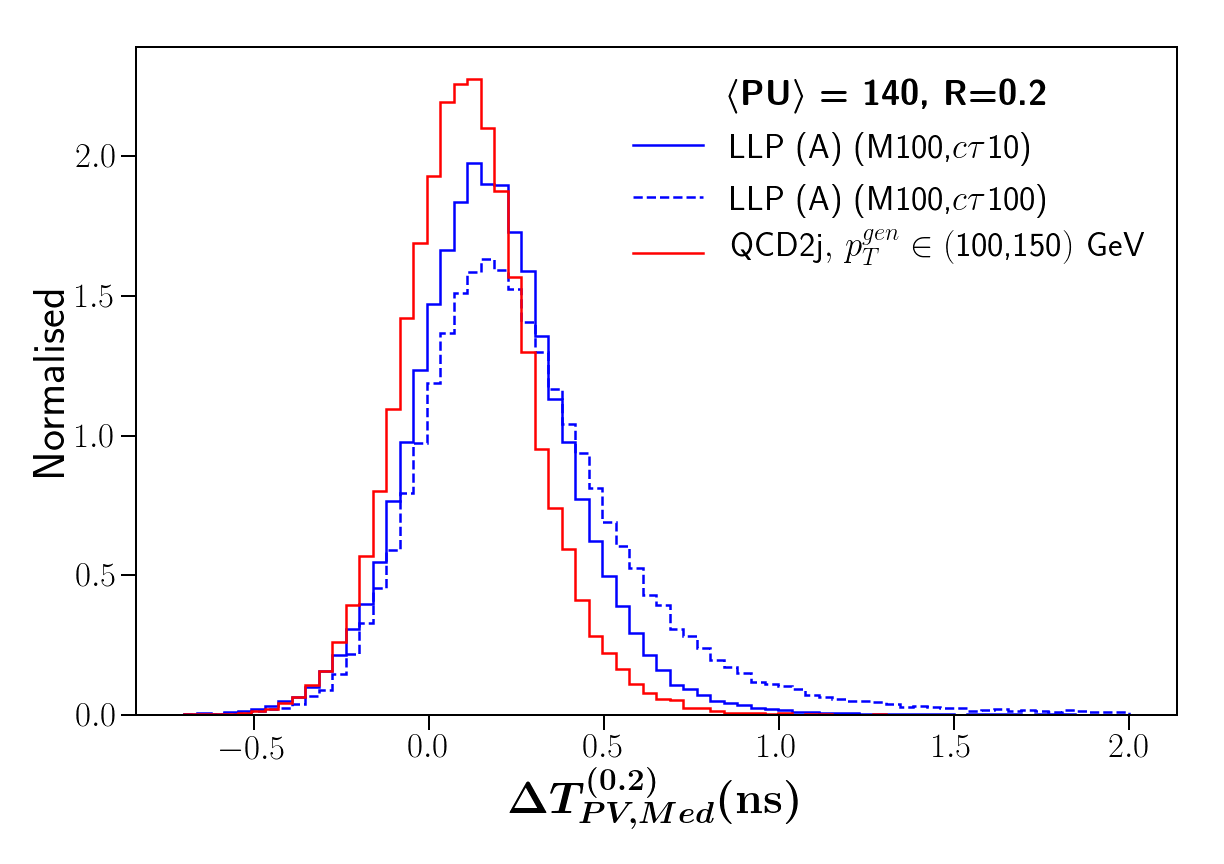}
\caption{Distributions of number of MTD hits ($N_{\text{MTD}}$) ({\it top left}), median of the absolute time ($T_{\text{Med}}^{(0.2)}$) of MTD hits associated with a jet ({\it top right}) and their time difference with respect to $z=0$ ($\Delta T_{\text{Med}}^{(0.2)}$) ({\it centre left}) for zero PU; Distributions of $N_{\text{MTD}}$ ({\it centre right}), $\Delta T_{\text{Med}}^{(0.2)}$ ({\it bottom left}) and time difference with respect to the primary vertex, PV, ($\Delta T_{\text{PV,Med}}^{(0.2)}$) ({\it bottom right}) are shown for events with 140 PU. The distributions are drawn for QCD prompt jets coming from dijet events with $p_T^{gen}\in(100,150){\rm~GeV}$ and those coming from LLP decays for scenario (A) with LLP mass 100 GeV having decay lengths of 10 cm and 100 cm.}
\label{fig:median_t_dist}
\end{figure}

From our discussion till now, we know that MTD hits associated with a jet will have significant PU contribution and 
these large number of MTD hits coming from PU will contaminate both the QCD and LLP hard processes equally, and therefore, the median timing of all hits associated with a jet will get biased. Let us now compare the timing variables of a jet in the zero and 140 PU scenarios to actually witness the PU contamination in the distributions of these variables. The {\it top panel} of fig.\ref{fig:median_t_dist} shows the median of the time values of all hits associated with a jet (within $\Delta R=0.2$ of the jet axis), after calibrating each $\eta$ bin with the time of prompt particles starting from $z=0$ ($\Delta T_{\text{Med}}^{(0.2)}$), and also the median of the absolute time of the hits ($T_{\text{Med}}^{(0.2)}$) for jets from QCD dijet events and LLP events with no PU. We find that the $\Delta T_{\text{Med}}^{(0.2)}$ distribution for jets coming from LLP decays have a significantly longer tail, which increases with the decay length of the LLP. 

However, with the addition of 140 PU vertices, the $\Delta T_{\text{Med}}^{(0.2)}$ distribution ({\it bottom left} plot of fig.\ref{fig:median_t_dist}), gets modified for both the signal (LLP events) and the background (QCD events), and the $\Delta T_{\text{Med}}^{(0.2)}$ distributions change, both dominated by MTD hits coming from PU events. The distribution for LLP hits does have a slightly longer tail than QCD and this tails increases with decay length of the LLP. However, the variable's discriminating power to separate prompt QCD jets from displaced jets decreases. Also, note there are more negative values of $\Delta T_{\text{Med}}^{(0.2)}$ after adding PU, since the time difference is calculated with the time of the prompt particle starting from $z=0$, and at larger $\eta$ values this time will be greater than time taken by particles originating from $z$ values from where that particular $\eta$ position on the MTD is closer. 

Another interesting thing to note is that the distributions with zero PU have many jets with zero MTD hits and therefore, zero median time or time difference, and this number increases with decay length. These correspond to events where the LLP decays after the MTD, and therefore, has energy depositions in the calorimeters but no hits in the MTD, as we discussed earlier. With addition of PU, every jet will have some number of associated MTD hits coming from PU processes, and therefore, the peak at zero goes away.

The effect of adding PU on the median of the time difference of all charged particles of a jet has two factors:
\begin{itemize}
\item MTD hits from PU processes getting associated with a jet
\item calculation of the time difference from the correct vertex
\end{itemize}
Till now, we have mostly discussed the first factor. The second factor would not affect the median of the absolute time of a jet. However, the time difference of a MTD hit will depend on where we consider the reference prompt particle to start from. We were calculating $T_{light}$ by considering the particle to start from $z=0$, i.e., we are calibrating each $\eta$ value on the MTD with the time of the prompt particle travelling at the speed of light from $z=0$ to that $\eta$ value on the MTD. However, this might not be a good measure of time difference since the production vertex of the LLPs might not be at $z=0$.

Instead of taking the time difference with respect to a prompt particle coming from $z=0$, we can, in principle, find out the primary vertex (PV) and then take the time difference with respect to the PV. PV is the vertex having the maximum value of $\sum p_T^2$ (or, $\sum p_T^2/n_{z_a}$) which are calculated for each vertex (at $z$-position $z_a$) with all the tracks coming from that vertex, $n_{z_a}$ (within a $z$ range of 1 mm)\footnote{We only calculate these quantities for vertices having at least two tracks.}.
If the PV vertex corresponds to the vertex of hard collision from where the LLPs are produced, then we can actually minimise the effect of the second factor which affects the median value of the time difference of MTD hits, if we calculate the difference from the PV. 
For LLPs, however, we don't have many tracks from the actual hard process at L1 since we are limited to $L_{xy}<1{\rm~cm}$ region at L1 and most of the displaced jets will have tracks beyond this region (see table \ref{tab:decay_frac_BP1}). Therefore, in most of the LLP events, we incorrectly assign a PU vertex as the primary vertex. Table \ref{tab:PV_eff_BP1} quotes the efficiency of identifying the primary vertex using both max($\sum p_T^2$) and max($\sum p_T^2/n_{z_a}$) for hard events involving pair production of LLPs of different mass and decay lengths, when merged with an average of 140 PU events. 

\begin{table}[hbt!]
\centering
\begin{tabular}{|c||c|c||}
\hline
Mass [GeV], & max($\sum p_T^2$) corresponds & max($\sum p_T^2$/$n_{z_a}$) corresponds\\
Decay Length [cm] & to hard collision & to hard collision \\
\hline\hline
50, 10 & 29.0\% & 36.1\%\\
50, 100 & 25.3\% & 32.3\%\\
100, 10 & 47.9\% & 51.0\%\\
100, 100 & 39.3\% & 43.8\%\\
200, 10 & 56.5\% & 59.0\%\\
200, 100 & 48.7\% & 51.4\%\\
500, 10 & 63.3\% & 63.5\%\\
500, 100 & 55.6\% & 56.0\%\\
\hline\hline
\end{tabular}
\caption{Efficiency for correctly identifying the vertex corresponding to the hard collision, from where LLPs are produced, as the primary vertex (PV) at L1 for scenario (A) using the vertex with the maximum $\sum p_T^2$ of tracks and maximum $\sum p_T^2/n_{z_a}$ where $n_{z_a}$ corresponds to the total number of tracks associated with that vertex.}
\label{tab:PV_eff_BP1}
\end{table}

The efficiency of identifying the hard collision vertex from where LLPs are produced as the primary vertex increases with increasing LLP mass and decreasing decay length. Decreasing decay length implies that decay is more probable within $L_{xy}=1{\rm~cm}$ and hence there will be more L1 tracks from the hard event. Increasing the LLP mass decreases the boost of the LLP and it is more likely to decay early. Also, production of heavy LLPs is accompanied by emission of more radiation, which makes it easier to identify the PV for such events. The {\it bottom right} plot of fig.\ref{fig:median_t_dist} shows the distribution of the median of time of all particles inside a jet calibrated with prompt particles, travelling at the speed of light, starting from the PV~\footnote{Here, we identify the PV as the vertex with the maximum value of $\sum p_T^2/n_{z_a}$) since it has slightly higher efficiency of correctly identifying the PV.}. We observe that this distribution has a slightly more prominent tail for the LLPs than that when only the difference with respect to $z=0$ is taken. However, it still is very different from the zero PU distributions, due to the fact that we can identify the PV correctly only half of the times even when the LLP has a mass of 100 GeV and decay length 10 cm (as seen from table \ref{tab:PV_eff_BP1}). 

At the high level trigger (HLT), one will have access to the full timing layer information and a 4D reconstruction of the vertices is possible, which can help reduce the PU contribution in the jets. This can also lead to the determination of the primary vertex correctly in most of the cases, and therefore, we can get closer to the distributions of median of time differences for the zero pile-up scenario.

\subsubsection{Association of MTD hits with L1 tracks}
\label{sssec:L1track_MTD_match}

We have now discussed how median timing (or time difference) of the MTD hits associated with a jet gets affected in the 140 PU scenario compared to the no pile-up one. The discriminating power of the median of time difference of all MTD hits associated with a jet decreases in the high PU LHC environments. The efficiency of identifying the correct primary vertex is not very high and also, even if one does identify the correct PV, the number of PU hits associated with a jet will affect the median timing adversely. In this section, we discuss how we can improve the performance of timing variables in order to discriminate between displaced and prompt jets. 

Jets from LLP signal processes are displaced and have very low probability to have L1 reconstructed tracks as discussed before, however, they will have MTD hits since most of them decay within tracker (see table \ref{tab:decay_frac_BP1}). If matching tracks with MTD hits is possible at L1, we can remove MTD hits which have associated tracks while calculating the median time. This will leave us with very low $p_T$ ($0.7{\rm~GeV}<p_T<2{\rm~GeV}$) MTD hits or hits from charged particles with ($L_{xy}>1{\rm~cm}$). The former category corresponds mostly to very soft particles coming from PU processes and the latter will be mostly particles from LLP decays, which don't have a track but have MTD hits. We denote the variables calculated using the set of MTD hits having no associated tracks with a superscript ``${\text{NT}}$''.

\begin{figure}[hbt!]
\centering
\includegraphics[scale=0.175]{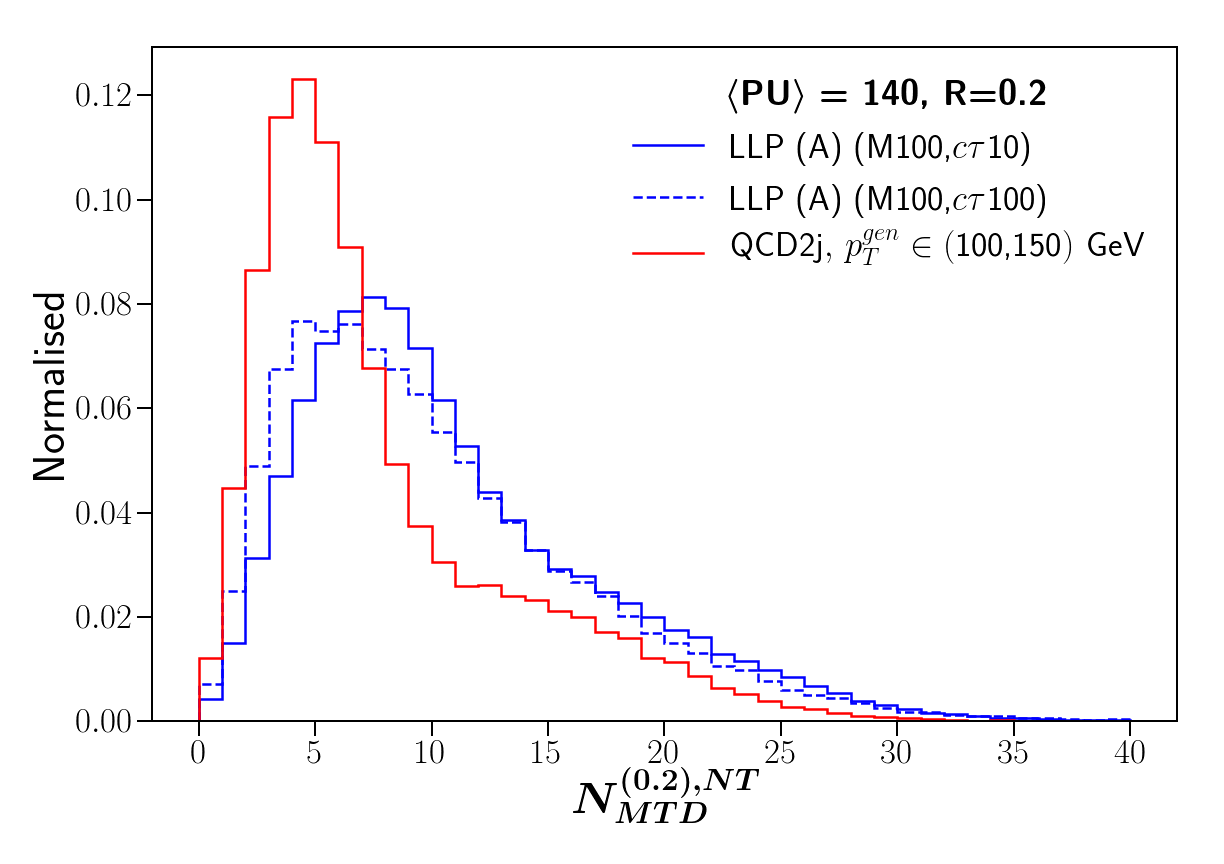}~
\includegraphics[scale=0.175]{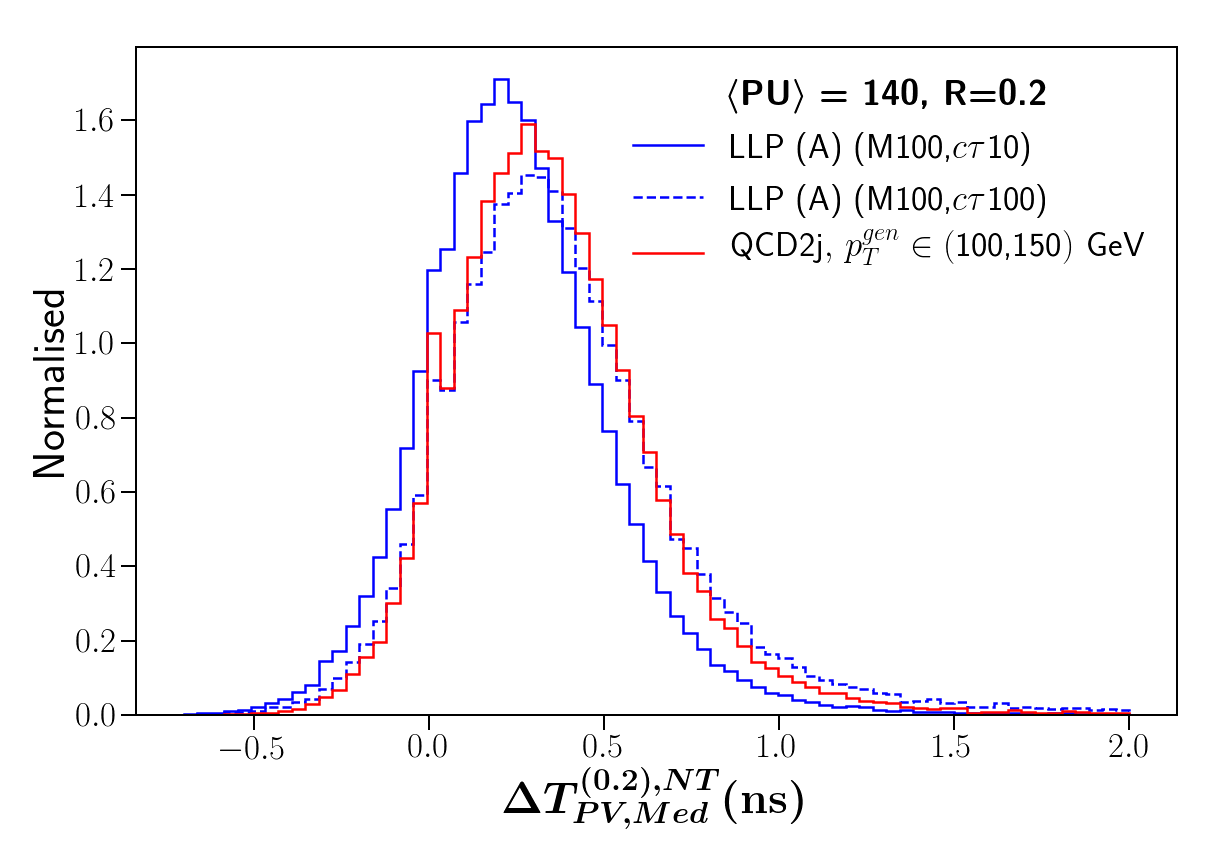}\\
\caption{Number of MTD hits and median of the time of hits (calibrated with respect to the PV) after removing the ones which have associated tracks $-$ $N_{\text{MTD}}^{(0.2),{\text{NT}}}$ ({\it left}) and $\Delta T_{\text{PV,Med}}^{(0.2),NT}$ ({\it right}).}
\label{fig:nt_vars}
\end{figure}

Fig.\ref{fig:nt_vars} shows the distributions of $N_{\text{MTD}}^{(0.2),NT}$ and $\Delta T_{\text{PV,Med}}^{(0.2),NT}$ with the ``${\text{NT}}$'' set of MTD hits, which don't have associated tracks, for the QCD prompt jets from dijet processes and displaced jets from LLPs for zero and 140 PU. The $N_{\text{MTD}}^{(0.2),NT}$ distribution shifts to lower values for the QCD jets. This is because for the QCD prompt jets, most of the MTD hits will have associated L1 tracks unlike displaced jets and soft PU jets where most of the tracks won't be reconstructed at L1 $-$ they being displaced in the former case and having $p_T<2{\rm~GeV}$ in the latter.

Another interesting trend that we observe in the {\it bottom right} plot of fig.\ref{fig:nt_vars} is that the $\Delta T_{\text{PV,Med}}^{(0.2),NT}$ distribution shifts to higher values for the QCD prompt jets than before removing hits with reconstructed tracks. This is due to the fact that after removing hits associated with reconstructed tracks, the remaining MTD hits will be characterized by lower $p_T$ and higher time values, which shifts the median time of the jet towards right. However, no such change is expected nor observed for jets coming from LLPs. Therefore, the correlation between these two variables $-$ $\Delta T_{\text{PV,Med}}^{(0.2)}$ and $\Delta T_{\text{PV,Med}}^{(0.2),NT}$ plays an important role in distinguishing the background and the signal in our case (see the correlation matrix in appendix \ref{app:corr_matrix}). They are more correlated for the signal displaced jets than the background prompt jets.

\subsubsection{Training based on regional timing}
\label{sssec:train_time}

We can see how well the timing variables can differentiate between the prompt QCD jets and the displaced ones. As we can see from figs. \ref{fig:median_t_dist} and \ref{fig:nt_vars}, none of the single variables have strong discriminating power after adding 140 PU, however, their correlations might help to discriminate between the signal and the background. Therefore, we train a BDT classifier using the following timing variables:
$$p_T,\, \eta,\, N_{\text{MTD}}^{(0.2)},\, T_{\text{Med}}^{(0.2)},\, \Delta T_{\text{Med, PV}}^{(0.2)},\, N_{\text{MTD}}^{(0.2),\text{NT}},\, \Delta T_{\text{Med, PV}}^{(0.2),\text{NT}}$$
The $p_T$ and $\eta$ of the jet are important inputs for the training because timing of a prompt jet will have some dependence on the $p_T$ and $\eta$ values as each charged constituent of a jet has as we have seen in the beginning of the timing section \ref{ssec:time_L1}.

\begin{figure}[hbt!]
\centering
\includegraphics[scale=0.195]{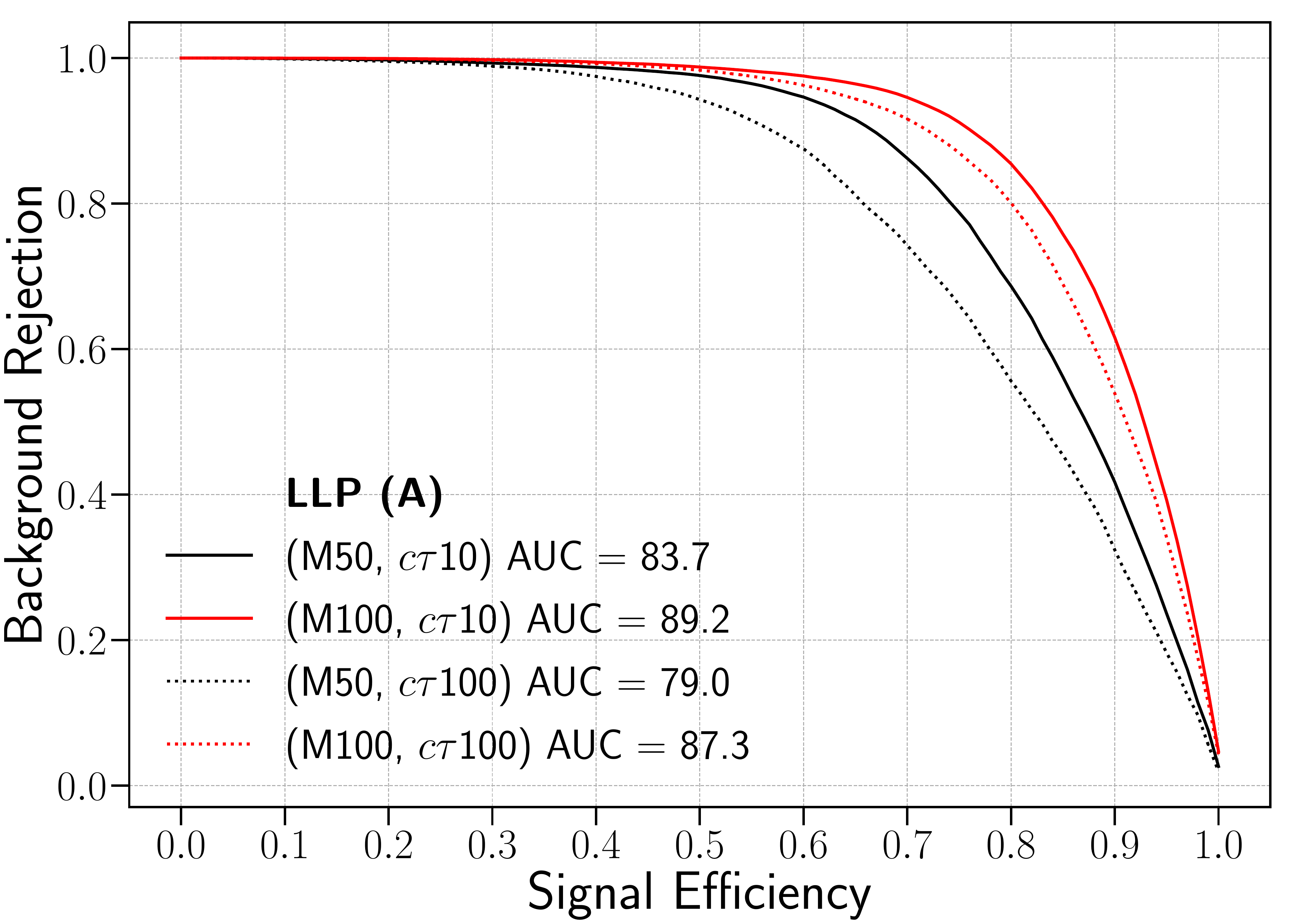}\\\vspace{0.5cm}
\includegraphics[scale=0.195]{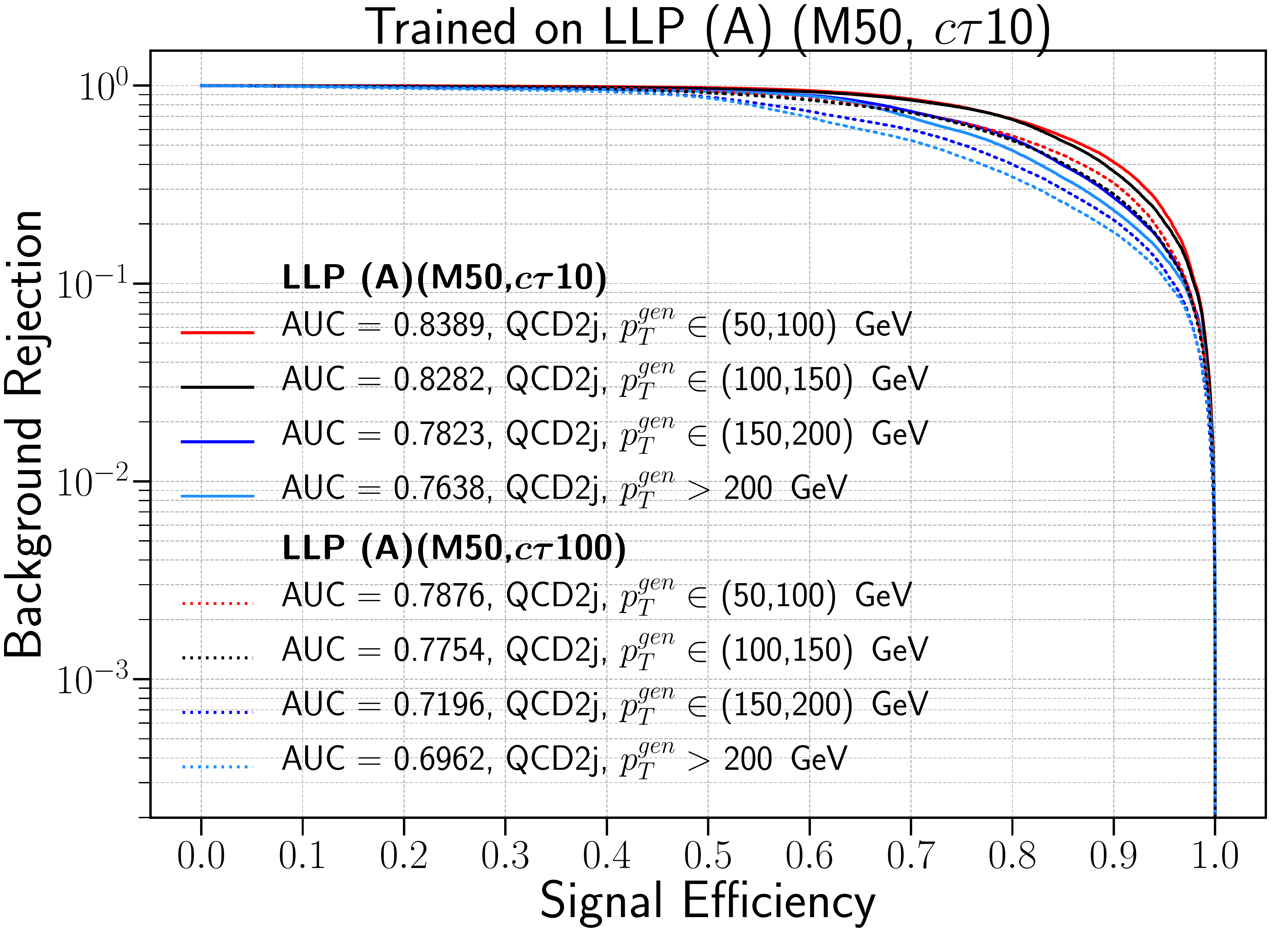}~
\includegraphics[scale=0.195]{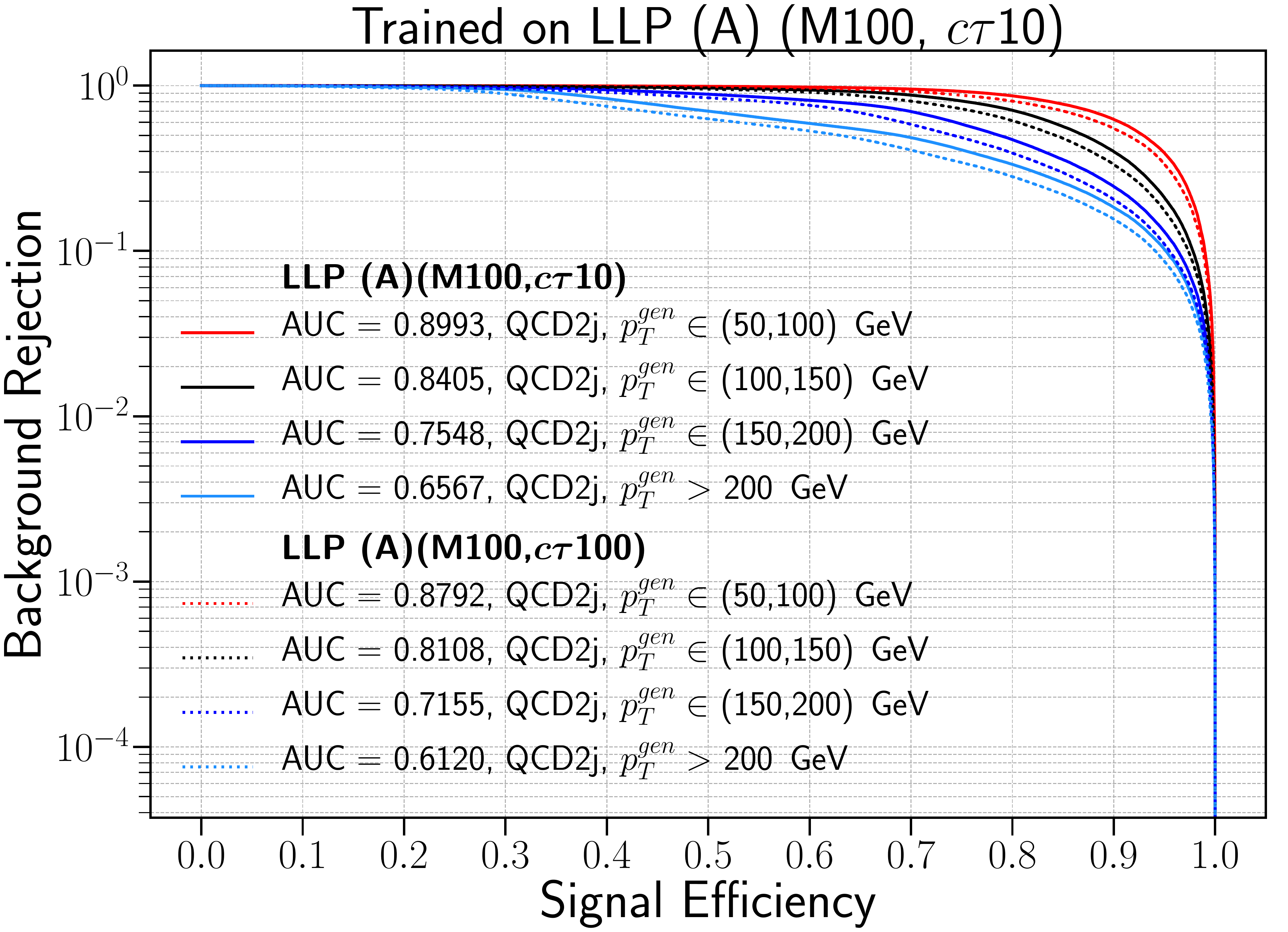}
\caption{ROC curves for identifying displaced jets from LLP decays (scenario (A)) from background QCD jets using timing variables.}
\label{fig:ROC_time_BP1}
\end{figure}

The {\it top panel} of fig.\ref{fig:ROC_time_BP1} shows the ROC curves for classifying the signal and background.
We now train the BDT on a particular mass value and decay length and apply it on that benchmark point as well as the benchmark point with the same mass and a different decay length. 
The {\it bottom} plots of fig.\ref{fig:ROC_track} show the application ROC curves for the LLP scenario (A) benchmark points when trained on (M50,$c\tau$10) ({\it left}) and (M100,$c\tau$10) ({\it right}) and applied on (M50,$c\tau$100) and (M100,$c\tau$100) respectively. The application ROCs are shown for jets from each of the $p_T^{gen}$ bins of the background QCD dijet events. 

The BDT performance is comparable to that when variables with tracking information are used, though the former is slightly weaker which can be seen from comparing the {\it area under the curve} (AUCs) of the ROCs. The reason for this is the low $p_T$ threshold of MTD, which results in more contribution from PU.
The track and time variables are not correlated with each other as can be seen from the correlation matrix in appendix \ref{app:corr_matrix}, and they individually give similar performance.

We find that the performance of the ROC degrades with increasing decay length and decreasing mass. The reason for the former is that there are many jets for higher decay lengths which are produced when the LLP decays outside the MTD and these jets will have all PU hits associated with them, and therefore, their timing is affected the most. For the lighter LLPs, the time delay caused by the LLP will be smaller and therefore, the performance based on timing variables decreases.

We have also trained the BDT classifier using both the tracking and timing variables together, however, the performance does not improve much. The performance of the timing variables mostly depends on the different correlations of the variables $\Delta T_{\text{Med, PV}}^{(0.2)}$ and $\Delta T_{\text{Med, PV}}^{(0.2),\text{NT}}$ for the signal and the background as we have discussed before while introducing the ``$\text{NT}$'' variables. This can also be seen from the correlation matrices of the variables for signal and background shown in appendix \ref{app:corr_matrix}. To recap, the ``NT'' variables are constructed from MTD hits which do not have any associated tracks at L1. When we use both tracking and timing variables together, tracking will mostly select jets with lesser number of L1 tracks as signal jets. The timing variables of these jets will be less affected if we use the ``$\text{NT}$'' set of MTD hits, because anyway they had lesser values of $N_{\text{trk}}$. Therefore, the $\Delta T_{\text{Med, PV}}^{(0.2)}$ and $\Delta T_{\text{Med, PV}}^{(0.2),\text{NT}}$ variables will be more correlated for such jets, irrespective of whether they are displaced jets or prompt jets.

\begin{figure}[hbt!]
\centering
\vspace{0.5cm}
\includegraphics[width=\textwidth]{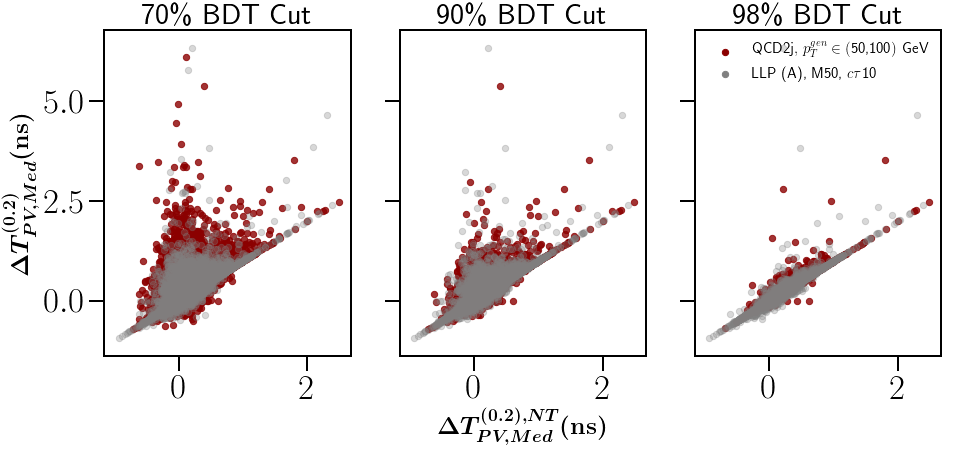}~
\caption{Correlation between the $\Delta T_{\text{Med, PV}}^{(0.2)}$ and $\Delta T_{\text{Med, PV}}^{(0.2),\text{NT}}$ variables for displaced (gray) and prompt (red) jets after a BDT cut based on tracking variables corresponding to 70\% ({\it left}), 90\% ({\it centre}) and 98\% ({\it right}) background rejections.}
\label{fig:corr_median_dt2_nt}
\end{figure}

Fig.\ref{fig:corr_median_dt2_nt} shows the correlation between the $\Delta T_{\text{Med, PV}}^{(0.2)}$ and $\Delta T_{\text{Med, PV}}^{(0.2),\text{NT}}$ variables for displaced jets (blue) and prompt jets (red) after a BDT cut based on tracking variables corresponding to 70\% ({\it left}), 90\% ({\it centre}) and 98\% ({\it right}) background rejections. We find that initially, jets from signal had more correlation in the values of the two variables than the background. However, when the BDT cut based on tracking variables is made more strict to reject 90\% or 98\% of the background, the remaining background jets are more signal-like, with less number of L1 tracks. Therefore, their median time difference values using all MTD hits associated with the jet and using only those which have no L1 tracks are highly correlated for both signal and background. Therefore, the inclusion of the timing variables will not improve the performance much here.

Still, for a signal efficiency of $\sim 70\%$, we get a background rejection of $\sim 95\%$ for the benchmark (M100,$c\tau$10) from scenario (A). Therefore, if regional timing is available at L1, we can define a trigger as we did in the previous section for tracking. The timing information of a jet having $p_T>60{\rm~GeV}$ can be used and putting a cut on the BDT score calculated from the various timing variables of that jet, can help in selecting LLP signals and rejecting QCD backgrounds. 

We can also use the timing training of the BDT as an added background rejection on the triggers based on track variables. The 70\% background rejection point of the previous section had very high background rates of $\sim 103$ kHz ($\mathcal{R}_B$ of $p_T^{gen}\in(50,100){\rm~GeV}$ for $T_2^2$). If we select only those events passing the $T_2^2$ trigger from the tracking section and then apply the BDT training using timing variables, we find that the trigger rate reduces to $\sim 26.9$ kHz from $\sim 103$ kHz and the signal efficiency becomes $15.47\%$ from $19.79\%$. Therefore, the timing information can give us an extra reduction factor of the background rate.

We have discussed many aspects of using the timing layer at L1. Let us summarise them once again:

\begin{itemize}
\item Timing of a jet is defined as some statistical measure of all the MTD hits associated with a jet. We have used the median time.
\item Statistical measures are mostly contaminated by PU hits getting associated with a jet.
\item The time difference of hits also depend on the starting $z$ position of the prompt particle travelling at the speed of light which we use as a reference. 
\item Identification of PV at L1 is not very efficient at L1.
\item Association of MTD hits with L1 tracks, and removal of these hits from the median calculation helps improve the performance of timing variables.
\item Regional timing can be used as a separate trigger or can be used in combination with the trigger based on L1 tracking, where in the latter it can give an extra factor of background rejection for similar signal efficiency.
\end{itemize}

In summary to our study of dedicated triggers for LLPs, we find that both tracking and timing information can be used to design some triggers for long-lived particles that can serve complementary to standard triggers at L1. In the next section, we study how these variables perform in some different LLP scenarios. 

We have also redefined some of the tracking and timing variables after removing outliers in two passes, like the jet vertex position and median time of MTD hits, where in the former outliers are removed based on the track's $z$ position and in the latter time of the MTD hits are used. We find no significant difference in the results.

There are also many pile-up removal strategies for the efficient removal of pile-up. In the current LHC runs, with about 30-50 PU interactions per bunch crossing, a donut PU subtraction method is used at level-1 of CMS as described in \cite{Zabi:2016ljo}. One needs to check the efficiency of this PU subtraction method at L1 in the busy environment of HL-LHC, where there will be around 140-200 PU events per bunch crossing. In the High-Level triggers, one can use the full timing information for a four-dimensional vertex reconstruction, which can reduce the PU number to the level of current LHC runs. In \cite{Kreis:2018job}, they also talk about using PUPPI (PileUp Per Particle Identification) at level-1 in CMS for HL-LHC runs. However, this algorithm is based on identification of the primary vertex, which is not very efficient for LLP benchmarks, as we have seen earlier.

Towards the end of the HL-LHC runs, the peak instantaneous luminosity is proposed to increase further (to a value $\sim 7.5\times10^{34}{\rm~cm^{-2}s^{-1}}$) which will lead to around 200 PU vertices per bunch crossing. As we can recall, the jet multiplicity increases drastically from around 10 in the 140 PU scenario to about 40 for 200 PU when we use a cone size of $R=0.4$, even when the $p_T$ threshold is 60 GeV (see fig.\ref{fig:jet_multiplicity_PU}). Fig.\ref{fig:200PU} in appendix \ref{app:200PU} shows that even in the 200 PU scenario, considering narrow jets with $R=0.2$ can really help to maintain the 0 PU jet $p_T$ distribution to some extent. The matching between the 0 PU and 140 PU $p_T$ distribution for $R=0.2$ jets was better than in the 200 PU scenario. However, we expect that the analyses presented in this work can be extended in the 200 PU environment of the LHC as well with some minute changes (like increasing the $p_T$ threshold).

We now discuss some other LLP scenarios and check the performance of the above discussed tracking and timing variables in the context of these different LLP scenarios.

\section{Performance of the classifiers based on tracking and timing for some other LLP scenarios}
\label{sec:other_bench}

We have discussed the prospects of using various tracking and timing variables to design some dedicated triggers at L1 for the LLP scenario where a pair of LLPs is directly produced and then decay to jets.
In this section, we extend our analysis to some other LLP scenarios and briefly discuss the findings. We have chosen each of these scenarios with some motivation to discuss slightly varying aspects of LLPs. 

\subsection{Scenario with direct pair-production of LLPs and their decay to jets and invisible particle}
\label{ssec:scenario_B}

We started with the direct production of LLPs in quark-initiated processes and their further decay into jets. Within the same production mechanism of LLPs, we now consider a different decay mode. The LLP can decay into jets and invisible particles. In cases of prompt decay, such events will be characterised by high missing transverse energies ($\met$) due to the presence of the invisible particles in the final state. However, if such decays are displaced, the $\met$ distribution might be different. Therefore, we are interested to study such a benchmark to have an idea of the efficiency of the standard $\met$ trigger at L1 for LLPs decaying into one or more invisible particles. Following are the simulation details for this benchmark, which will hereafter be referred to as LLP scenario (B).

$$\mathbf{{\rm~(B)~~~}p p \rightarrow X X,\ X \rightarrow Y j j}$$

We generate this using \texttt{PYTHIA6} similar to scenario (A) where a pair of LLPs are directly produced and they decay to an invisible particle and two quarks, which hadronize to give jets. We study two different mass points of the invisible particle $-$ $300{\rm~GeV}$, and $400{\rm~GeV}$, with the LLP mass fixed to $500{\rm~GeV}$, and having two different proper decay lengths of $10{\rm~cm}$ and $100{\rm~cm}$ each. The LLP is assumed to decay to the aforementioned final state with $100\%$ branching.

\begin{figure}[hbt!]
\centering
\includegraphics[scale=0.39]{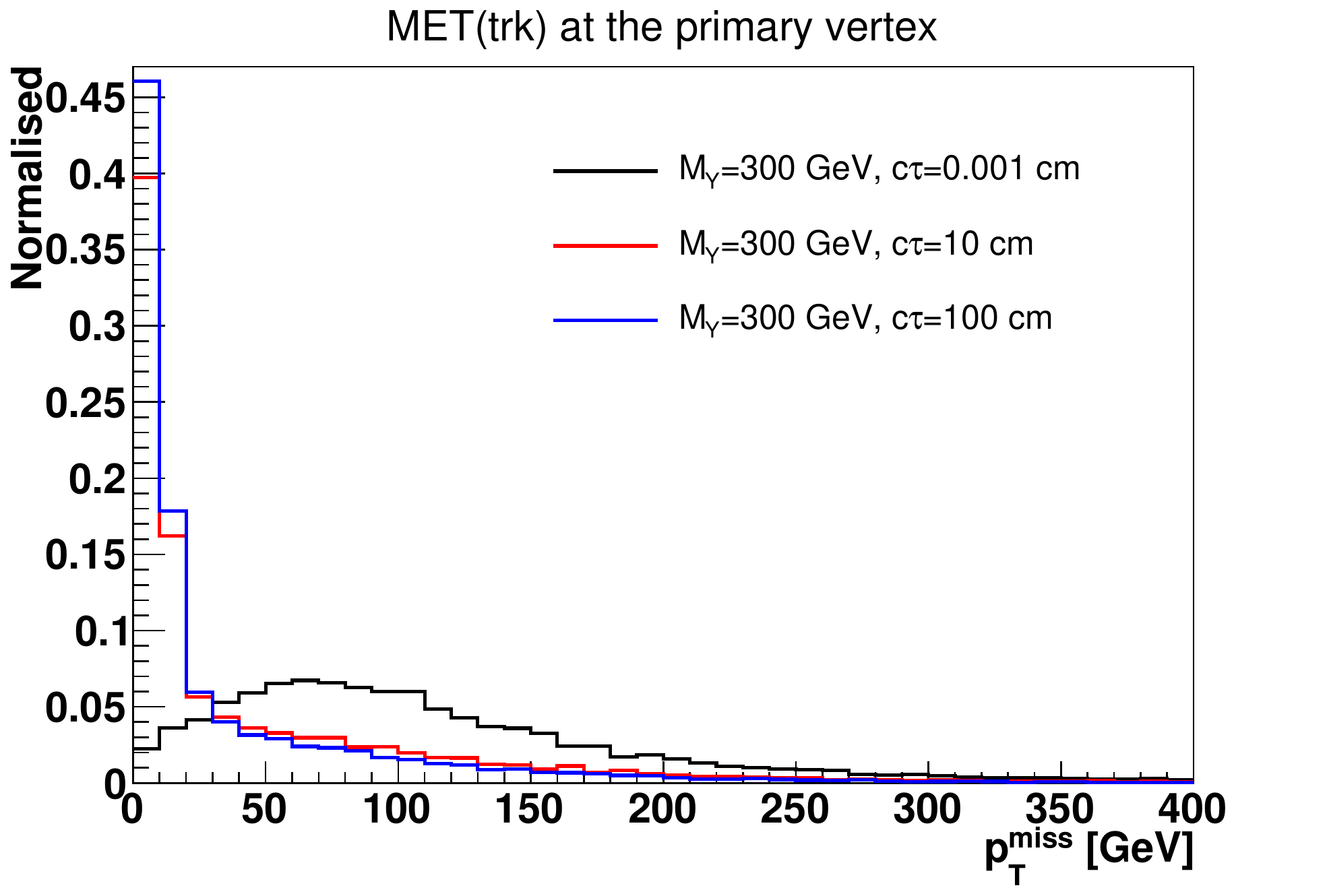}~
\includegraphics[scale=0.39]{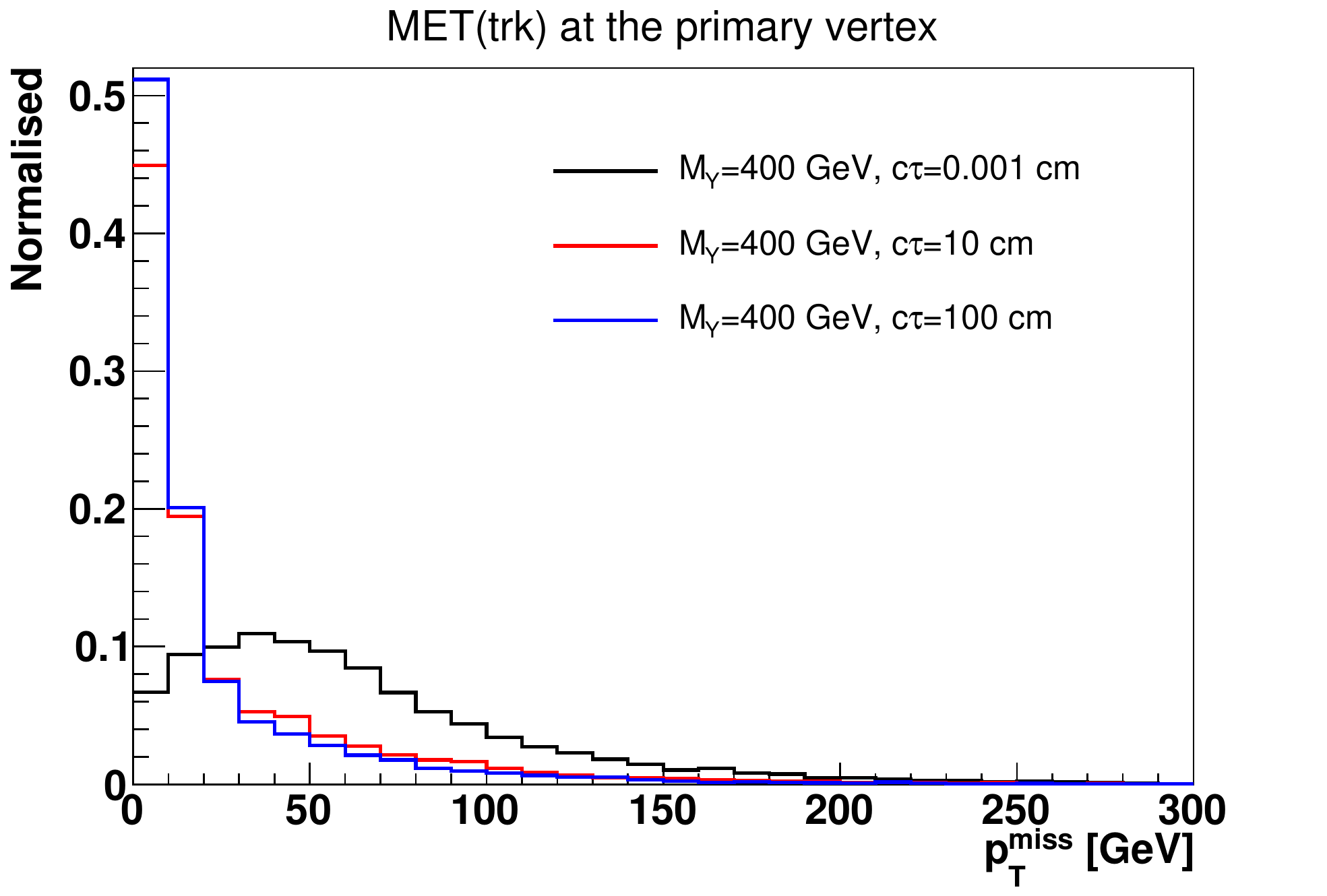}
\caption{Normalised distribution of $\met$(trk) calculated using L1 tracks from the identified primary vertex (PV) for prompt decay as well as for decay length values of 10 cm and 100 cm of particle $X$ of mass 500 GeV with the mass of the invisible particle ($M_Y$) being 300 GeV ({\it left}) and 400 GeV ({\it right}).}
\label{fig:ptmiss}
\end{figure}

Let us first start by looking at the $\met$ distribution of these benchmarks to see how it gets affected for displaced cases. We calculate $\met$ using tracks ($\met$(trk)) reconstructed at L1 which starting from the primary vertex~\footnote{The primary vertex is identified at L1, as discussed before, using the vertex with the maximum value of $\sum p_T^2/n_{z_a}$.}. 
Fig. \ref{fig:ptmiss} shows the distribution of $\met$(trk) for the particle $X$ having a mass of 500 GeV, decaying into jets and a missing particle of mass 300 GeV and 400 GeV.

From fig.\ref{fig:ptmiss}, we find that the $\met$(trk) distribution for prompt decay of $X$ does have a longer tail. However, when the LLP has a decay length of the order of few cm, there are no L1 tracks and as we have already discussed, the primary vertex identification is not correct. Therefore, most of the time, PU vertices get identified as primary vertex instead of the hard process and therefore, we find most of the events to have very low values of $\met$. Standard $\met$ triggers won't be efficient in selecting LLP events even if they decay into final states involving invisible particles.

\begin{figure}[hbt!]
\centering
\includegraphics[scale=0.195]{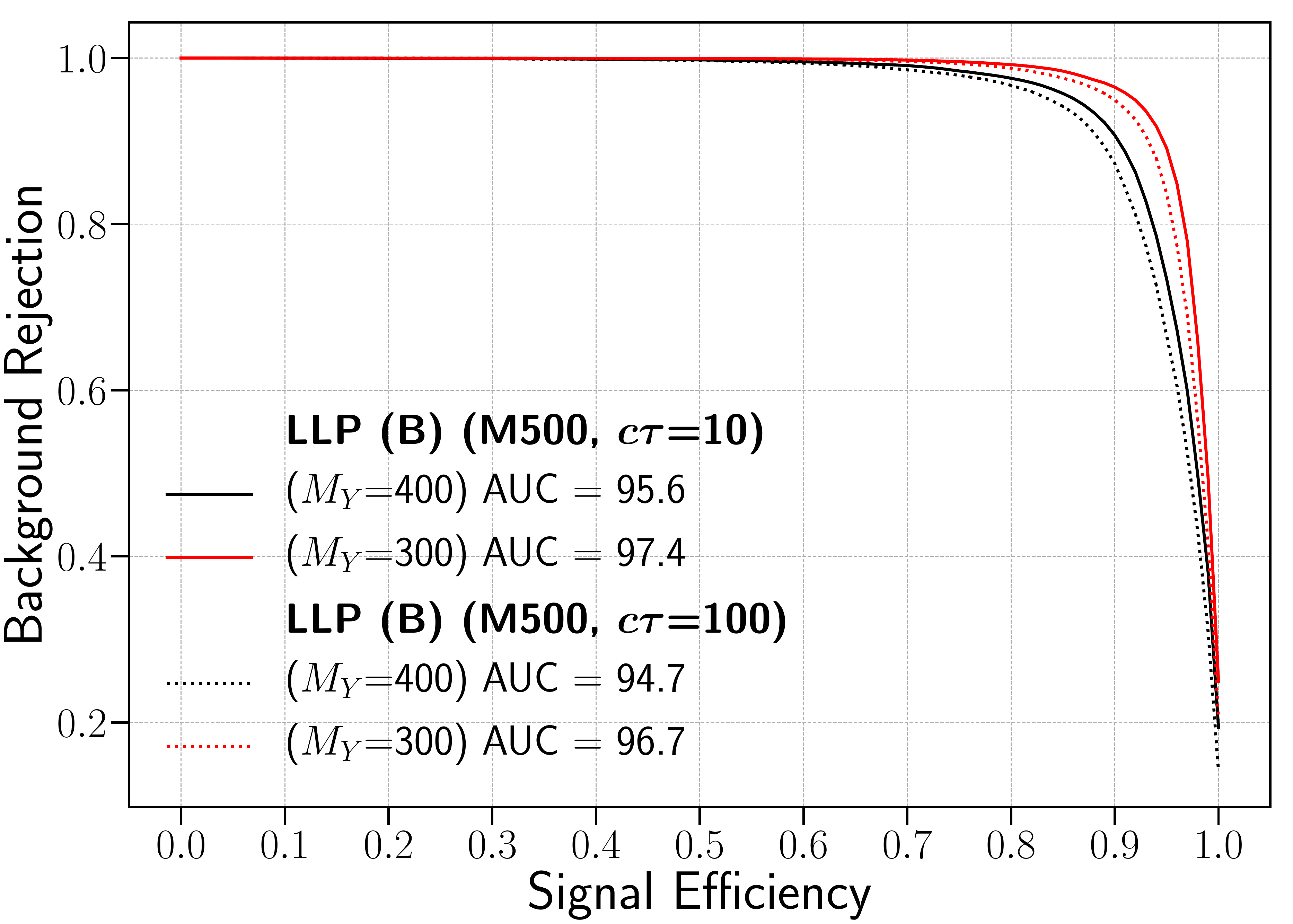}~
\includegraphics[scale=0.195]{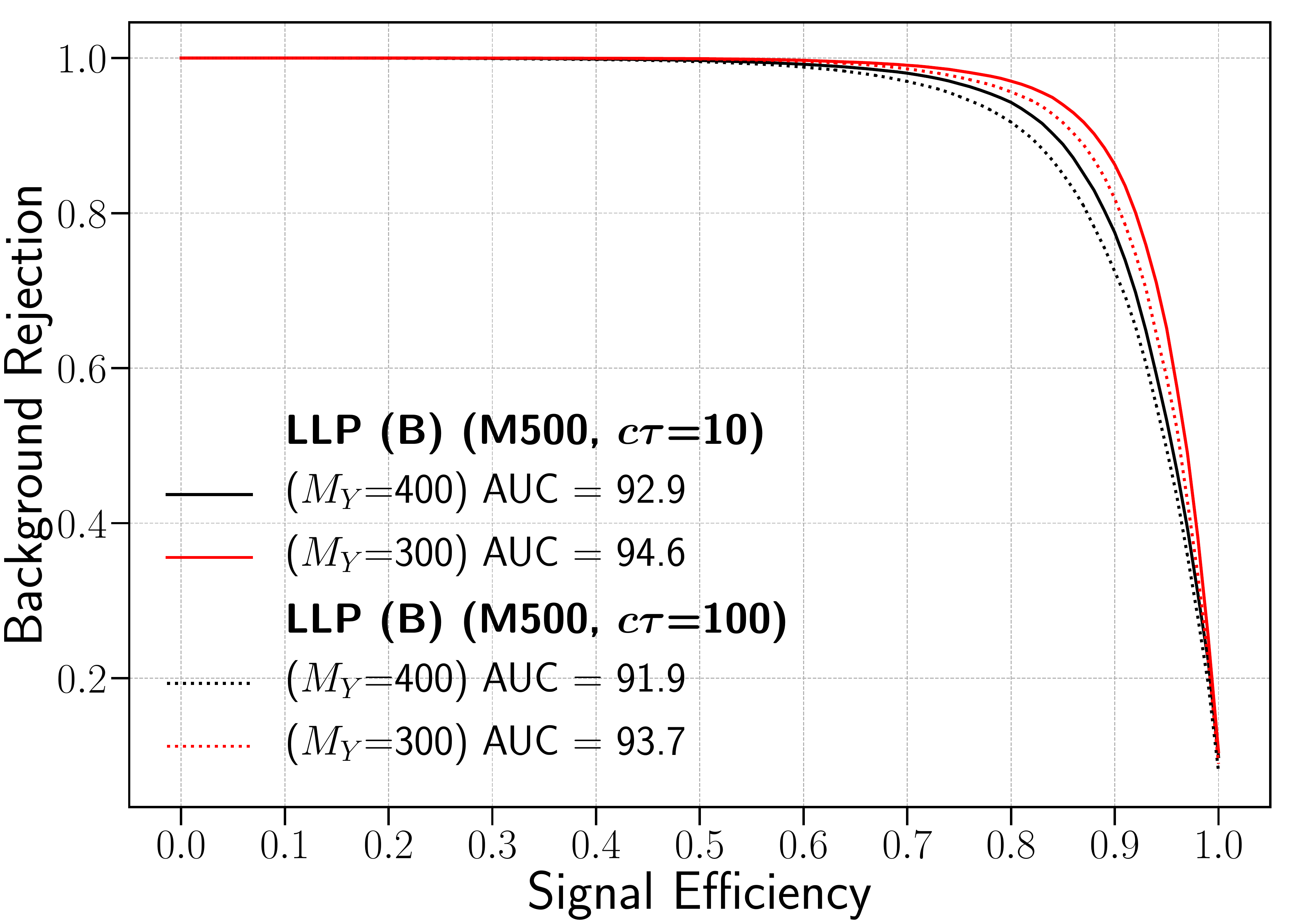}
\caption{ROC curves for selecting displaced jets coming from LLP decay from prompt QCD ones for four benchmark points from scenario (B) with mass of the LLP $M_X=500$ GeV having decay lengths 10 cm and 100 cm using tracking variables ({\it left}) and timing ({\it right}). Two different masses of the invisible particle $-$ 300 GeV and 400 GeV are considered.}
\label{fig:ROC_BP2_track_time}
\end{figure}

We now want to check how efficient L1 track and regional timing variables are in selecting events from this benchmark. The {\it left panel} of fig.\ref{fig:ROC_BP2_track_time} shows the classification ROC curve for jets coming from LLPs from different benchmark points of scenario (B) as signal and QCD prompt jets (merged sample of all $p_T^{gen}$ bins) as background using L1 track variables. The benchmark point from scenario (B) where mass of the invisible particle is 400 GeV (and $M_X=500$ GeV) is similar to the scenario (A) benchmark point where the LLP mass is 100 GeV $-$ in both cases the jets will have similar boost distribution. However, in this case, the boost of the LLP will be smaller since it has higher mass, and therefore, it decays more often within the tracker than in scenario (A). The ROC performance is, therefore, slightly better.
We expect the timing variables to perform slightly better (comparing scenario (A) where $M_X=100$ GeV with scenario (B) where $M_X=500$ GeV, $M_Y=400$ GeV) since the jets are now coming from the decay of a heavier particle, and hence, will take more time to reach the MTD. The {\it right panel} of fig.\ref{fig:ROC_BP2_track_time} shows the same classification ROC curve when timing information of these jets are used. 

\subsection{Scenario where LLPs are produced from the decay of Higgs boson and decay to jets}
\label{ssec:scenario_C}

LLPs can also be produced from the decay of an on-shell resonance, in which case the boost of the long-lived particles depend on the mass of the intermediate particle produced, in addition to its own mass. We now study this different type of production mode of LLPs, where they are produced from the decay of a SM Higgs boson, where the latter is produced in gluon-initiated process. We later also discuss briefly the case where the LLPs come from the decay of a heavy resonance, which can be a heavy Higgs boson or any other BSM particle. We consider the decay of LLPs into quarks only.

$$\mathbf{{\rm (C)~~~}p p \rightarrow h \rightarrow X X,\ X \rightarrow j j}$$

We generate the Higgs production, $p p \rightarrow h$, using \texttt{MG5\_aMC\_v2\_6\_6}\cite{Alwall:2011uj}, and then in \texttt{PYTHIA6} decay the Higgs to a pair of LLPs (with $100\%$ branching) and then these LLPs to quarks using similar coupling as in scenario (A). We study three different mass points, all having masses less than half the Higgs mass, and with two different proper decay lengths each $-$ $10{\rm~GeV}$, $30{\rm~GeV}$, and $50{\rm~GeV}$ having proper decay lengths of $10{\rm~cm}$ and $100{\rm~cm}$ each. The LLP is assumed to decay to light jets with $100\%$ branching.\\

Table \ref{tab:decay_frac_BP2} shows the decay fractions in various detector parts for benchmark points of scenario (C). In this benchmark, the LLPs are very light and their boost is controlled by the difference in their mass and the mass of the Higgs boson. Therefore, we find that the 50 GeV point has very less boost and decays mostly before the MTD.

\begin{table}[hbt!]
\centering
\begin{tabular}{|c||c|c|c|c|c|c|c||}
\hline
Mass [GeV], & Reco as & Before & Before & Before & Before & Inside & Outside \\
Decay Length [cm] & L1 tracks & MTD & ECAL & HCAL & MS & MS & detector\\
\hline\hline
10, 10 & 2.44 & 77.23 & 2.89 & 5.96 & 5.24 & 3.87 & 2.37\\
10, 100 & 0.25 & 18.28 & 1.90 & 6.14 & 16.06 & 18.40 & 38.98\\
30, 10 & 7.70 & 88.34 & 0.78 & 1.19 & 1.02 & 0.77 & 0.20\\
30, 100 & 0.85 & 44.58 & 3.56 & 9.71 & 16.28 & 12.32 & 12.71\\
50, 10 & 17.23 & 81.26 & 0.37 & 0.48 & 0.40 & 0.24 & 0.02\\
50, 100 & 1.97 & 64.92 & 3.53 & 7.27 & 8.20 & 7.70 & 6.41\\
\hline\hline
\end{tabular}
\caption{Fraction of decays for scenario (C) in various detector parts.}
\label{tab:decay_frac_BP2}
\end{table}

\begin{figure}[hbt!]
\centering
\includegraphics[scale=0.195]{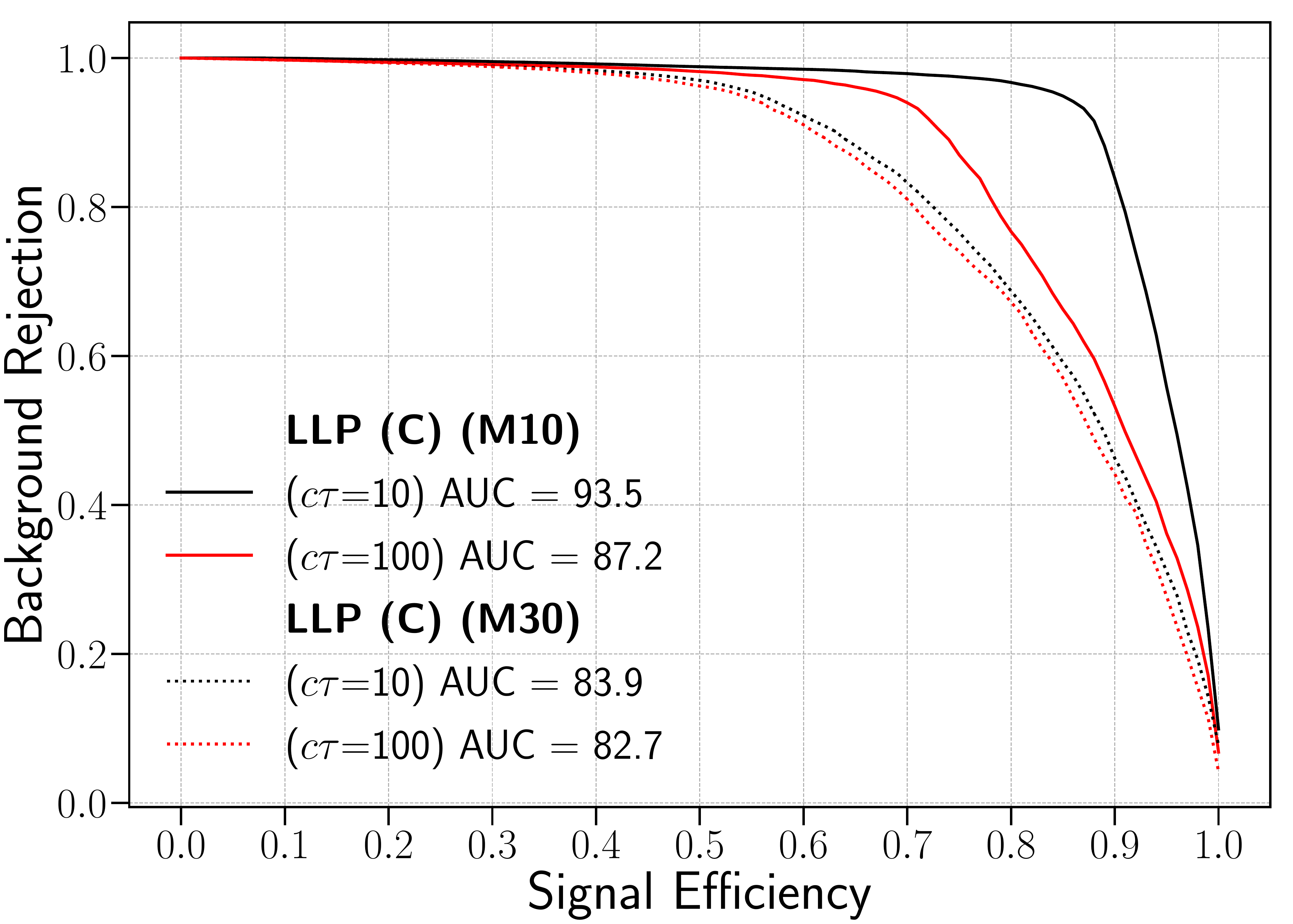}~
\includegraphics[scale=0.195]{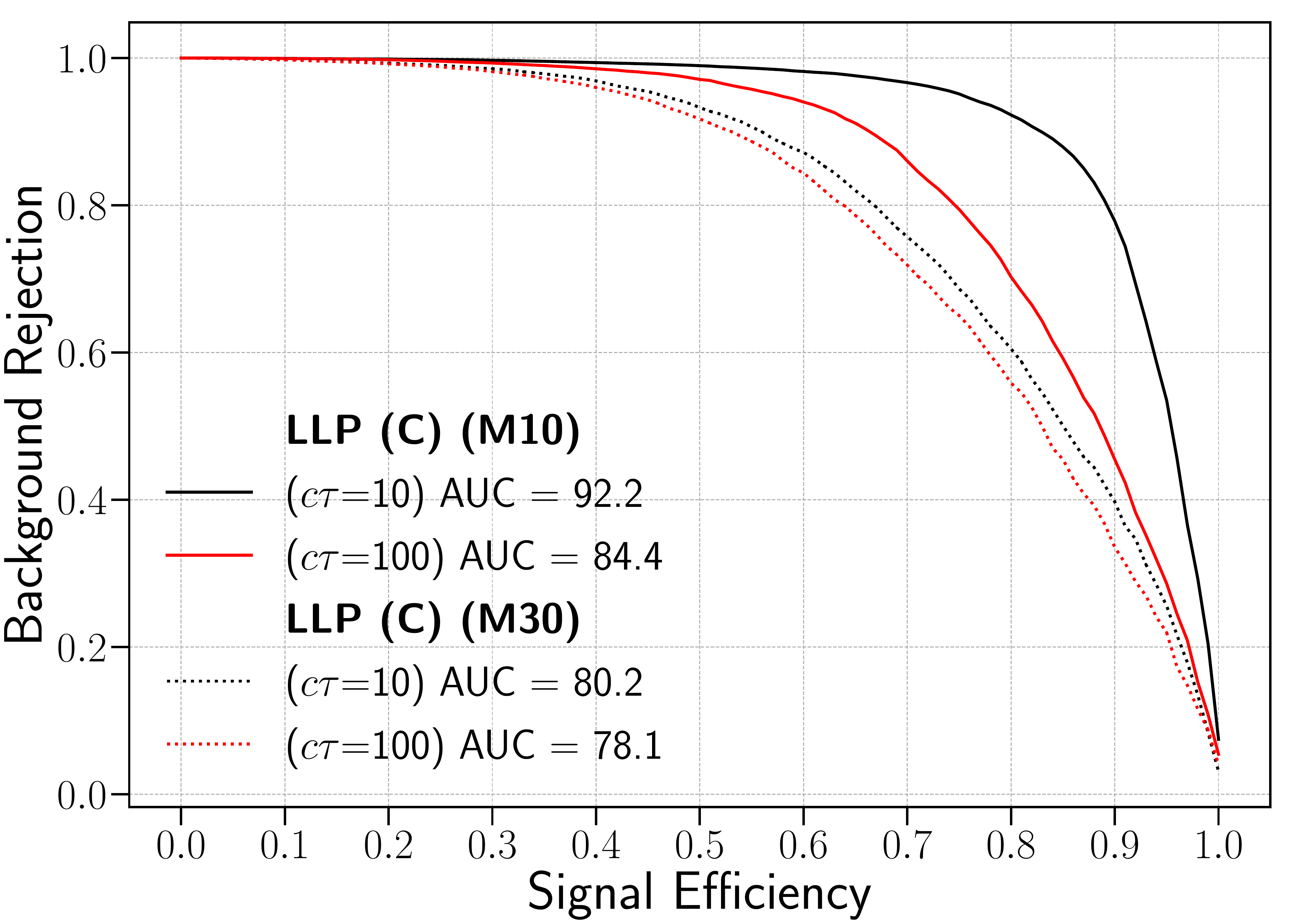}
\caption{ROC curves for selecting displaced jets coming from LLP decay from prompt QCD ones for four benchmark points from scenario (C) with LLP masses 10 GeV and 30 GeV and decay lengths 10 cm and 100 cm using tracking variables ({\it left}) and timing ({\it right}).}
\label{fig:ROC_BP3_track_time}
\end{figure}

We apply our BDT classification using track and time variables defined at L1 as discussed above and check their performance in classifying jets from benchmarks from scenario (C) and QCD background jets. The ROCs are shown in fig.\ref{fig:ROC_BP3_track_time}. The performance degrades than the scenario (A) and scenario (B) cases, as is expected, since these are very light LLPs and have very low jet multiplicities above our $p_T$ threshold of $60{\rm~GeV}$. Lowering the $p_T$ threshold requires taking into account QCD background with $p_T$ bins less than $50{\rm~GeV}$ at generation level, which again have huge cross sections and can increase the rates to very high values. Also, the pile-up rates are high at lower $p_T$ values. The ROCs are almost flat till a signal efficiency of about $70-80\%$ for LLP having a mass of 10 GeV, and we can get reasonable QCD rejection ($\sim 95\%$) at these values of signal efficiencies, which will help in increasing sensitivity to lower LLP masses. For LLP mass of 30 GeV, there are even less chances of having jets with $p_T>60{\rm~GeV}$, and the performance also degrades.

\begin{figure}
\centering
\includegraphics[scale=0.39]{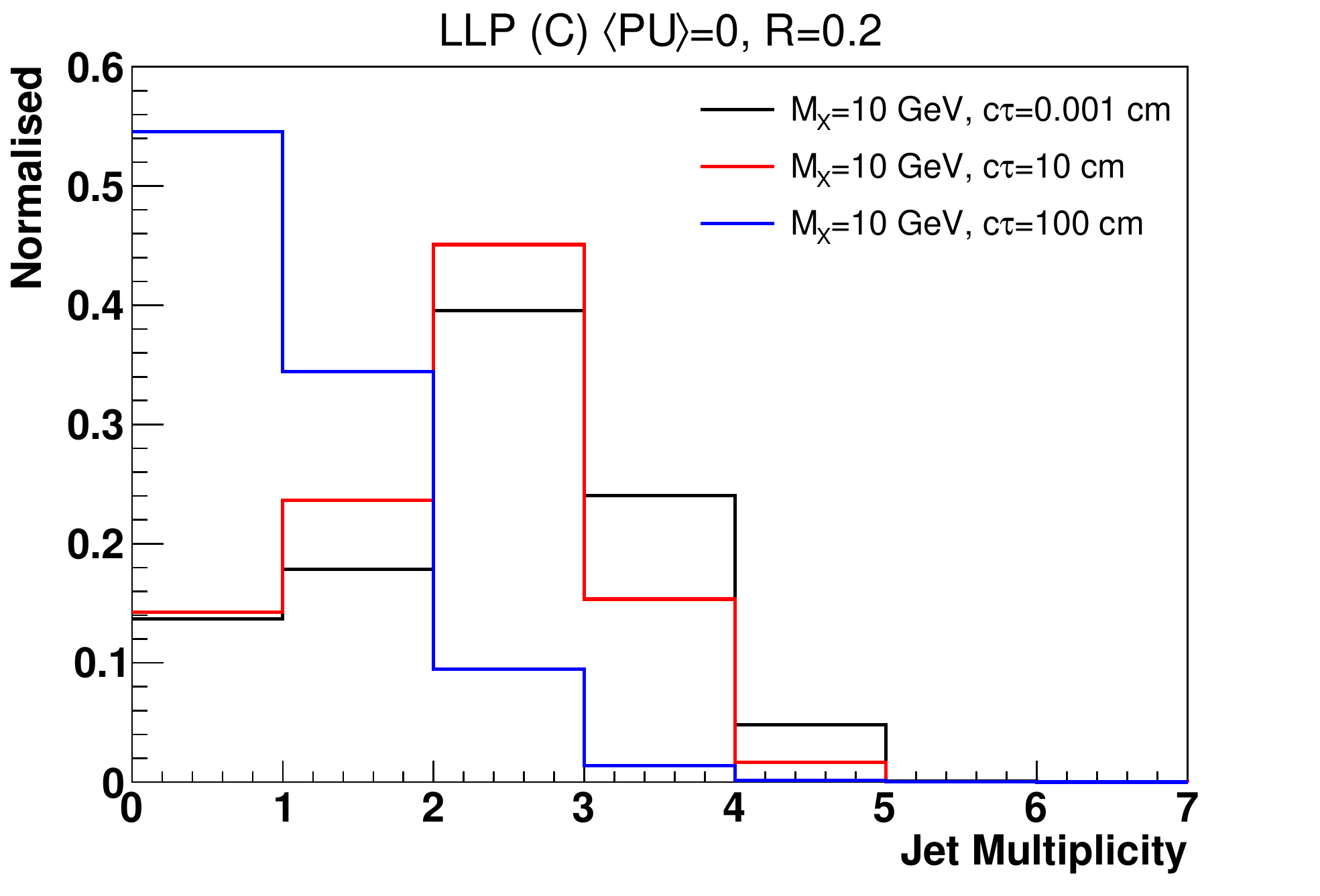}~
\includegraphics[scale=0.39]{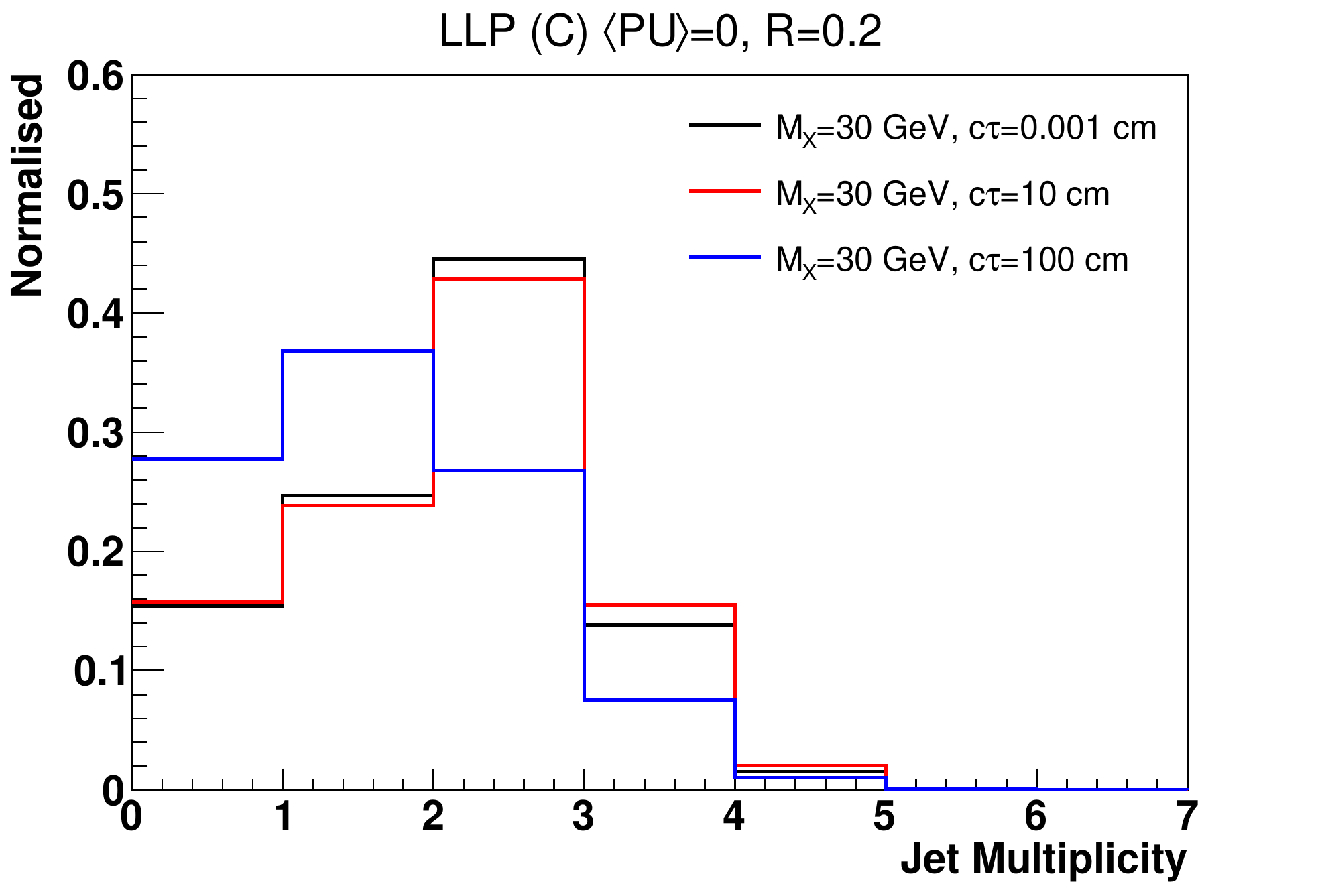}
\caption{Distributions of jet multiplicity for LLPs of mass 10 GeV ({\it left}) and 30 GeV ({\it right}) with varying decay length.}
\label{fig:jet_multiplicity_BP3}
\end{figure}

Another observation is that the performance degrades faster from decay length 10 cm to 100 cm when the mass of the LLP is 10 GeV than when it is 30 GeV. This is due to the fact that the 10 GeV mass LLP has very high boost values due to its smaller mass which when has a large decay length, will mostly decay outside the HCAL, and hence, the jet multiplicity will be low. We compare in fig.\ref{fig:jet_multiplicity_BP3} the jet multiplicity distribution of the 10 GeV ({\it left}) and 30 GeV ({\it right}) LLPs with varying decay lengths. We observe that the jet multiplicity changes drastically for the 10 GeV LLP from 10 cm decay length to 100 cm, however, this change is not so drastic in the 30 GeV LLP case. 


We now will discuss the effect when an ISR jet is present in such a process. The ISR jet increases the boost of the LLPs, and we can get more $p_T>60{\rm~GeV}$ jets. However, an increased boost will also mean increase in the decay length in the lab frame, and lesser number of jets within the first layer of the HCAL.  
Table \ref{tab:decay_frac_BP2_ISR} shows how these decay fractions get modified when the Higgs is produced with an ISR jet of $p_T>50{\rm~GeV}$. 

\begin{table}[hbt!]
\centering
\begin{tabular}{|c|c|c|c|c|c|c|c||}
\hline
Mass [GeV], & Reco as & Before & Before & Before & Before & Inside & Outside \\
Decay Length [cm] & L1 tracks & MTD & ECAL & HCAL & MS & MS & detector\\
\hline
10, 10 & 2.27 & 71.63 & 3.20 & 7.14 & 8.33 & 4.83 & 2.61\\
10, 100 & 0.24 & 16.55 & 1.72 & 5.39 & 14.83 & 17.45 & 43.82\\
30, 10 & 6.53 & 88.21 & 1.07 & 1.80 & 1.39 & 0.81 & 0.18\\
30, 100 & 0.70 & 39.67 & 3.28 & 9.23 & 17.34 & 14.48 & 15.31\\
50, 10 & 12.90 & 85.34 & 0.49 & 0.58 & 0.44 & 0.21 & 0.03\\
50, 100 & 1.40 & 55.70 & 3.59 & 8.86 & 13.47 & 9.63 & 7.35\\
\hline\hline
\end{tabular}
\caption{Fraction of decays for scenario (C) with an ISR jet of $p_T>50{\rm~GeV}$ in various detector parts.}
\label{tab:decay_frac_BP2_ISR}
\end{table}

The benefits of adding an ISR jet will be two-folds. First, if the ISR jet is hard enough ($p_T>173{\rm~GeV}$), it will help the event pass the single jet trigger. Also, it will increase the efficiency of identifying the correct primary vertex, which might improve the timing variables. However, it comes with the cost of reduced cross section. As you increase the $p_T$ of the ISR jet, the cross section falls off drastically. The second column of table \ref{tab:isr_PV_cs} shows how the efficiency of identifying the correct PV increases when the $p_T$ of the ISR jet is high, and the third column shows the decrease in the the production cross section. To pass the single jet trigger, we need an ISR jet having at least $p_T>173{\rm~GeV}$, and from the last row of table \ref{tab:isr_PV_cs}, we find that the cross section for such high $p_T$ jets is really small and we will only be able to trigger on 3.76\% of such events.

\begin{table}
\centering
\begin{tabular}{|c|c|c||}
\hline
	& max($\sum p_T^2/n_{z_a}$) corresponds &  \multirow{2}{*}{$\frac{\sigma(gg\rightarrow h+1j)}{\sigma(gg\rightarrow h)}$}\\
	&  to hard collision & \\\hline
$+0j$ & 14.14\% & $-$\\
$+1j$, $p_T^j>50{\rm~GeV}$ & 68.26\% & 25.73\%\\
$+1j$, $p_T^j>100{\rm~GeV}$ & 84.82\% & 8.81\%\\
$+1j$, $p_T^j>150{\rm~GeV}$ & 93.54\% & 3.76\%\\
\hline\hline
\end{tabular}
\caption{Efficiency for correctly identifying the vertex corresponding to the hard collision, from where LLPs are produced, as the primary vertex (PV) at L1 for scenario (C) without and with an ISR jet, with increasing the minimum $p_T$ of the latter. Also quoted are the production cross sections and the fraction of this cross section to the total cross section, without any hard ISR jet.}
\label{tab:isr_PV_cs}
\end{table}

\begin{figure}[hbt!]
\centering
\includegraphics[scale=0.195]{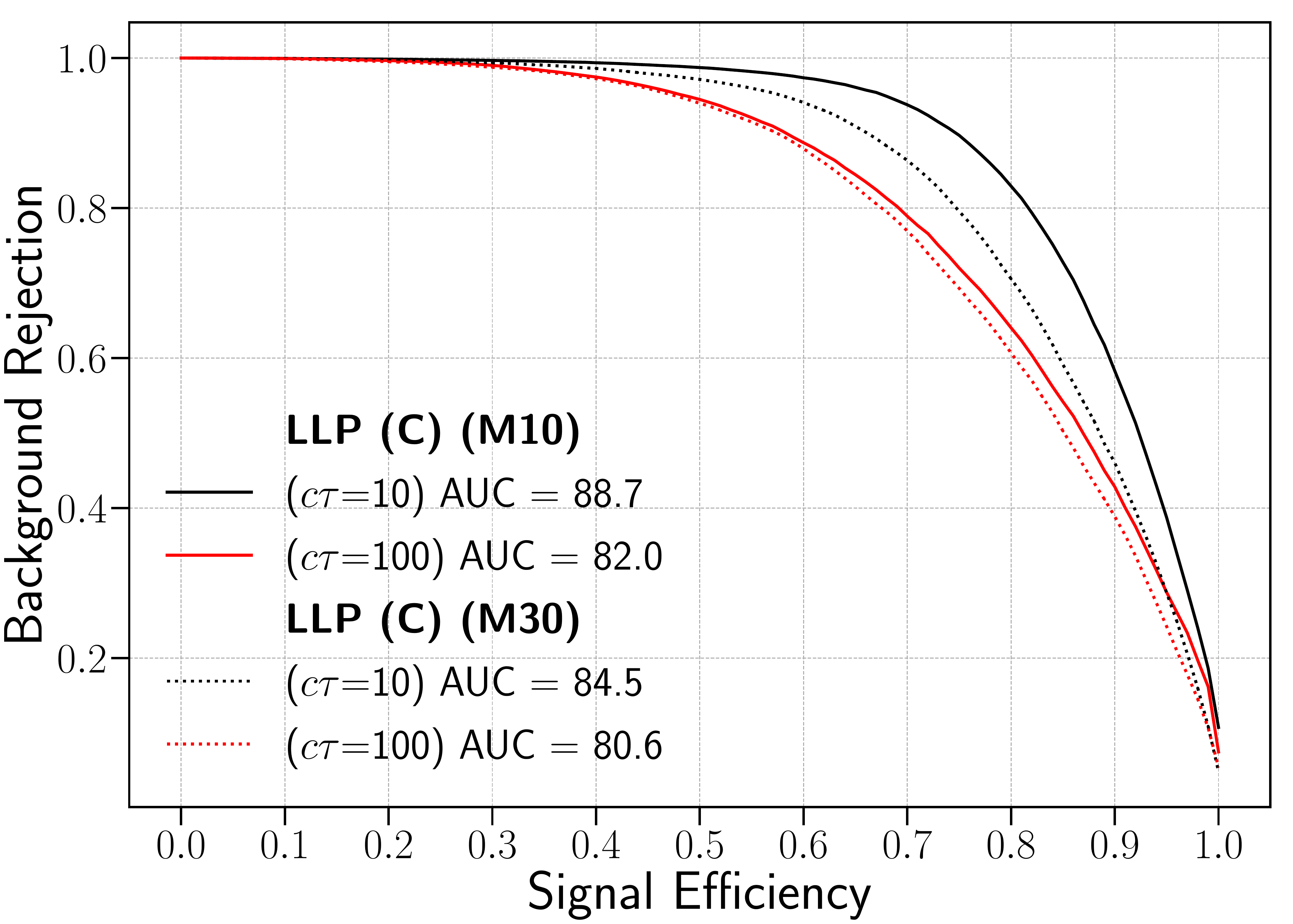}~
\includegraphics[scale=0.195]{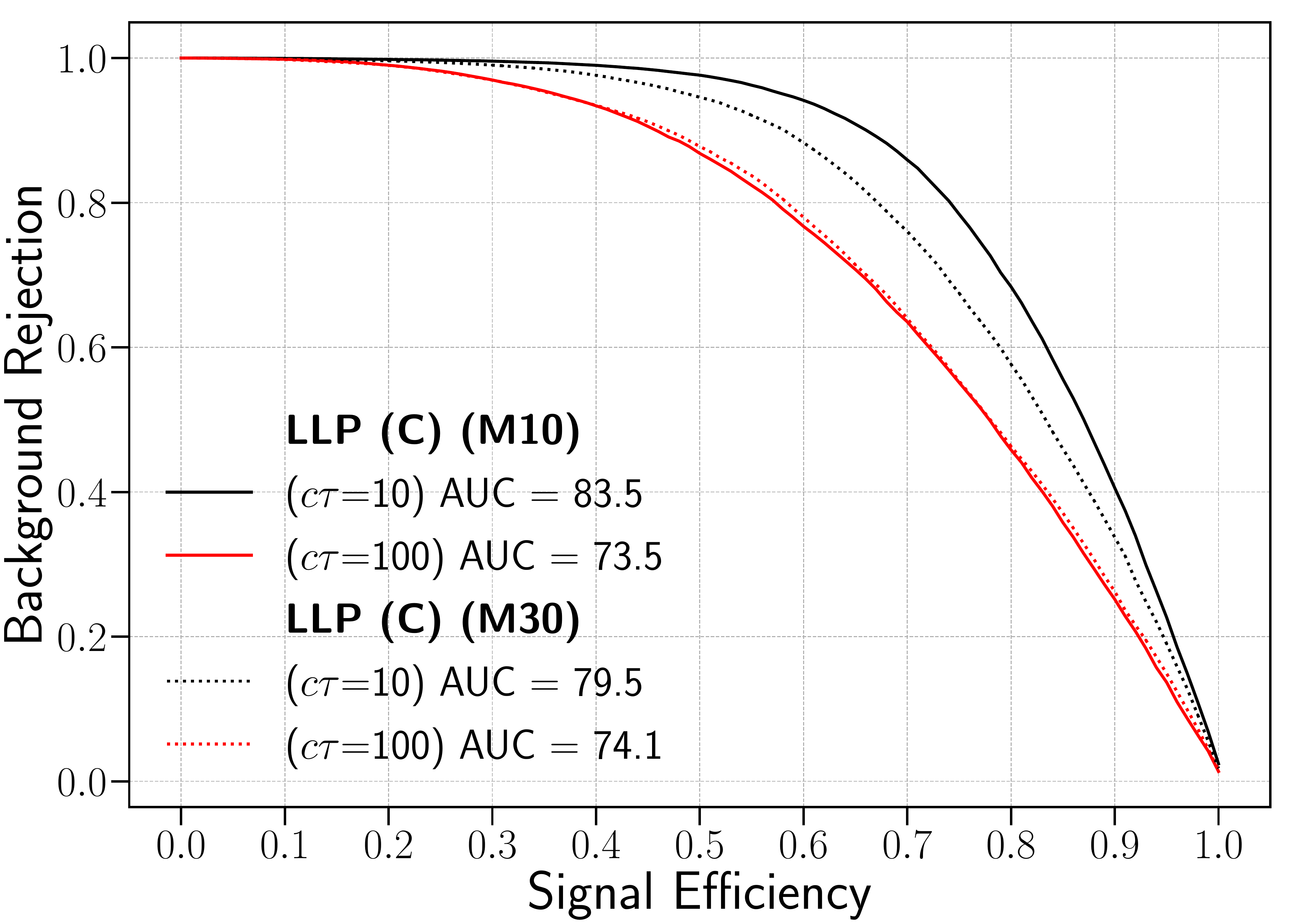}
\caption{ROC curves for selecting displaced jets coming from LLP decay from prompt QCD ones for four benchmark points from scenario (C) with LLP masses 10 GeV and 30 GeV and decay lengths 10 cm and 100 cm using tracking variables ({\it left}) and timing ({\it right}) when we generate the LLP processes with an initial state radiated jet having $p_T>50{\rm~GeV}$.}
\label{fig:ROC_BP3_ISR_track_time}
\end{figure}

\begin{figure}[hbt!]
\centering
\includegraphics[scale=0.185]{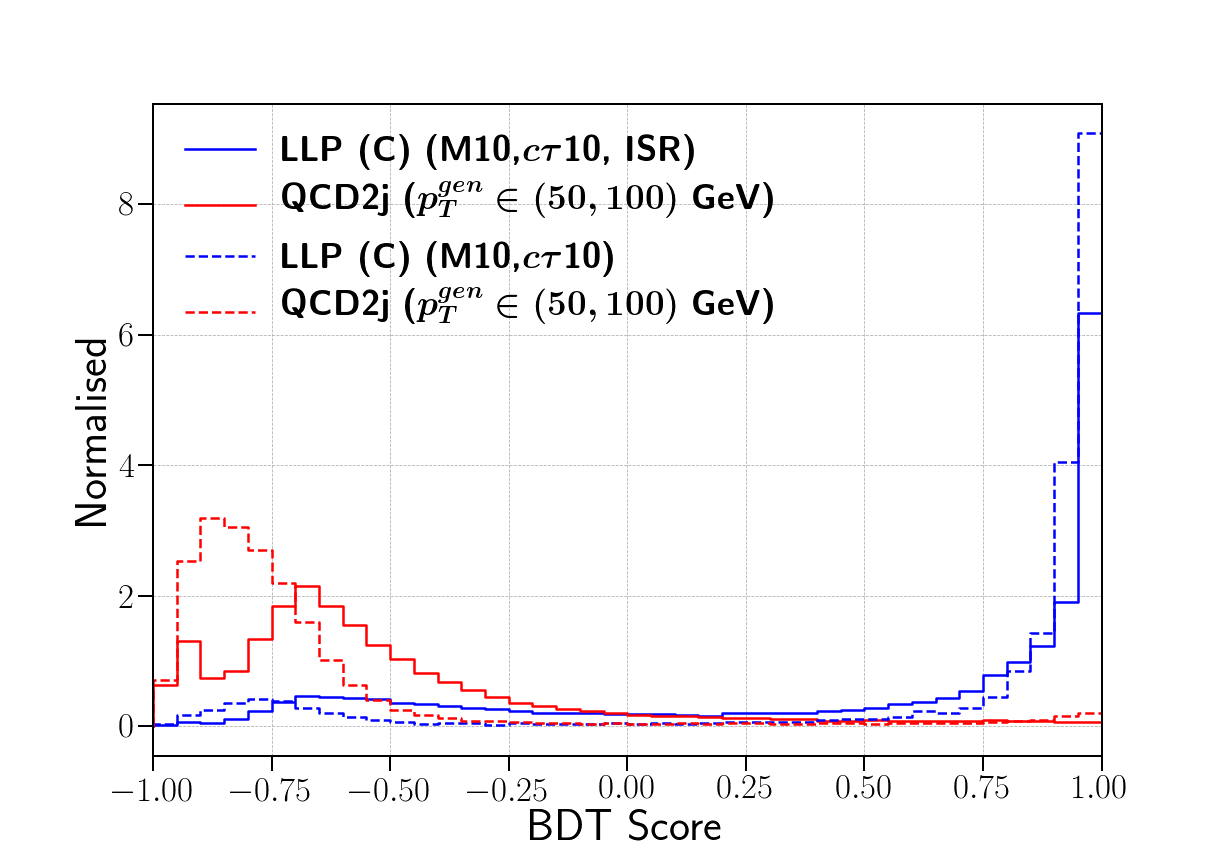}
\includegraphics[scale=0.185]{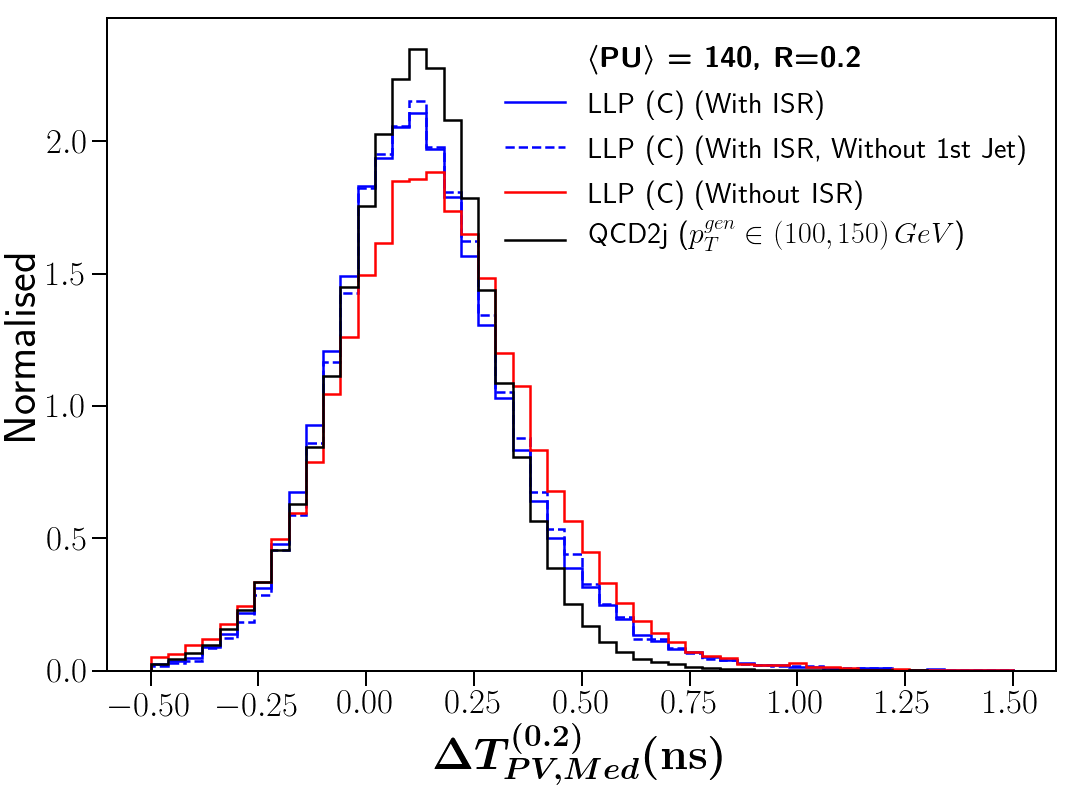}
\caption{Distribution of BDT scores when trained using track variables ({\it left}) and $\Delta T_{\text{PV,Med}}^{(0.2)}$ ({\it right}) for prompt and displaced jets from an LLP with mass 10 GeV and decay length 10 cm produced with and without an ISR jet (scenario (C)).}
\label{fig:BP3_BDT_Med}
\end{figure}

We use the sample with an ISR jet of $p_T>50{\rm~GeV}$ to check how including the ISR affects the BDT performance. Fig.\ref{fig:ROC_BP3_ISR_track_time} shows the BDT performances for the different benchmark points from scenario (C) when generated with an ISR jet using track ({\it left}) as well as timing ({\it right}) variables. We find that the performance degrades when the event has an ISR jet in both the track and timing ROCs, contrary to what we had expected since in these events we have better efficiency of identifying the primary vertex. The ISR jet resembles the prompt QCD jet, which is our background and therefore, the training is affected. This can be seen in the {\it left panel} of fig.\ref{fig:BP3_BDT_Med} where we have compared the BDT score distributions for the signal and background when trained with LLP benchmark point from scenario (C) where mass and decay length of the LLP is 10 GeV and 10 cm respectively, generated with and without the ISR. Also, the ISR increases the boost of the LLP and therefore, the timing also shifts to lower values, as can be seen in fig.\ref{fig:BP3_BDT_Med}.

\subsection{Scenario where LLPs are produced from the decay of a heavy resonance and decay to jets}
\label{ssec:scenario_D}

The LLPs can also be produced from the decay of a heavy resonance which could be a new BSM particle. We consider such a case where the mass of the intermediate resonance is 300 GeV and the LLPs which come from its decay have a mass of 10 GeV and decay length of 10 cm, and this scenario is denoted as scenario (D). The LLPs in this case will be highly boosted since they are light and come from the decay of a heavy particle.

\begin{figure}[hbt!]
\centering
\includegraphics[scale=0.2]{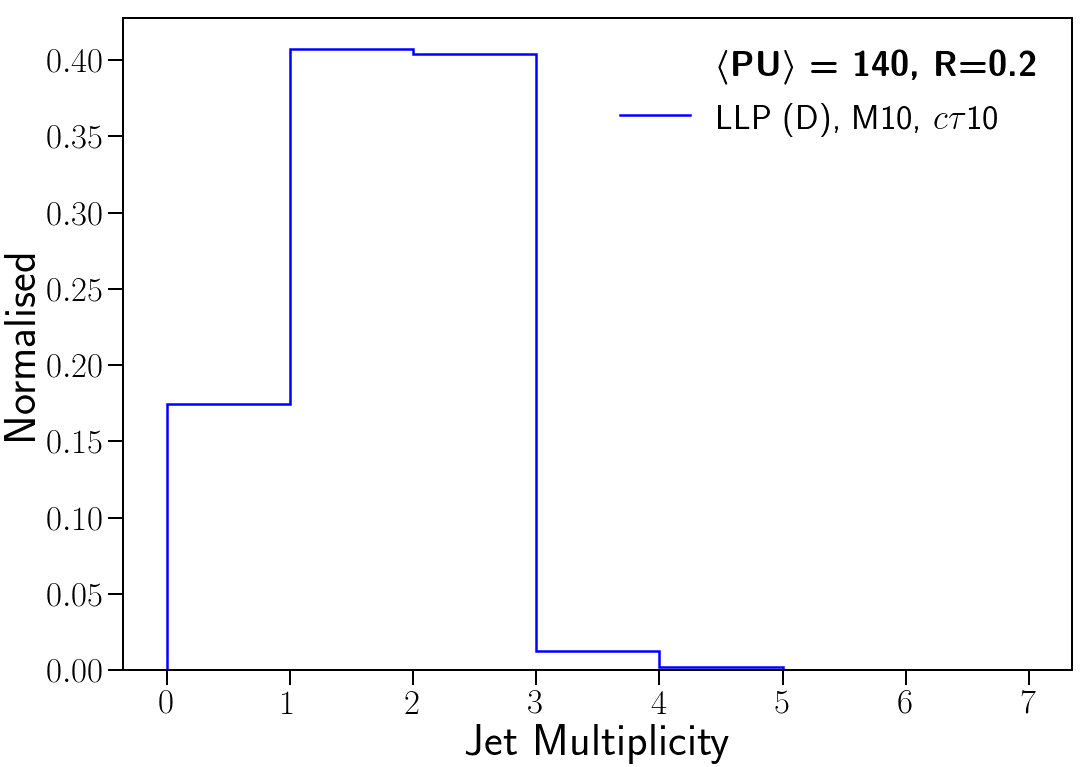}~
\includegraphics[scale=0.2]{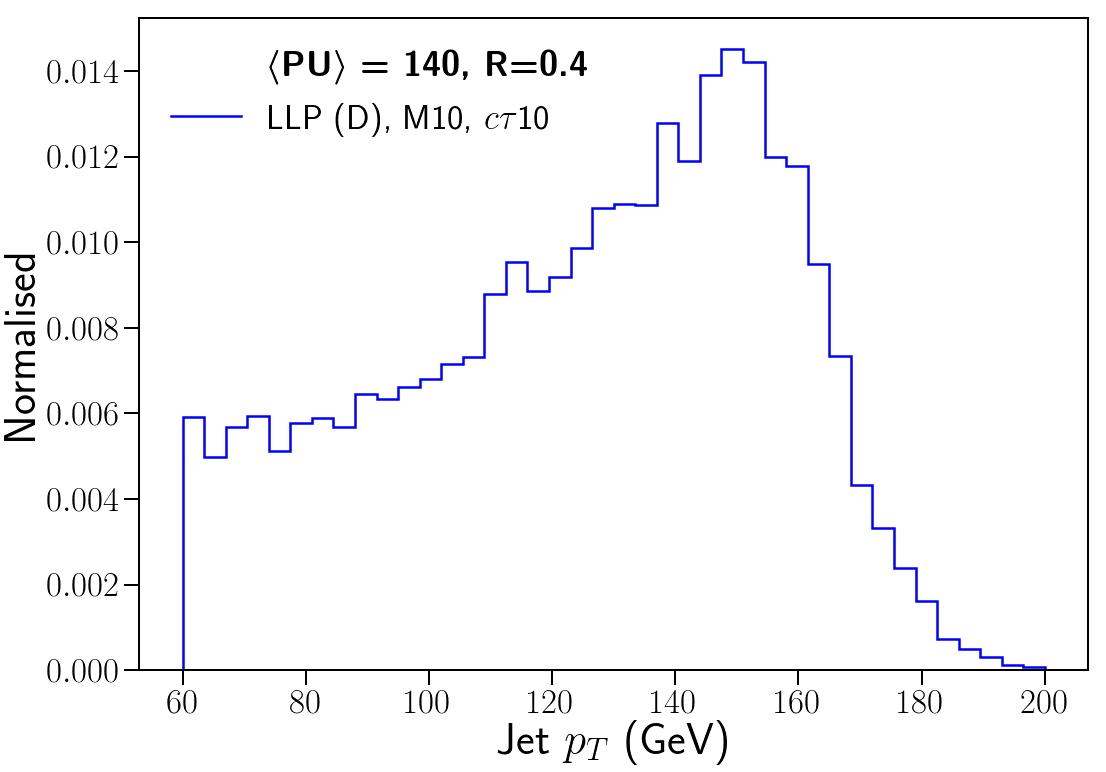}\\
\includegraphics[scale=0.2]{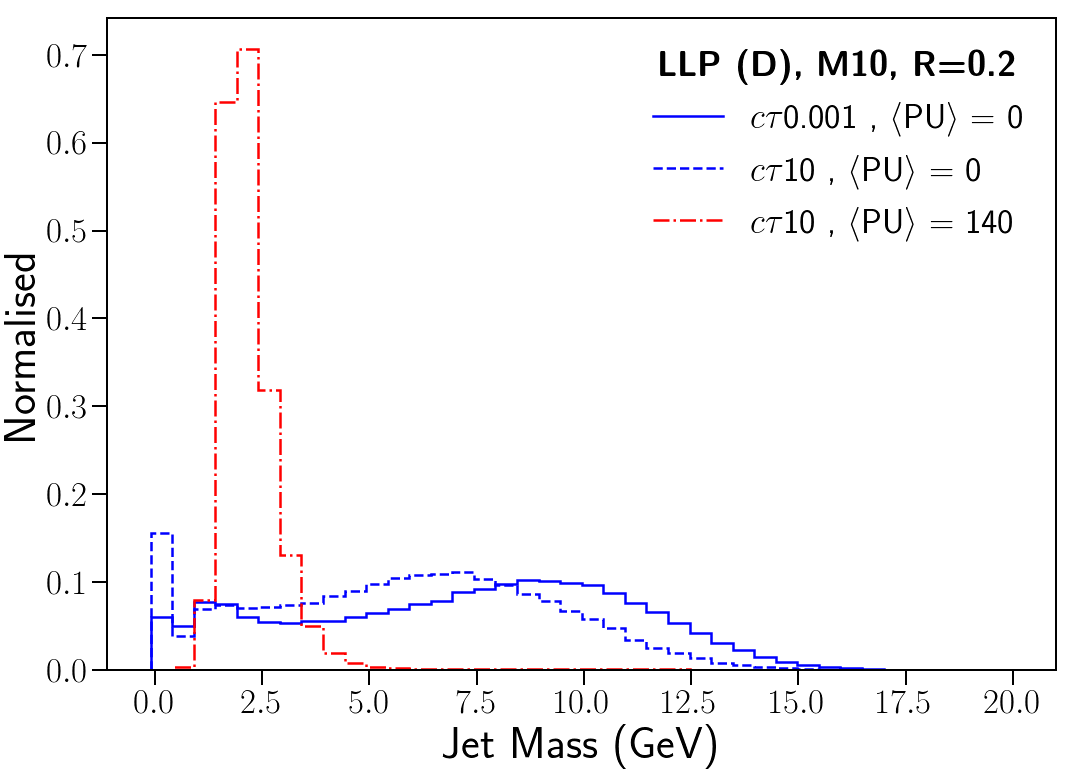}
\caption{{\it Top:} Jet multiplicity, jet $p_T$, and {\it bottom:} jet mass distribution for the case when LLPs (of mass 10 GeV and decay length 10 cm) come from the decay of a heavy resonance of mass 300 GeV.}
\label{fig:BP4_distributions}
\end{figure}

Fig.\ref{fig:BP4_distributions} shows the multiplicity, $p_T$ and jet mass distributions of jets coming from an LLP of mass 10 GeV and decay length 10 cm, which comes from the decay of a 300 GeV resonance. As we discussed earlier, in this case the LLPs will be highly boosted, which will increase their decay length in the lab frame. Therefore, the decay products can be close enough to get identified as a single jet. As we can see from the multiplicity plot, most of the time the multiplicity is one or two, which is due to increased probability of merging of two of the jets coming from the LLP decay. 

The jet $p_T$ also peaks around 150 GeV, which is half the heavy resonance mass, and in most cases the $p_T$ of the LLP will be close to this value. This also implies that both the jets coming from the LLP decay is identified as a single jet. 

We therefore, expect a peak around 10 GeV in the jet mass distribution. In case of no displacement, we do find a small peak around that value, since the jets are boosted enough that they can be contained within $R=0.2$ in some cases, even with very low $c\tau$ value. However, with increasing $c\tau$, we do not find any resonance peak around 10 GeV. This is due to the mismatch of the jets' actual $\eta-\phi$ with the one that is measured in the detector, which gives a wrong estimate of the jet mass. 

We also notice an important feature that with increasing $c\tau$ the jet mass is mostly close to zero, which happens when the LLP is displaced enough such that all its constituents fall within a single tower, and the mass of a single tower is close to zero.
However, with the addition of 140 PU events, this effect goes away. 

The jet $p_T$ for this scenario is mostly on the higher side and can therefore, pass the standard single jet trigger. We therefore, do not present our analysis based on tracking and timing at L1 for this scenario.


\section{Conclusions and Outlook}
\label{sec:concl}

Long-lived particles, strongly motivated from many BSM theories, that travel some distance in the detector before decaying to SM particles, provide
an compelling substitute to prompt hypothetical particles, especially, considering the absense of any discovery of BSM particles at the current LHC
experiments.
Models including LLPs could address important questions still unresolved by the SM.
Unique signatures of LLPs offer staggering scope for discovery of physics beyond the standard model in the LHC experiments, but simultaneously they require specific and dedicated triggering strategies since they can be easily missed. Since their production cross sections depend on various model parameters, they can also have really small values, and in that case, we need to ensure that the triggers are efficient enough so that we don't miss them even at the first level. 

The LHC experiments are going through a significant Phase II upgrade scheme to get ready for the gruelling conditions of HL-LHC. For CMS, a new timing sub-detector is being built to measure the timing of electrically charged particles with excellent precision. The time information will diminish the adverse effects of the elevated levels of PU anticipated at the HL-LHC. The timing information associated to each track will allow the use of 4-dimensional vertex reconstruction and will help recovering the current conditions. At the L1 trigger, it is expected that regional timing information will be available. Moreover, the increase of instantaneous luminosity at the HL-LHC will necessitate the introduction of tracker information, for the first time, in the level-1 trigger system to retain an acceptable trigger rate.

In this work we have shown that the usual
level-1 triggers targeting prompt particles, will be inefficient for
LLP searches in some scenarios, for example, when the light LLPs are pair-produced directly and decay into jets, in the high PU environment of HL-LHC. This points to the need to develop
dedicated L1 triggers for LLP searches. In our paper, we have presented the studies in the context of CMS detector and its upgrade
in HL-LHC, however, all the ideas should be qualitatively valid for ATLAS detector as well. 

We begin by studying how the jet distributions of both signal and background events change in the high PU scenarios, and how this effect can be minimised by considering narrow jets. For LLPs as signal, considering narrow jets do not affect the jet distributions as much as it affects the background jets, since displaced jets are physically more contained in smaller regions. However, this is not enough to suppress the huge QCD background and one has to use other features of displaced jets to differentiate them from the prompt ones.

In Phase-II the FPGAs used in L1 trigger hardware will be sophisticated enough to handle small scale machine learning(ML) applications.
In the context of MIP timing detector and L1 track triggers, the two important upgrades of CMS experiment in phase II of LHC, we have explored the performance of 
CMS in differentiating between prompt and displaced jets, the former coming from QCD dijet background and the latter from LLP decays. 
We have constructed several variables related to L1 tracks and timing, and used them to train two boosted decision trees, one for track variables, and another for timing variables. We found that the track and timing variables are
mostly uncorrelated and the individual BDTs produce similar performances. 
Our study has shown that it is possible to develop some dedicated LLP triggers based on the BDT classification which can help increase the trigger efficiency of events with LLPs and reasonably moderate background rates, even when we lower the $p_T$ threshold.
Also, the trigger based on timing can be used for an extra suppression of the background rate, after using the track based trigger. 

We have also presented a discussion on some other LLP scenarios. We have showed that when the LLP decays to some invisible particles in addition to jets, the $\met$ distribution as calculated using the L1 tracks will change completely when the decay length of the LLP is increased, and therefore, triggers based on $\met$ won't be efficient. However, the track and timing based triggers have promising results in selecting such events. We have also discussed the effect of an ISR jet produced in association with light LLPs on the trigger efficiencies.

We conclude that using variables related to track and timing will greatly improve the capability to trigger on LLP in L1 triggering stage in the high PU environment of HL-LHC. We have studied in detail how one can utilise L1 tracks and timing information to develop dedicated BDT based triggers for LLPs. One has to explore how more complex avenues of ML tools can be implemented at the L1 trigger and how they can improve the performance of dedicated LLP triggers, with our results as the starting point.

\subsection*{\textit{Acknowledgement}}
The work of B.B. is supported by the Department of Science and Technology, Government of India, under Grant No. IFA13- PH-75 (INSPIRE Faculty Award). 
The work of S.M. is supported by the German Federal Ministry of Education and Research BMBF. 
R.S. and P.S. would like to thank Rahool Kumar Barman and Amit Adhikary for useful discussions.

\newpage

\appendix

\section{Appendix}

\subsection{Distributions of tracking variables}
\label{app:track_vars}

\begin{figure}[hbt!]
\centering
\includegraphics[scale=0.1]{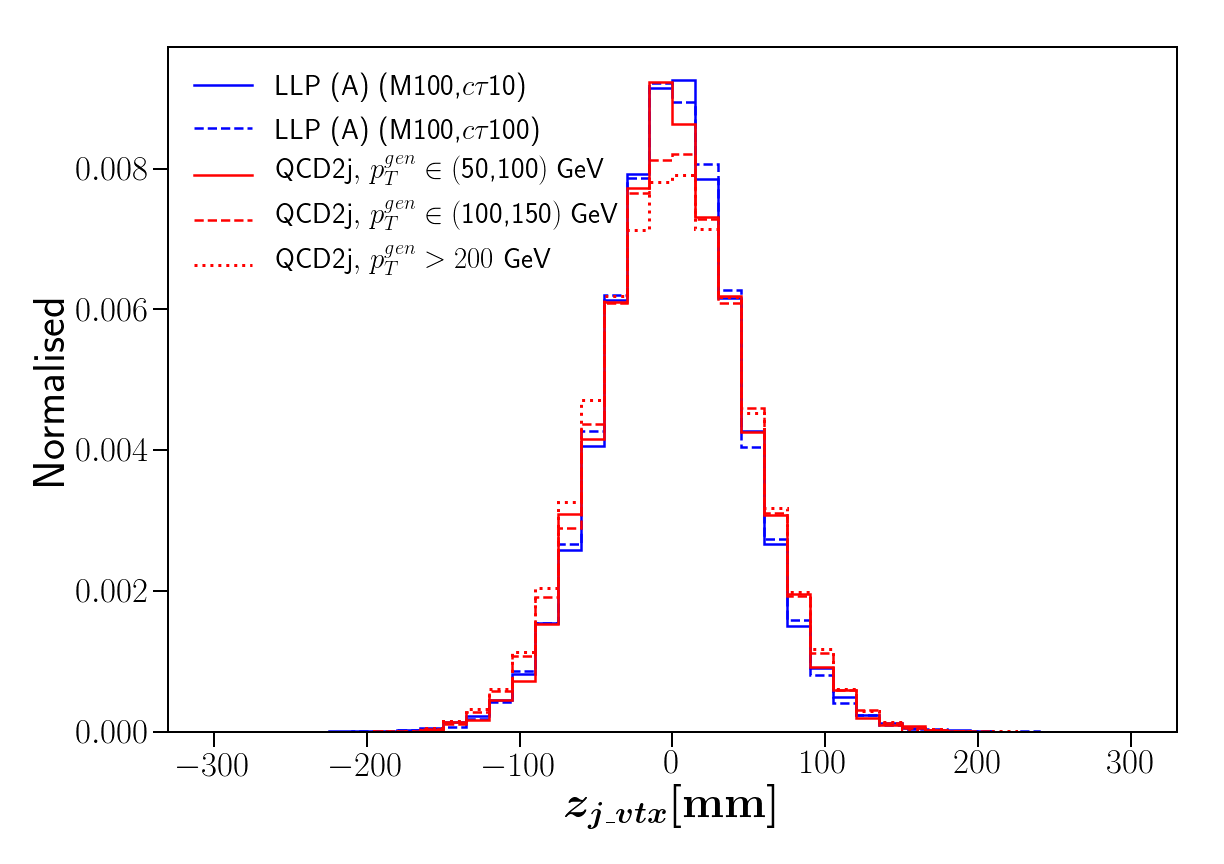}
\includegraphics[scale=0.1]{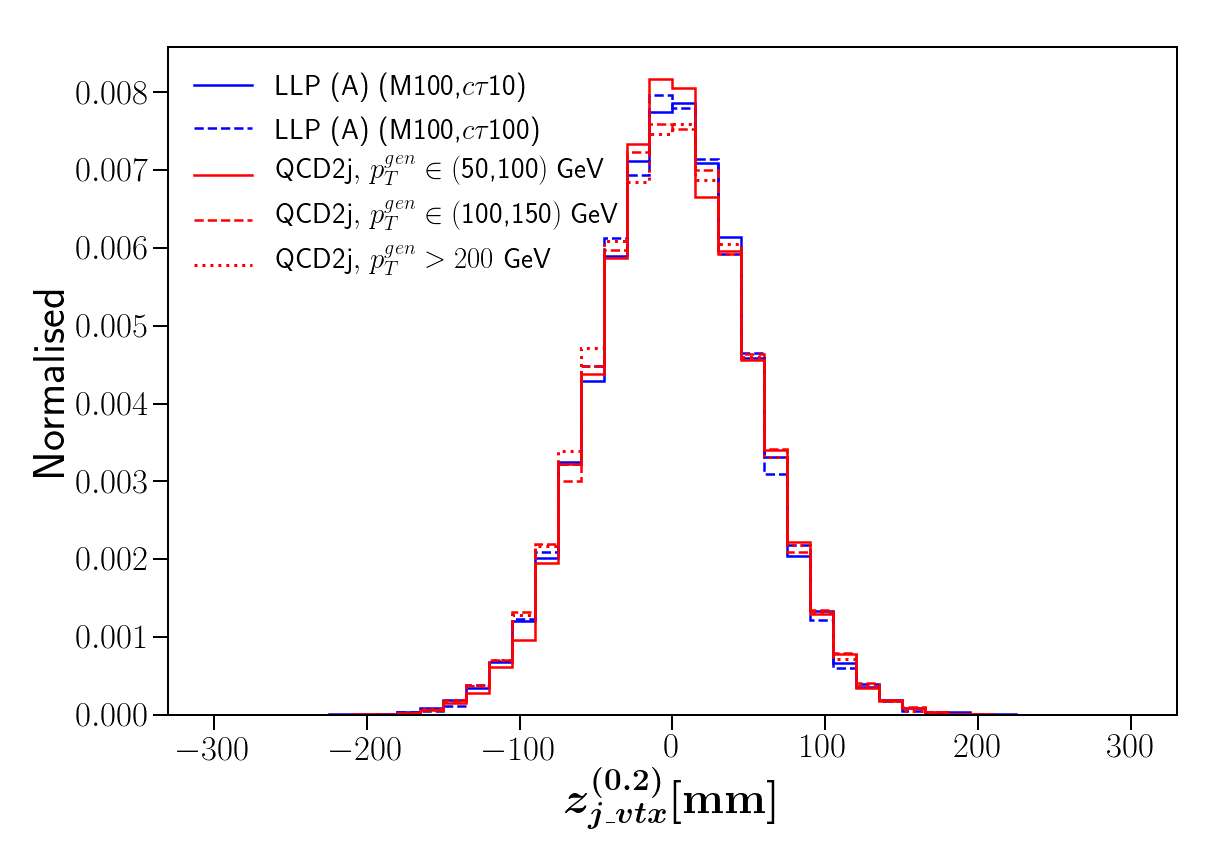}
\includegraphics[scale=0.1]{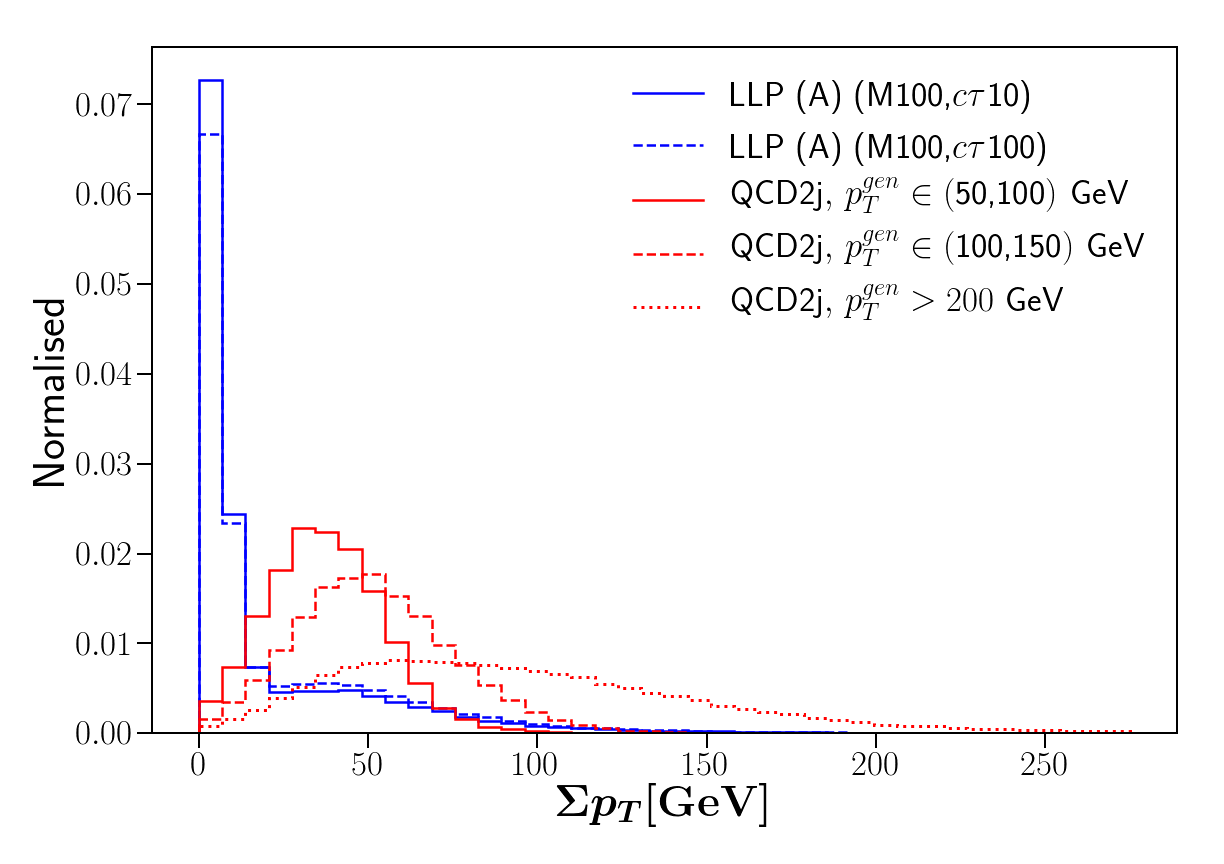}\\
\includegraphics[scale=0.1]{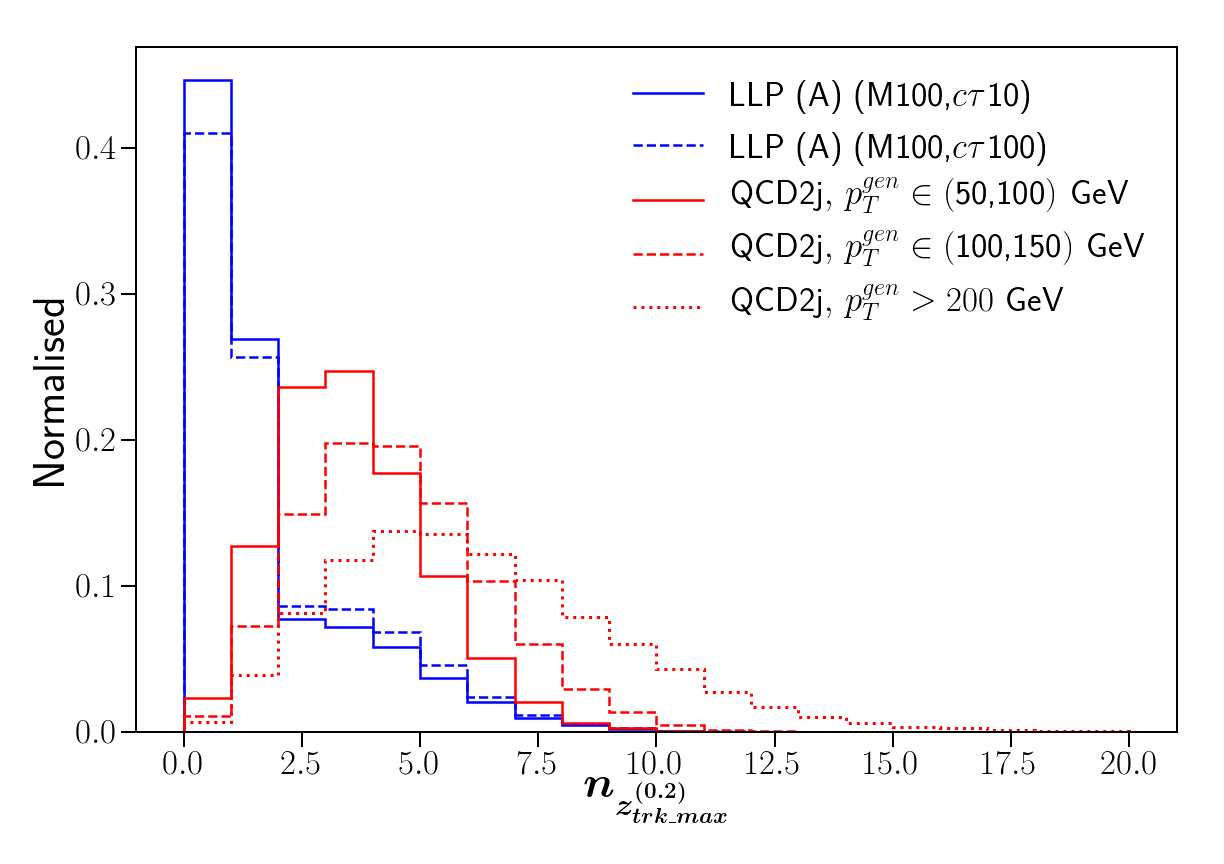}
\includegraphics[scale=0.1]{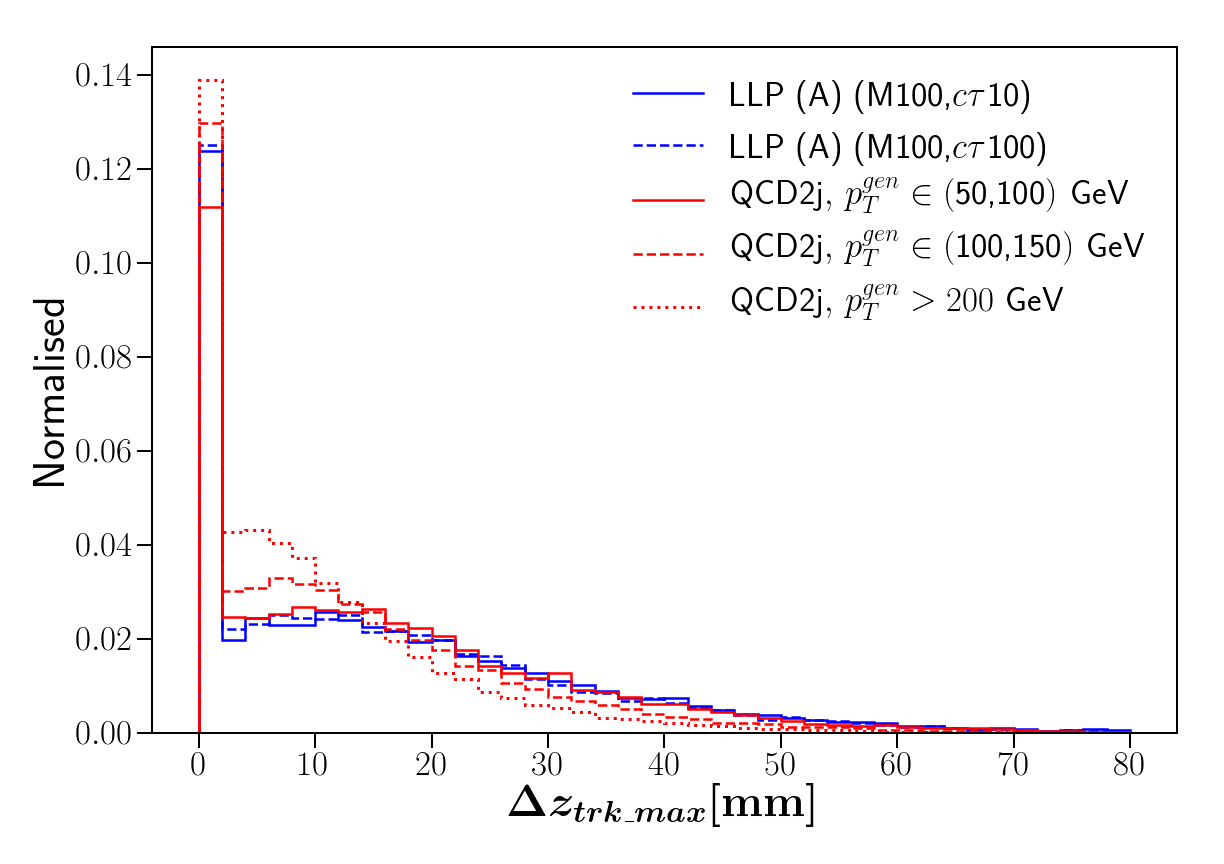}
\includegraphics[scale=0.1]{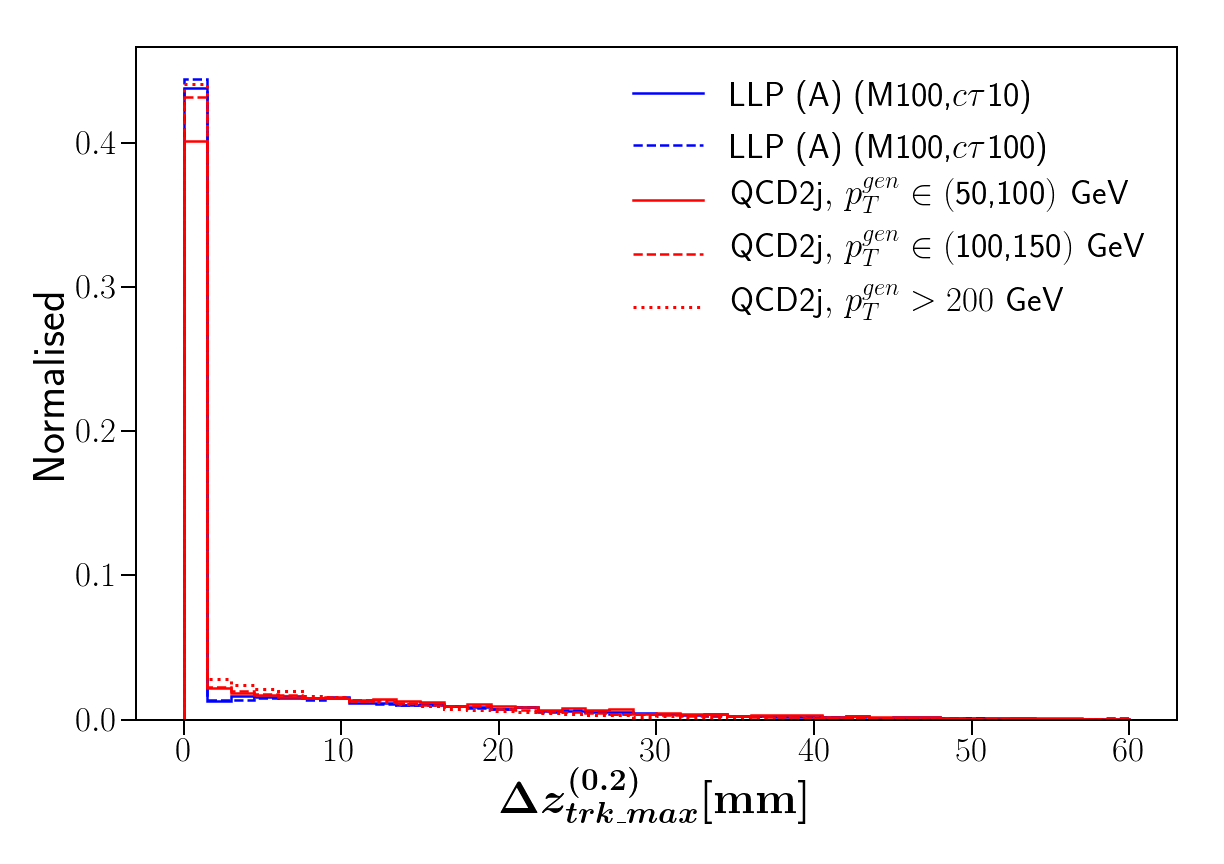}\\
\includegraphics[scale=0.1]{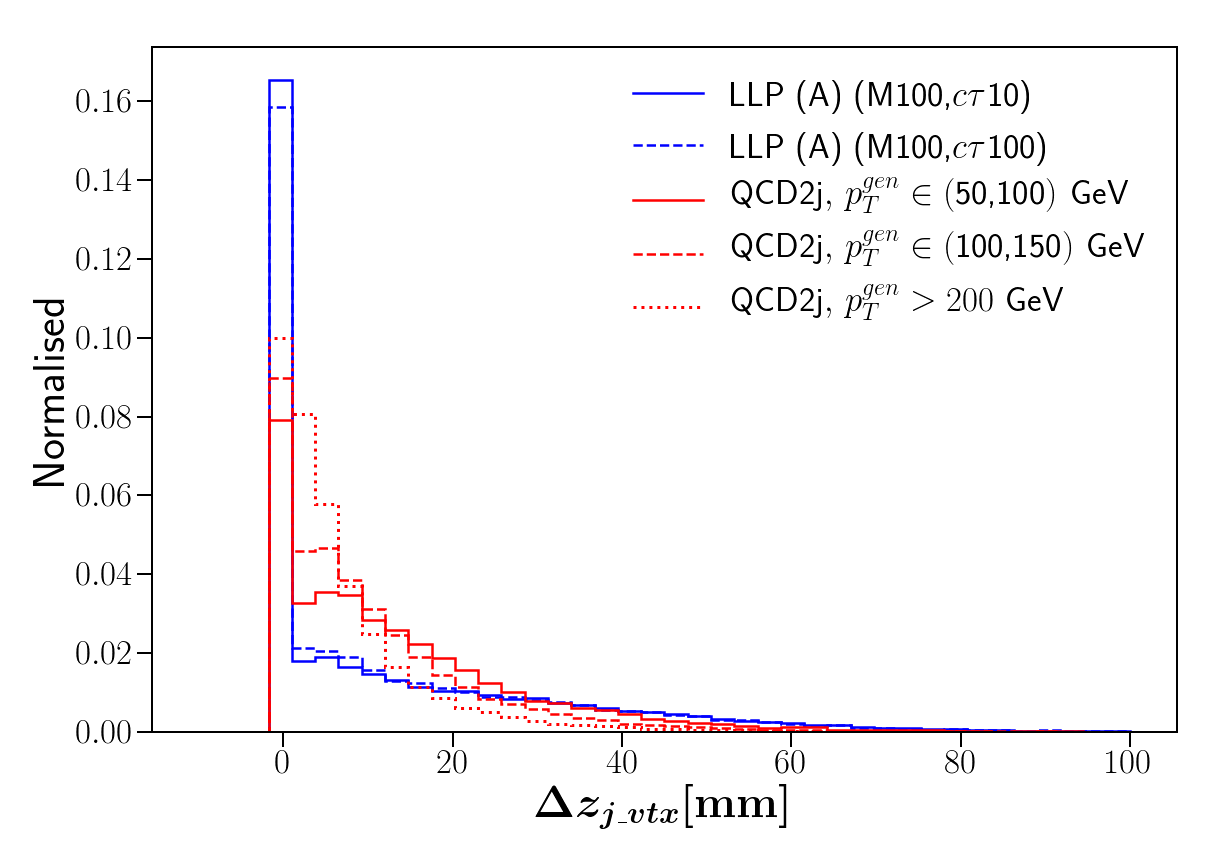}
\includegraphics[scale=0.1]{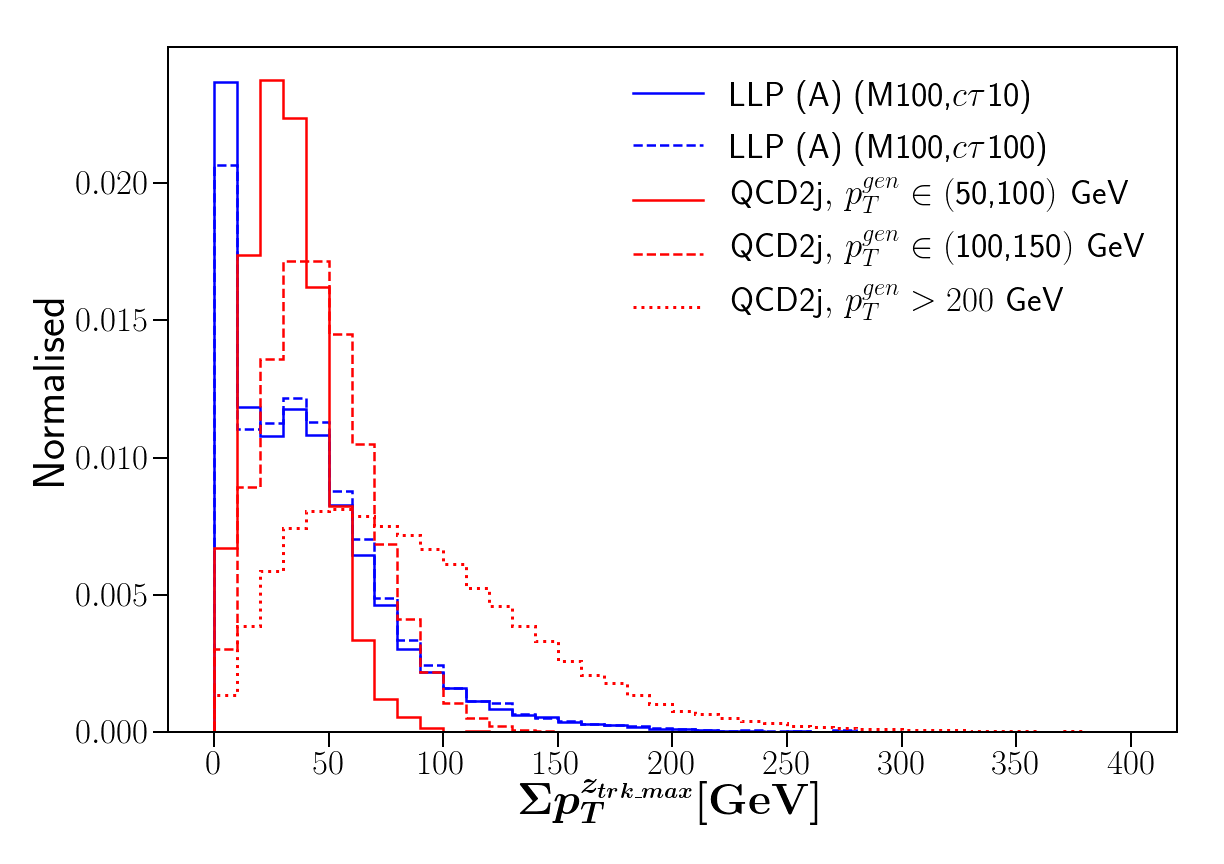}
\includegraphics[scale=0.1]{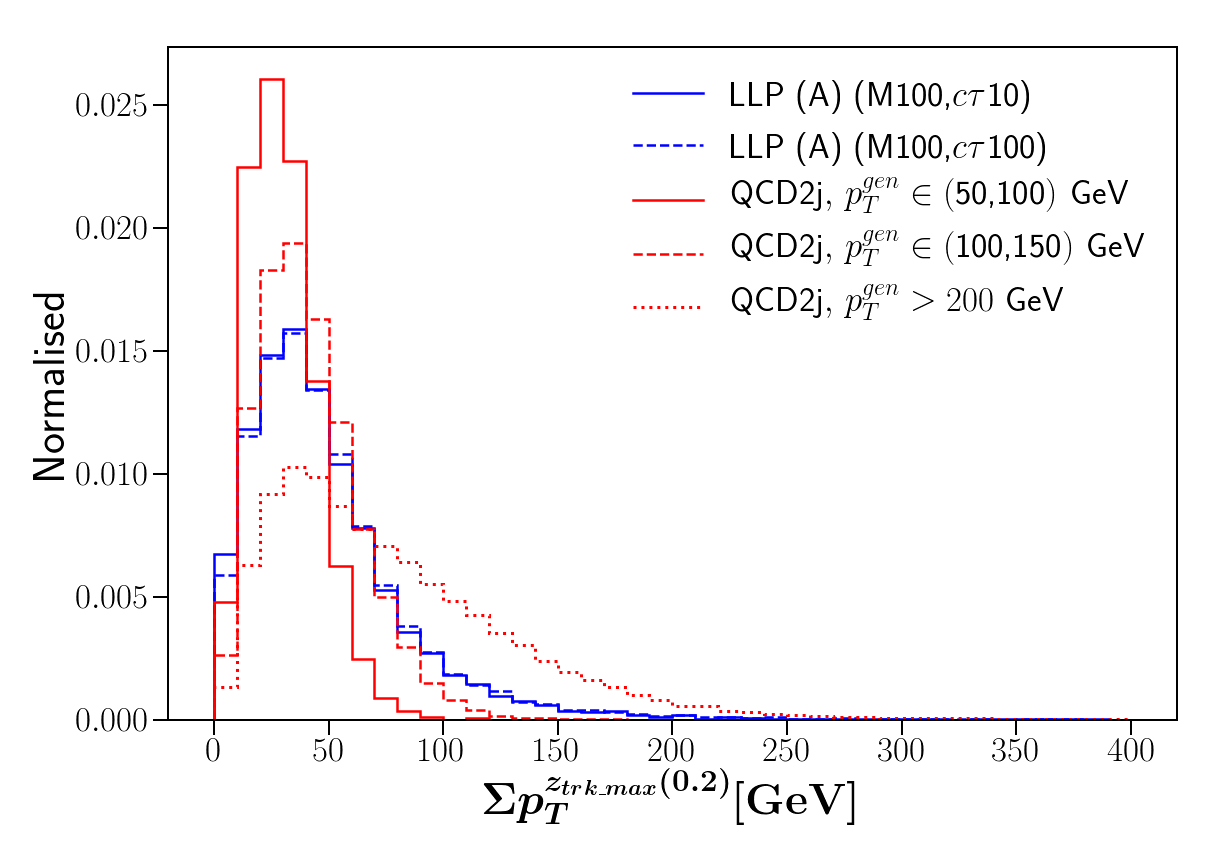}\\
\includegraphics[scale=0.1]{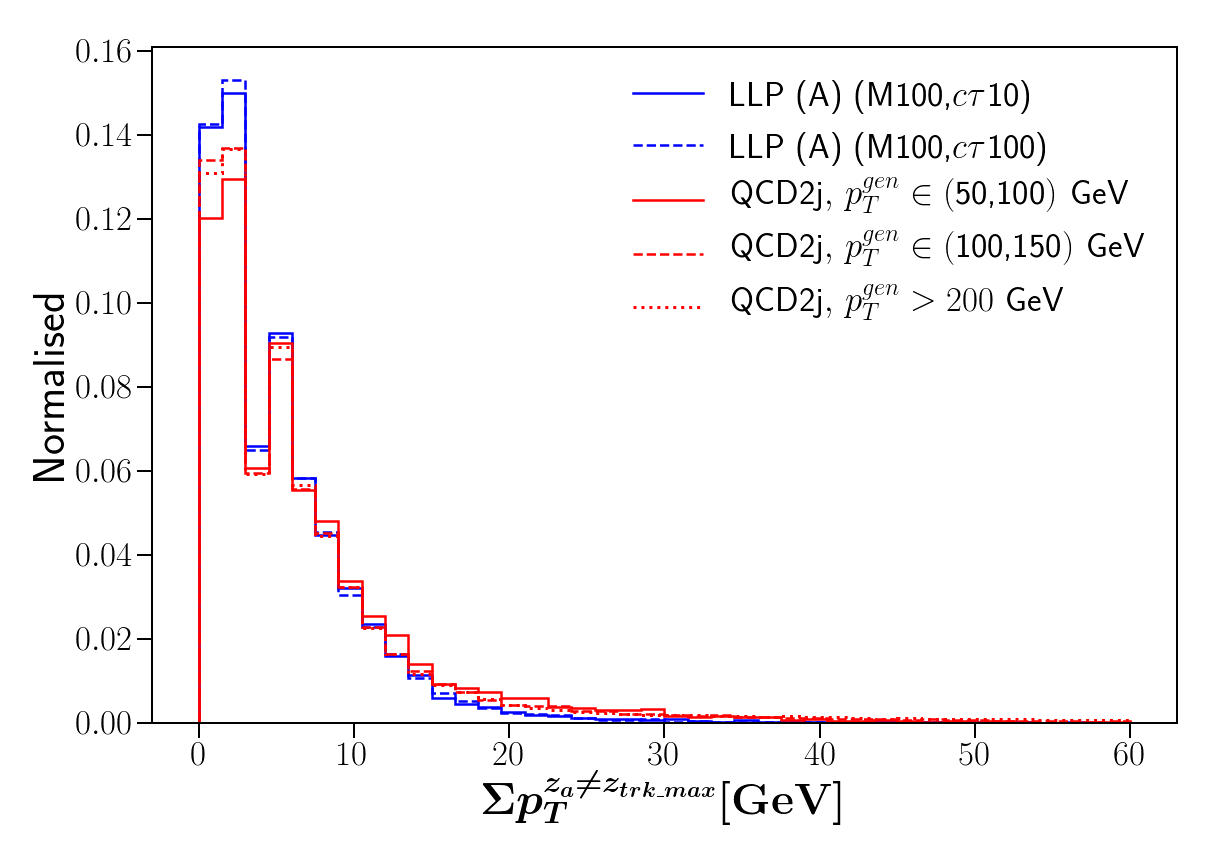}
\includegraphics[scale=0.1]{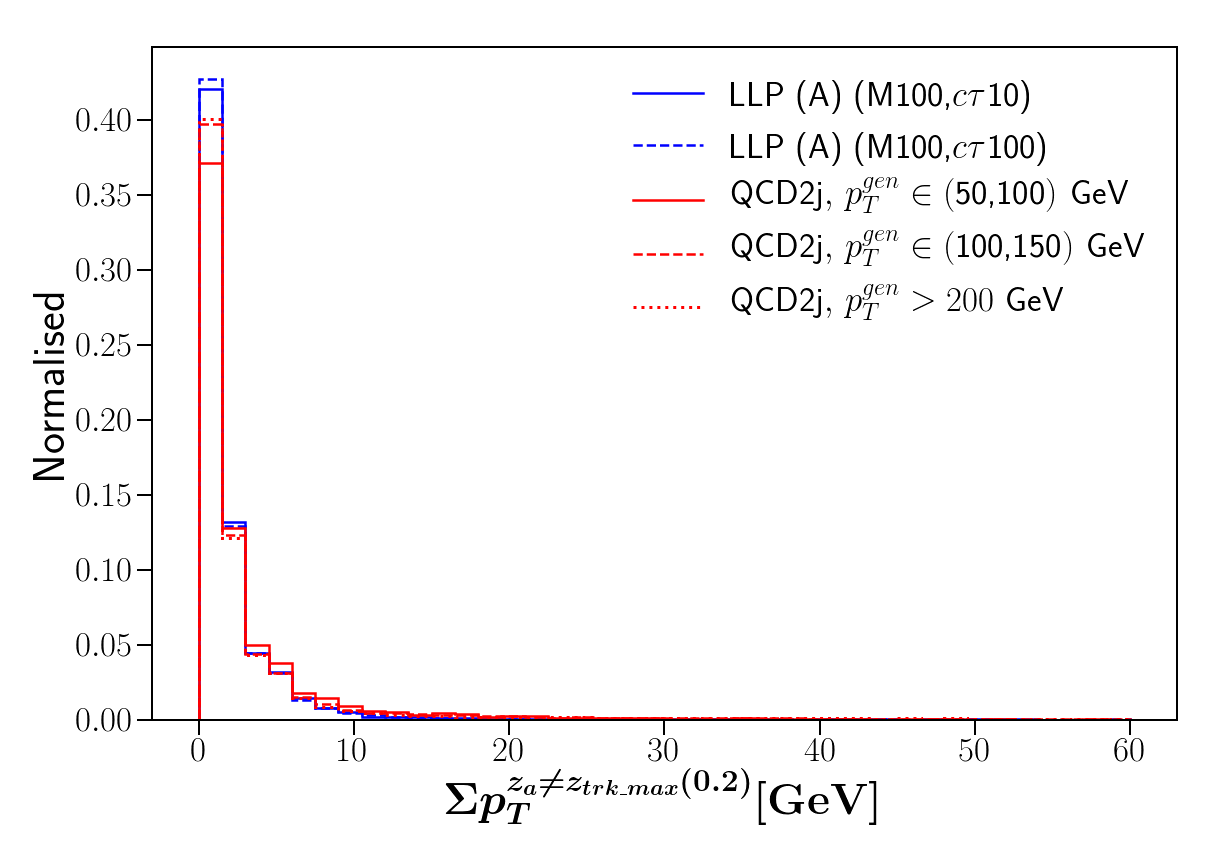}
\includegraphics[scale=0.1]{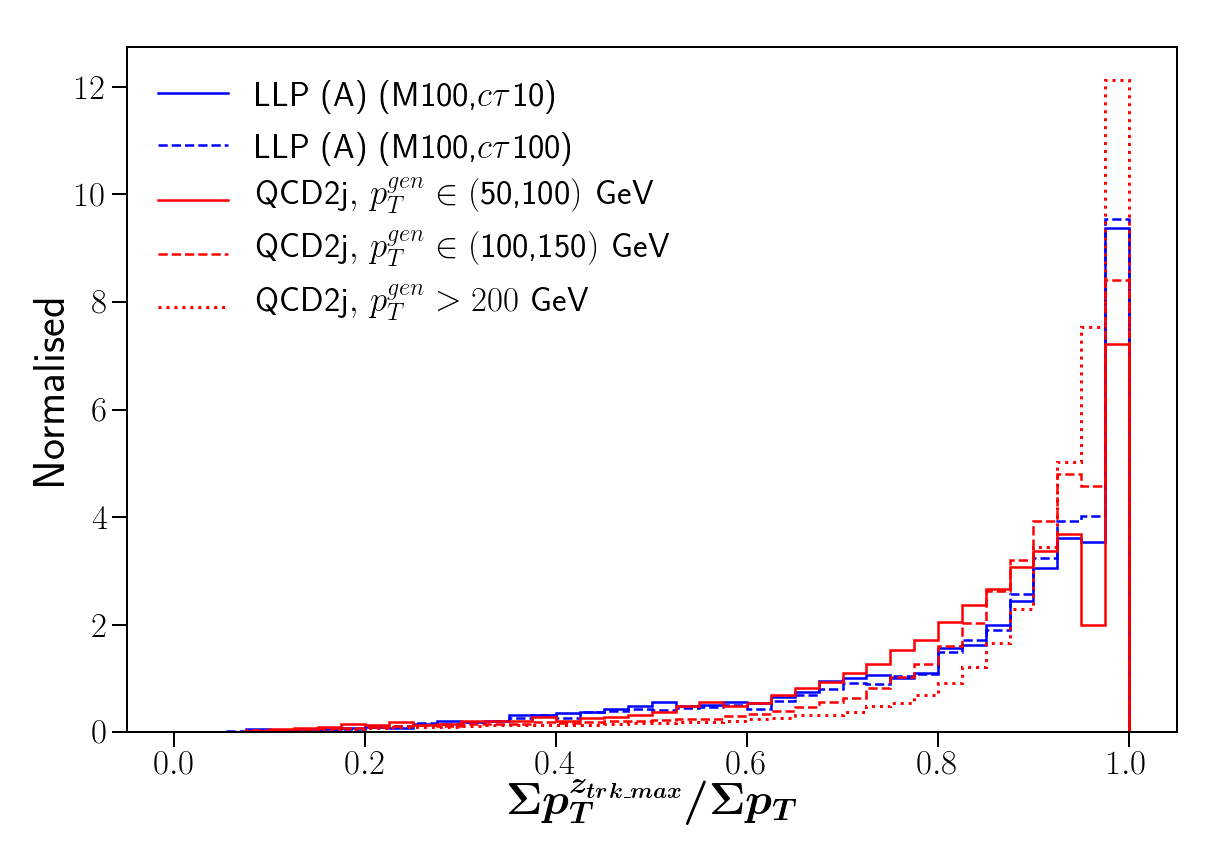}\\
\includegraphics[scale=0.1]{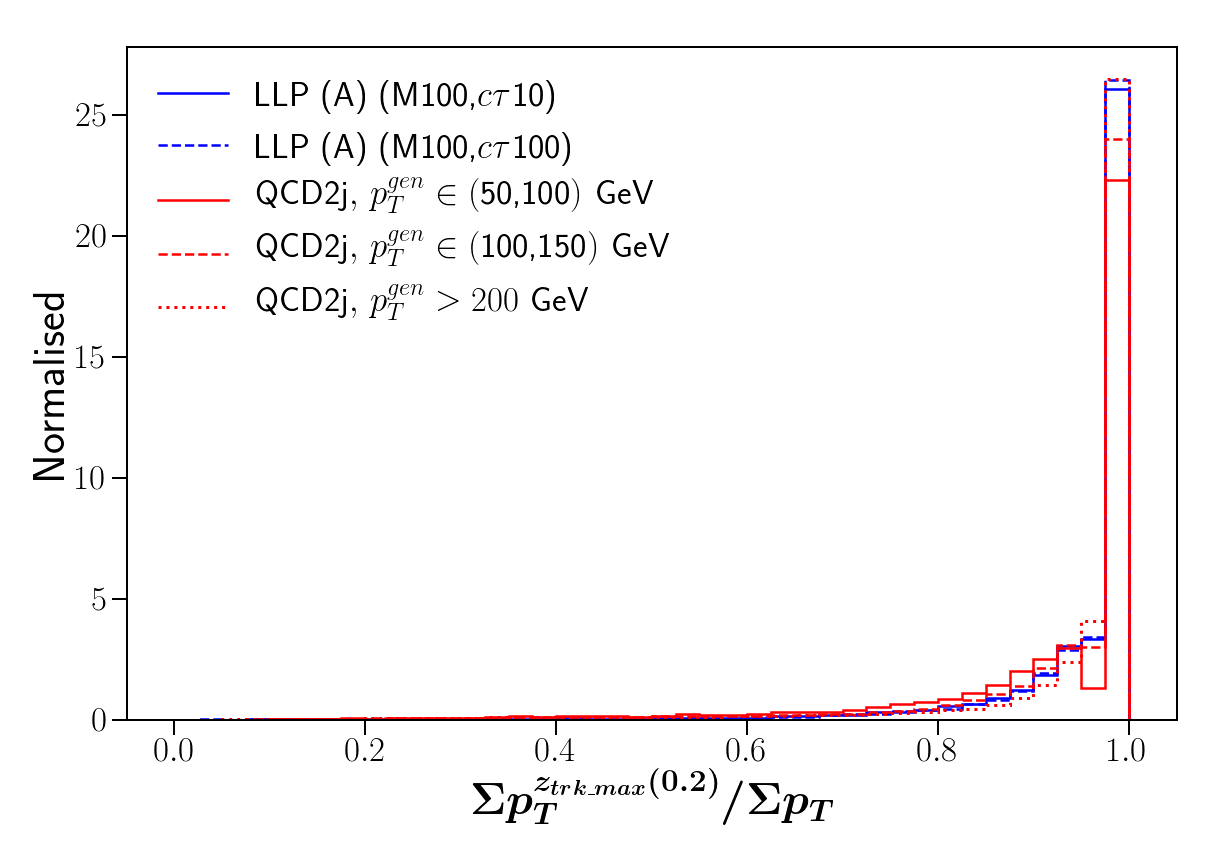}
\includegraphics[scale=0.1]{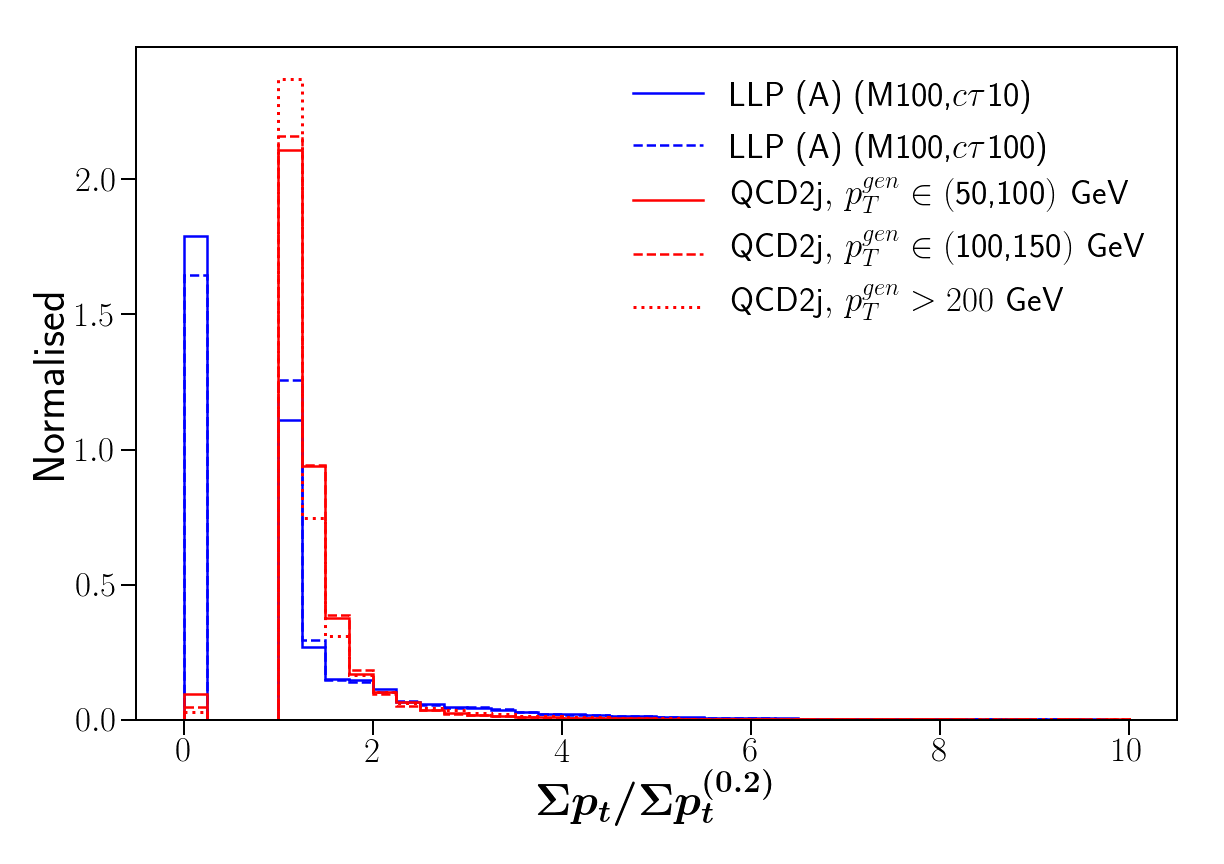}
\includegraphics[scale=0.1]{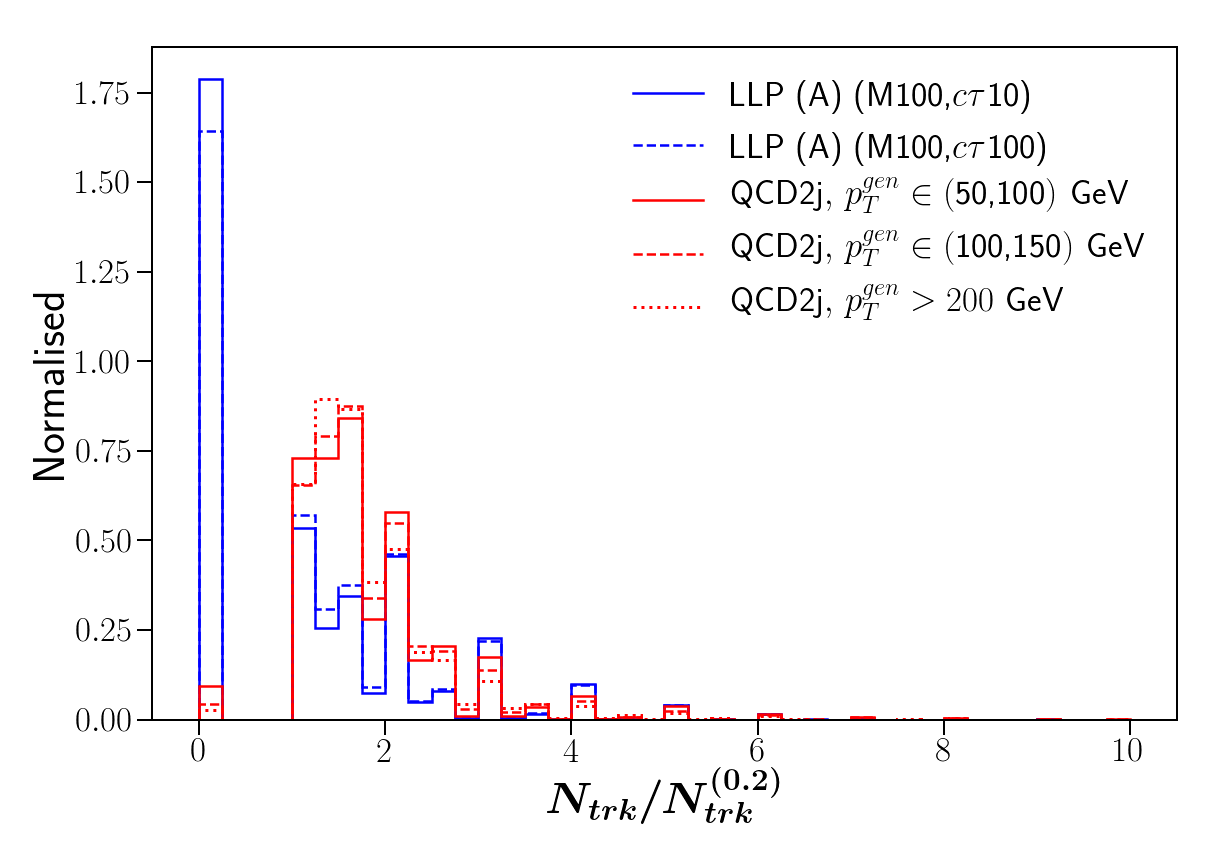}
\caption{Distributions for the rest of the track variables for two signal benchmark points from scenario (A) and QCD dijet processes with different $p_T$ cuts at the parton level.}
\label{fig:all_track_1}
\end{figure}

\begin{figure}[hbt!]
\centering
\includegraphics[scale=0.1]{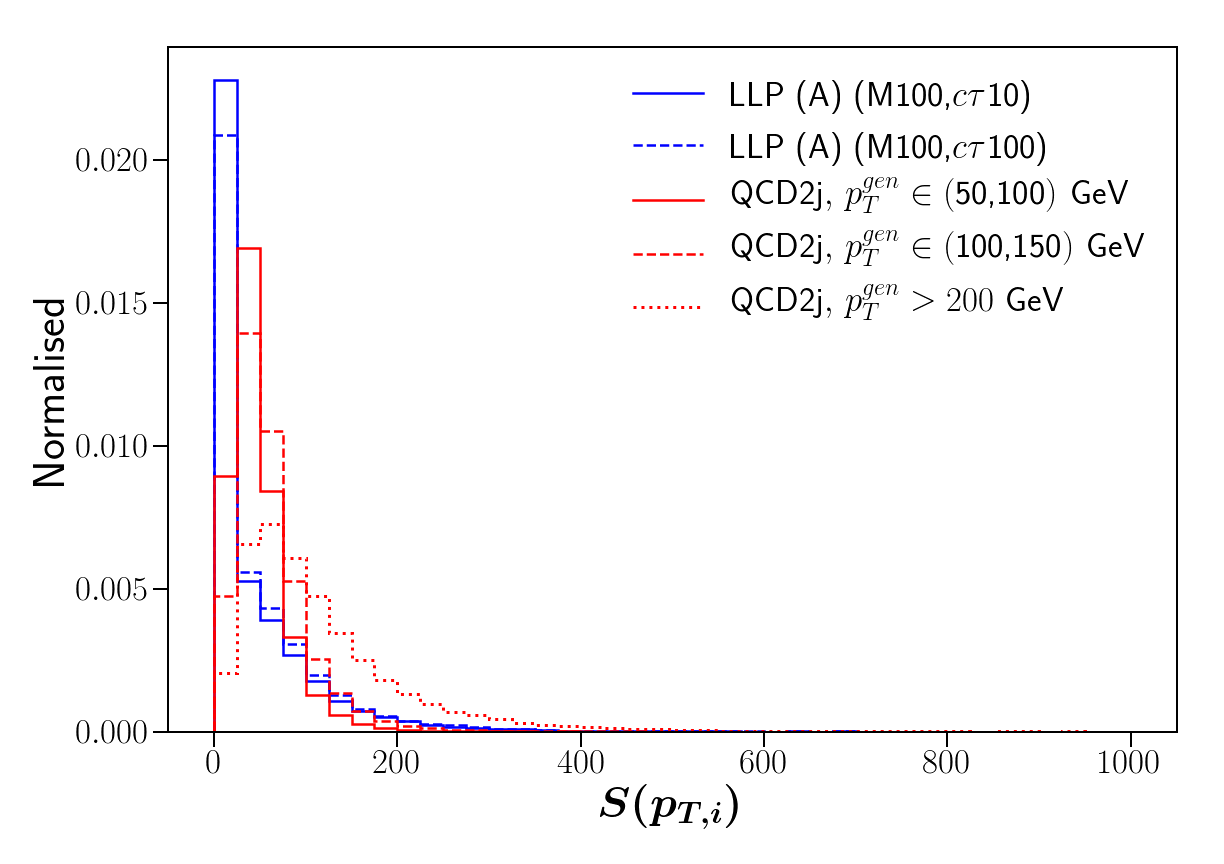}
\includegraphics[scale=0.1]{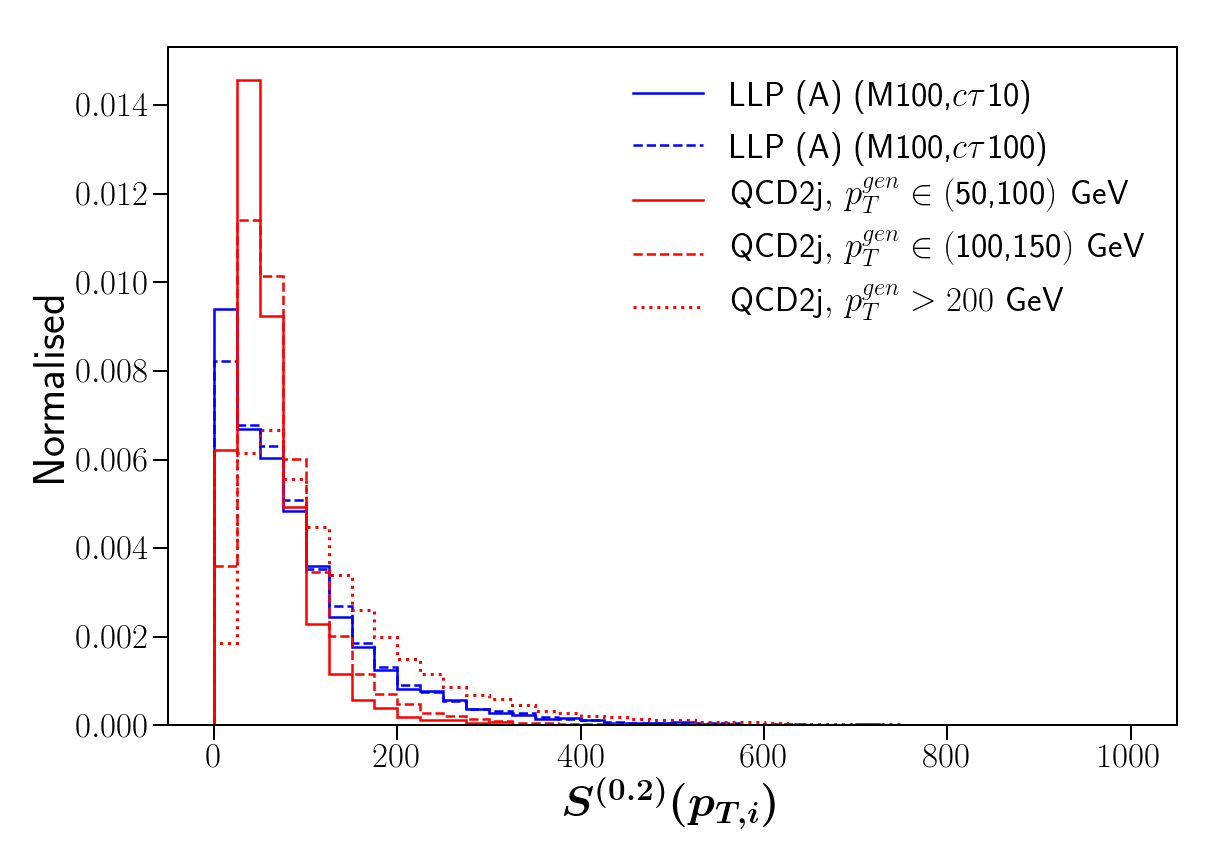}
\includegraphics[scale=0.1]{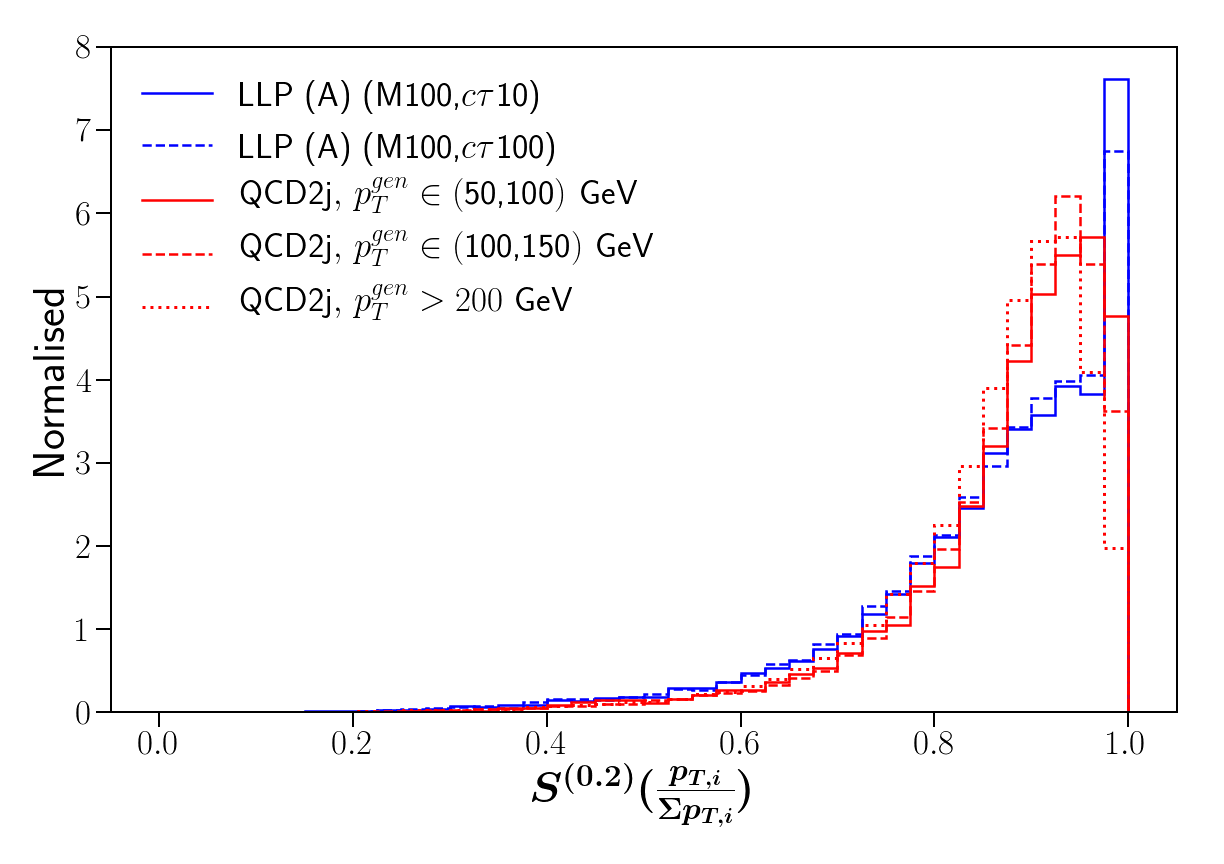}\\
\includegraphics[scale=0.1]{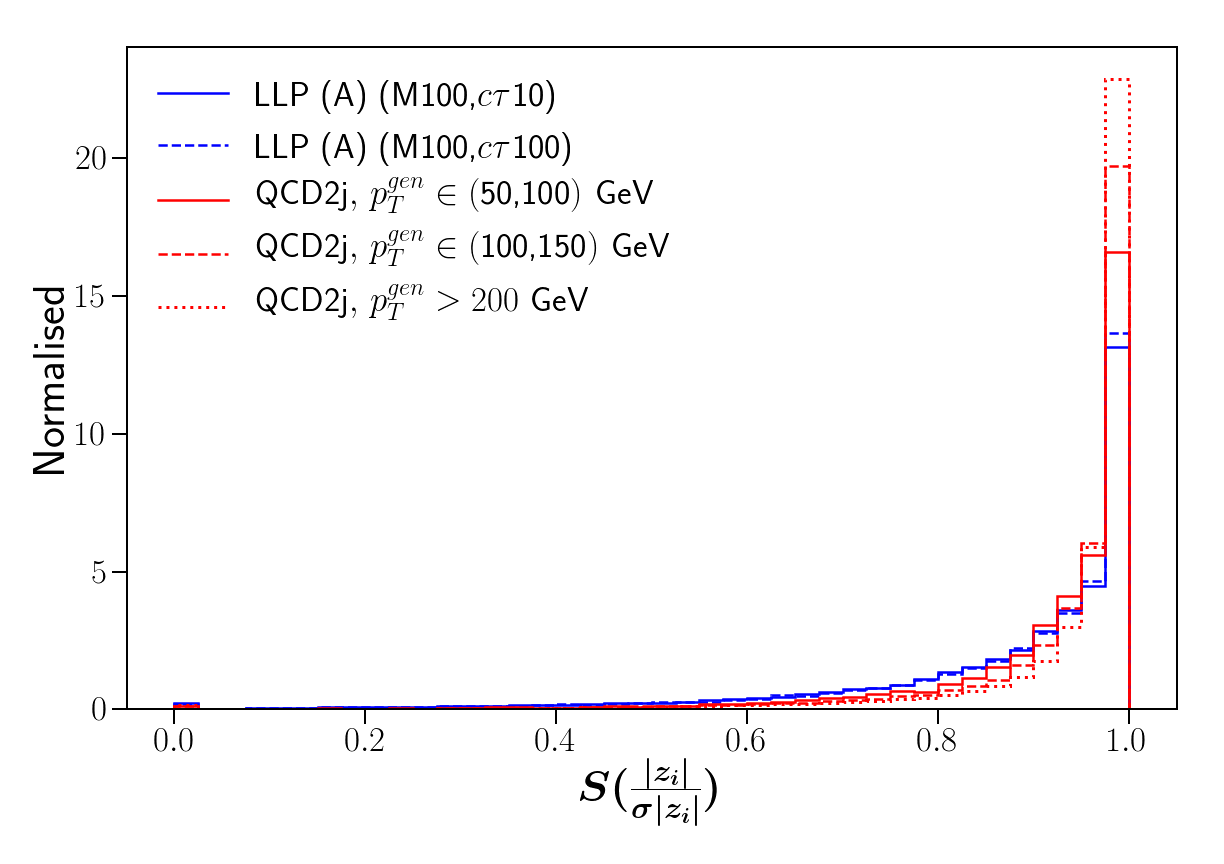}
\includegraphics[scale=0.1]{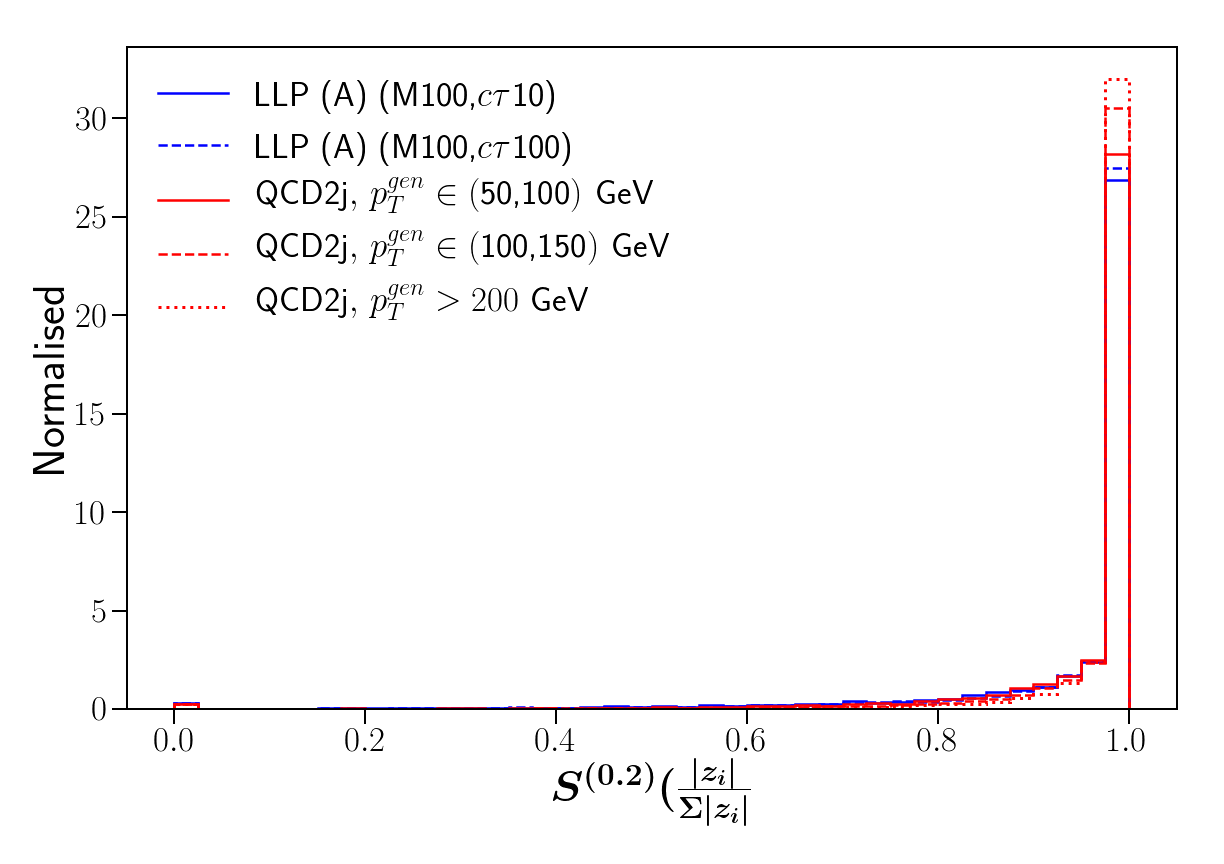}
\includegraphics[scale=0.1]{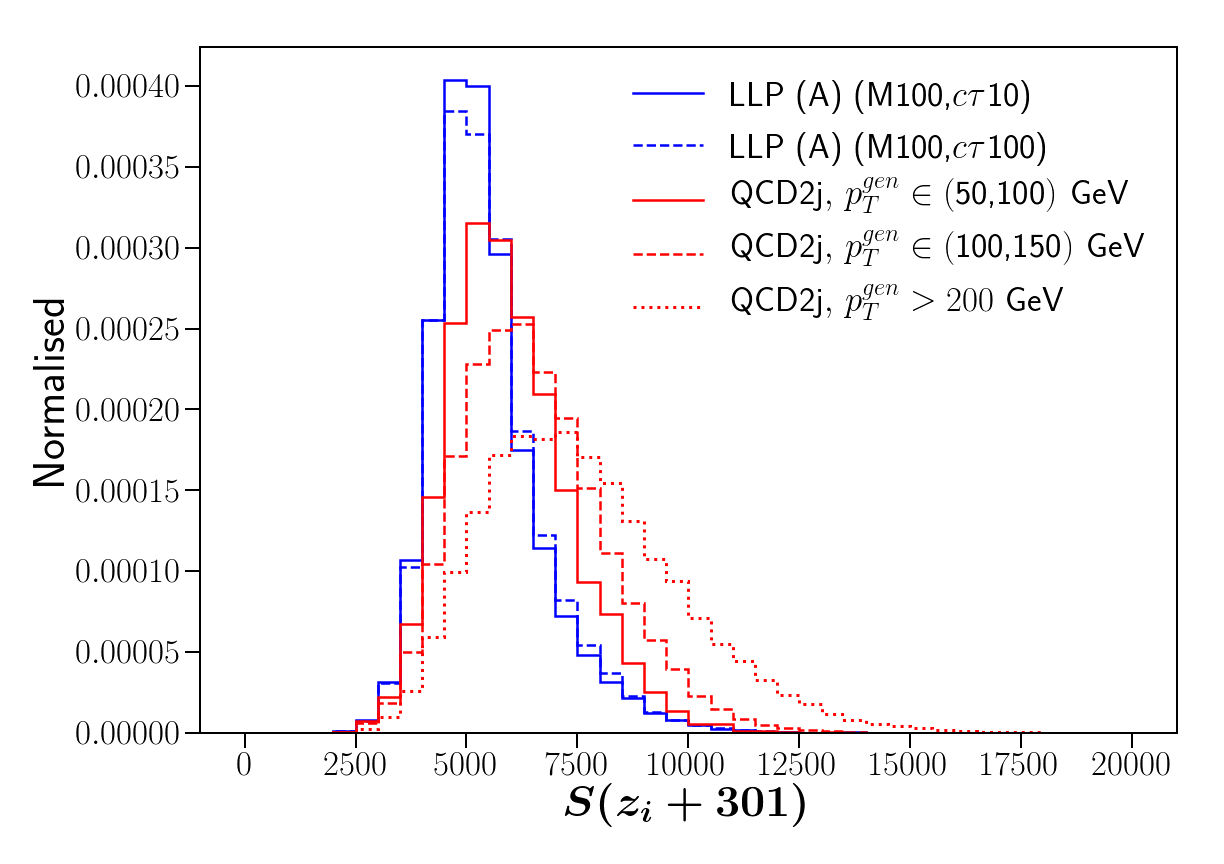}\\
\includegraphics[scale=0.1]{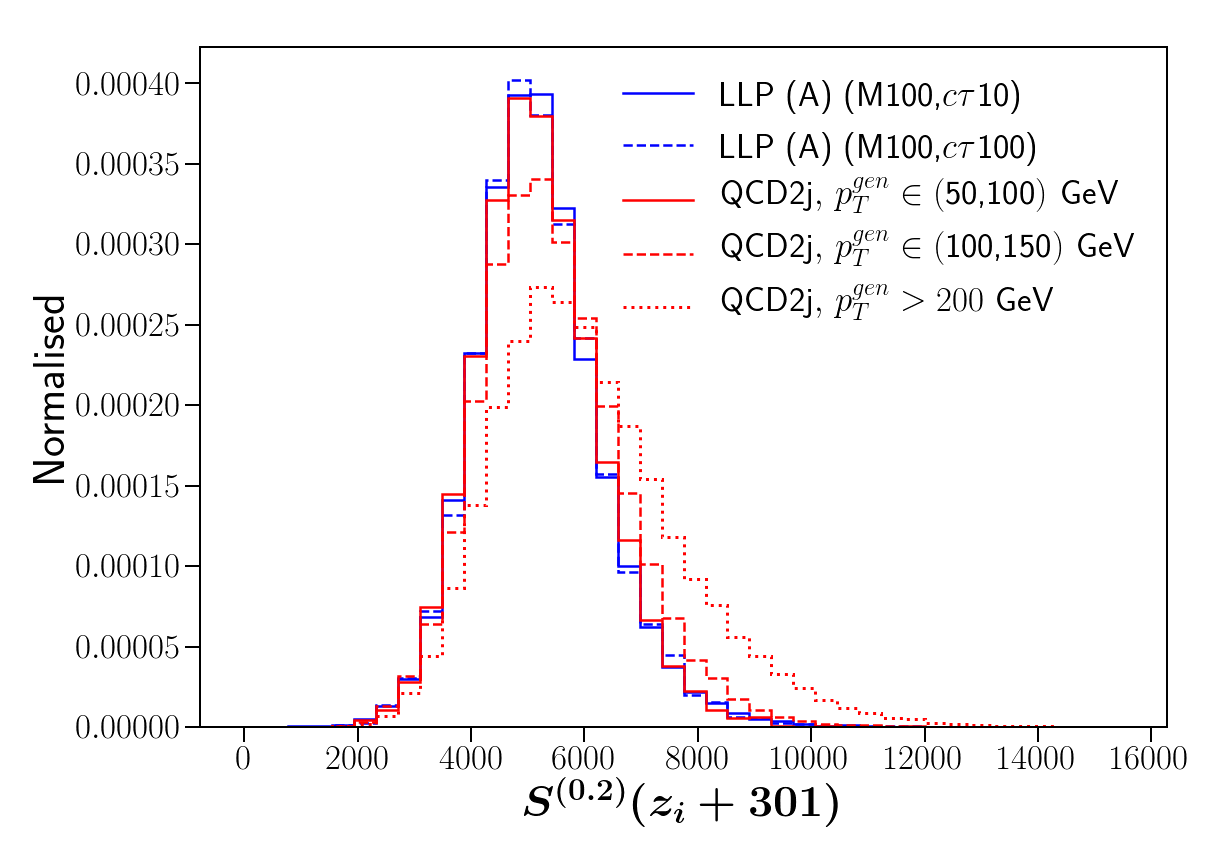}
\includegraphics[scale=0.1]{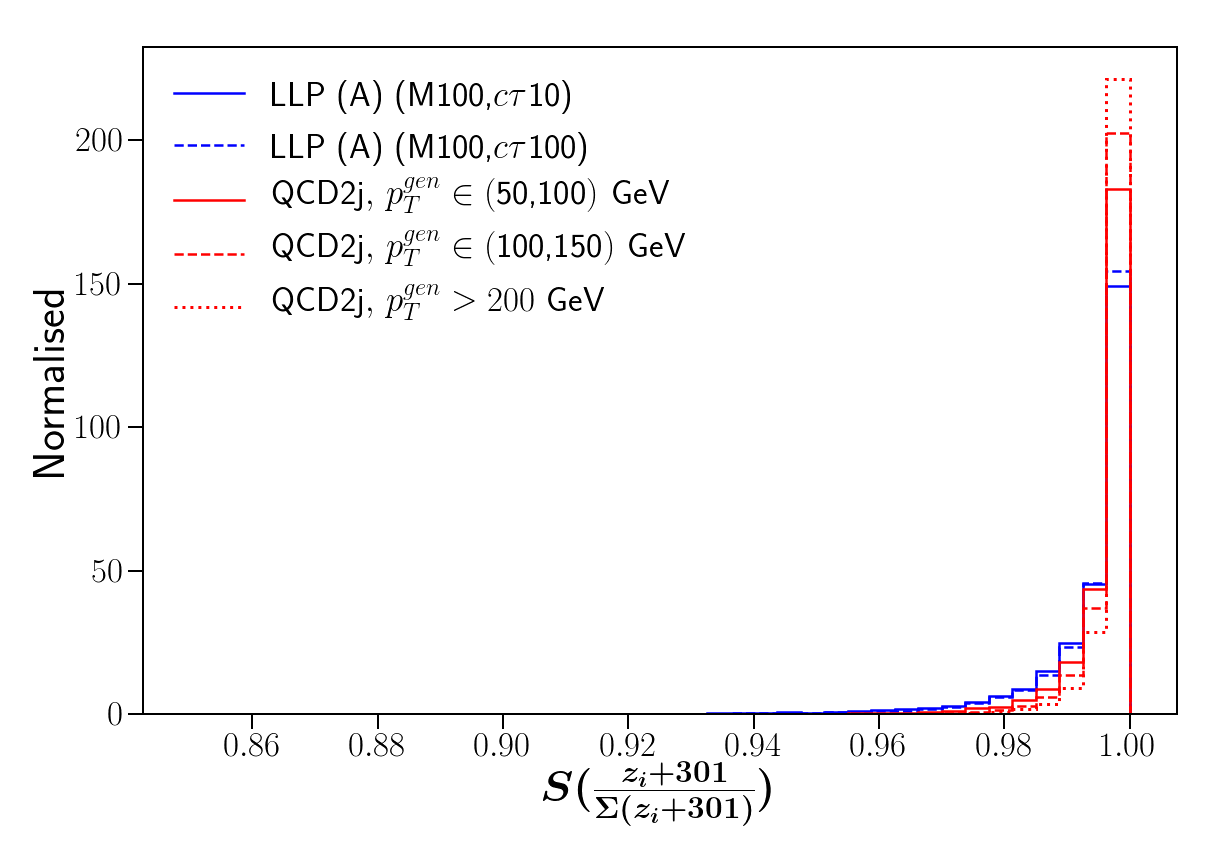}
\includegraphics[scale=0.1]{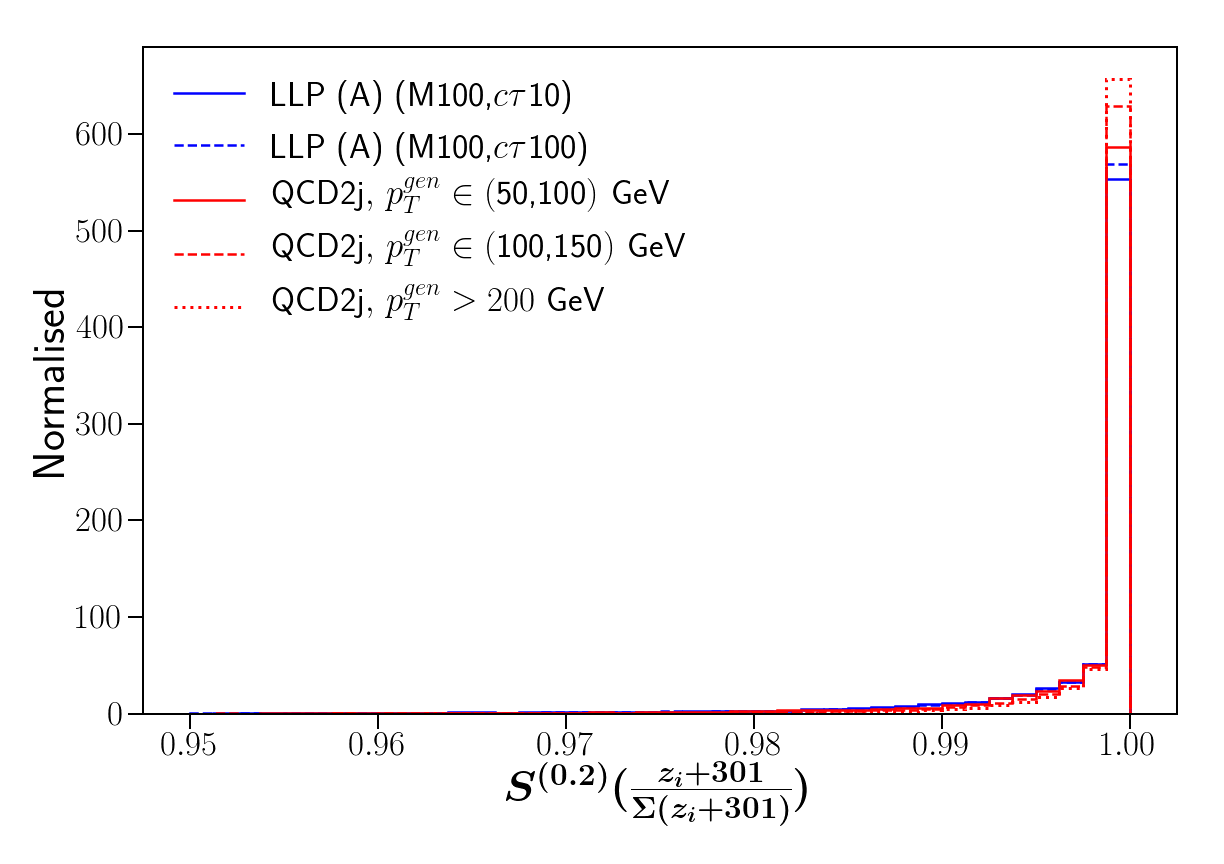}
\caption{Distributions for the rest of the track variables {\it(contd.)} for two signal benchmark points from scenario (A) and QCD dijet processes with different $p_T$ cuts at the parton level.}
\label{fig:all_track_2}
\end{figure}

\clearpage

\subsection{Propagation of a charged particle in magnetic field}
\label{app:magnetic}

The radius and angular frequency of the helical trajectory traversed by a charged particle having charge $q$ (in units of positron charge) and transverse momentum $p_T$ (in ${\rm~GeV}$) are given by

\begin{equation}
\begin{split}
\omega &= \frac{q\times B_z}{\gamma m}~~~[{\rm 89875518/s}]\\
r &= \frac{p_T\times10^9}{(\omega\times\gamma m)~c}~~~[{\rm m}]\\
\text{where,}~~\gamma m &= E\times10^9/c^2~~~[{\rm eV/c^2}]
\end{split}
\label{eq:helix_prop}
\end{equation}

Depending on which time among $t_r$ and $t_z$ is smaller, the particle will either exit from the radial sides or from the sides along the $z$-direction, where $t_r$ and $t_z$ are the respective times taken by the particle to reach the sides. The latter, $t_z$, is simply calculated using

\begin{equation}
t_z = \frac{\gamma m\times c}{p_z\times10^9}~(-z + z_{\rm max}\times {\rm sign}(p_z))
\label{eq:tz}
\end{equation}
where $p_z$ is the momentum of the particle along the $z$-direction, $z_{\rm max}$ is the half-length ($2.6{\rm~m}$) and ${\rm sign}(p_z)$ specifies the direction of motion of the particle in the $z$-direction.

\begin{figure}[hbt!]
\centering
\includegraphics[scale=0.1]{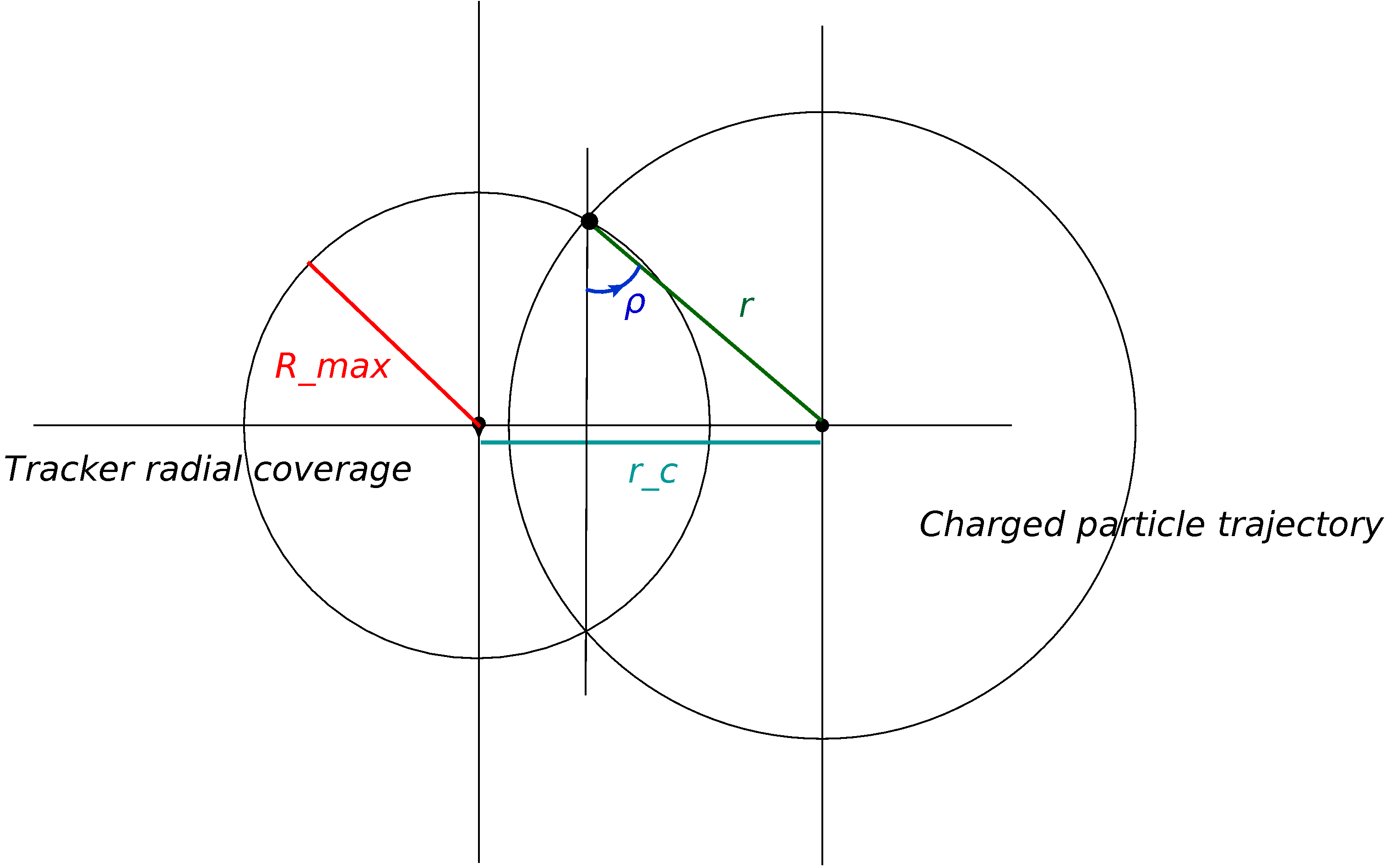}
\caption{Intersection of the charged particle track with the MTD.}
\label{fig:intersect}
\end{figure}

To find out the time when a particle exits the tracker from the sides of the cylinder,
i.e., cuts the cylinder in the radial direction, we need to find out the intersection of two circles in the transverse plane $-$ the circular path of the charged particle in the transverse plane with the circle describing the radial position of the MTD, as shown in fig.\ref{fig:intersect}. Equations of these two circles are as follows

\begin{equation}
\begin{split}
(x + r_c)^2 + y^2 = R_{\rm max}^2 ~~~\&~~~  x^2 + y^2 = r^2
\end{split}
\end{equation}
where $r_c$ is the distance of the PV from the helix axis, $R_{\rm max}$ is the radial position of the MTD, and $r$ is the radius of the helical trajectory of the charged particle given by eq.\ref{eq:helix_prop}. Here, the helix axis is the centre and all coordinates are measured from it.  

There are six possible path lengths that can be taken by a charged particle to exit from the radial sides of the cylinder \ref{eq:time_six}. They are as follows:

\begin{equation}
\begin{split}
t1 = (\delta + \rho) / \omega\\
t2 = (\delta + \pi - \rho) / \omega\\
t3 = (\delta + \pi + \rho) / \omega\\
t4 = (\delta - \rho) / \omega\\
t5 = (\delta - \pi - \rho) / \omega\\
t6 = (\delta - \pi + \rho) / \omega
\end{split}
\label{eq:time_six}
\end{equation}

\begin{figure}[hbt!]
\centering
\includegraphics[scale=0.06]{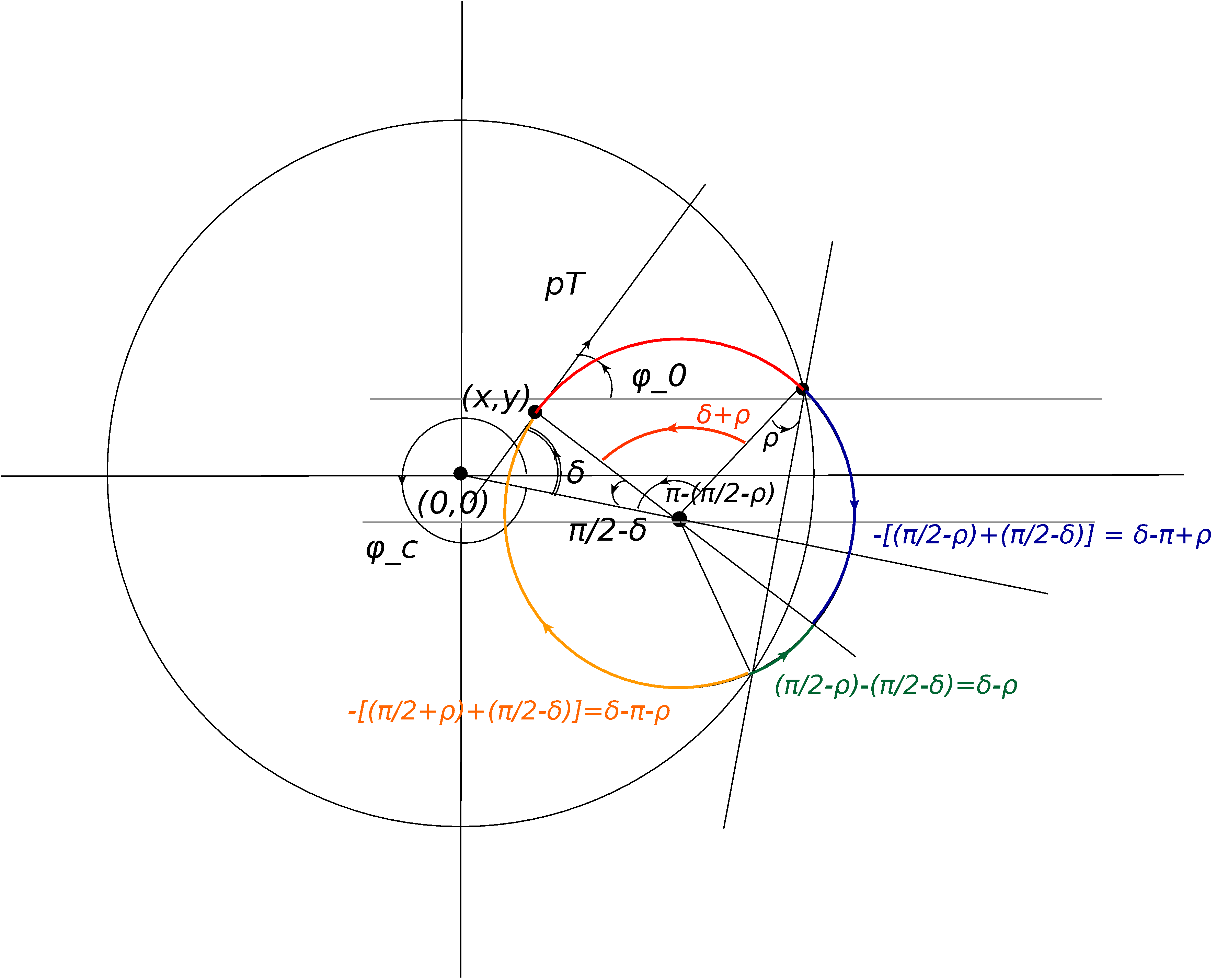}\hspace{0.5cm}
\includegraphics[scale=0.06]{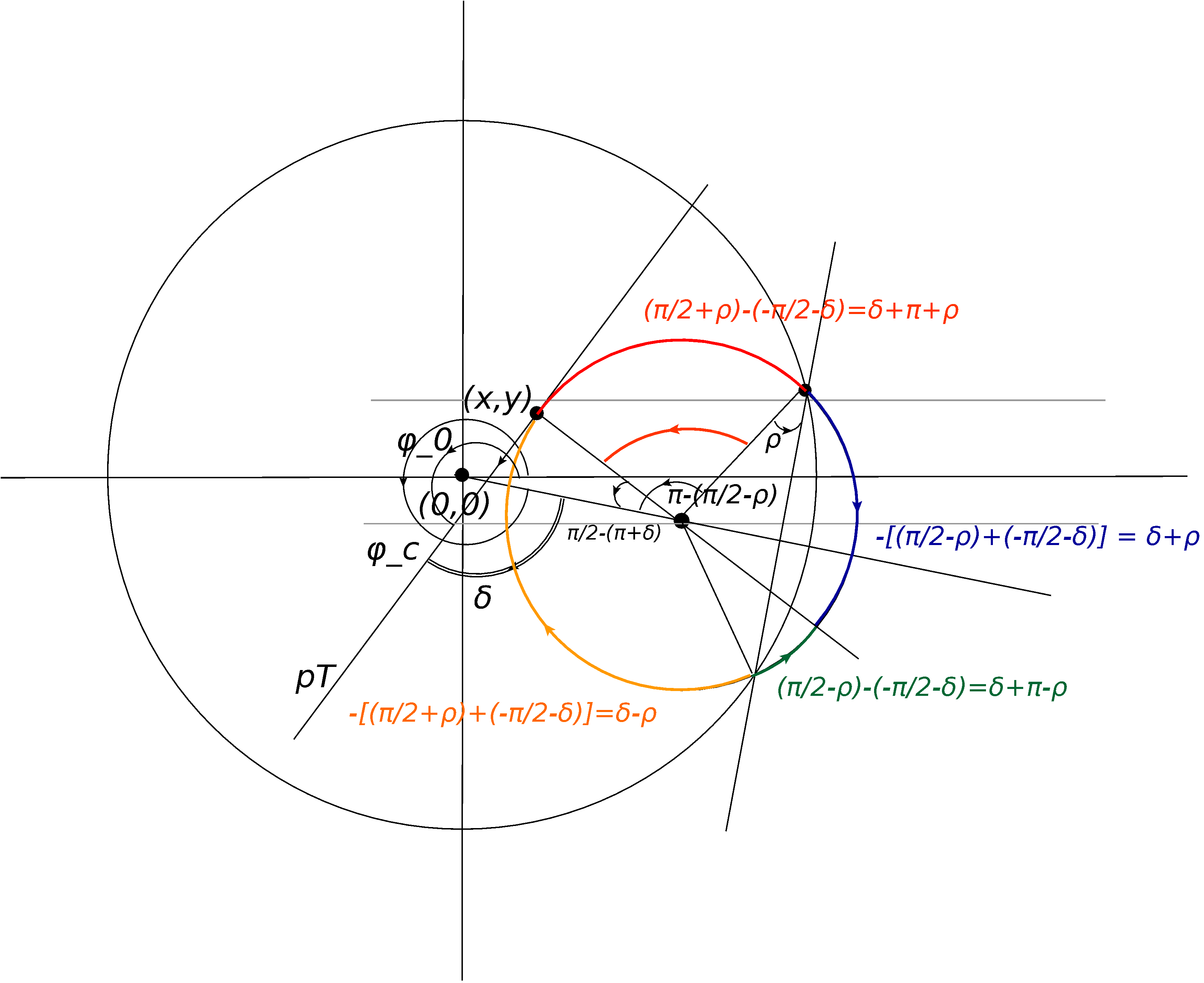}
\caption{Six possible path lengths that can be taken by a charged particle to exit from the radial sides of the cylinder.}
\label{fig:time_possibility}
\end{figure}

Fig. \ref{fig:time_possibility} shows how we land up with the above six possibilities. The two figures show two different directions of the transverse momenta of the particle, one with $\phi\_0<\pi$ (left) and other with $\phi\_0>\pi$ (right). The particle can move either along or opposite the direction of $p_T$ (depending on whether it has positive or negative charge) and also its $p_T$ can be such that $r_c^2+r^2<R_{\rm max}^2$ or $r_c^2+r^2>R_{\rm max}^2$. This gives us four possibilities in each case. But two of these possibilities give the same path length formula, and therefore, we have six possible path length values.

The present \texttt{ParticlePropagator} code selects the minimum time from the above six combinations. When $$r_c^2+r^2<R_{\rm max}^2,$$ the minimum time combination doesn't correspond to the correct path taken by the particle, and hence shows a much smaller time compared to the actual one. To select the correct path, we have added an additional condition that the radial distance of the point where the particle hits the MTD ($\sqrt{x_t^2+y_t^2}$) is close to $R_{\rm max}$.

\clearpage

\subsection{Correlation matrix of variables}
\label{app:corr_matrix}

\begin{figure}[hbt!]
\centering
\includegraphics[scale=0.27]{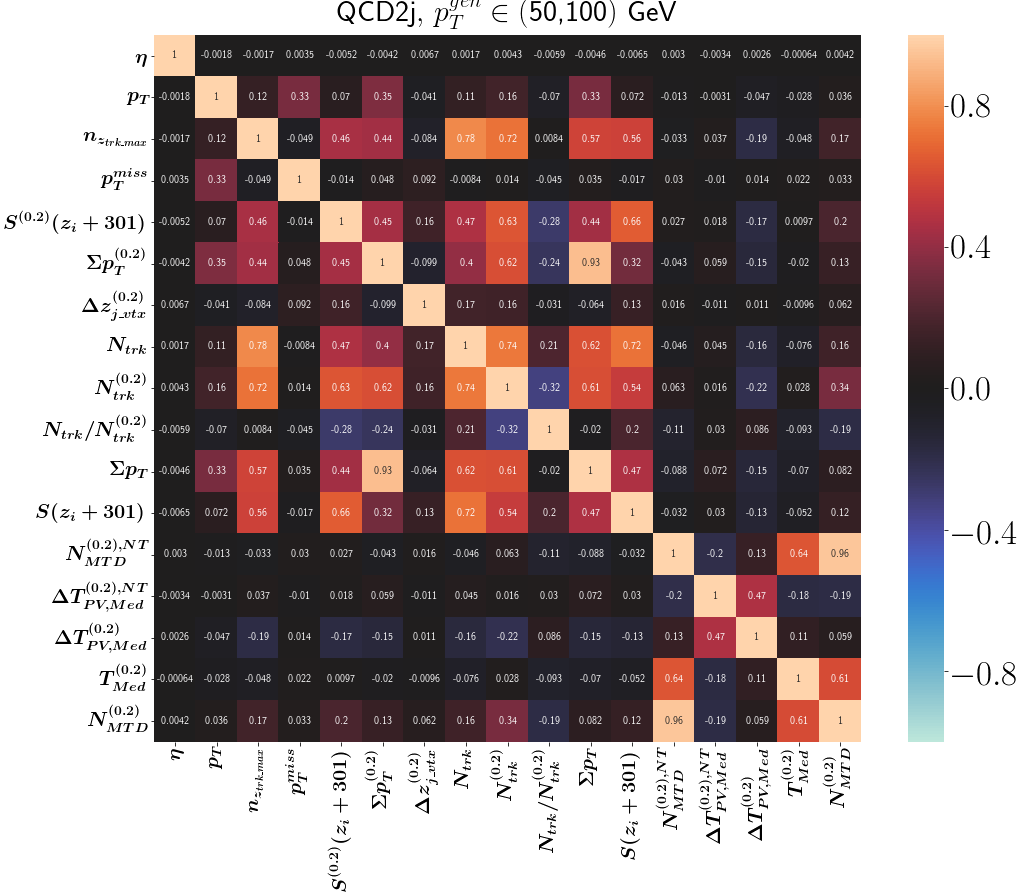}
\caption{Correlation matrices of background jets from QCD dijet events with $p_T\in(50,100){\rm~GeV}$ for variables using both tracking and timing informations.}
\label{fig:corr_track}
\end{figure}

\begin{figure}[hbt!]
\centering
\includegraphics[scale=0.27]{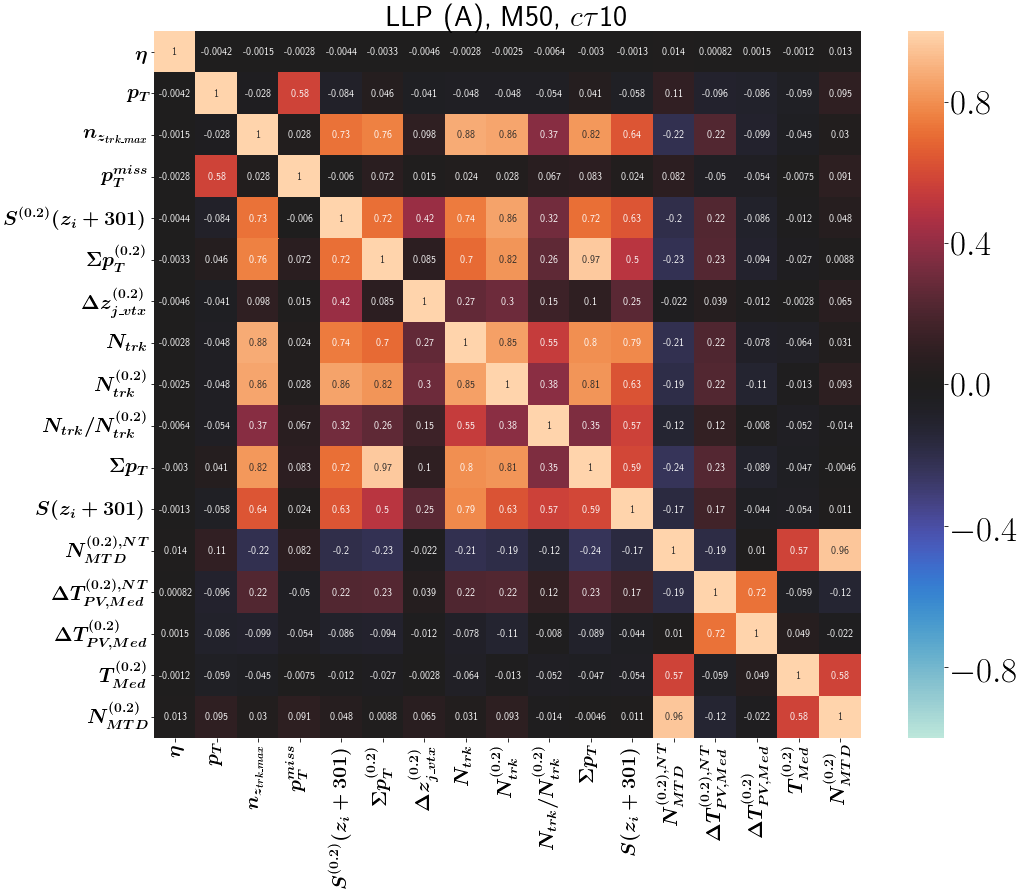}
\caption{Correlation matrices of signal jets from LLP benchmark point where mass and decay length of the LLP is 50 GeV and 10 cm respectively from scenario (A) for variables using both tracking and timing informations.}
\label{fig:corr_track}
\end{figure}

\clearpage

\subsection{Trigger efficiencies and rates for 90\% background rejection}
\label{app:trigger_eff}

\begin{table}[hbt!]
\centering
\resizebox{\textwidth}{!}{
\begin{tabular}{|c|c||c|c|c|c||}
\hline
\multirow{3}{*}{LLP (A)} & QCD2j & $T_2^1$ & $T_3^1$ & $T_{41}^1$ & $T_{42}^1$\\
 & $p_T^{gen}$ [GeV] & $\mathcal{R}_B$ [kHz] & $\mathcal{R}_B$ [kHz] & $\mathcal{R}_B$ [kHz] & $\mathcal{R}_B$ [kHz]\\
 & ($\mathcal{R}_B$ [kHz]) & $(\epsilon_S~[\%])$  & $(\epsilon_S~[\%])$ & $(\epsilon_S~[\%])$ & $(\epsilon_S~[\%])$\\
\hline\hline
& 50,100 (1046) & 34.7(17.41) & 31.6(17.07) & 25.2(14.47) & 29.8(15.84) \\
$M=50{\rm~GeV}$ & 100,150 (53.4) & 6.7(17.02) & 5.3(16.71) & 1.5(14.21) & 3.0(15.47)\\
$c\tau=10{\rm~cm}$ & 150,200 (7.5) & 1.2(16.28) & 0.9(16.01) & 0.1(13.61) & 0.14(14.77) \\
& $>$200 (2.7) & 0.5(15.76) & 0.2(15.53) & 0.02(13.12) & 0.03(14.29) \\
\hline\hline
& 50,100 (1046) & 34.7(12.32) & 31.6(11.98) & 25.2(9.63) & 29.8(10.71) \\
$M=50{\rm~GeV}$ & 100,150 (53.4) & 6.7(11.90) & 5.3(11.59) & 1.5(9.39) & 3(10.34)\\
$c\tau=100{\rm~cm}$ & 150,200 (7.5) & 1.2(11.14) & 0.9(10.89) & 0.1(8.89) & 0.14(9.72) \\
& $>$200 (2.7) & 0.5(10.73) & 0.2(10.51) & 0.02(8.57) & 0.03(9.39) \\
\hline\hline
& 50,100 (1046) & 34.4(72.44) & 30.3(68.91) & 22.1(46.69) & 28.3(57.03)\\
$M=100{\rm~GeV}$ & 100,150 (53.4) & 6.7(68.23) & 4.7(65.74) & 0.9(43.52) & 2.1(52.99) \\
$c\tau=10{\rm~cm}$ & 150,200 (7.5) & 1.2(60.95) & 0.8(59.39) & 0.05(36.21) & 0.08(45.70) \\
& $>$200 (2.7) & 0.5(51.47) & 0.2(50.33) & 0.0(25.37) & 0.01(35.45) \\
\hline\hline
& 50,100 (1046) & 34.4(58.99) & 30.3(55.49) & 22.1(34.22) & 28.3(42.70)\\
$M=100{\rm~GeV}$ & 100,150 (53.4) & 6.7(54.64) & 4.7(51.98) & 0.9(31.50) & 2.1(38.72) \\
$c\tau=100{\rm~cm}$ & 150,200 (7.5) & 1.2(47.58) & 0.8(45.68) & 0.05(25.55) & 0.08(32.40) \\
& $>$200 (2.7) & 0.5(38.98) & 0.2(37.51) & 0.0(17.12) & 0.01(24.16) \\
\hline\hline
\end{tabular}
}
\caption{Efficiency of selecting QCD and LLP events for scenario (A) benchmark points with the modified trackless trigger at L1 for Phase-II. BDT cut is applied considering 90\% background rejection. Quantity in parenthesis represents corresponding signal efficiency.}
\label{tab:mod1_eff_llp}
\end{table}

\clearpage

\subsection{The 200 PU scenario}
\label{app:200PU}
\begin{figure}[hbt!]
\centering
\includegraphics[scale=0.25]{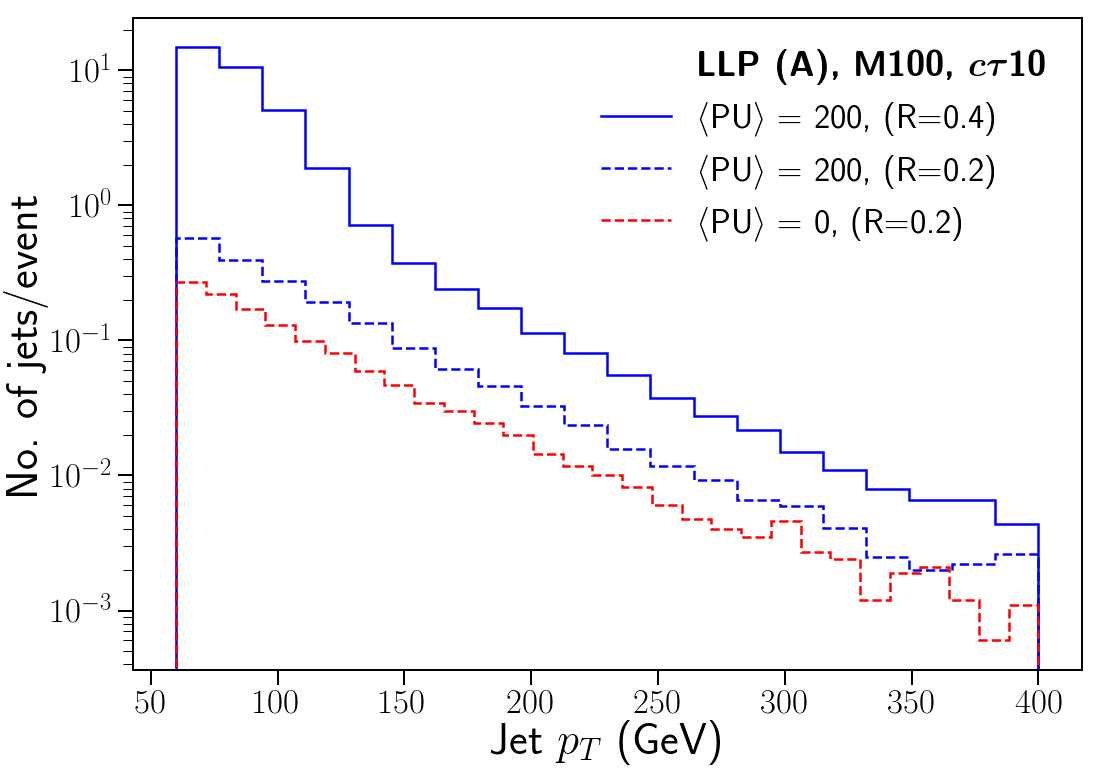}
\caption{Jet $p_T$ distribution for jets clustered using anti-$k_T$ with $R=0.2$ and $R=0.4$ coming from LLP benchmark ($M_X = 100{\rm~GeV}$ and $c\tau = 10{\rm~cm}$) from scenario (A) with 200 PU compared with the zero PU jet $p_T$ distribution.}
\label{fig:200PU}
\end{figure}

\end{document}